\PassOptionsToPackage{table}{xcolor}
\documentclass[12pt,a4paper,twoside]{report}

\usepackage[a4paper,
            top=3.1cm,
            bottom=3.1cm,
            inner=3.0cm,
            outer=3.0cm]{geometry}

\usepackage{tikz}
\usetikzlibrary{calc,shapes.geometric}
\definecolor{coveraccent}{RGB}{170,150,120}
\definecolor{covercream}{RGB}{236,226,205}
\definecolor{covercream2}{RGB}{214,198,170}
\usepackage{etoolbox}
\usepackage{tocloft}
\usepackage[font=small,labelfont=bf]{caption}
\usepackage{lmodern}
\usepackage{amsmath,amssymb}
\usepackage{txfonts}
\usepackage[utf8]{inputenc}
\usepackage[T1]{fontenc}
\usepackage{graphicx,float}
\usepackage{pdfpages}
\usepackage{xcolor}
\usepackage{adjustbox}
\usepackage{url}
\usepackage{subcaption}
\usepackage{titlesec}
\usepackage{xcolor}
\usepackage{fancyhdr}
\usepackage{ebgaramond}

\normalfont

\definecolor{citeblue}{RGB}{0,60,130}
\definecolor{chapterred}{RGB}{180,0,0}

\makeatletter
\pretocmd{\tableofcontents}{\setlength{\parskip}{0pt}}{}{}
\pretocmd{\listoffigures}{\setlength{\parskip}{0pt}}{}{}
\pretocmd{\listoftables}{\setlength{\parskip}{0pt}}{}{}
\makeatother

\usepackage{setspace}
\usepackage{comment}
\usepackage[
  backend=biber,
  style=authoryear,
  sorting=nyt,
  labelnumber=true,
  defernumbers=true,
  giveninits=true,
  natbib=true,
  maxcitenames=1,
  mincitenames=1,
  maxnames=1,
  minnames=1,
  uniquelist=false,
  uniquename=false,
  dashed=false
]{biblatex}

\DeclareFieldFormat{labelnumberwidth}{\mkbibbrackets{#1}\addspace}
\defbibenvironment{bibliography}
  {\list
     {\printtext[labelnumberwidth]{\printfield{labelnumber}}}
     {\setlength{\labelwidth}{\labelnumberwidth}%
      \setlength{\leftmargin}{\labelwidth}%
      \setlength{\labelsep}{\biblabelsep}%
      \addtolength{\leftmargin}{\labelsep}%
      \setlength{\itemsep}{\bibitemsep}%
      \setlength{\parsep}{\bibparsep}}}
  {\endlist}
  {\item}

\DeclareNameAlias{default}{family-given}
\AtEveryCitekey{\color{citeblue}}

\usepackage[
  colorlinks=true,
  linkcolor=black,
  citecolor=citeblue,
  urlcolor=citeblue
]{hyperref}

\definecolor{chapterred}{RGB}{180,0,0}
\titleformat{\chapter}[display]
 {\normalfont\bfseries\color{chapterred}}
 {\filleft\fontsize{80}{80}\selectfont\thechapter}
 {-8pt}
 {\filleft\Huge\bfseries\color{chapterred}}
 [\vspace{-20pt}\filleft\rule{0.92\textwidth}{0.5pt}]
 
\titlespacing*{\chapter}{0pt}{-20pt}{25pt}

\usepackage{lettrine}
\fancypagestyle{plain}{%
  \fancyhf{}
  \fancyfoot[C]{\small\thepage}
  
}

\definecolor{headergray}{RGB}{80,80,80}

\begin{document}
\raggedbottom

\begin{titlepage}
\thispagestyle{empty}

\begin{center}
\vspace{-0.25cm}

{\Huge\bfseries
\textsc{Universidad Complutense de Madrid}
}

\vspace{0.0cm}

{\huge
\textsc{Facultad de Ciencias Físicas}
}

\vspace{0.0cm}

{\Large
\textsc{Programa de Doctorado en Astrofísica}
}

\vspace{0.25cm}

\includegraphics[width=5.5cm]{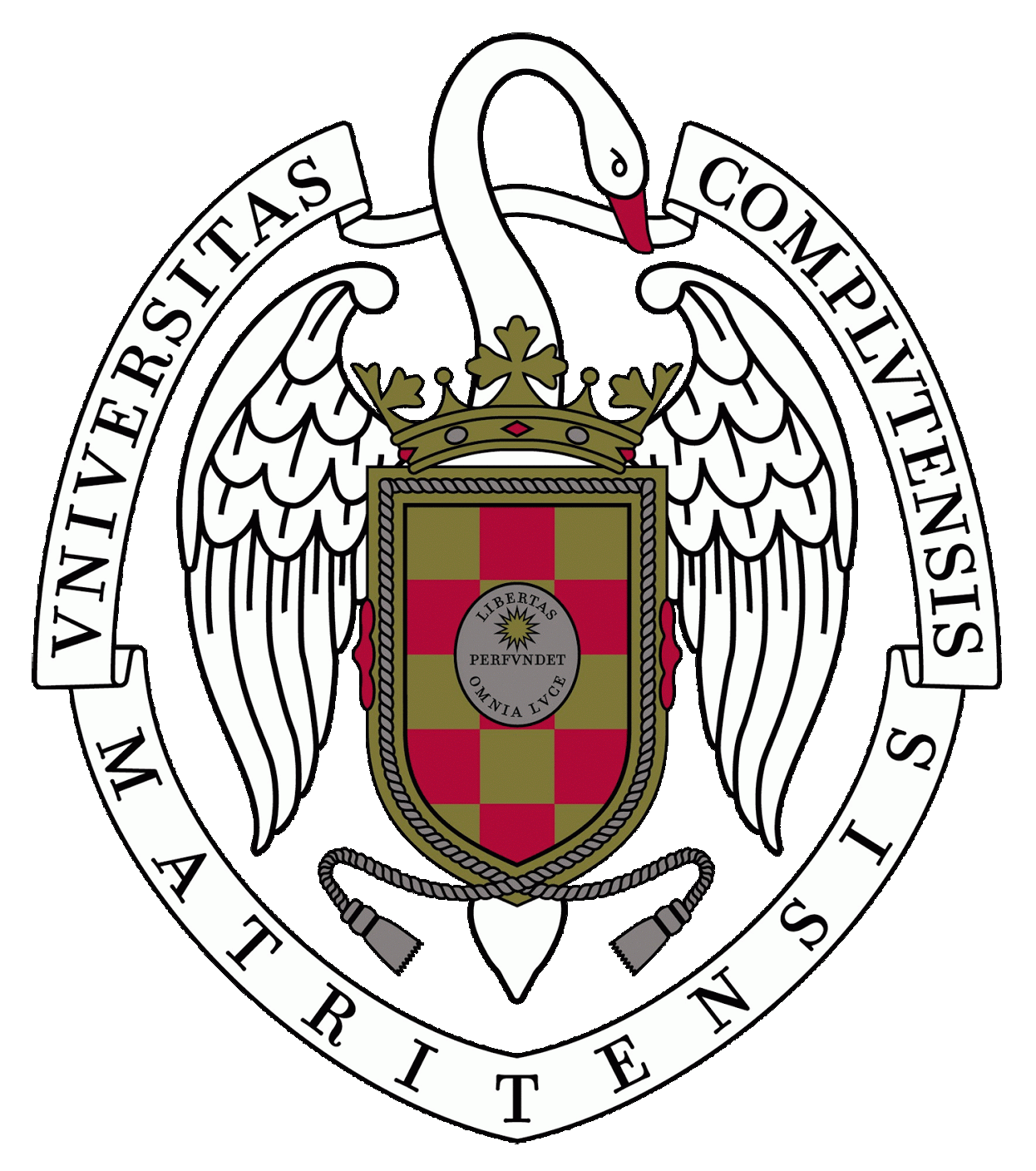}

\vspace{0.55cm}

{\LARGE\bfseries
\textsc{Tesis Doctoral}
}

\vspace{0.6cm}

{\Large
\textsc{Hacia la Utilización Astronómica de las Observaciones Geodésicas de Nueva Generación}
}

\vspace{-0.5cm}
{\large
\textsc{Towards an Astronomical Use of New-Generation Geodetic Observations}
}

\vspace{0.3cm}

{\large
\textsc{Memoria para optar al grado de Doctor}
}

\vspace{0.3cm}

{\large
\textsc{Presentada por}
}

\vspace{0.1cm}

{\Large
\textsc{Víctor Pérez Díez}
}
\vspace{0.1cm}

{\large
\textsc{Observatorio Astronómico Nacional}
}
\vspace{0.2cm}

{\large
\textsc{Dirigida por}
}

\vspace{-0.1cm}

{\large
\textsc{Iván Martí Vidal}
}

{\large
\textsc{Rafael Bachiller García}
}

\end{center}
\end{titlepage}

\clearpage
\thispagestyle{empty}
\null
\clearpage

\includepdf[
  pages=-,
  pagecommand={\thispagestyle{empty}}
]{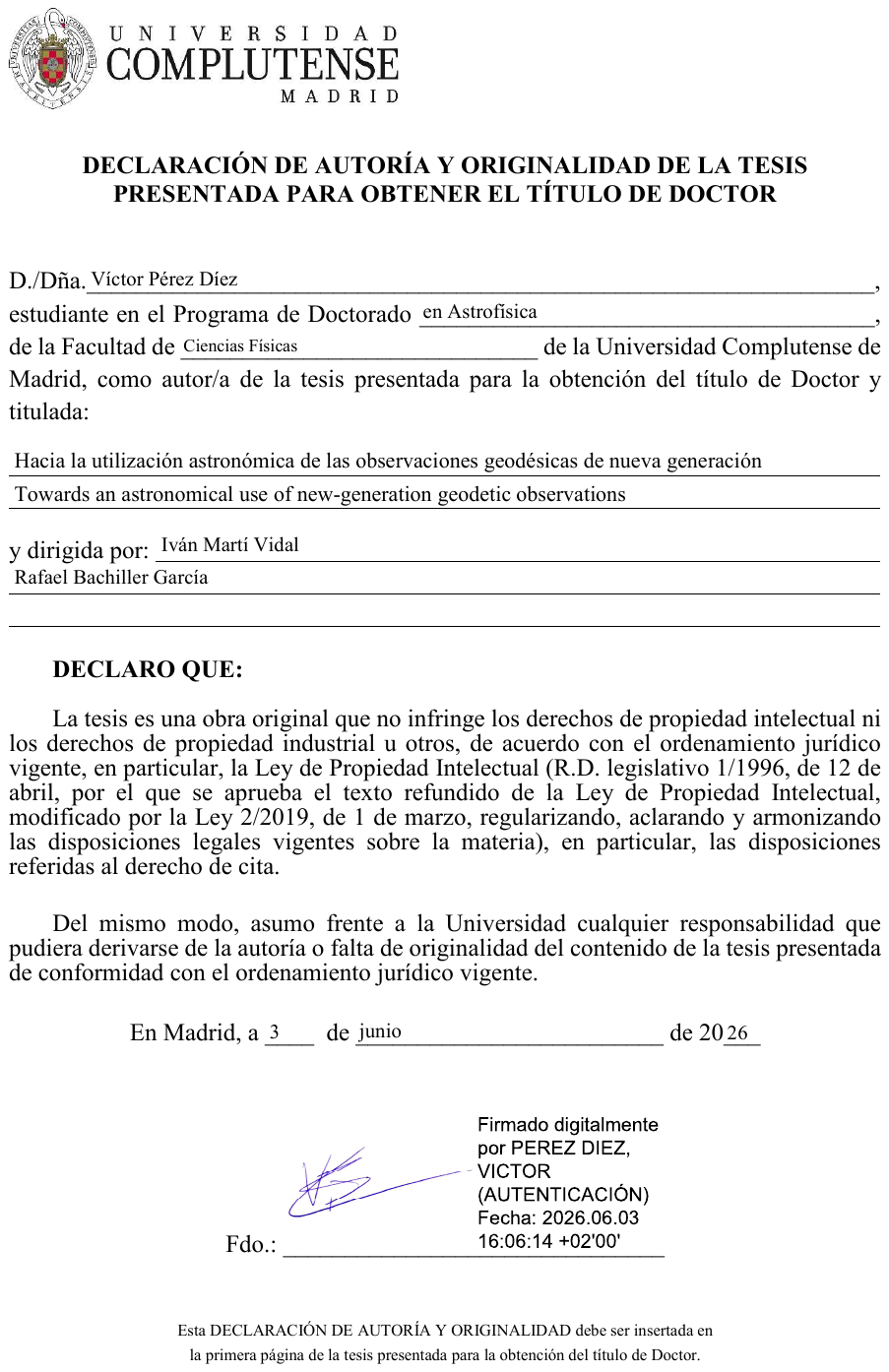}

\clearpage
\thispagestyle{empty}
\null
\clearpage

\pagenumbering{roman}

\phantomsection
\addcontentsline{toc}{chapter}{Agradecimientos}
\chapter*{\centering Agradecimientos}
\markboth{Agradecimientos}{Agradecimientos}

\lettrine{D}{  urante} todos estos años, he sido muy afortunado de tener cerca de mí a muchas buenas personas que me han ayudado a poder realizar este trabajo. En primer lugar, quiero agradecértelo a ti, abuela. Fuiste la primera en saber que iba a estudiar el Máster de Astrofísica, así que me hace mucha ilusión que seas la primera en aparecer en este manuscrito. Cada vez que veo la Luna desde la ventana de casa, me acuerdo de ti. Hace poco, toda la familia hemos estado viendo el eclipse desde tu casa, sé que te habría hecho muchísima ilusión.

No podría haber tenido una familia mejor. Gracias papá y mamá por haber hecho tanto por mí, por haber luchado para que yo tuviera todas las oportunidades que quisiera y por haber conseguido que pueda dedicarme a lo que más me gusta. Gracias por haberme enseñado y cuidado (y por seguir haciéndolo). Sentir que siempre tengo vuestro apoyo me ha ayudado muchísimo a conseguir mis metas.

También he tenido mucha suerte de tener a mis dos hermanos mayores. Gracias a los dos por haberme cuidado desde pequeño. Rebeca, gracias por llevarme a conciertos, a conocer tantos lugares y por algunos remineamientos. Dani, gracias por haberme dado ``ejemplo'', por haberme cuidado cuando llegué a Madrid (y en Mallorca) y por todas las partidas a tantos juegos.

Gracias también a mis sobrinos, por hacerme sentir tan querido, por querer pasar tanto tiempo conmigo y por jugar conmigo. Me lo paso muy bien con vosotros cada vez que voy a Burgos. Álvaro, gracias por todas las partidas de ajedrez (espero poder seguir plantándote cara mucho tiempo), por las preguntas de agujeros negros y por los regalos chulos. Max, gracias por los juegos y las preguntas sobre ciencia, seguro que en unos años aparezco en los agradecimientos de tu tesis científica. Leia, gracias por quererme tanto, por jugar conmigo y por querer venir a verme a Madrid siempre que puedes. Adrián, gracias por ayudarme en este final de tesis con tus sonrisas. Tus fotos y vídeos estos últimos meses me han ayudado mucho.

A Leo por ayudarme en tantos momentos, como a encontrar el Huevito o pegarse un viaje tan largo para buscarme en medio de una pandemia. A Marina por todas las tartas ricas (¡incluida la del agujero negro!), por prepararme tan bien mi primera habitación en Madrid y por los años viviendo juntos (y con Mafin). A mi tía Ascen, por acordarse siempre de mí y estar presente en todas las celebraciones.

Gracias también a mi familia de Caraquiz, Pedro y Conchita, por los fines de semana en los que estoy tan bien cuidado en la casa rural del Almendro y por la pensión Sierra Bermeja, que me ha venido tan bien estos años. Gracias también a Fran, Mariluz, Daniel y Juncal por los buenos momentos.

A mis amigos que me han llenado durante todos estos años. Gracias a todo Takoma: Caro, Poguero, Txus, Buddy, Violi y todos los demás que han pasado, por tantos grandes ratos con y sin instrumentos encima. También a mis compañeros de Blanco y Negro por los domingos de ajedrez y por otro año más manteniendo la categoría (con y sin necesidad de pandemias). A mis compañeros de DK, por las tardes de partidas. A Enrique, Marcos, Bego, Rull, Potter, Carmen, Luana, Sergio, Adri, Raquel y Quevedo por los juegos y las cenas. A mis amigos de La Vieja Escuela; Miguel, Rull, Andy, Diego, Mario y Pablo, por darme tantos millones en clausulazos y tenerme pendiente de los goles del domingo. A Elvira por estar siempre y por tanta música juntos.

En lo académico no he podido tener más suerte con mis directores. Es un gran privilegio poder estar a hombros de dos gigantes de la radioastronomía internacional: Iván y Rafael. Gracias por todo lo que me habéis enseñado estos años.

Desde mi llegada al observatorio, no he podido estar mejor acompañado. En primer lugar, gracias a Miguel e Iván por ayudarme tanto nada más llegué al observatorio. Por echarme una mano con las observaciones, con los papeleos, con los artículos, con el final de tesis... Siempre me he sentido muy acompañado. Gracias Marina por ser la mejor compañera de despacho que podría tener, por comprenderme y echarme una mano siempre que lo necesito, por las chuches y por las risas. Gracias también a los que habéis ido llegando después, mis amigos de la AstroCueva: Dani, Elisa y Paula Dones/Cutanda. Hacéis que cada día vaya con ganas al observatorio, para compartir los buenos y los malos momentos, los cafés y alguna buena comida. Entre todos, habéis conseguido mantenerme sano y salvo en todos los viajes (lo cual no es nada fácil).

Gracias también al resto de compañeros que me habéis acompañado y ayudado en todos los aspectos: Pablo, Javier, Miguel Santander, Miguel Querejeta, Santi, Jean-François, Belén, Nuria, Paula, María Jesús, Pedro Pablo, Alba, Gisela, María, Jose, Mario, Antonio, Miguel Izquierdo, Alexandra, Alejandro y Mariángeles. Espero seguir compartiendo conversaciones interesantes en las comidas durante mucho tiempo.

Muchas gracias también a todos los compañeros de Yebes, en especial a Javi, Cristina, Pablo, Candela y Santi. Gracias por haberme enseñado tanto de VLBI, de la antena, de las observaciones y haber confiado en mí para varios proyectos. Es un verdadero placer trabajar con vosotros y aprender de todo lo que sabéis.

Todo mi trabajo en VLBI ha sido también gracias a mis dos compañeros y amigos de Valencia. Alejandro, en mi primer viaje a Valencia no dudaste en ayudarme a aprender los primeros pasos del VLBI y a hacer mis primeros scripts de calibración. Gracias también por acompañarme por las tardes, por enseñarme la ciudad, buenos restaurantes y por compartir tantos momentos desde ese día (y todos los que quedan). Ezequiel, sin ti, todas las tardes calibrando datos de VGOS habrían sido mucho más duras y largas. Gracias por pelearte conmigo con esos datos tan rebeldes y por acompañarme por Valencia, Madrid, Yebes, Helsinki, Manchester, Dwingeloo y Potsdam.

I also want to thank my international colleagues. Thank you Frédéric, Minghui, Tuomas and Jong-Seo for trusting in my work and bringing me the opportunity to visit you and collaborate on your projects.

Nala, muchas gracias por toda tu compañía y por todo el amor que me das. Cada día que llego a casa y me saludas con toda tu alegría me llenas de motivación para seguir adelante y consigues hacer buenos los días malos. Gracias también por acompañarme en mi regazo mientras estoy teletrabajando y hacer el día más ameno.

Sobre todo, gracias a Alba. Me has acompañado a lo largo de todo este camino: desde los primeros días tocando Victor and His Demons o I Want to Hold Your Hand hasta haber montado nuestro propio estudio en casa. Hemos descubierto muchos lugares y comidas juntos y eres una gran rival a todo tipo de juegos. Empezamos a vivir juntos el mismo día que empecé este doctorado. Han sido los mejores días de mi vida, y en gran parte es gracias a ti. Me has ayudado y cuidado cuando estaba mal, hemos compartido los buenos momentos y me has hecho llevaderos los malos, me has hecho compañía las noches de observaciones y has convertido cada día en especial. Además de mi mejor amiga, eres mi modelo a seguir, no dejas de asombrarme con todo lo que haces. Tras todo este tiempo juntos, me ha quedado claro que quiero seguir viviendo aventuras y proyectos contigo. ¿Tú también quieres pasar la vida conmigo?

\vspace{18pt}

\thispagestyle{empty}
\pagebreak
\begingroup
\addtolength{\cftaftertoctitleskip}{\baselineskip}
\tableofcontents
\endgroup

\thispagestyle{empty}
\clearpage

\phantomsection
\addcontentsline{toc}{chapter}{List of Tables}
\listoftables
\thispagestyle{empty}
\clearpage

\phantomsection
\addcontentsline{toc}{chapter}{List of Figures}
\listoffigures
\thispagestyle{empty}

\clearpage
\thispagestyle{empty}
\null
\clearpage
\phantomsection
\addcontentsline{toc}{chapter}{Resumen}
\chapter*{\centering Resumen}
\markboth{Resumen}{Resumen}
\thispagestyle{empty}

\lettrine{E}{  sta} tesis estudia el uso astronómico de las observaciones del VLBI Global Observing System (VGOS), una red geodésica de nueva generación diseñada para mejorar la precisión de los productos geodésicos, como los parámetros de orientación de la Tierra (EOPs) y los sistemas de referencia terrestre y celeste. Aunque VGOS fue concebido para su uso en geodesia, su gran ancho de banda en frecuencia, sus numerosas observaciones y el seguimiento regular de núcleos galácticos activos (AGN) lo convierten en un instrumento valioso para su uso astrofísico. El objetivo principal de este trabajo es demostrar que los datos de la red VGOS pueden calibrarse con suficiente precisión para obtener imágenes multifrecuencia. Además, pueden utilizarse tanto para estudiar la física de los chorros de los AGN como para reducir los efectos de estructura de la fuente en los observables geodésicos.

La primera parte de la tesis desarrolla una serie de procedimientos para la calibración de observaciones VGOS. Los datos correlados se convierten de polarización lineal a una base circular y se calibran las fases y las amplitudes. Este procedimiento da como resultado unas visibilidades calibradas adecuadas para la deconvolución de imágenes. El elemento más importante de esta serie de algoritmos es un nuevo ajuste global de franjas para gran ancho de banda, que, tras sustraer una contribución a priori de la ionosfera, estima valores de retraso dispersivos y no dispersivos a lo largo de las frecuencias de VGOS. Los datos calibrados se procesan después mediante algoritmos de deconvolución de imágenes basados en técnicas de máxima verosimilitud con regularizadores (RML), produciendo imágenes de intensidad total y polarimetría en las cuatro bandas de VGOS, así como productos multifrecuencia como mapas de índice espectral y de rotación de Faraday.

Para verificar su validez, el procedimiento se aplica al blázar 1803+784, utilizado como fuente de referencia por ser muy brillante, observado frecuentemente en experimentos de VLBI geodésico y por haber sido estudiado con otras redes de VLBI astronómico anteriormente. Las imágenes recuperan la morfología esperada, formada por un núcleo compacto seguido de un chorro. La gran cobertura en frecuencia, desde aproximadamente 3 hasta 11 GHz, permite medir de forma simultánea la variación de la posición del núcleo en función de la frecuencia (desplazamiento del núcleo o \textit{core shift}). Este ajuste da un valor de $k_r=0.73^{+0.12}_{-0.19}$, por debajo del valor de equipartición, $k_r\simeq1$, lo que sugiere que el jet interno de 1803+784 no queda completamente descrito por un modelo cónico simple en equipartición de energía. Los mismos datos permiten también modelar la rotación de Faraday, separando una contribución externa de una componente interna que depende de la frecuencia. La comparación de la morfología, el comportamiento espectral y la polarimetría con estudios previos de otras redes VLBI confirma la fiabilidad del procedimiento.

A continuación, el análisis se extiende a una muestra de AGN: 3C418, 0059+581, 0613+570, 1849+670 y 2229+695. En todos ellos, los datos VGOS proporcionan imágenes multifrecuencia y medidas del desplazamiento del núcleo. La mayoría de los valores ajustados de $k_r$ son compatibles con la hipótesis de equipartición de energía, siendo 1803+784 el único caso de desviación. Los mapas de índice espectral muestran núcleos ópticamente gruesos y emisión del chorro ópticamente delgada, tal como se espera en chorros con emisión sincrotrón. Las medidas de rotación de Faraday dependen más de cada fuente, pero demuestran que las observaciones VGOS contienen suficiente información polarimétrica para recuperar medidas de rotación del mismo orden que las obtenidas con redes de VLBI astronómico.

La parte final de la tesis analiza la utilización geodésica de las imágenes obtenidas con este procedimiento. La misma estructura de las fuentes que traza la física de los chorros introduce también errores en el VLBI geodésico, ya que las distribuciones de brillo de las fuentes se suelen suponer puntuales. Por ello, los modelos obtenidos de las imágenes se utilizan para corregir la contribución de esta estructura en la fase de las visibilidades. Los datos corregidos se procesan con el mismo procedimiento de análisis geodésico que los datos originales y se observa que la mejora es mayor en fuentes estructuradas como 3C418, para la cual las distribuciones de clausuras de retraso pasan a ser más estrechas y los residuos post-ajuste se reducen. Las fuentes compactas, como 1849+670, muestran pocos cambios, como cabía esperar. Las pruebas con sesiones posteriores muestran que la corrección resulta más útil cuando los modelos son cercanos en el tiempo a las observaciones corregidas. Tras varios meses, la variabilidad de la fuente puede reducir el efecto beneficioso o incluso perjudicar los residuos del ajuste.

En conjunto, esta tesis muestra que las observaciones VGOS pueden utilizarse no sólo con un fin geodésico, sino que también se puede obtener una valiosa información astronómica. Permiten estudiar el desplazamiento del núcleo, la estructura y la rotación de Faraday en AGN, al mismo tiempo que proporcionan la información necesaria para mitigar los efectos de la estructura de la fuente en geodesia.

\pagebreak

\cleardoublepage
\phantomsection
\addcontentsline{toc}{chapter}{Abstract}
\chapter*{\centering Abstract}
\markboth{Abstract}{Abstract}
\thispagestyle{empty}

\lettrine{T}{  his} thesis explores the astronomical use of observations from the VLBI Global Observing System (VGOS), a next-generation geodetic network designed primarily to improve the precision of the geodetic products, such as the Earth Orientation Parameters and the terrestrial and celestial reference frames. Although VGOS was conceived for geodesy, its ultra-wide frequency coverage, high observing cadence and regular monitoring of compact radio-loud active galactic nuclei (AGN) make it a valuable instrument for astrophysics. The main goal of this work is to demonstrate that VGOS data can be calibrated and imaged with sufficient accuracy to obtain full-polarization, multi-frequency images, and that these images can be used both to study AGN jets and to reduce source structure effects in geodetic observables.

The first part of the thesis develops and validates a complete calibration pipeline for VGOS observations. Starting from the correlator output, the procedure converts the original linear polarization data into a circular polarization basis, calibrates instrumental phases, applies amplitude calibration, and produces calibrated visibilities suitable for imaging. A central element of the pipeline is a new wideband global fringe fitting algorithm, which estimates non-dispersive and dispersive delay terms across the VGOS frequency range after subtracting an a priori ionospheric contribution. The calibrated data are then imaged using regularized maximum likelihood techniques, producing total intensity and polarimetric images in the four VGOS bands, as well as multi-frequency products such as spectral index and Faraday rotation maps.

The pipeline is first applied in detail to the blazar 1803+784, used as a reference source because it is bright, frequently observed in geodetic VLBI experiments, and well studied with astronomical VLBI arrays. The VGOS images recover the expected compact core and downstream jet morphology. The simultaneous frequency coverage from approximately 3 to 11 GHz allows the frequency dependent position of the VLBI core to be measured simultaneously. The resulting core-shift fit gives $k_r=0.73^{+0.12}_{-0.19}$, below the canonical equipartition value of $k_r\simeq1$, suggesting that the inner jet of 1803+784 is not fully described by a simple conical jet in energy equipartition. The same data also allow the Faraday rotation to be modelled, separating an external contribution from a frequency-dependent internal component. The agreement between source morphology, spectral behaviour and polarization results with previous VLBI studies confirms the reliability of the calibration and imaging strategy.

The analysis is then extended to a sample of AGN: 3C418, 0059+581, 0613+570, 1849+670 and 2229+695. For all of them, VGOS data provide meaningful multi-frequency images and core-shift measurements. Most fitted values of $k_r$ are compatible with the equipartition expectation, while 1803+784 remains the clearest case of a significant deviation. The spectral index maps generally show optically thick compact cores and optically thin jet emission, as expected for synchrotron AGN jets. The Faraday rotation measurements are more source-dependent, limited in some cases by polarization signal-to-noise ratio (S/N) and angular resolution, but they demonstrate that VGOS observations contain enough polarimetric information to recover rotation measures of the same order as those measured by astronomical VLBI networks. For several sources, these results provide new astrophysical information obtained from data originally scheduled for geodesy.

The final part of the thesis examines the geodetic use of the VGOS images. The same source structures that trace the AGN jet physics also introduce systematic errors in geodetic VLBI, where source brightness distributions are often approximated as point-like. Image models are therefore used to correct the structure phase contribution directly at the visibility level, and the corrected data are processed through the same geodetic analysis chain as the original data. The improvement is strongest for structured sources such as 3C418, for which the closure delay distributions become narrower and the post-fit delay residuals are reduced. Compact sources, such as 1849+670, show little change, as expected. Tests with later sessions show that the correction is most useful when the image model is close in time to the corrected observations. After several months, source variability can reduce or even reverse the benefit.

Overall, this thesis shows that VGOS observations can be transformed from purely geodetic measurements into scientifically useful astronomical data. They allow the study of core shift, spectral structure and Faraday rotation in compact AGN, while at the same time providing the information needed to mitigate source structure effects in geodesy. VGOS therefore has a dual value: it is the best performing geodetic observing system and, with suitable calibration, a powerful multi-frequency VLBI instrument for AGN astrophysics.

\cleardoublepage

\pagenumbering{arabic}

\chapter{Introduction}
\label{ch:introduction}

\lettrine{T}{  his} thesis focuses on the development of a data calibration pipeline for wide-bandwidth Very Long Baseline Interferometry (VLBI) and its astrophysical and geodetic application to Active Galactic Nuclei (AGN) observations. AGN constitute one of the most important classes of extragalactic sources. They play an important role in astrophysics and geodesy. Their compact, bright, and highly polarized emission makes them ideal targets for high-resolution observations with VLBI, while their intrinsic structural complexity introduces challenges for precise astrometry and geodesy.

This thesis is structured as follows: Chapter \ref{ch:vgos_pipeline} describes the developed calibration pipeline. Its application and validation are presented in Chapter \ref{ch:1803+784}, while astrophysical and geodetic applications are shown in Chapters \ref{ch:astro} and \ref{ch:geodesy}, respectively.

This introduction provides the framework for the work presented in this thesis. It first introduces the basic properties of AGN and their relativistic jets, including their radio emission, polarization, and launching mechanisms. It then presents the principles of VLBI, together with the calibration and imaging techniques required for this type of observations. Finally, it discusses the role of VLBI in high-precision astrometry and geodesy, with particular emphasis on the capabilities of the VLBI Global Observing System (VGOS) and its potential for astrophysical and geodetic applications.

\section{Active Galactic Nuclei (AGN)}
\label{sec:AGN}
Active Galactic Nuclei (AGN) are among the most luminous sources in the Universe. They are powered by accretion of matter into a supermassive black hole (SMBH) located at the centre of a galaxy. The release of gravitational potential energy in the accretion flow produces radiation spanning the entire electromagnetic spectrum, from radio to gamma rays. The multi-wavelength study of these sources is necessary to explain the combination of physical processes occurring in different regions around the SMBH.

\begin{figure}
    \centering
    \includegraphics[width=0.7\textwidth]{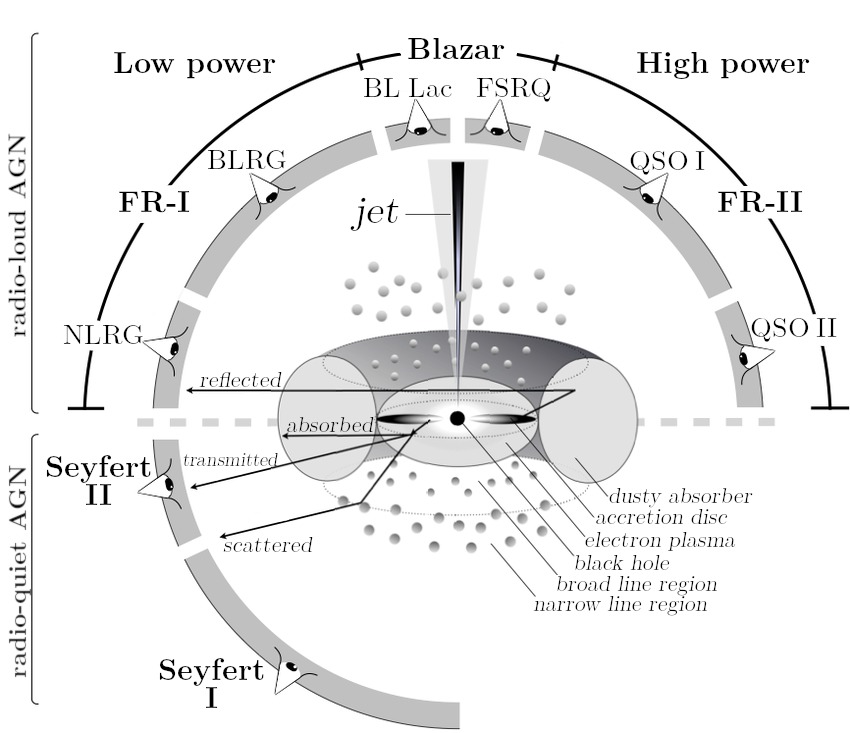}
    \caption[Schematic representation of the unified model of AGN.]{Schematic representation of the unified model of AGN. Adapted from \citet{Beckmann2012}.}
    \label{fig:UnifiedModelAGN}
\end{figure}

According to modern multi-wavelength studies, much of the apparent diversity of AGN can be interpreted as the result of variations in a small number of intrinsic physical parameters (black-hole mass, accretion rate, jet power) and geometric factors, mainly the inclination angle of the system with respect to the observer (see Fig. \ref{fig:UnifiedModelAGN}). This is called the Unified Model of AGN, proposed in \citet{Antonucci1993} and \citet{Urry1995}. This model has been tested and completed across the years (e.g. \citet{Heckman2014, Padovani2017}). It places a supermassive black hole (SMBH) at its centre. Gas falling toward the SMBH forms an accretion disc whose temperature rises steeply close to the centre, producing intense ultraviolet and X-ray radiation. Surrounding the disc, dense clumps of ionised gas form the broad line region (BLR), where orbital velocities of several thousand kilometres per second broaden the permitted emission lines. At larger distances, lower velocity clouds constitute the narrow line region (NLR), producing narrower optical and ultraviolet emission features. All these components are embedded in a torus of dust, whose orientation with respect to the observer determines how much of the inner engine is visible. Under appropriate conditions, large scale magnetic fields can accelerate plasma into two oppositely directed relativistic jets through magnetohydrodynamic processes associated with the accretion disc and the spinning black hole \citep{Blandford1977, Blandford1982}. These jets, when present, dominate the radio emission of the galaxy and can remain bright out to kiloparsec scales.

AGN may be broadly divided into non-jetted and jetted systems. Non-jetted AGN lack powerful relativistic outflows and their emission is dominated by thermal radiation associated with the accretion flow. Seyfert galaxies constitute the main class of these radio-quiet AGN. Seyfert I galaxies exhibit both broad and narrow emission lines, indicating a direct view of the BLR. Seyfert~II galaxies show only narrow lines, consistent with obscuration of the BLR by the dusty torus \citep{Khachikian1974}.

Jetted AGN include radio galaxies, radio-loud quasars, and blazars. Radio galaxies display extended synchrotron emission from large-scale jets and lobes. Their morphology separates them into Fanaroff–Riley classes \citep{Fanaroff1974}: FR I sources are edge-darkened and typically associated with lower power jets, while FR II galaxies are edge-brightened, hosting highly collimated, powerful jets terminating in hotspots.

Quasars, the most luminous AGN, are similarly subdivided into Type I (QSO I) and Type II (QSO II) sources. QSO I objects display broad permitted lines and a direct view of the accretion disc, while QSO II quasars exhibit only narrow lines, consistent with heavy obscuration. These two classes constitute the high-luminosity counterparts of Seyfert I and Seyfert II galaxies, respectively.

Radio-loud quasars are intrinsically similar to high-excitation FR II radio galaxies, but are viewed at relatively small inclination angles ($\leq 45^\circ$), exposing the BLR and central continuum \citep{Urry1995}. Their powerful jets contribute substantially to the observed emission, often outshining the host galaxy.

When a jet is oriented very close to the line of sight, relativistic Doppler boosting dramatically amplifies the non-thermal emission, producing a blazar \citep{Urry1995}. Blazars dominate the bright radio and gamma-ray sky and are subdivided into two populations: Flat-spectrum radio quasars (FSRQs), which show broad emission lines and strong external photon fields, and BL Lacertae objects (BL Lacs), defined by weak or absent emission lines and spectra dominated by beamed synchrotron and inverse Compton emission \citep{Padovani1995}.

\subsection{Radio Emission from AGN}
\label{RadioEmissionAGN}
AGN produce radiation in the entire electromagnetic spectrum, from low-frequency radio waves to high-energy gamma rays. Although different mechanisms dominate in different spectral bands, the overall emission is fundamentally non-thermal and requires the presence of relativistic particles embedded in strong magnetic fields.

\begin{figure}
    \centering
    \includegraphics[width=0.7\textwidth]{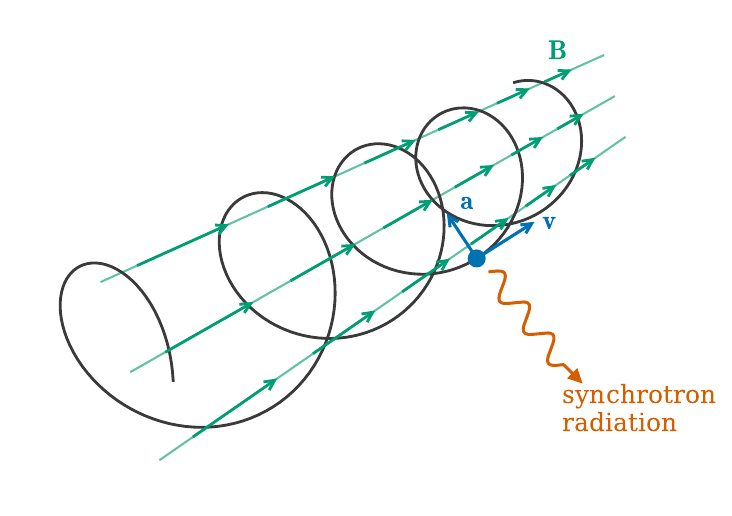}
    \caption{Synchrotron radiation generated by a particle with a helical motion across a uniform magnetic field.}
    \label{fig:synchrotron}
\end{figure}

At radio wavelengths, the dominant process is synchrotron radiation, produced by relativistic electrons moving along helical trajectories around magnetic field lines (see Figure \ref{fig:synchrotron}). This mechanism explains the smooth continuum spectra observed in most jetted AGN, as well as the high degrees of linear polarization, which trace the geometry and ordering of the magnetic field.

According to \citet{Pacholczyk1970}, synchrotron radiation from a population of relativistic electrons, $N(E)$, following a power law energy distribution, 
\[
N(E)\, dE = N_0\, E^{-\gamma}\, dE,
\]
where $N_0$ is a normalization constant related to the number density and energy density of the relativistic electrons and $\gamma$ is the electron energy index, produces a power law spectrum at high frequencies:
\[
I_\nu = S_\nu\, \tau_\nu \propto \nu^{-(\gamma - 1)/2} = \nu^{-\alpha},
\]
with the synchrotron spectral index:
\[
\alpha = \frac{\gamma - 1}{2}.
\]

At low frequencies, synchrotron self-absorption becomes important and the spectrum transitions to an optically thick regime, where:
\[
I_\nu = S_\nu \propto \nu^{5/2},
\]
a slope that is independent of $\gamma$. 

\begin{figure}[H]
    \centering
    \includegraphics[width=0.8\textwidth]{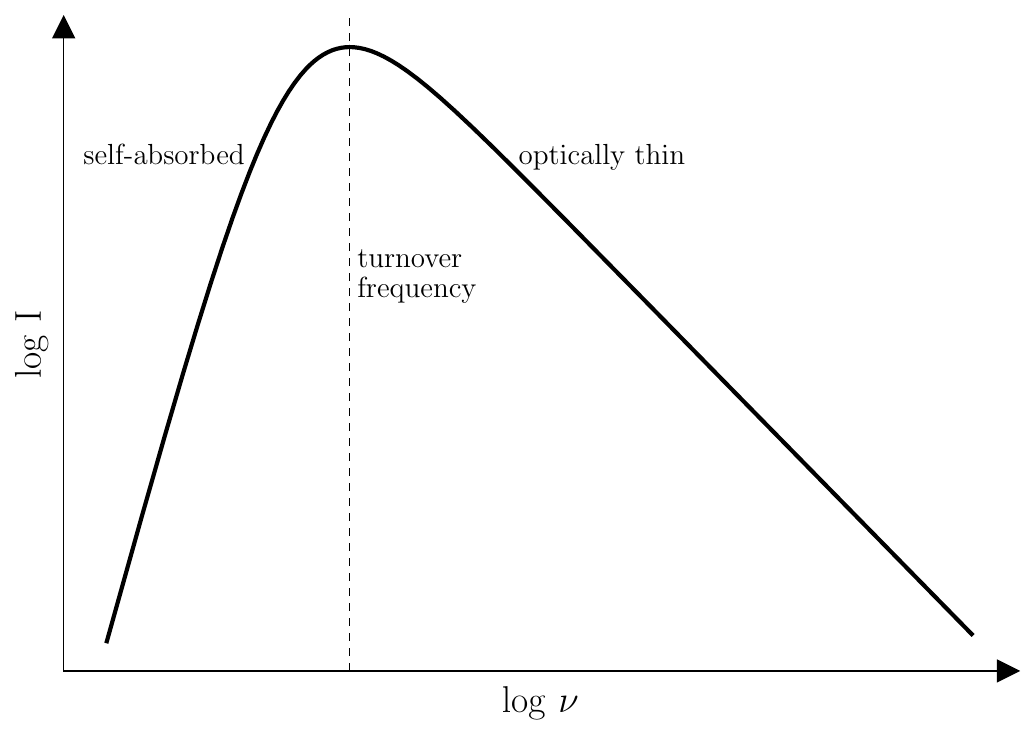}
    \caption[Idealized synchrotron spectrum of a homogeneous self-absorbed emitting region.]{Idealized synchrotron spectrum of a homogeneous self-absorbed emitting region. In real AGN jets, the compact VLBI core is generally inhomogeneous and the observed low-frequency spectrum is flatter than this idealized slope.}
    \label{fig:Spectrum}
\end{figure}

This spectrum exhibits a characteristic turnover at the frequency $\nu_t$, the \textit{turnover frequency}, which marks the transition between the optically thick and optically thin regimes. This prototypical synchrotron spectrum is illustrated in Figure~\ref{fig:Spectrum}.

\subsubsection{Relativistic Effects}

The observed appearance of AGN jets is strongly influenced by relativistic effects. Even if the intrinsic structure of a jet is symmetric, the combined action of relativistic aberration \citep{Rybicki1979}, Doppler boosting \citep{BlandfordKonigl1979}, and apparent superluminal motion \citep{Cohen1971} can alter its observed morphology and emission properties. These effects are essential for the interpretation of observations at very high spatial resolutions and underpin much of the observed diversity among radio-loud AGN \citep{Urry1995, Lind1985}.

Jets are believed to accelerate rapidly within the innermost regions close to the central engine, reaching bulk Lorentz factors of order
\[
\Gamma = \frac{1}{\sqrt{1 - \beta^2}} = \frac{1}{\sqrt{1 - (v/c)^2}} \sim 5 - 20,
\]
or even higher in some extreme blazars \citep{Jorstad2005, Hovatta2009}. At such velocities, the emission from the approaching jet is significantly amplified, while that from the receding jet is strongly suppressed. The degree of amplification is governed by the Doppler factor, $\delta$ , defined by \citet{Ghisellini2013}
\[
\delta = \frac{1}{\Gamma \left( 1 - \beta \cos \theta \right)},
\]
where $\beta = v/c$ is the bulk velocity of the flow in units of the speed of light and $\theta$ is the angle between the jet axis and the line of sight.

Small variations in $\theta$ can produce large changes in $\delta$, leading to big differences in the apparent brightness and the structure even among similar jets. Doppler boosting enhances not only the observed flux density of the approaching jet, but also affects the observed spectrum, brightness temperature, and variability time scales \citep{BlandfordKonigl1979}. Compact components exhibiting exceptionally high brightness temperatures or rapid variability are commonly interpreted as features with large Doppler factors. In contrast, the counter-jet generally remains undetected in VLBI images, except in nearby radio galaxies viewed at larger inclination angles, where both sides of the flow can sometimes be observed \citep{Giovannini2001}.

Relativistic aberration modifies the apparent directionality of the emitted radiation. Photons are preferentially beamed into a narrow cone of opening angle $\sim 1/\Gamma$ around the direction of motion. As a consequence, jets observed at small viewing angles appear unusually well collimated. Internal structures such as shocks, bends, or knots may appear compressed or distorted in projection \citep{Lind1985}.

One of the most striking observational manifestations of relativistic motion in AGN jets is the presence of apparent superluminal velocities, first identified through VLBI monitoring \citep{Whitney1971, Cohen1971}. When a jet component moves at relativistic speed at a small angle to the line of sight, its projected transverse velocity can exceed the speed of light without violating causality. The apparent velocity is given by
\[
\beta_{\mathrm{app}} = \frac{\beta \sin \theta}{1 - \beta \cos \theta},
\]
and can reach values of several times $c$. Measurements of $\beta_{\mathrm{app}}$ provide powerful constraints on both the intrinsic flow speed and the viewing angle. Long-term VLBI monitoring has revealed a wide range of kinematic behaviours, including accelerating components, quasi-stationary features downstream of the core, and trajectories suggestive of helical patterns.

Light travel time effects also play an important role in shaping the observed jet morphology. Emission from the counter-jet not only suffers Doppler deboosting, but also reaches the observer later than emission from the approaching side. As a result, the two sides of a jet do not correspond to the same intrinsic epoch, further enhancing their apparent asymmetry. This effect becomes particularly relevant during episodes of strong variability, when flares propagate along the jet and are observed at different apparent times depending on their direction of motion \citep{BlandfordKonigl1979}.

Taken together, these relativistic effects determine many of the defining observational characteristics of jetted AGN, including the one-sided jet seen in VLBI images, the presence of rapidly moving components, and the extreme luminosities and variabilities observed in blazars. They also have important implications for astrometry and geodesy, since Doppler boosting and structural evolution can shift the apparent centroid of emission, hence introducing changes in the estimated source positions.

\subsection{Polarization}
\label{sec:Polarization}

Polarization provides a compact description of the electromagnetic field beyond total intensity, and constitutes a key observable in radio astronomy. In interferometric observations, polarization is commonly described through the Stokes parameters \citep{Stokes1851}.

\begin{figure}[H]
    \centering
    \includegraphics[width=\textwidth, trim={0pt 0pt 0pt -1pt}, clip]{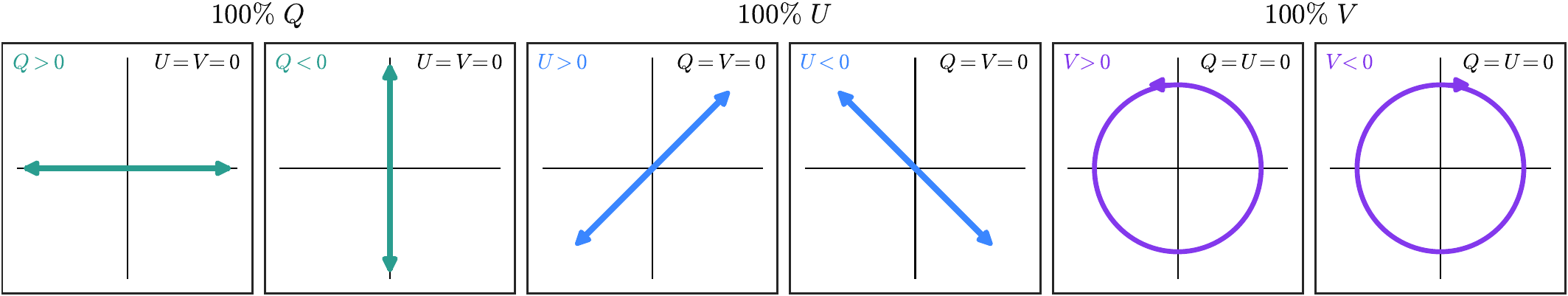}
    \caption{Schematic visualization of the Stokes parameters Q, U, and V, according to the IAU convention.}
    \label{fig:stokes}
\end{figure}

Let $e_x(t)$ and $e_y(t)$ be the complex signals of the electric field projected onto two orthogonal directions on the plane of the sky. Following the IAU convention, the $x$ axis is taken to point toward increasing declination (north), and the $y$ axis toward increasing right ascension (east) \citep{Hamaker1996}. The Stokes parameters, represented in Figure \ref{fig:stokes} are defined by \citet{Born1964, TMS1986} as:
\begin{align}
I &= \left\langle |e_x|^2 + |e_y|^2 \right\rangle, \\
Q &= \left\langle |e_x|^2 - |e_y|^2 \right\rangle, \\
U &= 2\left\langle \,|e_x| |e_y| cos \delta \right\rangle, \\
V &= 2\left\langle \,|e_x| |e_y| sin \delta \right\rangle,
\end{align}
where $\langle \cdot \rangle$ denotes an average over timescales long compared to the wave period, and $\delta$ represents the phase difference between the two oscillating components.

The total intensity is given by $I$, while $Q$ and $U$ describe the linear polarization and $V$ the circular polarization. It is often convenient to define the linearly polarized intensity and its fractional polarization,
\begin{equation}
\label{eq:linearlypolarizedintensity}
P = \sqrt{Q^2 + U^2}, 
\qquad 
m_{\mathrm{L}} = \frac{P}{I},
\end{equation}
together with the fractional circular polarization,
\begin{equation}
m_{\mathrm{C}} = \frac{V}{I}.
\end{equation}
The electric vector position angle (EVPA), denoted by $\chi$, is defined from the linear Stokes parameters as:
\begin{equation}
\chi = \frac{1}{2} \arctan{\left(\frac{U}{Q}\right)}.
\label{eq:evpa}
\end{equation}

\begin{figure}[h]
    \centering
    \includegraphics[width=0.75\textwidth, trim={30pt 100pt 30pt 120pt}, clip]{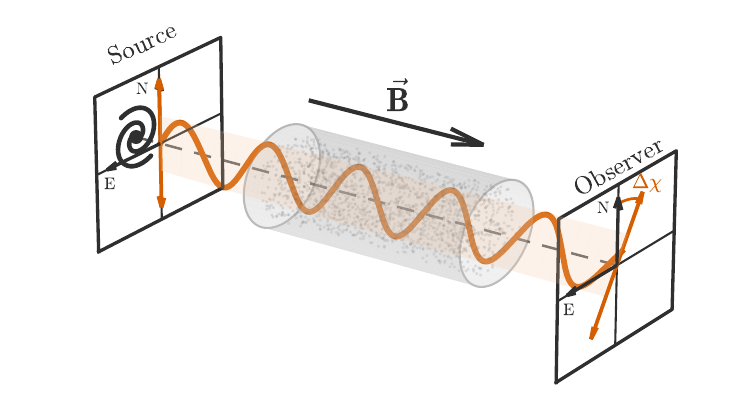}
    \caption[Schematic illustration of Faraday rotation in a magnetized plasma.]{Schematic illustration of Faraday rotation. Linearly polarized radiation emitted by the source propagates through a magnetized plasma. As the wave travels through the plasma, the plane of linear polarization rotates by an angle $\Delta\chi$ that increases with $\lambda^2$. The observed EVPA differs from the intrinsic one, with a rotation proportional to the rotation measure (RM) and $\lambda^2$ (Eq.~\ref{eq:faraday_rotation}).}
    \label{fig:faradayrotation}
\end{figure}

As polarized radiation propagates through a magnetized plasma, its polarization state can be modified. The most widely observed propagation effect at radio wavelengths is Faraday rotation (illustrated in Figure \ref{fig:faradayrotation}), where the plane of linear polarization rotates by an amount proportional to the wavelength squared $\lambda^2$,

\begin{equation}
\chi(\lambda) = \chi_0 + \mathrm{RM}\,\lambda^2,
\label{eq:faraday_rotation}
\end{equation}
where $\chi_0$ is the intrinsic EVPA in the limit $\lambda \rightarrow 0$ and RM is the rotation measure. \citet{Brentjens2005} defines the rotation measure in practical cgs-derived units by:
\begin{equation}
\mathrm{RM}\;[\mathrm{rad\,m^{-2}}] = 0.812 \int n_e\;[\mathrm{cm^{-3}}]\; B_{\parallel}\;[\mu\mathrm{G}]\; dl\;[\mathrm{pc}],
\label{eq:rm_definition}
\end{equation}
where $n_e$ is the thermal electron density, $B_{\parallel}$ is the line-of-sight component of the magnetic field, and $dl$ is the differential path length. Although Faraday rotation is dominated by thermal electrons, relativistic electrons can also contribute under certain conditions (although with an effect attenuated by the Lorentz factor as $\log(\Gamma)/\Gamma^2$), introducing additional complexity in compact jet regions.

Faraday rotation can lead to several forms of depolarization. Unresolved RM structure within the synthesized beam causes beam depolarization, while averaging over a finite receiver bandwidth leads to bandwidth depolarization when the EVPA rotates significantly across a frequency channel. Furthermore, Faraday dispersion can produce a frequency dependent decrease of the observed fractional polarization. It depends on whether the rotating plasma is mixed with the emitting region or lies in a foreground screen \citep[e.g.,][]{Burn1966}.

A related propagation effect is \emph{Faraday conversion}, in which linear
polarization is partially converted into circular polarization as the radiation
propagates through a magnetized relativistic plasma \citep{Jones1977, Homan2009}.
This process can generate small but measurable levels of circular polarization
even when the intrinsic synchrotron circular component is weak. It is expected
to be particularly relevant in compact, optically thick regions.

\subsubsection{Polarization of AGN Jets}
\label{sec:polarization_theory}
Polarization is a fundamental observable of synchrotron AGN jets. It provides direct insight into the geometry and the magnetic field within the relativistic plasma. Synchrotron radiation is intrinsically polarized, and both the polarization fraction and the orientation of the polarization vectors carry valuable information about the physical conditions in the emitting region.

For an optically-thin synchrotron source with a power-law distribution of relativistic electrons and in the absence of Faraday effects, the maximum degree of linear polarization is given by \citet{Pacholczyk1970}
\begin{equation}
\Pi = \frac{\gamma + 1}{\gamma + \tfrac{7}{3}},
\end{equation}
where $\gamma$ is the power law index of the electron energy distribution. This maximum value is independent of the frequency. In practice, the observed polarization fraction is typically much lower due to beam depolarization, turbulence, and the superposition of multiple magnetic field components along the line of sight. In the optically thin regime, the EVPA is orthogonal to the projected magnetic field direction, allowing EVPA maps to trace ordered structures such as shocks or large scale helical magnetic fields within the jet.

At lower frequencies, synchrotron self-absorption becomes important and the emission enters an optically thick regime. In this case, the degree of linear polarization is also frequency-independent, as indicated in \citet{Pacholczyk1970}

\begin{equation}
\Pi = \frac{3}{6\gamma + 13}.
\end{equation}

In the optically thick regime, the EVPA rotates by \(90^\circ\) relative to the optically thin case and becomes parallel to the projected magnetic field direction. This EVPA rotation provides a powerful diagnostic of synchrotron self-absorption and is routinely observed in multi-frequency VLBI polarization studies (e.g. \citet{Zavala2004, Gomez2011, OSullivan2009}).

As introduced in the previous subsection, Faraday rotation arises when linearly polarized radiation propagates through magnetized plasma and produces a frequency-dependent rotation of the EVPA proportional to $\lambda^2$. In the context of AGN jets, this effect is often particularly pronounced in the vicinity of the VLBI core, where dense magnetized material is present. Indeed, exceptionally high rotation measures have been reported in the inner jet regions of AGN, indicating strong magnetic fields and enhanced thermal electron densities close to the central engine \citep{Jorstad2007, Hovatta2012, IMV2015}. Moreover, systematic transverse gradients in RM across the jet width have been interpreted as observational signatures of large scale helical magnetic fields, consistent with jet launching and collimation scenarios \citep{Gabuzda2015}.

Circular polarization, although typically much weaker than linear polarization, has also been detected in a number of AGN jets at levels of $\sim 0.1$--$1\%$ (e.g. \citet{Homan2001, Wardle2001, Vitrishchak2008}). Its origin remains a subject of debate, with possible contributions from intrinsic synchrotron circular polarization and from Faraday conversion of linear to circular polarization in a relativistic plasma \citep{Homan2004}. Measurements of circular polarization therefore provide additional constraints on the particle content and magneto-ionic properties of relativistic jets. More recently, circular polarization has been detected in the immediate vicinity of the supermassive black hole in M87 by the Event Horizon Telescope \citep{EHT2023}, demonstrating that Stokes V emission is not limited to parsec scale jets, but is also present in the jet launching region itself.

\subsection{Jet Launching and Collimation}
\label{sec:jetlaunching}

Relativistic jets consist of highly collimated outflows of plasma launched from the immediate vicinity of the central supermassive black hole (SMBH), and can propagate from sub-parsec regions near the black hole to hundreds of kiloparsecs \citep{Pearson1988}. Radio observations have historically provided the clearest view of their morphology and kinematics.

The first evidence of jets came from optical observations of the nearby radio galaxy M87 \citep{Curtis1918}. The advent of radio astronomy \citep{Jansky1933, Reber1940} and the development of radio interferometry \citep{Ryle1946, Ryle1960}, revealed that many bright extragalactic radio sources exhibit extended double-jet morphologies.

\begin{figure}[h!]
    \centering
    \includegraphics[width=0.66\textwidth, trim={0pt 0pt 0pt 0pt}, clip]{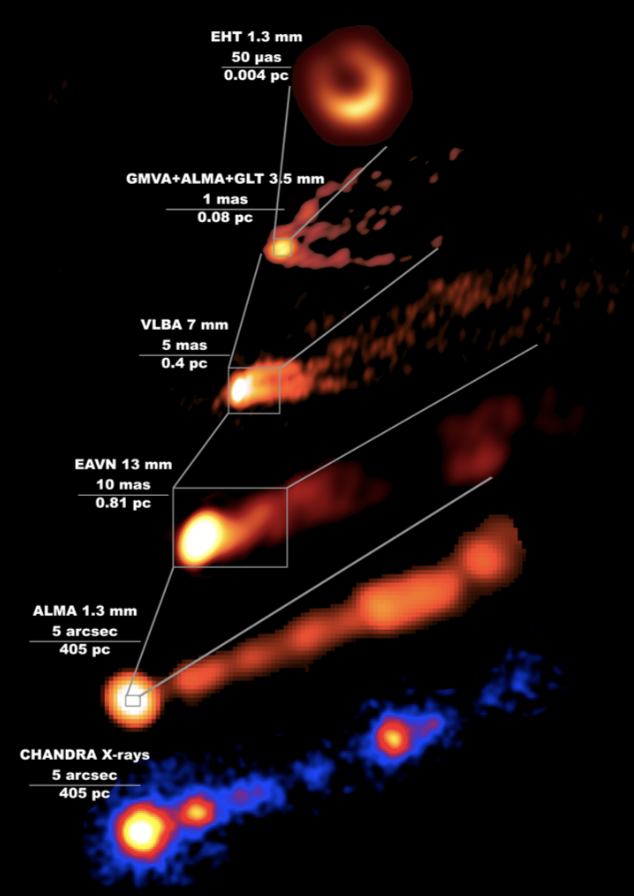}
    \caption[Composite multi-wavelength view of M87 at different spatial scales.]{Composite multi-wavelength view of M87 at different spatial scales, combining radio and X-ray observations. The colour scale in each image has been adjusted to enhance the visibility of the relevant morphological features at the corresponding scale. Taken from \citet{Algaba2024}.
}
    \label{fig:multiscaleM87}
\end{figure}

Jets can be studied over a wide range of spatial scales, as seen in Figure \ref{fig:multiscaleM87}. On parsec scales, VLBI observations resolve the innermost jet regions, revealing a bright compact core, and a series of downstream components, called knots. On kiloparsec scales, connected interferometers study the large-scale jets (e.g. \citet{Laing2008}), where the outflow interacts with the surrounding interstellar and intergalactic medium. The transition from the compact jet to the extended radio structure is commonly seen as a gradual expansion and bending of the flow, accompanied by the formation of shocks and brightness enhancements along the jet. 

Relativistic jets provide a unique window into the central engine of AGN. Their structure and kinematics encode information about the launching region, the collimation and acceleration processes, and the interaction with the ambient medium. In addition, their compact, bright cores are ideal targets for high-resolution VLBI studies, which will be discussed in the following sections in the context of radio emission mechanisms and geodetic applications.

The origin of relativistic jets in AGN is closely linked to the presence of strong magnetic fields in the immediate vicinity of the central supermassive black hole (SMBH). While the detailed physics of jet launching remains an active area of research, two theoretical frameworks have usually been used for understanding how rotational energy can be extracted and channelled into powerful and collimated outflows. One class of models attributes the jet power primarily to the rotational energy of the black hole itself. In the Blandford-Znajek (BZ) mechanism \citep{Blandford1977}, large-scale magnetic field lines threading the event horizon of a spinning Kerr black hole extract energy electromagnetically from the black hole spin. In this scenario, the jet power scales with both the dimensionless spin parameter and the magnetic flux accumulated near the horizon, a result that has been extensively explored in analytical studies and numerical simulations \citep[e.g.,][]{Komissarov2001, Tchekhovskoy2011}. General relativistic magnetohydrodynamic (GRMHD) simulations have demonstrated that the BZ mechanism can efficiently produce highly relativistic jets under conditions of strong magnetic flux accumulation, such as in magnetically arrested discs.

A complementary class of models associates the jet launching process with the accretion disc rather than the black hole spin. In the Blandford-Payne (BP) mechanism \citep{Blandford1982}, large-scale magnetic field lines anchored in the rotating accretion disc act as centrifugal accelerators. Plasma loaded onto field lines inclined by less than $60^\circ$ with respect to the disc surface can be flung outward and collimated by magnetic stresses. This will form a magnetic outflow that may transition into a relativistic jet at larger distances.

These two mechanisms are not mutually exclusive. GRMHD simulations suggest that both processes may operate simultaneously, with the BZ mechanism dominating the energy extraction and acceleration close to the black hole, while the BP mechanism contributes to mass loading and collimation on disc scales \citep{McKinney2006}. This combined action naturally leads to a stratified jet structure, consisting of an inner relativistic spine surrounded by a slower sheath.

Recent high-resolution observations have begun to provide observational constraints on these theoretical models. In particular, polarimetric imaging of M87 by the Event Horizon Telescope (EHT) has revealed organized magnetic field structures on horizon scales \citep{EHT2021VIII} consistent with these processes.

\subsection{Jet Emission Model}
\label{sec:jetemissionmodel}
High-resolution observations have demonstrated that relativistic jets are not simple flows. They exhibit a stratified and evolving morphology, in which physical quantities such as magnetic field strength, particle density, and bulk velocity vary systematically with distance from the central engine \citep{Lister2009, Lister2013}. These longitudinal and transverse gradients give rise to frequency dependent opacity effects that strongly influence the apparent location of the jet base observed in VLBI images.

In VLBI observations, the compact and bright feature commonly identified as the \emph{core} does not correspond to a fixed physical structure. Instead, the VLBI core is understood as the surface along the jet where the optical depth to synchrotron self-absorption is approximately unity, $\tau_\nu \simeq 1$ \citep{BlandfordKonigl1979, Lobanov1998}. At a given observing frequency, emission from regions closer to the central engine is self-absorbed, and the observed core marks the transition between optically thick and optically thin synchrotron emission. As a consequence, the apparent position of the VLBI core depends on the observing frequency. This frequency-dependent displacement of the core position is known as the \emph{core shift}. The effect was proposed by \citet{BlandfordKonigl1979}, first detected observationally by \citet{Marcaide1984} and later formulated quantitatively within a physical jet framework by \citet{Lobanov1998}. In the standard model of a freely expanding conical jet, the magnetic field strength and the density of relativistic particles decrease as power laws with distance $r$ from the central engine. Under the assumption of equipartition between particle and magnetic energy densities, synchrotron self-absorption leads to a frequency-dependent location of the $\tau_\nu \simeq 1$ surface, such that the distance of the observed core follows a power-law dependence on observing frequency,
\begin{equation}
r_{\mathrm{core}}(\nu) \propto \nu^{-1/k_r},
\label{eq:coreshift}
\end{equation}
where the parameter $k_r$ depends on the gradients of the magnetic field strength and particle density along the jet. For a conical jet in equipartition, $k_r \approx 1$, implying a systematic downstream shift of the core position at lower observing frequencies.

Measurements of the core shift provide a powerful diagnostic of the physical conditions in the innermost regions of relativistic jets. By combining multi-frequency observations, it is possible to infer the magnetic field strength at a given distance from the black hole, estimate the location of the jet relative to the observed core, and constrain the jet opening angle \citep{Lobanov1998, Hada2011, Hirotani2025}. However, observational evidence indicates that the assumptions of a purely conical geometry and equipartition do not always hold simultaneously. Multi-frequency core-shift studies have revealed significant deviations from $k_r = 1$ in several sources, suggesting departures from equipartition or changes in jet geometry \citep{Chamani2021, Chamani2023, Bartolini2025}. In particular, values of $k_r < 1$ have been interpreted as indicating jets where particles dominate, while larger values may reflect magnetically dominated flows or additional opacity contributions. These deviations can also be understood in the context of jets that present a transition from an inner parabolic collimation regime to an outer conical expansion. High-resolution studies of nearby AGN have shown that jets often exhibit a parabolic geometry close to the black hole, gradually transitioning to a conical shape \citep{Kovalev2020, Boccardi2021}. When the core shift is measured across frequency ranges that sample both regimes, the resulting value of $k_r$ naturally deviates from unity \citep{Ricci2022}. In such cases, fitting the core shift separately in the parabolic and conical regions yields different indices, with values closer to $k_r \simeq 1$ recovered when only the conical region is considered.

\begin{figure}[htpb]
  \centering 
  \includegraphics[width=1.0\textwidth]{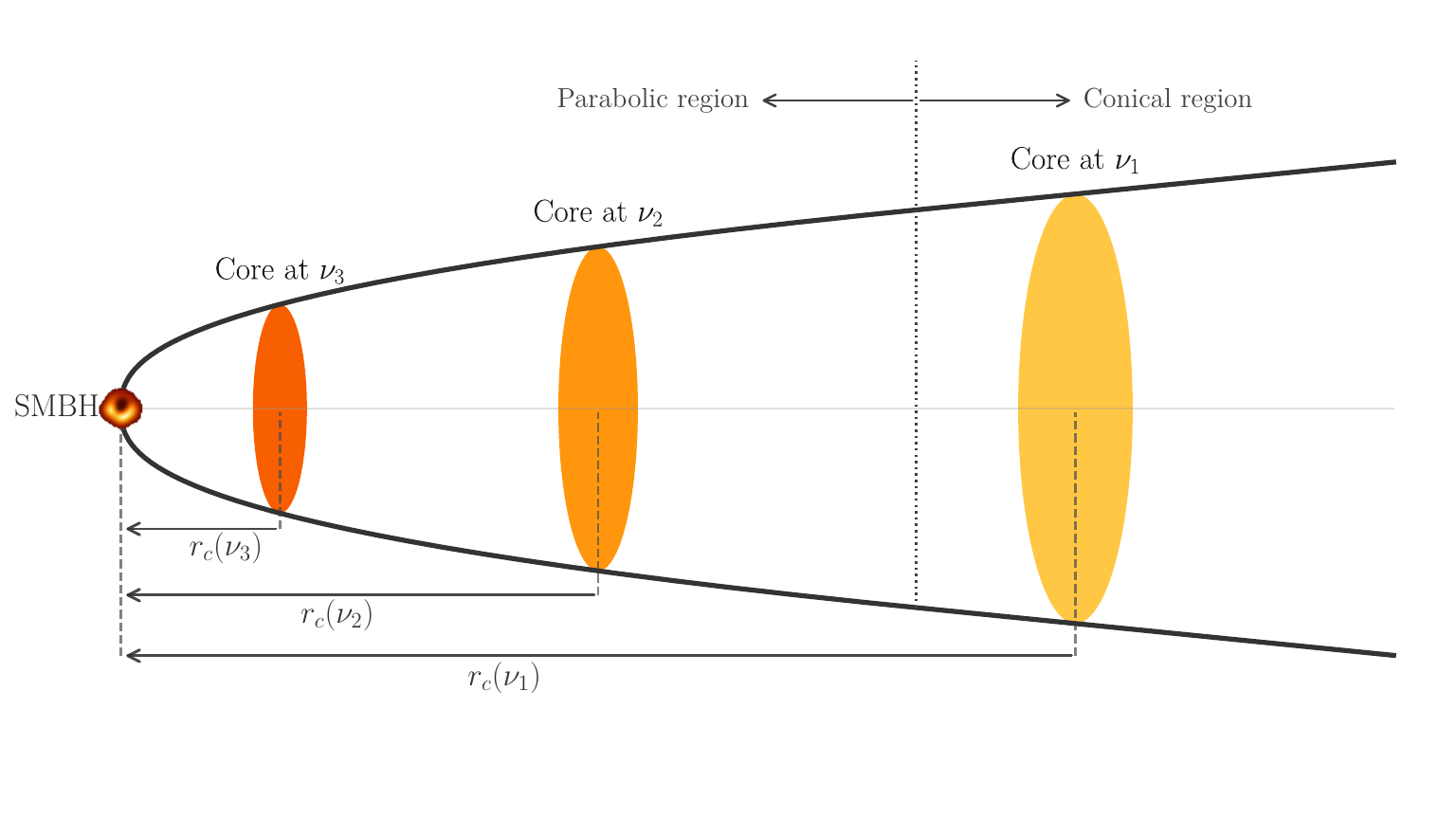}
  \caption[Schematic representation of a relativistic jet and radio core structure in the core-shift model.]{Schematic representation of a relativistic jet and the radio core structure in the framework of the core-shift model. Elliptical regions indicate the observed radio cores at different observing frequencies ($\nu_1 < \nu_2 < \nu_3$).}
  \label{fig:JetModel}
\end{figure}

A schematic representation of the core-shift phenomenon and the underlying jet geometry is shown in Fig.~\ref{fig:JetModel}. The figure illustrates how the observed VLBI core moves downstream at lower frequencies, tracing the $\tau_\nu \simeq 1$ surface along the jet, and highlights the transition between an inner parabolic region and an outer conical jet.

Beyond its astrophysical relevance, the core-shift effect has important implications for high-precision astrometry and geodesy. Since the apparent position of a compact radio source depends on both observing frequency and intrinsic source structure, uncorrected core shifts can introduce systematic errors in the realization of celestial reference frames \citep{Charlot2010, Xu2021}. In particular, when the frequency dependence of the core position deviates from the ideal $k_r = 1$ case, the group delay position measured in wideband VLBI observations no longer corresponds to a unique physical location in the source \citep{Porcas2009}. This effect is especially relevant for modern broadband VLBI systems and geodetic applications requiring sub-milliarcsecond accuracy.

\subsection{Physical Parameters of a Jet}
\label{sec:physicaljet}

Jet models typically assume equipartition between the energy densities of relativistic particles and magnetic fields \citep[e.g.][]{Plavin2019, OSullivan2008, OSullivan2009}. This assumption has been tested in some sources and found to be consistent with observations, as shown in \citet{Hada2011} and \citet{Sokolovsky2011}. The equipartition hypothesis implies a specific frequency dependence for the apparent core position, characterized by a power-law shift with exponent $k_r \approx 1$ (Eq. \ref{eq:coreshift}).

Following the standard conical jet model \citep{BlandfordKonigl1979, Lobanov1998}, the magnetic field strength and the relativistic particle density are assumed to decrease with distance $r$ from the central engine as power laws:
\begin{equation}
B(r) = B_1 \left( \frac{r_1}{r} \right)^m,
\label{eq:magnetic_field_r}
\end{equation}
\begin{equation}
N(r) = N_1 \left( \frac{r_1}{r} \right)^n,
\end{equation}
where $r_1 = 1\,\mathrm{pc}$, and $B_1$ and $N_1$ denote the magnetic field strength and particle density at that reference distance. The indices $m$ and $n$ describe the radial stratification of the jet.

The parameter $k_r$ in Eq.~\ref{eq:coreshift} is related to the magnetic field and particle density gradients, and to the optically thin spectral index $\alpha$, through
\begin{equation}
k_r = \frac{(3 - 2\alpha)m + 2n - 2}{5 - 2\alpha}.
\end{equation}
In the case of equipartition between magnetic and particle energy densities \citep{BlandfordKonigl1979}, the choice $m = 1$ and $n = 2$ leads to $k_r = 1$, independently of $\alpha$. This regime is frequently adopted as a standard description of compact AGN jets.

For two observing frequencies $\nu_1$ and $\nu_2$ ($\nu_1 < \nu_2$), the measured core shift $\Delta r_{\mathrm{mas}}$ can be expressed in terms of the core position offset $\Omega_{r\nu}$, which measures how the core shifts on the sky (in $mas$) as we change the observing frequency (in $GHz$):
\begin{equation}
\Omega_{r\nu} =
4.85 \times 10^{-9}
\frac{\Delta r_{\mathrm{mas}} D_L}{(1+z)^2}
\frac{\nu_1^{1/k_r}\nu_2^{1/k_r}}
{\nu_2^{1/k_r} - \nu_1^{1/k_r}},
\end{equation}
where $D_L$ is the luminosity distance in pc, $z$ is the redshift, and the observing frequencies are given in GHz. For an ideal conical jet with constant physical parameters, $\Omega_{r\nu}$ should be independent of the selected frequency pair.

The absolute distance between the observed core at frequency $\nu$ and the origin of the jet can then be written as
\begin{equation}
r_{\mathrm{core}}(\nu) =
\frac{\Omega_{r\nu}}{\sin\theta}
\nu^{-1/k_r},
\end{equation}
where $\theta$ is the jet viewing angle. This expression depends only on measured quantities ($\Delta r_{\mathrm{mas}}$, $k_r$) and $\theta$, so it provides a direct estimate of the deprojected core location.

This formalism provides a direct link between the observed core shift and the physical properties of the ultracompact jet. Furthermore, when $k_r = 1$, the frequency-dependent shift in the core position does not introduce a group delay. Therefore, group-delay VLBI astrometry is not affected by frequency-dependent core shifts \citep{Porcas2009}. However, when $k_r \neq 1$, the group-delay position becomes frequency-dependent and deviates from the physical jet origin. This can introduce systematic errors in astrometric VLBI solutions. \citet{Kovalev2008A} showed that such core shifts can bias the alignment between radio and optical reference frames, compromising the precision of celestial reference frame realizations such as the International Celestial Reference Frame \citep[ICRF;][]{Ma1998, Charlot2020} and the \textit{Gaia} catalogue \citep{Gaia2016, Lindegren2018, Gaia2023}. Verifying whether AGN cores adhere to the expected $k_r = 1$ regime is not only important for understanding jet physics, but is also crucial for ensuring the accuracy of geodetic measurements \citep[e.g.][]{Porcas2009, Ming2022} and for aligning radio and optical reference frames \citep{Kovalev2008A, Ming2021b}.

Assuming equipartition ($m=1$, $n=2$, $k_r=1$), the magnetic field at 1 pc can be estimated as
\begin{equation}
B_1 \approx
2.92 \times 10^{-9}
\left[
\frac{\Omega_{r\nu}^3 (1+z)^3}
{k_e \delta_j^2 \phi \sin^5\theta}
\right]^{1/4},
\label{eq:magnetic_field}
\end{equation}
where $\delta_j$ is the Doppler factor, $\phi$ is the half intrinsic jet opening angle, and $k_e$ accounts for deviations from strict equipartition. In this expression, $B_1$ is obtained in $G$ when $\Omega_{r\nu}$ is expressed in $pc\,GHz$.

Once $B_1$ is known, the particle density at 1 pc can also be estimated under the same equipartition assumption. Following \citet{Lisakov2017}, the relation between the magnetic field strength and the relativistic particle density is
\begin{equation}
N_1 =
\frac{B_1^2}
{8\pi \ln(\gamma_{\max}/\gamma_{\min})m_ec^2},
\label{eq:N1}
\end{equation}
where $m_e$ is the electron mass, $c$ is the speed of light, and $\gamma_{\min}$ and $\gamma_{\max}$ are the lower and upper limits of the electron Lorentz factor distribution. Here, $N_1$ is expressed in $\mathrm{cm^{-3}}$ when cgs units are used. A value $\gamma_{\max}/\gamma_{\min}=100$ is usually adopted, as in \citet{Lobanov1998} and \citet{Lisakov2017}. With this choice,
\begin{equation}
N_1 \simeq
1.06 \times 10^4
\left( \frac{B_1}{1\,\mathrm{G}} \right)^2
\,\mathrm{cm^{-3}}.
\label{eq:N1_B}
\end{equation}

The magnetic field strength and particle density at the core position observed at a given frequency can then be obtained from the radial scalings

\begin{equation}
B_{\mathrm{core}}(\nu) =
B_1
\left( \frac{1\,\mathrm{pc}}{r_{\mathrm{core}}(\nu)} \right),
\end{equation}
\begin{equation}
N_{\mathrm{core}}(\nu) =
N_1
\left( \frac{1\,\mathrm{pc}}{r_{\mathrm{core}}(\nu)} \right)^2.
\label{eq:N1_r}
\end{equation}

\pagebreak
\section{Very Long Baseline Interferometry (VLBI)}
\label{sec:vlbi}

In modern astrophysics, to better understand the different processes occurring in the Universe, it is necessary to use various techniques. This concept is known as \textit{multi-messenger astronomy}. Each of these techniques provides information about a range of the electromagnetic spectrum or even about other particles that carry information to the Earth.

This thesis focuses on the radio range of the electromagnetic spectrum. More specifically, it uses the VLBI technique, which provides the highest possible resolution of all astronomical methods so far.

\subsection{Historical Background}

Improving the angular resolution has been one of the most pursued goals in astronomy since its early days. For a single dish telescope, the resolution is inversely proportional to the aperture diameter as $\theta \sim \lambda/D$, so that even very large radio dishes are fundamentally limited by their long operating wavelengths \citep{TMS1986}. This limitation motivated the development of interferometric techniques, where the effective aperture is set by the separation between antennas rather than by the size of a single reflector \citep{TMS1986}.

\begin{figure}[h!]
    \centering
    \includegraphics[width=0.9\textwidth]{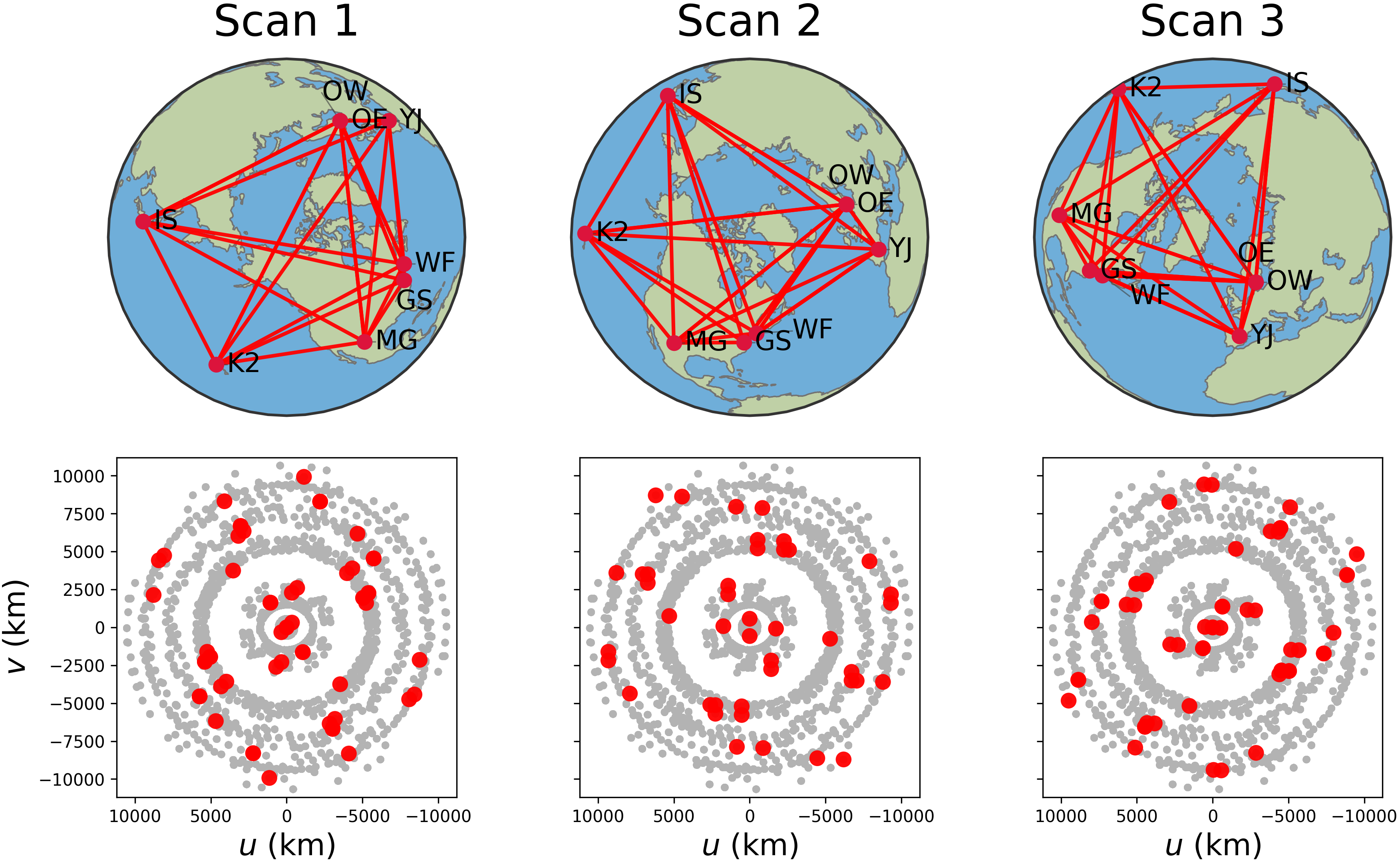}
    \caption[Earth rotation synthesis for 1803+784 in the VGOS experiment VO2187.]{Earth rotation synthesis illustrated with the observations of the AGN 1803+784 in the VGOS experiment VO2187 in 2022. The upper panel shows the observing stations and all the baselines of three different scans. The bottom panel shows the full UV coverage, highlighting in red the points observed in each respective scan.}
    \label{fig:earthrotation}
\end{figure}

A major step forward came with the formulation and implementation of the \emph{aperture synthesis} \citep{Ryle1960}, illustrated in Fig.~\ref{fig:earthrotation}. By exploiting the rotation of the Earth, a pair of antennas can sample many projected baselines over time, enabling image reconstruction. This idea led to the development of the first connected interferometers.

Taking this idea to an even more extreme, the technique known as Very Long Baseline Interferometry (VLBI) emerged. Here, the antennas are not physically connected. In VLBI, the signals are recorded independently at each station and, later, brought to a correlator, where the products are formed. Since observatories may be in quite distant places, with different local times, this approach becomes feasible only thanks to ultra-stable frequency standards (hydrogen masers) at each site, which provide accurate times of the data for successful post-correlation \citep{TMS1986}.

Modern VLBI arrays routinely achieve baselines comparable to the Earth's diameter, delivering angular resolutions that reach milliarcsecond scales \citep{Cooper2007} and, at millimetre wavelengths, even tens of microarcseconds, allowing radioastronomers to obtain images of supermassive black holes \citep{EHT2019IV, EHT2022I}.

\subsection{Principles of VLBI}

The theoretical foundation of radio interferometry is provided by the Van Cittert-Zernike theorem, which establishes a direct relationship between the spatial coherence of an electromagnetic wavefront and the brightness distribution of the source that emitted it \citep{vanCittert1934, Zernike1938}. In practical terms, the theorem states that the mutual coherence function measured by an interferometer is proportional to the Fourier transform of the sky brightness distribution.

\begin{figure}
    \centering
    \includegraphics[width=0.5\textwidth]{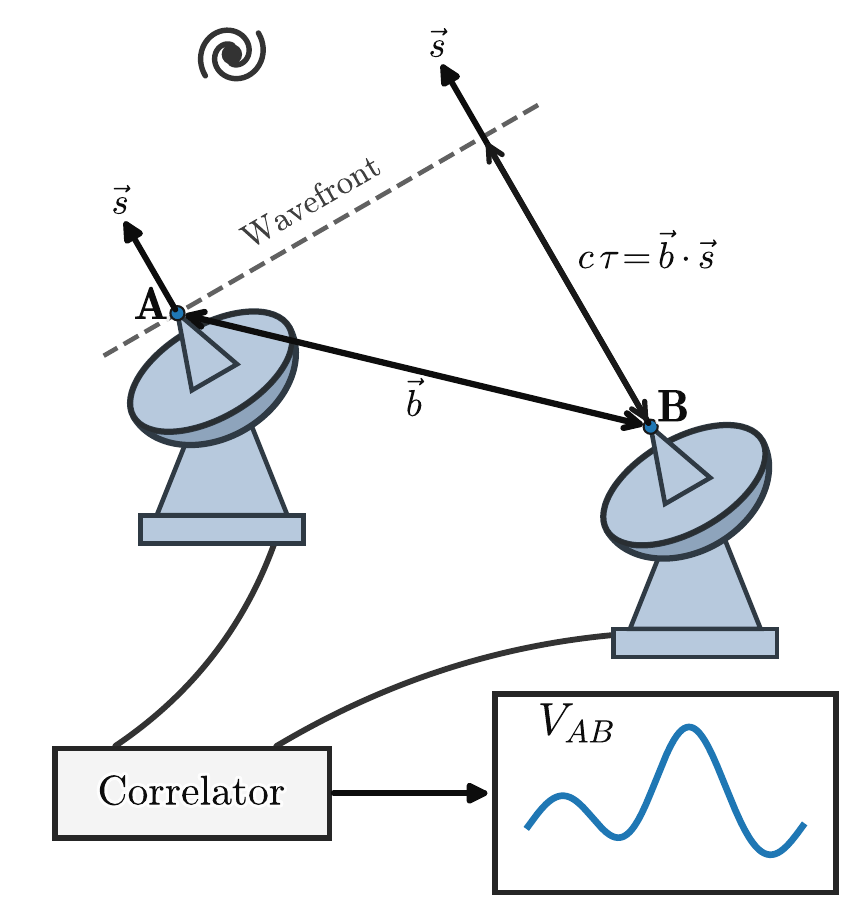}
    \caption[Schematic representation of a two-element radio interferometer.]{Schematic representation of a two-element radio interferometer. A plane wave emitted by a distant source arrives at the two antennas with a geometric delay $\tau_{\mathrm{g}} = (\overrightarrow{b}\cdot\overrightarrow{s})/c$, determined by the baseline vector $\overrightarrow{b}$ and the source direction $\overrightarrow{s}$. The correlated signals measure the spatial coherence of the incoming wavefront, which is related to the Fourier transform of the source brightness distribution.}
    \label{fig:2elementsinterferometer}
\end{figure}

The basic concept of any interferometric array can be explained by the two-element interferometer, shown in Figure \ref{fig:2elementsinterferometer}. Consider two antennas separated by a vector baseline $\overrightarrow{b}$ observing a distant radio source in the direction of the unit vector $\overrightarrow{s}$ (Fig.~\ref{fig:2elementsinterferometer}).

Because of the finite separation between the antennas, the wavefront arrives at each telescope at slightly different times. The corresponding geometric delay is given by
\begin{equation}
\tau_{\mathrm{g}} = \frac{\overrightarrow{b}\cdot\overrightarrow{s}}{c},
\label{eq:geometricdelay}
\end{equation}
where $c$ is the speed of light. Each antenna converts the incoming electric field into a voltage signal. For a monochromatic wave of frequency $\nu$ and amplitude $A$, the voltages measured at the two antennas can be written as
\begin{align}
V_1(t) &= A\, e^{2\pi j\nu t}, \\
V_2(t) &= A\, e^{2\pi j\nu (t-\tau_g)}.
\end{align}

The fundamental observable of an interferometer is obtained by multiplying these voltages and averaging the product over time, a process known as correlation, that gives as a result the interferometric fringe:
\begin{equation}
R = \langle V_1(t)V_2^{*}(t)\rangle
   = \frac{A^2}{2}\, e^{2\pi j \nu \tau_g}.
\end{equation}

This expression shows that the interferometer response depends on the relative phase delay between the two signals, that is set by the source direction and the baseline length. Introducing the observing wavelength $\lambda = c/\nu$, the baseline length in wavelength units $u = b/\lambda$, and the direction cosine $x = \sin\alpha$, the response, known as \textit{visibility}, can be written as
\begin{equation}
V(u) = I e^{2\pi jux}.
\end{equation}
where $I = A^2/2$ represents the received power from the source.

For an extended astrophysical source, the emission can be regarded as spatially incoherent, and the total signal measured by each antenna is the superposition of contributions arriving from all directions within the source solid angle. In this case, the visibility becomes an integral over the sky brightness distribution $I(\overrightarrow{s})$,
\begin{equation}
V(\overrightarrow{b}) =
\int I(\overrightarrow{s})\,\exp\!\left(-\frac{2\pi i}{\lambda}\overrightarrow{b}\cdot\overrightarrow{s}\right)\,d\Omega.
\label{eq:visibility}
\end{equation}

The visibility is a complex quantity, characterized by an amplitude and a phase, which encode the contrast and relative position of structures within the source. This result is a direct statement of the Van Cittert-Zernike theorem \citep{vanCittert1934, Zernike1938}, which establishes that the spatial coherence measured by an interferometer is the Fourier transform of the sky brightness
distribution.

By introducing angular coordinates $(x,y)$ on the sky and the corresponding baseline coordinates $(u,v)$ orthogonal to the line of sight, the relationship between visibilities and brightness can be written in its standard two dimensional form:
\begin{align}
V(u,v) &= \int\!\!\int I(x,y)\,e^{-2\pi i(ux+vy)}\,dx\,dy, \\
I(x,y) &= \int\!\!\int V(u,v)\,e^{+2\pi i(ux+vy)}\,du\,dv,
\end{align}
which forms the mathematical basis of aperture synthesis imaging in radio interferometry.

\subsubsection{Fundamental VLBI Observables}
\label{sec:VLBIobservables}
The complex visibility measured by a VLBI interferometer is usually described with a small number of fundamental observables.

The interferometric phase (shown in Eq. \ref{eq:visibility}) is defined for each baseline and observing time as the relative phase between the signals received at the two antennas. It contains
contributions from the geometric delay, atmospheric propagation effects, instrumental terms, and the intrinsic source structure:

\begin{equation}
\phi = \phi_{geo} + \phi_{tro} + \phi_{ion} +\phi_{str} + \phi_{ins} + 2\pi n
\label{eq:phase}
\end{equation}

One estimator of the relative delay between two antennas is the phase delay, defined as:
\begin{equation}
\tau_{\phi} = \frac{\phi}{2\pi\nu_0}
\end{equation}
where $\nu_0$ is the reference observing frequency. This quantity is affected by an inherent $2\pi$ ambiguity and by all non-geometric phase contributions.

Another robust observable is the group delay, defined as the slope of the phase as a function of frequency:
\begin{equation}
\tau_{\mathrm{g}} = \frac{1}{2\pi}\frac{\partial\phi}{\partial\nu}
\label{eq:groupdelay}
\end{equation}
The group delay is free from phase ambiguities and constitutes the primary observable in geodetic VLBI, where it enables extremely precise measurements of station positions and Earth orientation parameters.

The phase of the visibility also varies with time. This temporal variation, commonly referred to as the fringe rate, is defined as:
\begin{equation}
\dot{\phi} = \frac{\partial\phi}{\partial t}.
\end{equation}
Residual fringe rates arise from unmodelled Doppler effects, clock drifts, and source structure, and they limit the coherent integration time of the data. The associated delay rate describes the temporal variation of the delay and is directly related to the residual relative motion of the antennas with respect to the correlator model.

In addition to baseline-based observables, VLBI makes extensive use of closure quantities, which are immune to antenna calibration errors. The most widely used of these is the closure phase, defined for a closed triangle of antennas (pqr) as the sum of the interferometric phases around the closed loop:
\begin{equation}
\phi_{pqr} = \phi_{pq} + \phi_{qr} + \phi_{rp}.
\label{eq:closurephases}
\end{equation}
Antenna-dependent phase errors cancel out in this combination, leaving a quantity that depends only on the intrinsic source structure.

Similarly, closure amplitudes can be constructed from the visibilities measured on four antennas (pqrs) in such a way that antenna-based gain errors are eliminated:
\begin{equation}
C_{pqrs}
=
\frac{|V_{pq}|\,|V_{rs}|}{|V_{pr}|\,|V_{qs}|}.
\label{eq:closureamplitudes}
\end{equation}

An equivalent way of expressing this observable is through the logarithmic closure amplitude (log-closure amplitude):
\begin{equation}
c_{pqrs}
=
\ln C_{pqrs}
=
\ln\left(
\frac{|V_{pq}|\,|V_{rs}|}{|V_{pr}|\,|V_{qs}|}
\right)
=
\ln |V_{pq}|+\ln |V_{rs}|-\ln |V_{pr}|-\ln |V_{qs}|.
\label{eq:logclosureamps}
\end{equation}

The logarithmic closure amplitude therefore contains the same gain-independent information as the closure amplitude itself, but in an additive form. This is useful because multiplicative amplitude gain errors
become additive terms in logarithmic space and cancel in the same way as in the standard closure amplitude. In addition, the logarithmic form treats the numerator and denominator symmetrically, avoiding the strongly asymmetric behaviour that can arise from ratios involving low signal-to-noise visibility amplitudes. For this reason, log-closure amplitudes usually have better behaved statistical properties and are commonly used in closure imaging \citep[e.g.][]{Chael2018,Blackburn2020}.

Closure quantities therefore provide reliable constraints on the brightness distribution of the source, even in the presence of large and rapidly varying instrumental or atmospheric effects.

A fundamental limitation of closure observables is that they do not preserve absolute astrometric information. While they encode the relative positions and structure of emission components within a source, the absolute position on the sky is lost. Despite this limitation, closure quantities form the backbone of VLBI self-calibration and hybrid imaging techniques, enabling reliable image
reconstruction under challenging observational conditions (see Sect. \ref{sec:imaging}).

More recently, a more general class of closure observables, known as closure traces, has been introduced by \citet{Broderick2020} as an extension of classical closure quantities. Closure traces are constructed from full visibility matrices and remain invariant under arbitrary antenna-based Jones matrices. As a result, they provide calibration-independent constraints on the polarized source structure, even in the presence of complex and poorly known instrumental effects, including the D-terms (see Sect. \ref{sec:rime}). Their properties and potential applications to VLBI polarimetry have been further explored and developed by \citet{Albentosa2023} and \citet{Muller2024}, demonstrating their usefulness for robust polarimetric analysis and imaging.

\subsubsection{Radio Interferometer Measurement Equation (RIME)}
\label{sec:rime}

The Van Cittert-Zernike theorem, introduced in the previous section, provides a scalar relationship between the spatial coherence of the electromagnetic field and the brightness distribution of the source. In this formulation, only total intensity (Stokes I) is considered. However, real astrophysical sources are generally polarized, and interferometric receivers are sensitive to the full vector nature of the electric field. A complete description of radio interferometric measurements requires a formalism that incorporates polarization (see Section \ref{sec:Polarization}) and instrumental effects.

This generalization is provided by the Radio Interferometer Measurement Equation (RIME), originally formulated by \citet{Hamaker1996} and later developed by \citet{Smirnov2011}. The RIME expresses the measured visibilities in terms of matrices related to the source brightness, allowing a unified description of calibration effects.

In the RIME formalism, the electric field is represented by a two component complex vector, and the brightness distribution of the source is encoded in the \emph{brightness matrix}, which, in the base of linear polarization receivers, is given by:
\begin{equation}
\mathbf{B} =
\frac{1}{2}
\begin{pmatrix}
I + Q & U + iV \\
U - iV & I - Q
\end{pmatrix},
\label{eq:brightness_matrix}
\end{equation}
where $I$, $Q$, $U$, and $V$ are the Stokes parameters. This matrix representation contains the complete second order statistical description of the polarized emission.

For a baseline formed by antennas $p$ and $q$, the fundamental RIME for a point source can be written as \citep{Smirnov2011}
\begin{equation}
\mathbf{V}_{pq} = \mathbf{J}_p \, \mathbf{B} \, \mathbf{J}_q^{\dagger},
\label{eq:RIME_point}
\end{equation}
where $\mathbf{V}_{pq}$ is the $2\times2$ visibility (or coherency) matrix measured on the baseline, $\mathbf{J}_p$ and $\mathbf{J}_q$ are the Jones matrices describing the signal propagation and instrumental response of antennas $p$ and $q$, respectively, and the superscript $\dagger$ denotes the Hermitian transpose.

For an extended source, the observed visibility is obtained by summing the contribution from all sky directions weighted by the beam response of the antennas. This leads to a Fourier relationship between the brightness distribution and the measured visibilities:
\begin{equation}
V_{pq}
=
G_p
\left(
\iint B_{pq}(l,m)\,
e^{-2\pi i \left(u_{pq} l + v_{pq} m\right)}
\, dl \, dm
\right)
G_q^{H},
\label{eq:Vpq}
\end{equation}
\begin{equation}
B_{pq}(l,m) = E_p(l,m)\, B(l,m)\, E_q^{H}(l,m).
\label{eq:Bpq}
\end{equation}
where $l, m$ are the directions, the $G$ Jones matrices represent the antenna direction-independent effects and the $E$ Jones matrices represent the direction-dependent effects.

A key feature of the RIME is that the Jones matrix of each antenna can be factorized into a product of matrices, each representing a specific physical effect along the signal path,
\begin{equation}
\mathbf{J}_p = \mathbf{J}_{p,n}\, \mathbf{J}_{p,n-1}\, \cdots \mathbf{J}_{p,1}.
\label{eq:jones_chain}
\end{equation}
This factorization provides a natural framework for calibration, as each effect can be modeled and corrected independently.

In VLBI observations, signals are commonly recorded in a circular polarization basis. One practical advantage of this choice is that several important Jones matrices are diagonal (and hence commutable), which simplifies calibration. These two features lead to an easier independent calibration of each effect, as in this model:
\begin{equation}
\mathbf{J}_p =
\mathbf{P}_{p}\,
\mathbf{B}_{p}\,
\mathbf{G}_{p}\,
\mathbf{D}_{p}\,
\mathbf{E}_{p}\,
\mathbf{T}_{p}\,
\mathbf{Z}_{p}\,
\mathbf{F}_{p}.\label{eq:jones_chain_model}
\end{equation}

$\mathbf{P}_{p}$ models the parallactic angle, the orientation of the sky coordinate system with respect to the antenna frame. In a linear polarization basis, it has the form \citep{Hales2017}:
\begin{equation}
\mathbf{P}^{\rm LF}(\psi)=
\begin{pmatrix}
\cos\psi & \sin\psi\\
-\sin\psi & \cos\psi
\end{pmatrix},
\end{equation}
while in a circular polarization basis:
\begin{equation}
\mathbf{P}_p =
\begin{pmatrix}
e^{-i\psi_p} & 0 \\
0 & e^{+i\psi_p}
\end{pmatrix},
\end{equation}
where $\psi_p$ is the parallactic angle at antenna $p$. Because this matrix is diagonal (only in a circular polarization basis), it commutes with other diagonal Jones terms, allowing the parallactic angle correction to be applied at any stage of the calibration process.

The bandpass matrix $\mathbf{B}_p$ models frequency-dependent amplitude and phase variations introduced by the receiver chain. In the absence of significant instrumental polarization the bandpass response is assumed to be diagonal \citep{Smirnov2011b},
\begin{equation}
\mathbf{B}_p(\nu) =
\begin{pmatrix}
B_{p,R}(\nu) & 0 \\
0 & B_{p,L}(\nu)
\end{pmatrix}.
\end{equation}

The matrix $\mathbf{G}_p$ represents time-dependent and frequency-independent complex gains, accounting for variations in receiver amplification and electronic phase. As for the bandpass, $\mathbf{G}_p$ is typically assumed to be diagonal in circular polarization \citep{Hamaker1996}:
\begin{equation}
\mathbf{G}_p(t) =
\begin{pmatrix}
g_{p,R}(t) & 0 \\
0 & g_{p,L}(t)
\end{pmatrix}.
\label{eq:gainjones}
\end{equation}

The matrix $\mathbf{D}_p$ describes polarization leakage, modelled with the so-called D-terms, between the nominally orthogonal polarization channels. Unlike the previous terms, $\mathbf{D}_p$ is generally non-diagonal and is responsible for coupling between right- and left-circular polarization (or $x$ and $y$ in the case of receivers in a linear basis) \citep{Hales2017},
\begin{equation}
\label{eq:dterms}
\mathbf{D}_p =
\begin{pmatrix}
1 & D_{p,RL} \\
D_{p,LR} & 1
\end{pmatrix}.
\end{equation}

Accurate estimation of $\mathbf{D}_p$ is essential for reliable polarimetric calibration.

The tropospheric term $\mathbf{T}_p$ accounts for non-dispersive atmospheric
propagation effects. Since these effects are non-polarizing, the corresponding Jones term is scalar \citep{Smirnov2011} and may be written as:
\begin{equation}
\mathbf{T}_p(\nu,t) = T_p(\nu,t)\,\mathbf{I},
\end{equation}
where $T_p(\nu,t)=a_p(\nu,t)e^{i\phi^{\mathrm{trop}}_p(\nu,t)}$. Because $\mathbf{T}_p$ is scalar, it commutes with all other Jones terms.

Finally, ionospheric propagation can be represented by Jones terms as well. The dispersive ionospheric phase delay is commonly written as a scalar Jones term,
\begin{equation}
\mathbf{Z}_p(\nu,t) = e^{i\zeta_p(\nu,t)}\,\mathbf{I},
\end{equation}
where $\zeta_p$ depends on the total electron content along the line of sight and scales approximately as $\nu^{-1}$ in phase \citep{Smirnov2011b}. Since this term is scalar, it commutes with the other Jones matrices and can often be absorbed into the antenna-based complex gains.

In polarimetric observations, the ionosphere can also introduce Faraday rotation. Nevertheless, this extra effect is small for the Earth's magnetic field. In a circular polarization basis, the effect becomes a differential phase between the two circular hands,
\begin{equation}
\mathbf{F}^{\odot}_p(\beta_p)=
\begin{pmatrix}
e^{+i\beta_p} & 0\\
0 & e^{-i\beta_p}
\end{pmatrix}.
\end{equation}
The ionospheric Faraday contribution is expected to be small, although it remains conceptually distinct from the scalar dispersive ionospheric phase term.

In summary, having a diagonal Jones matrix for the parallactic angle, implies that most instrumental and propagation effects in VLBI are diagonal when expressed in a circular polarization basis. This property allows the majority of calibration terms to be solved independently for each polarization channel, while non-diagonal effects, such as polarization leakage, can be isolated and treated separately within the RIME framework.

\subsection{Data Calibration}
\label{sec:data_calibration}

The interferometric observables delivered by a VLBI correlator do not represent the intrinsic source coherence directly, but are instead affected by a variety of instrumental, atmospheric, and geometric effects that vary between antennas, with time, and with observing frequency, as described in the previous subsection. Data calibration consists of estimating and correcting these terms in order to recover visibilities only related to the Fourier transform of the source brightness distribution.

In VLBI, calibration plays a particularly critical role. The absence of a common reference clock among stations, together with rapidly varying propagation delays in the troposphere and ionosphere, introduces large variations of the visibilities that we should correct. Thanks to the diagonal Jones matrices, these calibrations can be estimated and applied independently.

\subsubsection{PolConvert}
\label{sec:polconvert}

Modern wideband VLBI systems increasingly rely on receivers that record the incoming radiation in a linear polarization basis. This choice is motivated by the superior polarization purity and bandwidth performance of linear feeds. Quarter-wave plates, used to generate circular polarization, may introduce some leakage and frequency-dependent distortions. Sometimes, mixed-polarization VLBI observations arise when stations with linear receivers are correlated together with antennas recording circular polarization. In other VLBI networks, all stations have linear receivers. In such cases, the VLBI visibilities are obtained on a mixed (linear–circular) or full linear polarization basis. This is not convenient for standard VLBI calibration and imaging (see \ref{sec:rime}).

The algorithm PolConvert \citep{PolConvert} is designed to post-process VLBI data obtained with linear polarization receivers (or with a mixture of linear and circular) and to produce calibrated visibilities as if circular polarization feeds had been used at all stations. This conversion is performed entirely after correlation. It ensures a consistent treatment of the instrumental response of each antenna \citep{IMV2016}.

The principal advantage of this approach is that it enables a robust recovery of all four Stokes parameters ($I$, $Q$, $U$, and $V$) while keeping instrumental polarization at a low level. In PolConvert, the full cross-polarization bandpasses of the linear feed stations are accounted for in the conversion from linear to circular. As a result, polarization leakage introduced by uncalibrated X/Y gain differences is largely avoided, which is essential for high-fidelity polarimetric VLBI.

Since its original development, PolConvert has been successfully applied to a wide range of VLBI experiments in different frequencies and arrays. These include millimeter and submillimeter VLBI observations with the Event Horizon Telescope (EHT) and the Global Millimeter VLBI Array (GMVA), as well as centimeter-VLBI observations with arrays such as the Long Baseline Array (LBA) and the European VLBI Network (EVN) (e.g. \citet{EHT2019III, EHT2019IV, Chanapote2019, Lu2023, Ross2023, Bhandari2023}).

PolConvert was originally developed in the context of ALMA–VLBI observations, where the use of linear feeds and their varying orientations with respect to the antenna mount introduced additional complexity in the polarization response. Depending on observing frequency, ALMA receivers exhibit different feed orientations, which motivated the adoption of a generic polarization channel nomenclature. For this reason, PolConvert consistently uses the labels $X$ and $Y$ for the orthogonal linear receptors, rather than $H$ and $V$, which is strictly valid only in the special case where one receptor is aligned with the antenna axis. In the following, and for consistency with the PolConvert code and its associated scripts, the $X$ and $Y$ labels are therefore used to denote the two linear polarization channels.

The conversion from linear to circular polarization is primarily controlled by the relative complex bandpass between the $X$ and $Y$ channels at each antenna. Both the phase difference and the amplitude ratio between these channels are generally frequency-dependent and must be characterized across the full observing band. PolConvert estimates these quantities using a Global Cross-Polarization Fringe-Fitting algorithm (GCPFF), which exploits the redundancy of mixed-polarization visibilities to solve coherently for the cross-polarization gains \citep{IMV2016}.

In the estimation of the $X/Y$ bandpasses, PolConvert incorporates a priori phase information derived from the phase-calibration (phase-cal) tones. Specifically, the phase difference between the extracted $X$ and $Y$ phase-cal tones is computed for each spectral window and integration time, and is fitted as a cross-polarization tone delay. This delay term is then added to the $X/Y$ bandpass model to be solved. Once determined, these cross-polarization bandpasses are applied to the remaining scans of the experiment. Any temporal drift in the relative $X/Y$ phase is tracked by the phase-cal tones, which are used by PolConvert throughout the entire dataset.

From a physical perspective, the quantities estimated by the GCPFF can be interpreted as differences in the optical path length of the $X$ and $Y$ polarization channels between the receiver frontend and the point at which the phase-cal tones are injected.

After the application of PolConvert, the visibilities are expressed in a pure circular polarization basis. Residual antenna-based phase, delay, and rate offsets are subsequently solved by using standard VLBI fringe-fitting techniques, which are described in the following sections.

\subsubsection{Phase Calibration}
\label{sec:phasecal}

Unlike connected-element interferometers, VLBI phases can vary significantly within short timescales due to residual clock offsets and atmospheric fluctuations. In addition, imperfect a priori delay models produce residual phase slopes with frequency. Both effects reduce coherence: averaging data over time or frequency without correcting these slopes leads to decorrelation and a
loss of signal-to-noise ratio (S/N).

After the correlation (and pol-conversion, if needed), phases can still have some residual rates and delays. In order to permit data averaging, these residuals must be removed. This is done by \emph{Fringe Fitting}, the method that estimates residual phase, delay, and rate parameters that maximize coherence. It searches for the peak of the fringe amplitude in the delay-rate space. A baseline-based solution can be converted into a consistent antenna-based calibration by solving for a set of station parameters that best fit all baselines simultaneously, an approach commonly referred to as \emph{Global Fringe Fitting}, introduced by \citet{Schwab1983}. Once applied, these solutions flatten the phase as a function of frequency and time, allowing longer coherent integrations and improving the robustness of subsequent self-calibration and imaging (Sect.~\ref{sec:imaging}).

In VLBI, the interferometric phase is the most rapidly varying observable. The measured baseline phase contains contributions from geometry, propagation through the troposphere and ionosphere, station clock behaviour, receiver electronics, and intrinsic source structure (see Eq. \ref{eq:phase}). Even after applying a fringe fitting with an a priori correlator model, residual phase errors are usually large enough to prevent long coherent integrations unless they are corrected.

For a given baseline $pq$, and restricting for clarity to a single polarization hand, the observed complex visibility can be written schematically as \citep{Schwab1983}:
\begin{equation}
V^{\mathrm{obs}}_{pq}(\nu,t) \;=\; G_p(\nu,t)\,G_q^{*}(\nu,t)\,V^{\mathrm{true}}_{pq}(\nu,t),
\end{equation}
where $G_p$ (see Eq. \ref{eq:gainjones}) is a complex, antenna-based gain that includes residual phase errors. Writing $G_p = |G_p|e^{i\varphi_p}$, the observed baseline phase becomes:
\begin{equation}
\phi^{\mathrm{obs}}_{pq}(\nu,t) \;=\; \phi^{\mathrm{true}}_{pq}(\nu,t) + \varphi_p(\nu,t) - \varphi_q(\nu,t).
\label{eq:baseline_phase_antenna}
\end{equation}

This antenna-based structure implies that certain combinations of phases, the closure phases (Eq. \ref{eq:closurephases}), are immune to station-dependent errors. They cancel the antenna-based terms, leaving a quantity that depends only on the intrinsic source structure (Sect.~\ref{sec:phasecal}). This invariance is the fundamental reason why self-calibration and hybrid mapping can recover reliable source images even when the station phases are strongly corrupted \citep{Cornwell1981,Rogers1974}.

Self-calibration solves for antenna-based complex gains by minimizing the difference between observed visibilities and a model visibility set derived from an assumed brightness distribution. In its simplest form, the starting model is a point source at the phase center, whose ideal visibility has zero phase on all baselines. The solved antenna phases are then applied to the data, improving the coherence and enabling imaging. The resulting image can be used to update the source model, iterating between calibration and imaging in the \emph{hybrid mapping} procedure, initiated by \citet{Rogers1974} and \citet{Readhead1978}. In practice, this iterative approach is essential in VLBI because the true source structure often introduces non-zero intrinsic phases $\phi^{\mathrm{true}}_{pq}$, and a simple model would bias the gain solutions (see Sect.~\ref{sec:imaging}).

\subsubsection{Amplitude Calibration}
\label{sec:ampcal}

Not only the visibility phases, but also the amplitudes, must be calibrated. Amplitude calibration converts correlator units into an absolute flux density scale and removes antenna dependent sensitivity variations with time and elevation. In VLBI, this step is typically performed in two stages: an \emph{a priori} calibration based on measured system temperatures and antenna gain curves, followed by an \emph{a posteriori} refinement through amplitude self-calibration. The overall goal is to recover visibilities whose amplitudes reflect the intrinsic source coherence, while minimizing systematic errors due to receiver noise, atmospheric attenuation, and instrumental gain changes.

The instantaneous sensitivity of an antenna is commonly characterized by its system equivalent flux density (SEFD), which depends on the system temperature $T_\mathrm{sys}$ and on the elevation-dependent antenna gain \citep{TMS1986}. In practice, variations in $T_\mathrm{sys}$ arise from changes in receiver noise, spillover, and atmospheric opacity, and can be monitored continuously, while the antenna gain variation with elevation is corrected using measured gain curves.

Even after an a priori calibration, residual gain errors may remain due to imperfect $T_\mathrm{sys}$ measurements, unmodelled opacity variations, pointing offsets, or time-variable instrumental gain. These residuals can be reduced by amplitude self-calibration, which solves for antenna-based amplitude gains by minimizing the difference between observed and model visibilities, analogously to the phase case described in the previous section.

Antenna-based gain errors cancel in appropriate closure combinations, as the closure amplitudes (see Eq.~\ref{eq:closureamplitudes}). These data retain calibration-independent information that constrains the source structure and prevents arbitrary distortions of the brightness distribution during self-calibration \citep{TMS1986}.

In VLBI imaging, amplitude self-calibration is therefore applied iteratively. It requires adequate SNR per solution interval, and a sufficiently accurate source model to avoid biasing the gains. In typical hybrid imaging pipelines, phase solutions are stabilized first. Then, amplitude
self-calibration is introduced on progressively shorter timescales as the source model improves.

\subsubsection{Polarization Calibration}
\label{sec:polcal}

Polarization calibration aims to recover the intrinsic polarized emission of the source by correcting instrumental effects that couple the nominally orthogonal polarization channels. In the framework of the RIME, these effects are primarily described by the polarization leakage (D-term) Jones matrices (see Sec~\ref{sec:rime}), together with residual cross-hand phase and delay offsets. Accurate polarimetric calibration is essential for reliable measurements of Stokes $Q$, $U$, and $V$, whose amplitudes are typically much smaller than total intensity.

After the application of PolConvert (Sect.~\ref{sec:polconvert}), the VLBI visibilities are expressed in a circular polarization basis. Here, most instrumental effects are diagonal, while polarization leakage appears as a non-diagonal Jones term (Eq. \ref{eq:dterms}). It describes the coupling between right and left circular polarization at antenna $p$. Typical leakage amplitudes are of order a few percent or less, but their impact on polarimetric observables can be significant if not properly corrected \citep{Roberts1994}.

In VLBI, polarization calibration is particularly challenging for several reasons. First, the parallactic angle coverage of individual antennas may be limited, reducing the ability to disentangle source polarization from instrumental leakage. Second, the absence of a connected array prevents the use of standard absolute polarization references. Finally, source structure and
opacity effects can introduce baseline-dependent polarization signatures that bias simple calibration approaches. As a result, polarimetric calibration in VLBI relies heavily on antenna-based modeling within the RIME framework and on the use of calibration-independent observables, such as the closure traces (explained in Sect. \ref{sec:VLBIobservables}).

A commonly adopted strategy is to estimate the D-terms by fitting the observed cross-hand visibilities to a source polarization model, assuming that the instrumental leakage is stable over the duration of the experiment, as in \citet{Leppanen1995,Hales2017}. This approach requires an initial estimate of the source polarization structure and needs an iterative calibration and imaging scheme, analogous to hybrid mapping for total intensity.

More recently, advanced calibration algorithms have been developed to exploit the full information content of wideband polarimetric VLBI data. In particular, the PolSolve algorithm \citep{PolSolve} implements a global solution for instrumental polarization within the RIME formalism, fitting simultaneously for D-terms and source polarization parameters. It corrects for residual leakage effects that remain after the initial polarization conversion. PolSolve has been successfully applied to both synthetic and real VLBI datasets \citep{EHT2024, Paraschos2024, Kim2025}. Using PolSolve ensures a reliable full-Stokes calibration across the entire frequency bandwidth.

\subsection{Imaging}
\label{sec:imaging}

The reconstruction of the sky brightness distribution from interferometric data is an inverse problem fundamentally limited by the incomplete sampling of the Fourier plane (the uv-coverage). As discussed in Sect.~\ref{sec:vlbi}, a VLBI array samples only a discrete set of spatial frequencies $(u,v)$, so the measured visibilities do not provide a complete representation of the source brightness distribution. As a consequence, direct Fourier inversion leads to an
image that is distorted by the sampling function of the array.

The inverse Fourier transform of the measured visibilities produces the so-called \emph{dirty image}:
\begin{equation}
I_D(l,m) = I(l,m) * B_D(l,m),
\end{equation}
where $I(l,m)$ is the true sky brightness distribution, $*$ denotes convolution, and $B_D(l,m)$ is the \emph{dirty beam}, i.e., the point spread function (PSF) of the interferometer. The dirty beam is given by the Fourier transform of the sampling function in the $(u,v)$ plane and typically exhibits strong sidelobes due to the sparse and anisotropic baseline coverage of VLBI arrays \citep{TMS1986}. The presence of these sidelobes implies that bright features in the source generate spurious responses throughout the image, making the dirty image difficult to interpret directly. Imaging therefore requires a \emph{deconvolution} process to separate the intrinsic source structure from the instrumental response of the array.

The most widely used deconvolution algorithm in radio interferometry is CLEAN, originally introduced by \citet{Hogbom1974}. CLEAN models the sky brightness distribution as a collection of point sources and operates iteratively. At each iteration, the algorithm identifies the peak of the dirty image that corresponds to a point-like component, and subtracts a scaled version of the dirty beam centered at that location.  The process is repeated until the residual image approaches the noise level.

\begin{figure}[h!]
    \centering
    \includegraphics[width=0.93\textwidth, trim={70pt 6pt 60pt 11pt}, clip]{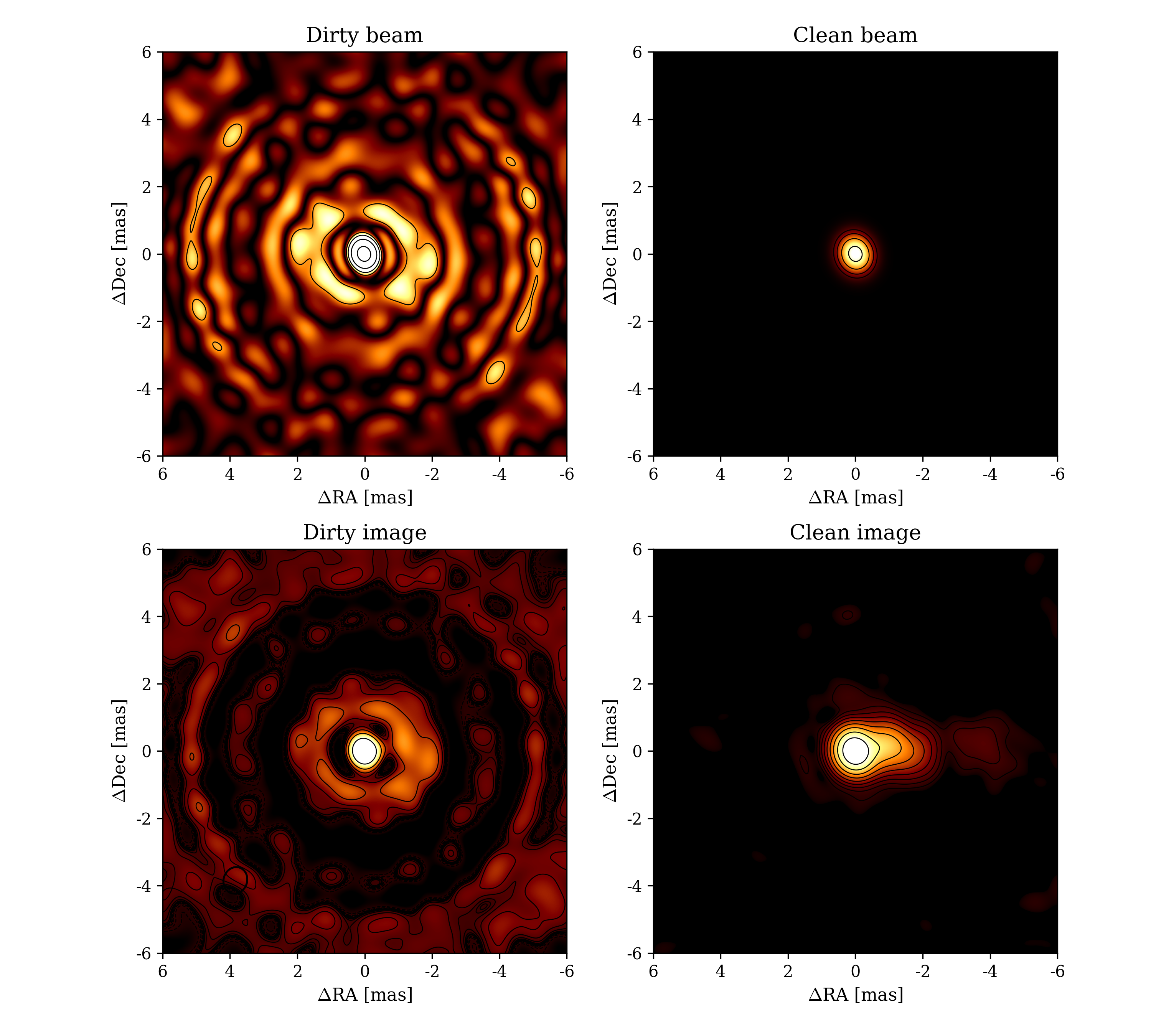}
    \caption[Main terms of CLEAN imaging for 1803+784 in the VGOS experiment VO2187.]{Main terms of CLEAN imaging. Imaging of the source 1803+784, observed during the VGOS experiment VO2187 in C band ($\simeq 6.75$~GHz). \textit{Top left:} dirty beam (PSF), the Fourier transform of the $(u,v)$ sampling of the array. \textit{Top right:} CLEAN beam, an elliptical Gaussian fitted to the central lobe of the dirty beam. \textit{Bottom left:} dirty image, the Fourier inversion of the calibrated visibilities. \textit{Bottom right:} restored CLEAN image.}
    \label{fig:clean}
\end{figure}

The collection of point components identified during the CLEAN iterations constitutes the \emph{CLEAN model}. To obtain a physically meaningful image, these components are convolved with an idealized restoring beam, usually a Gaussian fit to the central lobe of the dirty beam, and added to the final residual image. The result is the \emph{CLEAN image}, which represents an estimate of the diffraction-limited true sky brightness distribution with the instrumental response largely removed. All these terms are illustrated in Fig. \ref{fig:clean}.

In VLBI, imaging with CLEAN is tightly coupled to calibration through the process of self-calibration (Sect.~\ref{sec:phasecal}). Within the RIME framework (Sect.~\ref{sec:rime}), CLEAN provides a model of the source brightness matrix $\mathbf{B}$, which is used to solve for
antenna-based Jones matrices that describe residual phase and amplitude errors. These calibration solutions are applied to the data, improving coherence and enabling a correct CLEAN deconvolution. Iterating between imaging and calibration constitutes the \emph{hybrid imaging} procedure \citep{Rogers1974,Readhead1978}. It is an essential procedure for VLBI observations, where initial calibration is often insufficient.

While CLEAN imaging has proven successful, it relies on direct visibility data and is therefore sensitive to residual station-based calibration errors. Moreover, CLEAN implicitly assumes a sparse representation of the source brightness distribution in terms of point components, which may not be optimal for extended or smoothly varying emission.

An alternative class of imaging techniques is provided by the \emph{regularized maximum likelihood} (RML) methods \citep{Narayan1986}. In RML imaging, the image is obtained by minimizing an objective function composed of a data fidelity term and one or more regularization terms. The data term quantifies the agreement between the model image and the observed interferometric observables, while the regularizers encode prior information about the expected source structure, such as smoothness, compactness, or entropy:
\begin{equation}
\mathcal{J}(\mathbf{I})
=
\chi^2(\mathbf{I})
+
\sum_k \lambda_k\, \mathcal{R}_k(\mathbf{I}),
\label{eq:rml_objective}
\end{equation}
where $\chi^2$ quantifies the agreement between the model image $\mathbf{I}$ and the interferometric observables, $\mathcal{R}_k$ are regularization terms encoding prior assumptions on the image, and $\lambda_k$ are hyperparameters controlling their relative weight.

A key advantage of RML methods is their flexibility in the choice of observables. In addition to complex visibilities, RML algorithms can use other observables, such as closure quantities (see Sect. \ref{sec:VLBIobservables}), as input data. These quantities are invariant to antenna-based
calibration errors, which simplifies the imaging process in VLBI and reduces sensitivity to residual calibration errors.

RML methods can be applied to full-Stokes imaging within the RIME formalism, imaging independently all Stokes parameters ($I$, $Q$, $U$ and $V$). This makes it possible to obtain all the polarimetric information, such as the degree of linear or circular polarization and the EVPA (see Sect.~\ref{sec:Polarization}).

RML imaging also enables effective superresolution by exploiting prior information and the global structure of the data, allowing the reconstruction of features on angular scales smaller (up to the 33\%) than the nominal synthesized beam \citep{Chael2018}. This capability has been demonstrated in a variety of VLBI applications, including imaging with the Event Horizon Telescope \citep{EHT2019IV}.

Recent developments have extended RML techniques to wideband interferometric data through multi-frequency synthesis. In this approach, the image intensity is modeled as a smooth function of frequency, allowing the simultaneous reconstruction of spatial and spectral information.

One of the most used RML imaging software packages is \texttt{eht-imaging} \citep{Chael2016, Chael2018}. In its multi-frequency implementation \citep{Chael2023}, the logarithm of the image intensity is expanded as a Taylor series in $\log(\nu/\nu_0)$ around a reference frequency $\nu_0$:
\begin{equation}
\log I_i =
\log I_0
+ \alpha \log\!\left(\frac{\nu_i}{\nu_0}\right)
+ \beta \log^2\!\left(\frac{\nu_i}{\nu_0}\right)
+ \dots,
\end{equation}
where $I_0$ is the total intensity image at $\nu_0$, $\alpha$ is the spectral index map, and $\beta$ represents the spectral curvature. This formulation allows information from all frequency bands to be combined coherently, producing images that are naturally aligned across frequency and well suited for studies of opacity effects and core shift.

\pagebreak
\section{High-Precision Astrometry and Geodesy}

Geodesy is the science that determines the Earth’s geometry, gravity field, and rotation in space. Modern geodesy underpins a wide range of scientific and societal applications, including global navigation and positioning \citep{Teunissen2017}, monitoring of sea level rise \citep{Cazenave2004}, tectonic plate motion \citep{DeMets2010}, Earth orientation \citep{Schuh2012}, and the establishment of stable terrestrial and celestial reference frames \citep{Charlot2020}.

\begin{figure}[h!]
    \centering
    \includegraphics[width=0.7\textwidth, trim={0pt 0pt 0pt 0pt}, clip]{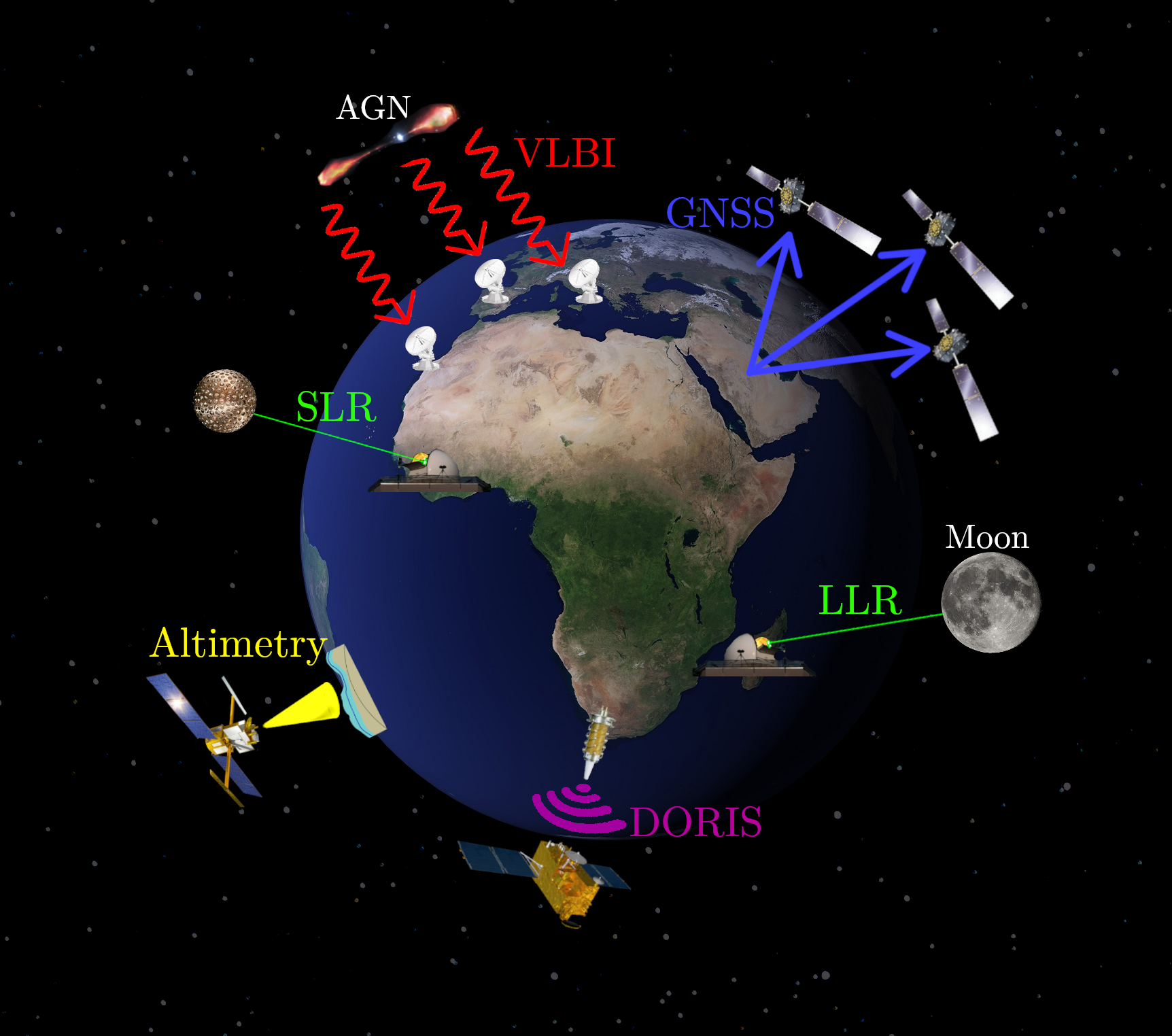}
    \caption[Schematic overview of the main space geodetic techniques.]{Schematic overview of the main space geodetic techniques: Global Navigation Satellite Systems (GNSS), Very Long Baseline Interferometry (VLBI), Satellite Laser Ranging (SLR), Lunar Laser Ranging (LLR), DORIS and satellite altimetry.}
    \label{fig:spacegeodesy}
\end{figure}

High precision geodesy relies on a combination of space geodetic techniques, each of which is sensitive to different physical quantities (see Figure \ref{fig:spacegeodesy}). Global Navigation Satellite Systems (GNSS) provide continuous measurements of station positions and velocities with millimeter-level precision, making them indispensable for regional and global deformation studies. Satellite Laser Ranging (SLR) and Lunar Laser Ranging (LLR) measure the distance to artificial satellites and the Moon, respectively. They are particularly important for the realization of the geocenter and the terrestrial scale. The DORIS system provides precise orbit determination and station positioning. Satellite altimetry enables global measurements of sea level and ocean topography. A comprehensive overview of these techniques and their roles in modern geodesy is given in \citet{Plag2009} and \citet{Luzum2015}.

\begin{figure}[h]
    \centering
    \includegraphics[width=\textwidth, trim={0pt 0pt 215pt 0pt}, clip]{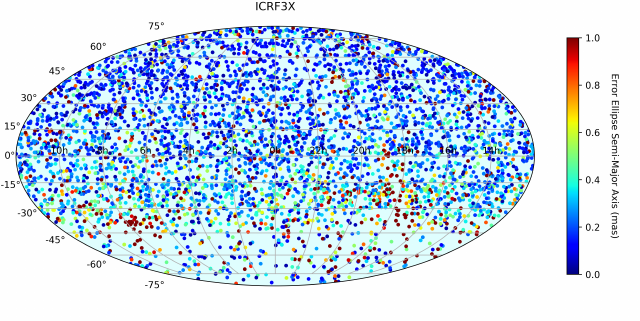}
    \caption[Distribution of the 4536 sources included in the ICRF3.]{Distribution of the 4536 sources included in the ICRF3. Each source is plotted as a dot color-coded according to its position uncertainty (defined as the semi-major axis of the error ellipse in position) \citep{Charlot2020}.}
    \label{fig:icrf3}
\end{figure}

A central concept in space geodesy is the definition of reference frames. The International Terrestrial Reference Frame (ITRF) provides a realization of the Earth-fixed coordinate system, defined by station positions and velocities on the rotating Earth. The International Celestial Reference Frame (ICRF) is a quasi-inertial reference frame defined by the positions of distant extragalactic radio sources, primarily compact active galactic nuclei observed with VLBI \citep{Ma1998,Charlot2020}. These sources are located at cosmological distances, so they exhibit no measurable global rotation, making the ICRF the most inertial reference frame currently available (see Fig. \ref{fig:icrf3}).

The link between the terrestrial and celestial frames is provided by the Earth Orientation Parameters (EOPs), which describe the rotation of the Earth in space as a function of time. The EOPs include polar motion, the Earth rotation angle (UT1-UTC), and precession and nutation parameters. These quantities are fundamental because they enter directly into the modeling of all space geodetic observables. Accurate knowledge of the EOPs is required to transform measurements between terrestrial and celestial frames.

\subsection{Geodetic and Astrometric VLBI}

Very Long Baseline Interferometry (VLBI) is the only one of the space geodetic techniques capable of determining all EOPs independently and simultaneously \citep{Schuh2012}. It is unique in providing Universal Time (UT1). Furthermore, it provides celestial pole movements (precession and nutation) over longer time ranges. Geodetic VLBI uses the direct observation of extragalactic radio sources, which define the celestial frame, combined with measurements performed by stations fixed to the rotating Earth. As a result, VLBI plays a central role in the realization of both the ICRF and the EOP time series used by all other geodetic techniques.

The fundamental VLBI observable in geodesy is the geometric group delay (see Eq. \ref{eq:geometricdelay}). This simple parameter highlights the direct sensitivity of VLBI delays to both station positions (through $\overrightarrow{b}$) and source directions (through $\overrightarrow{s}$).

The extreme precision of VLBI can be illustrated with a simple estimation. A delay uncertainty of $\Delta\tau = 1$~ps corresponds to a path length difference of
\[
c\,\Delta\tau \simeq 0.3~\mathrm{mm}
\]
For a baseline of length $|\overrightarrow{b}| \simeq 7000$~km, this delay
precision translates into an angular sensitivity of
\[
\Delta\theta \sim \frac{c\,\Delta\tau}{|\overrightarrow{b}|}
\sim 4 \times 10^{-11}\,\mathrm{rad}
\sim 8~\mu\mathrm{as},
\]
demonstrating that a level of picoseconds in delay measurements enables microarcsecond astrometry and millimeter accuracy in the stations positioning. This level of precision is essential for improving the detection of tectonic plate motions \citep{Feigl1993}, loading effects \citep{Petrov2004}, and small variations in Earth rotation \citep{Nothnagel2008}.

Achieving such precision requires accurate estimation of the group delay, defined as the derivative of the interferometric phase with respect to frequency (see Eq.~\ref{eq:groupdelay}). For a given phase uncertainty $\Delta\phi$, the corresponding delay uncertainty scales inversely with the effective bandwidth $\Delta\nu$. Consequently, wideband VLBI observations provide a better precision in delay estimations, making these systems particularly well suited for geodetic applications.

In summary, high-precision astrometry and geodesy rely on VLBI observations. By measuring group delays with picosecond accuracy over intercontinental baselines, VLBI enables an extremely accurate realization of the celestial reference frame, the determination of Earth orientation parameters, and the precise monitoring of the dynamic Earth system.

\pagebreak
\section{VLBI Global Observing System (VGOS)}
\label{sec:vgos}

The VLBI Global Observing System (VGOS) is the next generation realization of geodetic VLBI. It is designed to meet the accuracy requirements of modern Earth system monitoring. VGOS is developed and coordinated within the International VLBI Service for Geodesy and Astrometry (IVS)\footnote{https://ivscc.gsfc.nasa.gov/} \citep{Nothnagel2017}, which provides the international framework for geodetic and astrometric VLBI observations, correlator operations, analysis standards, and the delivery of geodetic products (EOPs, ITRF, ICRF) to the International Earth Rotation and Reference Systems Service (IERS)\footnote{https://www.iers.org/}.

The design drivers of VGOS are tightly connected to the goals of the Global Geodetic Observing System (GGOS)\footnote{https://geodesy.science/ggos/}, an initiative of the International Association of Geodesy (IAG)\footnote{https://geodesy.science/iag/} aiming at a consistent and long-term monitoring of the geodetic ``Earth system'' (geometry, gravity field, and rotation) by integrating space geodetic techniques. GGOS has articulated requirements at the level of millimeters for station coordinates and the geocenter \citep{Gross2009}, together with high temporal resolution and low latency products to capture rapid Earth system variability.

Historically, the geodetic VLBI network has operated with the Legacy dual S/X system, which has delivered the long-term EOP series and contributed decisively to the realization of the celestial reference frame \citep{Schuh2012, Ma1998, Charlot2020}. However, the accuracy requirements of modern geodesy, together with the need for improved robustness against radio-frequency interference (RFI), faster observing strategies, and higher precision group delays, motivated the development of a new VLBI standard. This transition was formalized in the ``VLBI2010'' concept \citep{Petrachenko2009}, which laid out the fundamental system architecture that later evolved into VGOS \citep{Hase2012}.

\subsection{VGOS Technical Characteristics}

In order to achieve the goals of 1~mm in the position of the stations and 0.1~mm/yr in their velocities, VGOS needs to use wideband observing frequency ranges. It improves the precision of the geodetic group delay, since the delay uncertainty scales inversely with the effective bandwidth (see Eq. \ref{eq:groupdelay}). Ultra-wideband receivers spanning several GHz are used to reach an accuracy of picoseconds in the group delay. In particular, VGOS uses frequency bandwidths from $\sim$2 to $\sim$14~GHz.

VGOS stations are typically based on smaller antennas with high slew rates, enabling short scan lengths and rapid source switching. Specifically, the diameters of VGOS antennas are $\sim 13\,\mathrm{m}$, with a slew speed of $360^\circ/\text{min}$. This strengthens the estimation of tropospheric parameters, clocks, and EOPs, and improves the sky coverage and the temporal resolution of geodetic products.

\begin{figure}[h!]
    \centering
    \hspace{-2.1cm}
    \begin{minipage}{0.43\textwidth}
        \centering
        \includegraphics[
            height=4.5cm,
            trim=0.3cm 0cm 0.3cm 0cm,
            clip
        ]{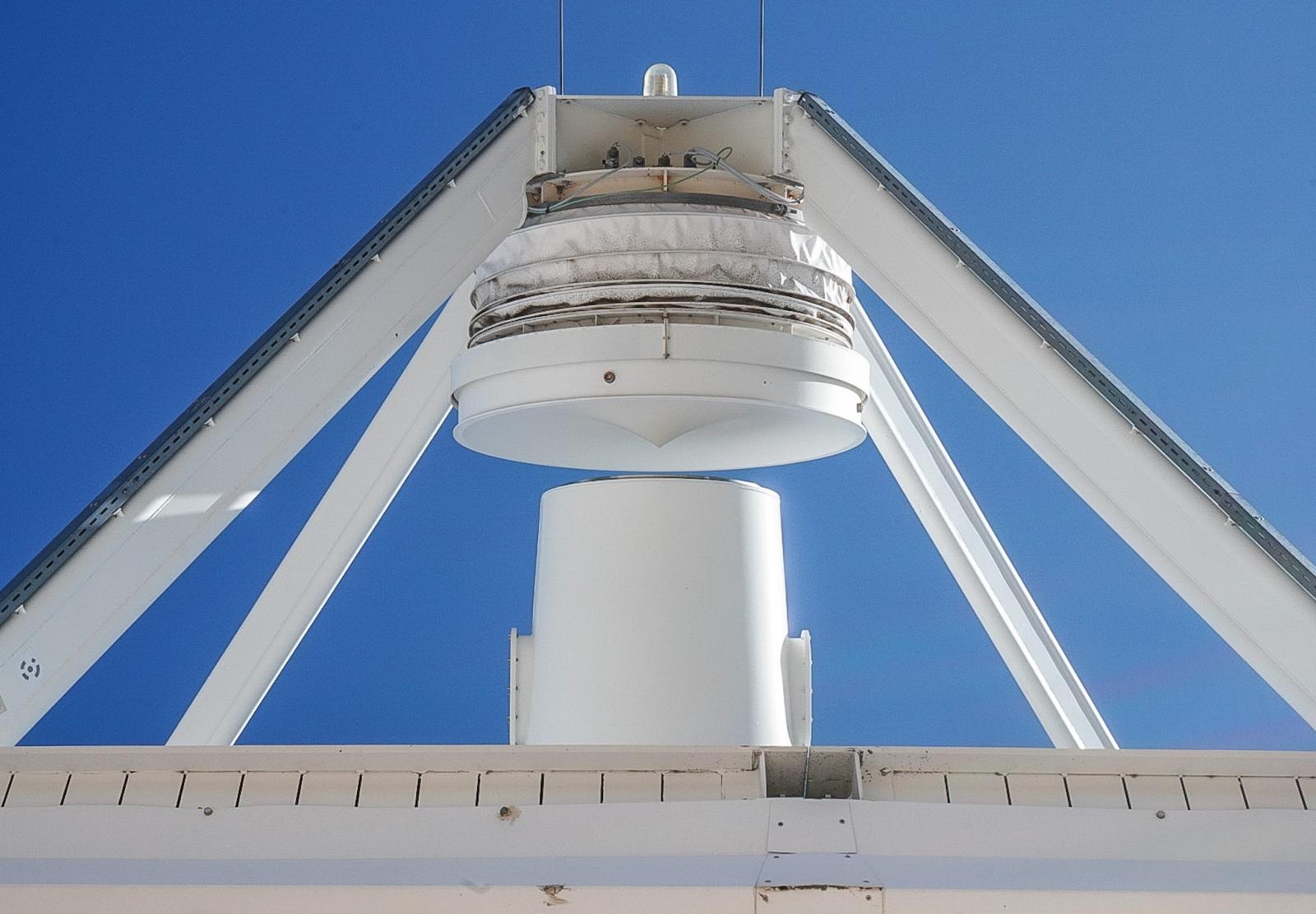}
\end{minipage}
    \begin{minipage}{0.43\textwidth}
        \centering
        \includegraphics[height=4.5cm]{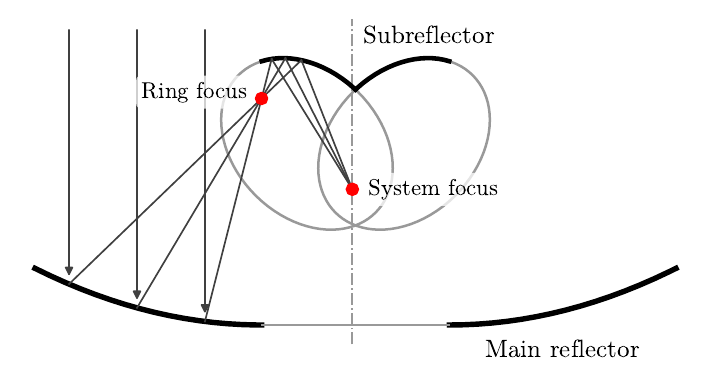}
    \end{minipage}
    
    \caption[Ring-focus geometry of the 13.2\,m VGOS antenna at Yebes Observatory.]{Left: The subreflector of the 13.2\,m VGOS antenna at the Yebes Observatory (Spain), where ring-focus geometry can be seen. Right: Geometrical optics scheme of a ring-focus antenna. Incoming rays are reflected by the main paraboloidal reflector and converge at the ring focus. After reflection on the subreflector, the radiation is redirected toward the system focus, where the feed is located.}

    \label{fig:ringfocus}
\end{figure}

Most VGOS antennas adopt a ring-focus design (see Figure \ref{fig:ringfocus}), which provides a compact structure and a flatter gain–elevation curve compared to classical large S/X antennas \citep{Nothnagel2017}. It also improves gain stability with elevation and reduces spillover losses.

The antennas of the VGOS array have a similar System Equivalent Flux Density (SEFD) of $\sim 2500\,\mathrm{Jy}$, making the system more homogeneous than the Legacy S/X system.

The VGOS architecture includes broadband digitization, flexible channelization, and standardized data transport and correlation pipelines, enabling faster product delivery in comparison to legacy geodetic VLBI. Signals are recorded at rates of 16 Gbps or even up to 32 Gbps in advanced configurations \citep{Niell2018}.

The digital backend allows flexible channelization optimized for geodetic delay estimation, maximizing the effective bandwidth $\Delta\nu_\mathrm{eff}$ and minimizing delay uncertainty. Data are recorded using modern Mark~6 systems and correlated with software correlators such as DiFX \citep{Deller2011}. 

A practical and defining feature of the VGOS design is the use of linear polarization receivers at all stations. This choice was primarily motivated by the need to achieve very large fractional bandwidths (2–14\,GHz) while maintaining high polarization purity and stable instrumental response across the band.

For ultra-wideband systems, the implementation of circular polarization using quarter wave plates becomes increasingly challenging. The phase shift introduced by a quarter wave plate is inherently frequency dependent, and maintaining an accurate $90^\circ$ phase offset across a huge bandwidth is technically difficult. Residual frequency-dependent phase and amplitude mismatches introduce instrumental polarization leakage and bandpass distortions. In contrast, linear feeds provide a more uniform polarization response over broad frequency ranges and avoid the need for dispersive polarization conversion hardware. This makes them better suited for broadband delay measurements, which are central to geodetic VLBI performance \citep{Petrachenko2009, Niell2018}.

However, this design choice introduces additional complexity for polarization calibration and for compatibility with the traditional circular polarization basis commonly used in VLBI calibration and imaging (see Sect.~\ref{sec:rime}). This conflict between broadband instrumental design and calibration simplicity is resolved in practice through post-correlation polarization conversion. The PolConvert algorithm (Sect.~\ref{sec:polconvert}) performs a software transformation from linear to circular polarization after correlation, preserving the wideband advantages of linear feeds while enabling calibration and imaging to proceed in the circular basis.

Beyond software solutions, hardware approaches for realizing broadband circular polarization are also being investigated. One promising option is the use of microwave 3\,dB/90$^\circ$ hybrid couplers, which can combine two orthogonal linear signals with a controlled phase shift to synthesize right and left circular polarization. In this scheme, the polarization conversion is performed at the frontend, prior to the low noise amplifiers, thereby avoiding the use of frequency sensitive quarter wave plates.

The potential applicability of such hybrid couplers to VGOS systems has been evaluated in \citet{GarciaPerez2018}. This report analyzes their broadband performance, polarization isolation, and phase stability, and discusses their possible implementation in next-generation wideband receivers. While these devices offer an elegant analog solution to polarization synthesis, practical considerations must be carefully assessed. At present, polarization conversion via software (Sect.~\ref{sec:polconvert}) remains the operational solution in VGOS \citep{Jaron2024}.

\subsection{Global Network and Observing Strategy}

\begin{figure}[h!]
    \centering
    \includegraphics[width=\textwidth, trim={0pt 0pt 0pt 0pt}, clip]{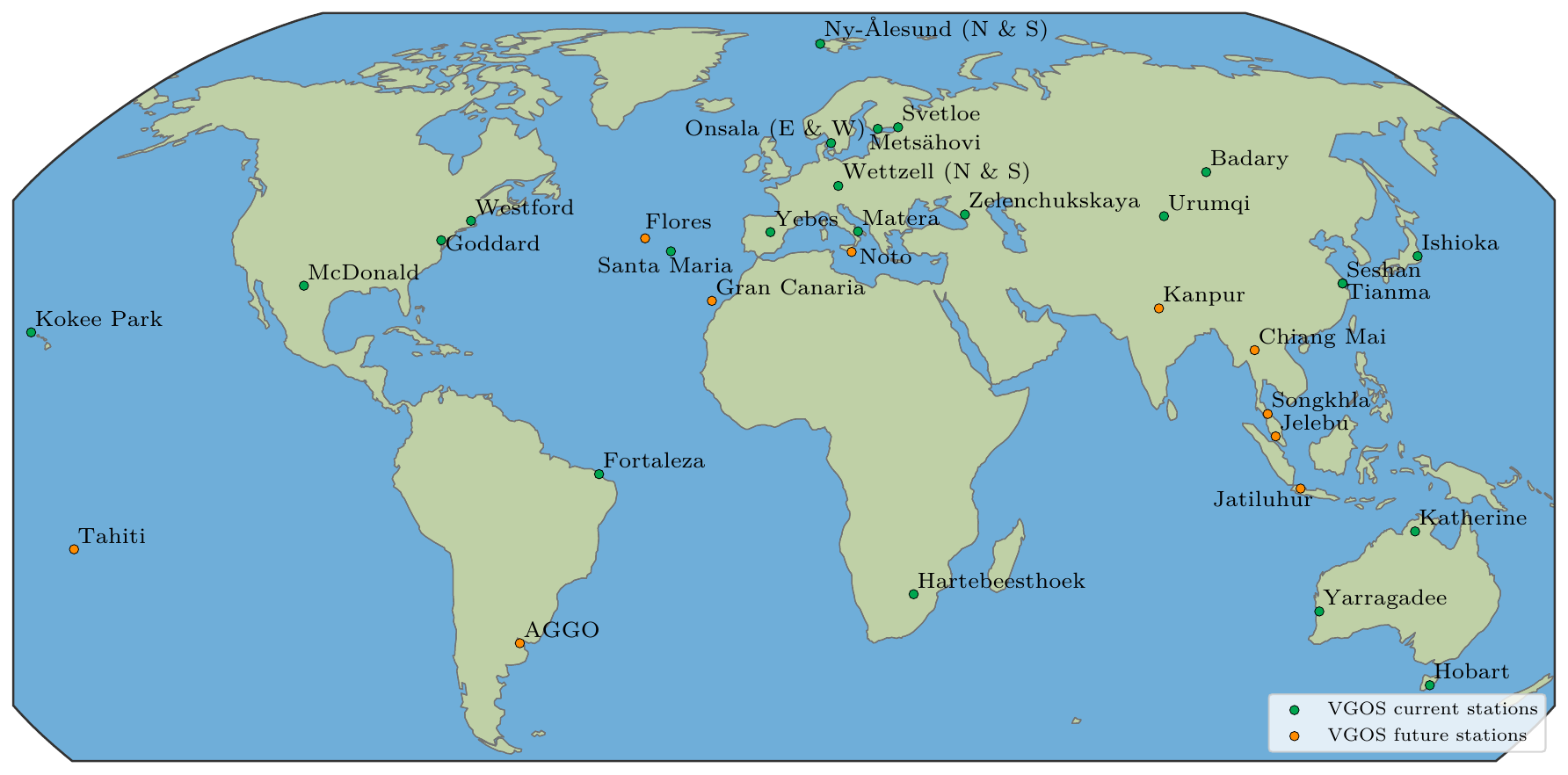}
    \caption[Geographic distribution of the VGOS network.]{Geographic distribution of the VGOS network. Operational broadband VGOS antennas at 2026 are shown in green. In project or future stations, in orange.}
    \label{fig:vgosstations}
\end{figure}

The current VGOS network consists of approximately 25 radio telescopes distributed worldwide (see Fig.~\ref{fig:vgosstations}), providing almost global coverage. This geographical distribution is essential for achieving long intercontinental baselines, which maximize sensitivity to Earth orientation parameters and station coordinates, and improving image resolution. 

The network geometry plays a critical role in geodetic performance. Long north-south baselines improve sensitivity to polar motion, while east-west baselines are particularly important for the determination of UT1-UTC. The inclusion of high latitude stations, such as those in the Arctic region, enhances the stability of celestial frame realization and improves the separation of tropospheric and geometric effects. Southern hemisphere stations are especially important for reducing hemispheric imbalance in the reference frame and strengthening global scale determination.

Furthermore, there are some stations with two twin close telescopes (as in the cases of Onsala, Wettzell and Ny-Ålesund). These observatories are quite useful at calibrating data, since they have the shortest baselines and share the same sky conditions.

The VGOS network is designed for continuous and regular geodetic monitoring of the Earth system. VGOS observations are, currently, scheduled weekly. Two main types of sessions are currently conducted: standard 24-hour sessions and rapid UT1-intensive sessions.

The 24-hour VGOS sessions are dedicated to the simultaneous determination of station coordinates, tropospheric parameters, clocks, and the full set of Earth Orientation Parameters (EOPs). These sessions provide the backbone of the long-term geodetic time series and contribute directly to the realization of the International Terrestrial Reference Frame (ITRF) and the celestial reference frame \citep{Nothnagel2017}.

Intensive sessions are usually also carried out weekly. They focus primarily on the rapid determination of UT1--UTC, the difference between UT1 (Universal Time 1), which is derived from the Earth's rotation angle, and UTC (Coordinated Universal Time), the international atomic time scale used for civil timekeeping. These sessions use a limited number of stations observing over shorter time spans (usually about one hour), optimized to maximize sensitivity to Earth rotation angle. Rapid UT1 estimates are essential for operational applications, including satellite navigation, spacecraft tracking, and precise orbit determination \citep{Schuh2012}.

The long-term goal of the VGOS concept is to transition to an almost continuous 24/7 observing mode, enabling significantly improved temporal resolution and reduced latency in geodetic products \citep{Petrachenko2009}. A key objective of VGOS is the eventual delivery of geodetic products in near real time, reducing latency from days to hours or even minutes. Achieving this requires not only continuous observing but also automated correlation pipelines, fast data transport, and streamlined analysis procedures.

Source structure effects constitute one of the dominant error sources in geodetic VLBI. While geodetic analysis traditionally models AGN as ideal point-like sources, high resolution astronomical VLBI imaging has demonstrated that many of them exhibit extended jet structures. These asymmetric brightness distributions introduce phase variations that translate into systematic biases in the group delay observable. 

\citet{Xu2021, Jaron2025} quantified the impact of source structure on broadband VGOS group delays and showed that structure contributions can reach several tens of picoseconds, a level comparable to the combined magnitude of residual station errors. This makes source structure one of the leading contributors to the geodetic error budget in modern broadband VLBI.

Consequently, the synergy between astronomical VLBI imaging and geodetic analysis has become increasingly important. Accurate multi-frequency brightness models are required not only for celestial frame realization but also for achieving the millimeter level station accuracy and picosecond delay precision targeted by GGOS. In this context, detailed astrophysical characterization of geodetic radio sources is no longer optional, but a fundamental component of next-generation geodetic analysis.

At the same time, VGOS observations represent a unique opportunity for astrophysics. The network routinely observes hundreds of compact AGN with global baselines, broadband frequency coverage (2–14\,GHz), and full polarization capability. These observations provide high angular resolution, large fractional bandwidth, and regular temporal sampling, enabling detailed studies of jet morphology, spectral properties, polarization structure, and variability. The periodic nature of VGOS sessions allows long-term monitoring of structural evolution, core-shift behavior, and magnetic field properties in a huge sample of AGN.

VGOS joins geodesy and astrophysics: the same datasets that define the terrestrial and celestial reference frames also contain rich information about relativistic jet physics. Exploiting this dual potential requires advanced calibration, wideband imaging, and polarimetric analysis techniques, which constitute the main focus of this thesis.

\cleardoublepage

\chapter{Calibration Pipeline for VGOS Data}
\label{ch:vgos_pipeline}

\vspace{-0.5cm}

\noindent
\hfill
\begin{minipage}{0.8\textwidth}
\small
\raggedleft
\textit{This chapter is mainly based on:}\\[0.3em]
\textit{``Towards an astronomical use of new-generation geodetic observations.\\
I. From the correlator to full-polarization images''}\\[0.3em]
\textit{\textbf{V. Pérez-Díez}, I. Martí-Vidal, E. Albentosa-Ruiz, J. González-García,\\
F. Jaron, T. Savolainen, M. H. Xu and R. Bachiller.}\\[0.3em]
\textit{Astronomy \& Astrophysics, 688, A151 (2024).}\\
\end{minipage}

\vspace{0.8cm}

\lettrine{T}{  he} VLBI Global Observing System (VGOS) was designed as the next-generation framework for geodetic VLBI. Its ultra-wideband receivers make VGOS a powerful resource not only for geodesy, but also for astrophysical studies of compact active galactic nuclei (AGN). However, this astronomical potential has remained largely unexplored because VGOS observations are acquired and processed primarily for geodetic purposes.

This chapter presents a complete calibration pipeline that enables the astronomical exploitation of VGOS observations. Starting from the raw correlator outputs, the pipeline performs polarization conversion from a linear to a circular basis, phase calibration with a novel wide band global fringe fitting algorithm that includes dispersive ionospheric terms, and amplitude calibration.

The resulting pipeline allows the recovery of full Stokes calibrated visibilities suitable for high fidelity imaging and polarimetric analysis. It therefore demonstrates that VGOS data can be transformed from purely geodetic observables into valuable astronomical datasets, while also providing source structure information that can help improve geodetic observables.

\pagebreak
\section{Introduction}

As already discussed in Chapter \ref{ch:introduction}, circular polarization receivers have been commonly employed in Very Long Baseline Interferometry (VLBI) for various reasons. VLBI observations involve extremely long baselines, which introduce significant differences in the parallactic angles of the different antennas. Correcting for the effects of parallactic angle in the interferometric phase is easier while observing in a circular polarization basis (i.e., dividing the signal into right-hand, RCP, and left-hand, LCP, circular polarizations), as explained in Sect. \ref{sec:rime}. On the contrary, if linear polarization feeds are used in the VLBI observations, the gain differences (both phases and amplitudes) between the polarization channels at each antenna (and for each frequency, i.e. a delay) must be known before the parallactic angle correction can be applied. Therefore, the calibration and fringe fitting strategy for linear polarization feeds may become significantly more complex and computationally expensive, since such a pre-calibration is needed prior to the fringe fitting, as shown in \citet{IMV2016}.

However, circular polarization receivers also have certain disadvantages. On the one hand, the extra hardware required for the conversion into circular polarization (e.g., quarter-wave plates at the receiver frontend) may add additional noise and degrade the polarization purity of the receiver, resulting in a higher instrumental polarization and a slightly lower S/N. On the other hand, the fractional bandwidth available is narrower, since the circular polarization hardware converters are designed for an optimum performance at a specific frequency, worsening their polarization purity as we depart from that frequency. Due to this last drawback, linear polarization receivers may be the most convenient choice for VLBI antennas with huge wide band receivers (e.g., with fractional bandwidths of the order of 100\%).

This is actually the case of VGOS. In order to achieve the goal of 1 mm accuracy in station position, it requires extremely wide bandwidths to allow the measurement of group delays with a precision of the order of a picosecond. In particular, VGOS demands the use of ultra-wideband receivers covering a broad frequency range from 2\,GHz to 14\,GHz, as it was originally planned \citep{Petrachenko2009} and recently confirmed \citep{Niell2018}.

Various algorithms have been proposed to calibrate the data from VGOS observations. The one officially used for its processing is found in \citet{Cappallo2014}. It is based on the use of {\em pseudo-Stokes I} (henceforth, pI), a quantity obtained by combining the four linear polarization correlation products (i.e., $XX$, $XY$, $YX$, and $YY$), accounting for instrumental cross-polarization phase (not amplitude) bandpasses and parallactic angle differences between antennas. The main drawback of this method for its astronomical exploitation is that it only handles total intensity, so it is not possible to obtain information about the polarized brightness distribution of the observed sources. Furthermore, deriving the instrumental cross-polarization phase bandpasses of all the antennas is usually an expensive procedure in the pI approach, which may take several hours of processing for a typical global VGOS observation.

In order to address these drawbacks and find ways to improve the calibration and analysis of VGOS, a European project called {\em EU-VGOS} was initiated as reported in \citet{EuVGOS}, and the current status has been reported in  \citet{EuVGOS23}. The EU-VGOS project decided to use the PolConvert software \citep{PolConvert}, explained in Sect.~\ref{sec:polconvert} as a way to convert the VGOS data to a circular polarization basis just after correlation, in such a way that legacy VLBI calibration algorithms (all based on observations with circular polarization receivers) could be used in VGOS. We refer to the conversion from linear to circular polarization using the tool PolConvert as \textit{polconversion}. 

In this Chapter, we present a detailed description of the complete calibration process of the data obtained during a global VGOS 24-hr session by using an approach based on PolConvert. All steps will be described, starting from correlation to polconversion with existing software. Then, we will present our new global fringe fitting algorithm, as well as the amplitude calibration and multi-frequency imaging of sources at full polarization.

In Sect. \ref{sec:Observations}, we summarize the observational setup (participating antennas, scheduling, frequency setup, correlation strategy, etc.). In Sect. \ref{sec:paperdata_calibration}, we describe our pipeline, emphasizing the steps related to polconversion, global wide-band fringe fitting, and amplitude calibration. Finally, Sect. \ref{sec:mfimaging} describes the imaging procedure.

\pagebreak
\section{Observations}
\label{sec:Observations}
The observations used for testing our pipeline correspond to the VGOS IVS experiment with code VO2187, which was observed on 6-7 July 2022. The eight participating antennas (with their code names) were Goddard (GS), Ishioka (IS), Kokee (K2), McDonald (MG), the twin Onsala telescopes (OE and OW), Westford (WF) and Yebes (YJ). The antenna coordinates used in the correlation are listed in Table \ref{tab:antennas}.

\begin{table}[ht]
\centering
\caption[Antennas used in experiment VO2187.]{Antennas used in experiment VO2187, together with their reference Geocenter coordinates used in the DiFX correlation.}

\resizebox{\textwidth}{!}{%
\begin{tabular}{l c| r r r}
\multicolumn{2}{c|}{Antenna} & \multicolumn{3}{c}{GC Coords (m)} \\ 
\hline
IVS Component Name & Code & X & Y & Z \\
\hline 
Goddard Geophysical and Astronomical Observatory & GS & 1130729.877 & -4831245.972 & 3994228.300 \\
Ishioka VLBI Station & IS & -3959636.203 & 3296825.448 & 3747042.571 \\
Kokee Park Geophysical Observatory & K2 & -5543831.745 & -2054585.590 & 2387828.974 \\
McDonald Geodetic Observatory & MG & -1330788.462 & -5328106.593 & 3236427.492 \\
Onsala Space Observatory East & OE & 3370889.298 & 711571.199 & 5349692.048 \\
Onsala Space Observatory West & OW & 3370946.779 & 711534.507 & 5349660.925 \\
Westford Antenna, Haystack Observatory & WF & 1492206.223 & -4458130.552 & 4296015.629 \\
IGN Yebes Observatory & YJ & 4848831.021 & -261629.388 & 4122976.576 \\
\hline
\end{tabular}
}
\label{tab:antennas}
\end{table}

The total recorded bandwidth was 1\,GHz, divided into 32 spectral windows (spw) of 32\,MHz each. The spw are distributed across a wide frequency coverage, ranging between 3\,GHz and 11\,GHz, and are arranged in subsets of 4 bands (i.e., 8 spw each). The bands, centered around 3.25, 5.5, 6.75 and 10.5\,GHz, respectively, are labelled as A, B, C, and D, following the convention of the VGOS community. 

Phase-calibration tones equally spaced in frequency (hereafter phase-cal tones) were inserted at each station, covering the full band in intervals of 5\,MHz, which ensures a total of 6-7 phase-cal tone detections per spw, for a robust determination of the instrumental delay and phase at each frequency. The exception is the Yebes antenna (YJ), for which the phase-cal tones are spaced in intervals of 10\,MHz (so there are only 3-4 usable tone detections per spw). This limitation at YJ may degrade the quality of its instrumental phase calibration, as discussed in Sect. \ref{sec:instrumentalPhases}.

The correlation was processed at the MPIfR in Bonn using DiFX-2.5.4 \citep{Deller2011} in full polarization mode, i.e., the four polarization correlations, $XX$, $XY$, $YX$ and $YY$, were produced. A total of 160 spectral channels, with a correlator integration time of 1\,s, were generated for each spw. DiFX refers the correlation times to the geocenter, which allows the computation of closure quantities from the correlated outputs \citep[e.g.,][]{TMS1986}. These closures are unaffected by antenna gain effects (i.e., they only depend on the brightness distributions of the sources), which allows the imaging of VGOS data if a Global Fringe-Fitting algorithm \citep{Schwab1983} is used (see Sect. \ref{sec:GFF}).

There was a total of 74 sources observed (radio-loud AGN; mostly blazars), spanning an overall observing time of 24 hours in total. The sources were observed in short interleaved scans with a typical duration of 30\,s. The number of scans is different for each source, ranging from just one scan (source 0847-120) up to 66 scans (source 1803+784). In Table \ref{tab:sources}, we list the 10 sources on which more observing time was spent, and for which an astronomical study has been done (see Chapter \ref{ch:astro}).

\begin{table}
    \centering
    \small
    \caption{Sources most commonly observed in experiment VO2187.}
    \begin{tabular}{r | r | c}
    Source & $N_{obs}$ & $N_{scan}$ \\
    \hline
    1849+670 & 1004 & 60 \\ 
    2229+695 &  987 & 61 \\ 
    1803+784 &  836 & 66 \\ 
    0059+581 &  692 & 59 \\ 
    0613+570 &  625 & 46 \\ 
    3C418  & 532 & 56 \\ 
    0955+476 &  392 & 52 \\ 
    3C274 &  203 & 39 \\ 
    DA426 &  177 & 43 \\ 
    OJ287 &  168 & 37 \\ 
    \hline
    \end{tabular}
    \label{tab:sources}
    \caption*{Sources are ordered by the number of observations ($N_{obs}$, which is the sum of the number of baselines per scan, over all scans). $N_{scan}$ is the number of scans.\\}
\end{table}

\pagebreak
\section{Data Calibration}
\label{sec:paperdata_calibration}
As explained in Sections \ref{sec:rime} and \ref{sec:data_calibration}, the visibilities delivered by the correlator do not directly represent the intrinsic source coherence, but are affected by instrumental, geometric, and propagation effects that vary with antenna, time, and frequency. In the case of VGOS, the use of ultra-wideband receivers (2-14\,GHz) and linear polarization feeds introduces additional complexity in the calibration process, particularly for polarization conversion and phase alignment across the full bandwidth.

The goal of the calibration procedure described in this chapter is to transform the raw correlator outputs into fully calibrated, full-Stokes visibilities that  are suitable for astronomical analysis. In the following subsections, we describe each step in detail and demonstrate their performance for experiment VO2187.

\subsection{PolConvert Cross-Polarization Bandpass}
\label{sec:paperpolconvert}

In order to convert the linear-based visibilities to a circular basis, the first step of the calibration uses the software PolConvert (see Sect. \ref{sec:polconvert}) \citep{PolConvert}. The conversion from linear to circular polarization is primarily affected by the cross-polarization ($X/Y$) bandpass of each antenna in phase and amplitude. These quantities can be estimated in PolConvert using the algorithm called global cross-polarization fringe-fitting (GCPFF) \citep{IMV2016}.

To apply this algorithm, the first step is to select one or more calibrator scans. In order to choose them, we performed a coarse baseline-based preliminary fringe fitting on a selected set of spw for all sources in the four correlation products ($XX$, $YY$, $XY$, and $YX$), and we calculated the S/N of the fringes. In this way, we selected suitable calibration scans with a good S/N. It is preferable to have either a long calibration scan and/or a set of scans covering some parallactic-angle changes in the antennas, which are needed to break a 180-degree global phase ambiguity in the $X/Y$ bandpass. This ambiguity is equivalent to exchanging $R$ and $L$ polarizations at all antennas. In short, we required a source that was bright, was observed with a sufficient parallactic-angle coverage in any antenna, and had (preferably) a low fractional linear polarization.

In the process of estimating the $X/Y$ bandpasses, PolConvert applies a priori phase information computed from the difference of the phases of the extracted phase-cal tones between $X$ and $Y$ for each spw and at each integration time. This difference was fit as a cross-polarization tone delay for each spw, which was then added to the $X/Y$ bandpass to be solved.

These cross-polarization bandpasses were then applied to the remaining scans of the experiment because any drift in the $X/Y$ phase differences was tracked by the phase-cal tones that are used by PolConvert in all scans. In essence, the quantities estimated by the GCPFF can be interpreted as related to the difference in the optical path of the $X$ and $Y$ polarization channels from the receiver frontend until the point at which the phase-cal tones are injected (located after the polarization splitter in the waveguides). In our procedure, we therefore assumed that the part of the hardware that causes the GCPFF cross-polarization gains was stable at the timescale of the duration of the experiment.

\begin{figure}
    \includegraphics[width=\textwidth, trim={2pt 0pt 2pt 25pt}, clip]{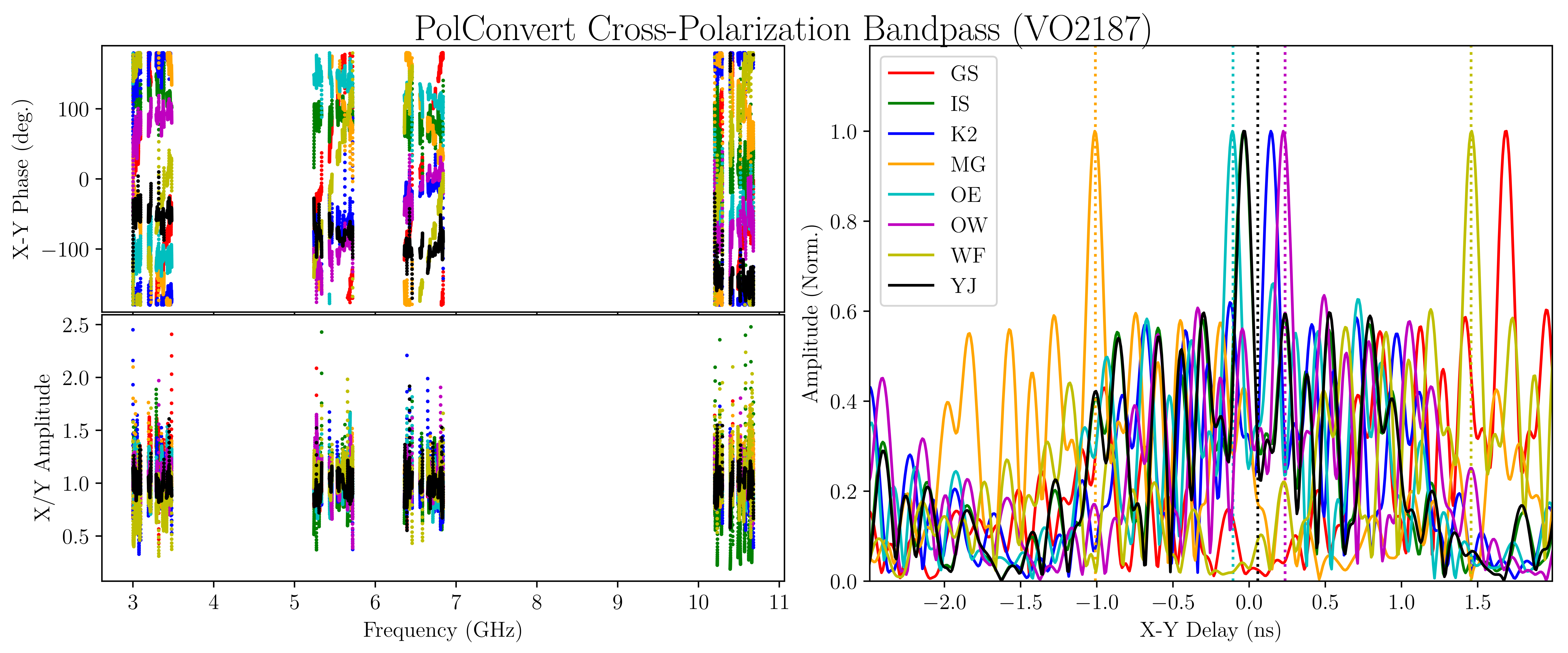}
    \caption[Cross-polarization bandpasses and multiband delays for experiment VO2187.]{Cross-polarization bandpasses for experiment VO2187 (left; phases at top; amplitudes at the bottom), as estimated with PolConvert (see text). Multiband cross-polarization delays (right; computed from the values shown at the left). The dotted lines (same colors) mark the cross-polarization multiband delays obtained 63 days later from experiment ER2201 \citep{Jaron2024}.}
    \label{fig:XPBandpass}
\end{figure}

In Fig. \ref{fig:XPBandpass}, we show the GCPFF cross-polarization bandpasses obtained with PolConvert using the first two scans of VO2187 (observations of sources 0059$+$581 and 1741$-$038). The phases and amplitudes of the $X/Y$ bandpasses are shown in the left panel (each antenna with a different color). The bandpasses in delay space are shown in the right panel, where the dotted lines mark the delays reported for another VGOS epoch \citep{Jaron2024} for the subset of antennas that also participated in VO2187.

Several conclusions can be drawn from this figure. First, the $X/Y$ phases for each antenna can be well connected across the whole band (from 3 to 11\,GHz), without jumps or discontinuities. This indicates that the $X/Y$ bandpass effects introduced by the receiver components before the injection of the phase-cal tones can be predominately described by a single delay between the polarization channels. All $X/Y$ amplitude ratios were around unity, which indicates a similar electronic gain for the two polarization channels at all antennas. There are some exceptions, however. For instance, IS in band D shows a low $X/Y$ ratio, as does WF in band A.

The quality of the alignment of the $X/Y$ phases across the VGOS band appears clear in the fringe plots shown in Fig. \ref{fig:XPBandpass} (right), which show clean peaks that also remain stable (differences smaller than 10\,ps) across a time baseline of more than two months when we compare the peak positions to those reported in \citet{Jaron2024} for epoch ER2201 (observed 63 days after VO2187). The exception is the YJ antenna, for which a shift of about 100\,ps is seen between the $X/Y$ delay in VO2187 and ER2201. The shift can be explained by changes in the phase-cal system at YJ between the two epochs because some amplifiers at the end of the analog signal chain were replaced due to failure.

\subsubsection{PolConverted Fringes}
After the GCPFF cross-polarization bandpasses were obtained, we executed PolConvert to apply them (together with the polarization difference of the phase-cal tones) to the whole experiment.

One way to assess the quality of the polconversion is to compare the fringe amplitudes in the parallel-hand correlations (i.e., $RR$ and $LL$) to those of the cross-hand correlations (i.e., $RL$ and $LR$). In particular, the former should be remarkably higher than the latter under normal conditions (i.e., as long as the fractional polarization of the sources is low and their polarized structures are not much more compact than the Stokes I brightness distribution). This condition should hold for all sources at all observing times.

\begin{figure}
    \hspace*{-6mm}
    \includegraphics[width=0.57\textwidth]{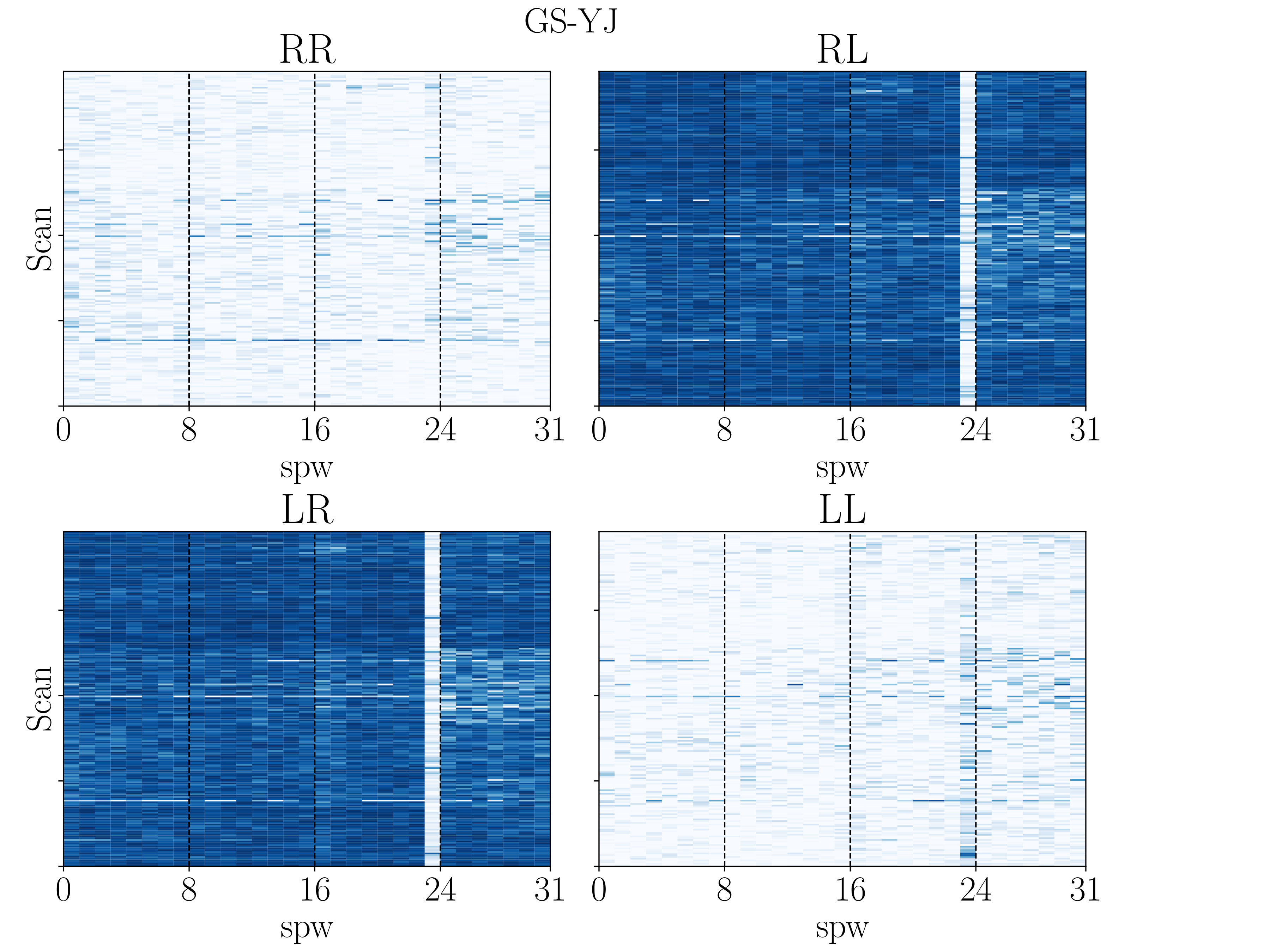}
    \hspace*{-13mm}
    \includegraphics[width=0.57\textwidth]{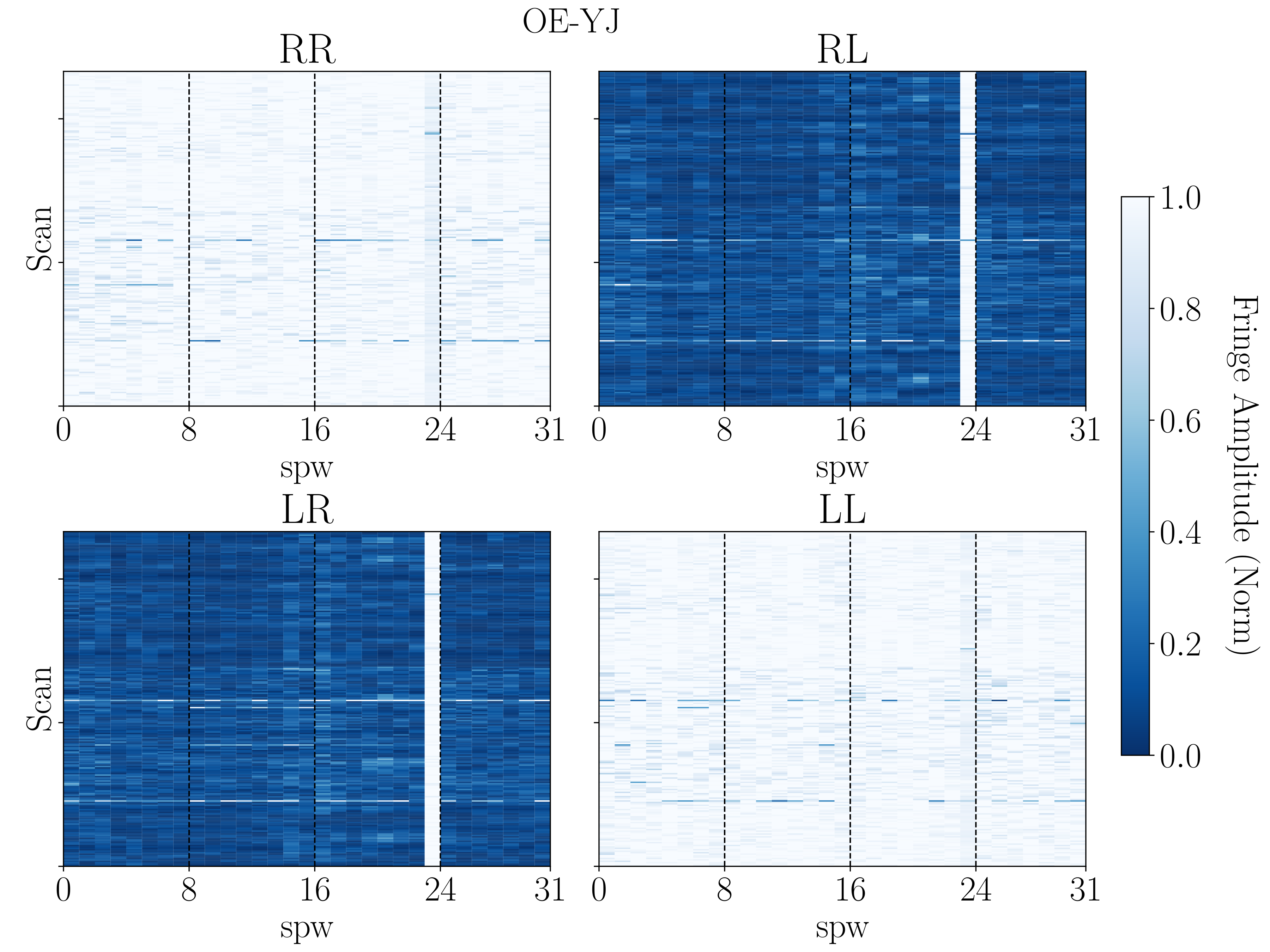}
    \caption[Fringe peaks for selected baselines as a function of spectral window and scan.]{Fringe peaks for a selection of baselines as a function of spw (horizontal axis) and scan (vertical axis), and for the set of sources listed in Table \ref{tab:sources}. All four correlation products are shown. Left, baseline GS to YJ. Right, baseline OE to YJ.}
    \label{fig:FringePeak}
\end{figure}

Fig. \ref{fig:FringePeak} shows the fringe amplitude peaks of baselines GS-YJ and OE-YJ in all spw (and all four polconverted correlation products) for all the scans of the seven most frequently observed sources (see Table \ref{tab:sources}). The different scans (a total of 197 and 234 for GS-YJ and OE-YJ, respectively) are arranged on the vertical axes of the plots, and the color palette for each scan is normalized to its peak. For all spw, the parallel-hand amplitudes of all these scans are significantly higher than the cross-hand amplitudes. The only exception is spw 23, which corresponds to the highest frequency of band C. In this spw, similar amplitudes are obtained in all four correlation products, which indicates that the polconversion has not worked properly at that particular frequency window. This failure may be related to a degraded quality in the phase-cal tones at YJ (see Sect. \ref{sec:instrumentalPhases}).  

From Fig. \ref{fig:FringePeak}, a clear conclusion can be drawn: The polconversion has produced much higher fringe amplitudes in the parallel-hand products of different sources at times separated by several hours, covering a large region of the sky with very distinct parallactic angles. Therefore, the polconversion is not only restricted to the calibration scans, but can be extrapolated to the whole experiment.

\subsubsection{Absolute EVPA Calibration}
\label{sec:EVPA}
In an ordinary VLBI observation using circular-polarization receivers, any instrumental phase offset between the $R$ and $L$ polarization channels, uniformly added to all the VLBI antennas of the interferometer, cannot be distinguished from a global rotation of the electric vector position angle (EVPA) of a linearly polarized source. This limitation in the EVPA calibration using circularly-polarized antenna feeds has traditionally been a major issue in VLBI polarimetric studies. 

Usually, in full-polarization VLBI, the use of data from other arrays (like the VLA) was required at similar epochs, in order to force a match between the EVPAs observed with VLBI and those from the other array. Needless to say, such an EVPA calibration approach suffers from important potential biases, related to the extended (polarized) scales that may affect the EVPAs measured with, e.g., the VLA, that may be completely resolved out at the VLBI resolution. 

However, as discussed in \citet{IMV2016} and \citet{Goddi2019}, an important advantage of the use of linear polarizers in VLBI, via the algorithm PolConvert, is that the absolute phase between the polconverted $R$ and $L$ polarizations (i.e., in essence, the absolute EVPA) is automatically calibrated, as long as a polconverted antenna is used as the reference station in the process of GFF. In the case of VGOS, all antennas have linear-polarization feeds (i.e., all of them are being polconverted), so the condition for an absolute EVPA calibration is always satisfied.

Having the phases between $R$ and $L$ absolutely calibrated allows us to combine the $RR$ and $LL$ visibilities right after the polconversion, so that the fringes corresponding to Stokes I can be obtained. This improves the S/N by $\sqrt{2}$, as compared to the independent processing of $RR$ and $LL$. However, for a correct combination of $RR$ and $LL$, the different orientations of the two antenna feeds (i.e., basically, the difference between the parallactic angles of the antennas) have to be accounted as well. Once the parallactic angle effects are removed from $RR$ and $LL$, both visibility products should have the same phases, assuming that Stokes V is negligible compared to Stokes I. 

We have tested this statement by studying the phases of the $RR/LL$ visibility ratio as a function of frequency (i.e., spw) and time. In order to be sensitive to the (small) potential deviations in the phases of the $RR/LL$ ratio, we have restricted the analysis to the scans with the highest S/N; in particular, we have selected those scans with S/N$>$10 in all spw (only 120 scans out of 1919 passed this restriction). 

\begin{figure}[ht!]
    \centering
    \includegraphics[width=0.66\textwidth]{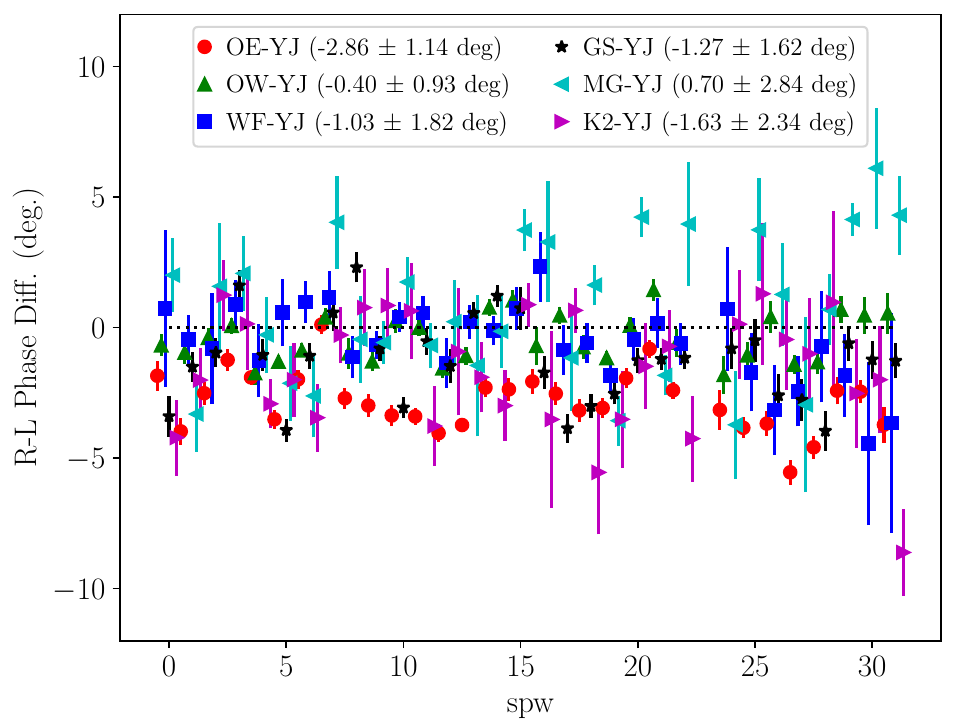}
    \caption[Scan-averaged $RR/LL$ phase across spectral windows for baselines to YJ.]{Scan average of the $RR/LL$ phase for all spw, using the scans with highest S/N (see text) for the baselines to YJ. The spw 23 has been removed from the analysis. Different baselines are slightly shifted in the spw axis, for clarity. The complete statistics for each baseline, averaged over spw, are included in the figure labels.}
    \label{fig:RLDiff}
\end{figure}

In Fig. \ref{fig:RLDiff}, we show the phases of $RR/LL$ (after accounting for the parallactic angles), averaged over all the detections with S/N$>$10 and referenced to the YJ antenna. The error bars shown in the figure correspond to the scatter of the $RR/LL$ phases across the scans (which involve different sources under different parallactic angles). It is remarkable the very little scatter in the $RR/LL$ phases across scans of baseline OE-YJ (error bars of the order of only one degree), which happens to be a relatively short baseline. Actually, for that baseline, the average $RR/LL$ phase over scans and spw is different from zero at a $\sim2-3\sigma$ level ($-2.9\pm1.1$\,deg). The unexpected (though small) phase offset between RR and LL in this baseline could be explained, for instance, by a small rigid rotation of the orthogonal X--Y feed pair with respect to its nominal orientation in the antenna mount.

For the rest of the baselines, all $RR/LL$ phases are compatible with zero and take typical values lower than 10 degrees, which is indicative of a proper conversion of the data into a circular polarization basis. Otherwise, the phases of $RR$ and $LL$ would not be rotating according to the parallactic angle of each scan, and the parallactic-angle-corrected phases of the $RR/LL$ ratios would not be compatible with zero.

\subsection{Instrumental Phase Calibration}
\label{sec:instrumentalPhases}
As explained in Sec. \ref{sec:paperpolconvert}, PolConvert takes the difference in phase-cal tone phases between the $X$ and $Y$ polarization channels (i.e., $X/Y$ phase-cals) into account when it estimates the cross-polarization gains (using the GCPFF algorithm) and also when it converts the data. This instrumental difference between polarizations was added to the cross-polarization bandpasses estimated with the GCPFF algorithm at each integration time. Therefore, because the $X/Y$ tone difference is already accounted for in the polconversion, the instrumental phases affecting the polconverted $RR$ and $LL$ visibilities are described by exactly the same phase-cal phases, which correspond to those in the $Y$ polarization channel (i.e., the reference polarization used in the GCPFF).

These remaining instrumental phases were estimated in a way similar to the so-called multitone mode of the fourfit program, which is the option used in the official VGOS calibration pipeline\footnote{\url{https://www.haystack.mit.edu/wp-content/uploads/2020/07/docs_hops_011_multitone_phasecal.pdf}}. The multitone mode estimates a tone delay, $\tau_{pc}$, from the phase-cals, $\phi_i$, found in each spw, by obtaining the peak of the tone fringe. Then, the instrumental tone phase, $\phi_{pc}$, is computed as the average of the tone residual phases centered at the reference spw frequency, $\nu_0$, when the tone delay has been subtracted. The equation used is:

\begin{equation}
    \phi_{pc} = \left< \phi_i - 2\pi\tau_{pc} (\nu_i - \nu_0)\right>.
    \label{eq:pcal}
\end{equation}

\begin{figure}[h]
  \centering
  \includegraphics[width=\textwidth]{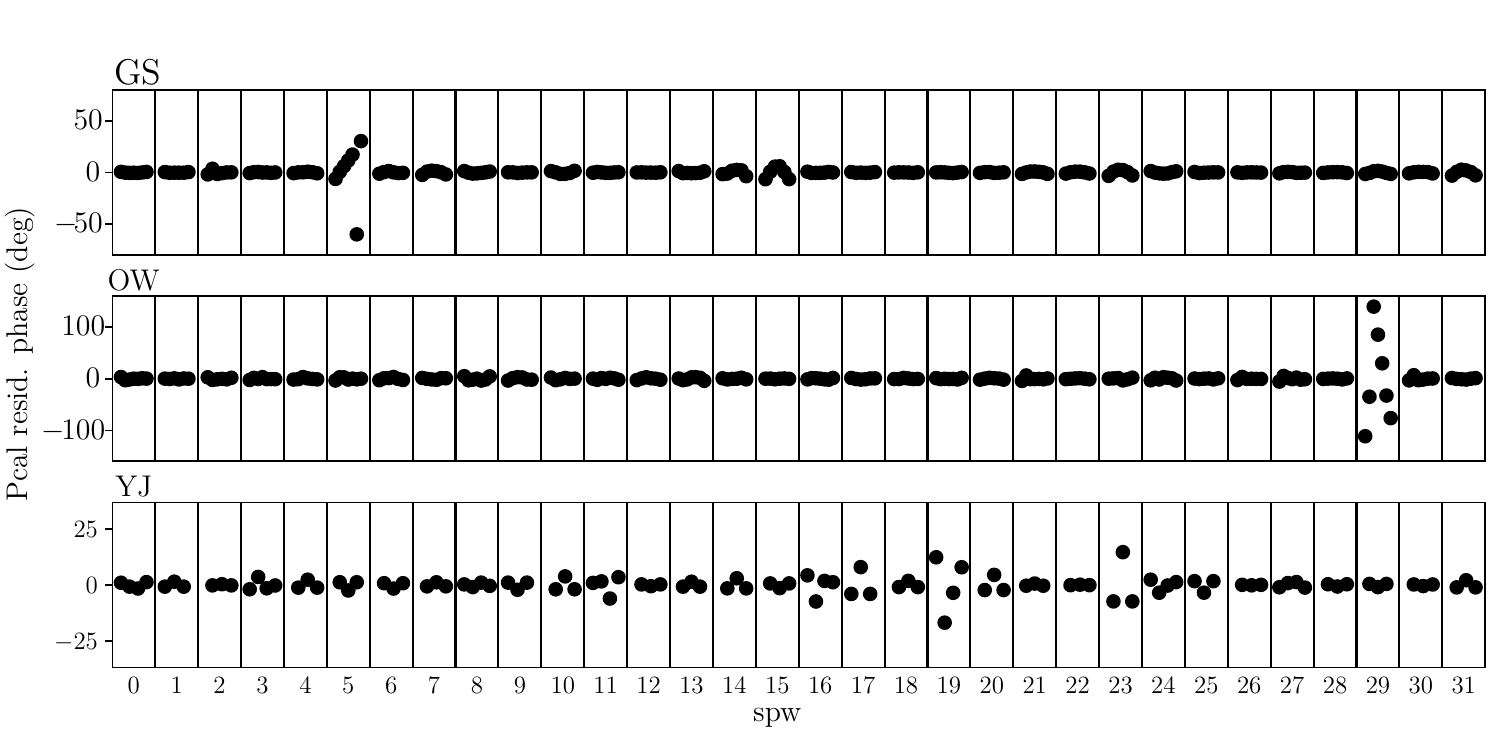}
  \caption[Residual phase-cal tones after subtraction of the estimated instrumental delays.]{Residual phase-cal tones (at 18:00 UT) after subtraction of the estimated instrumental delays, without removal of the bad entries. Clear outliers are seen (e.g., for GS, OW, and YJ in spw 5, 29, and 19, respectively). However, it is not possible to identify some of the outliers at YJ (e.g., in spw 17 and 23; see text).}
  \label{fig:phasecals}
\end{figure}

In Fig. \ref{fig:phasecals}, we show the residual phase-cal tone phases of the scan starting at 18\,UT for three representative antennas and after subtraction of the tone phases estimated using Eq. \ref{eq:pcal}. No outlier tones were removed. Some of the outliers are clearly seen, for instance, at spw 5 in GS or spw 29 at OW. Outlier tones result in incorrect estimates of the tone delays, $\tau_{pc}$, which in turn produce biased estimates of the instrumental tone phases, $\phi_{pc}$. Because several (6$-$7) tones are available within each spw, removing the outliers in these cases allowed us to retrieve the correct tone phases. However, for cases such as spw 23 of YJ, where only three tones are available for the whole spw, it is not possible to identify which tone is the actual outlier. This prevents a correct calibration of the tone delay, and as a consequence, the instrumental tone phase. This limitation clearly degrades the instrumental phase calibration at YJ for all the spw showing a behavior similar to that of spw 23.

\subsubsection{Manual Phase Alignment Across the VGOS Band}
\label{sec:ManualPhasecal}
After the phases were corrected using the phase-cal tones, as described in the previous section, we expected the visibility phases to be properly aligned across all the spw. However, this was not the case and was partially due, for instance, to bandpass effects from the receiver components prior to the injection of the phase-cal tones. In practice, the additional correction needed to properly align the phases across all the spw (quantities known as additive phases) was computed from the observation of a calibrator source. In the VLBI jargon, this procedure is known as manual phase-cal.

The ionosphere over each station may introduce strong effects in the visibility phases across a band as wide as that of VGOS. If the ionosphere is not accounted for in the manual phase-cal, strong biases in the estimate of dispersive and nondispersive delays from the global fringe fitting (see next section) may affect the final geodesy products.

In order to minimize the contamination from the ionosphere in our manual phase-cal (and in the visibilities, in general), we used estimates of the total electron content (TEC) along the line of sight of each telescope, computed from GNSS IONEX models. In particular, we employed the algorithm PyPhases (included in PolConvert), which is based on the TECOR task implemented in the Astronomical Image Processing System (AIPS)\footnote{http://www.aips.nrao.edu}. Our algorithm reads ionospheric models produced by Jet Propulsion Laboratory (JPL) \footnote{https://cddis.nasa.gov/archive/gnss/products/ionex} that provide maps with the TEC at each point on the Earth's surface every 2 hours, assuming a thin ionospheric layer. A simple geometric model of the ionosphere was then employed to calculate the vertical TEC at the point at which the signal passes the thin layer. Then, spherical trigonometry was used to map this vertical TEC to the line of sight. Finally, we eliminated the ionospheric contribution from the observed phases at each integration time according to the following equation:

\begin{equation}
\label{eq:tec}
\phi_{obs} = \phi_0 + 2\pi \left(\tau_{gr}(\nu - \nu_0) + \dot{\tau}(t - t_0) \right) + \frac{\kappa \cdot \text{dTEC}}{\nu},
\end{equation}

\noindent where $\phi_{obs}$ is the observed phase, $\phi_0$ is the fringe phase, $\tau_{gr}$ is the nondispersive delay, $\nu$ is the frequency, $\nu_0$ is the fringe reference frequency, $\dot{\tau}$ is the fringe rate, $t$ is the visibility time, $t_0$ is the fringe reference time, and dTEC is the difference of the line-of-sight TEC (sTEC) between the two antennas of the baseline ($\kappa = 40.3 \, m^3s^{-2}$ is a constant, relating the TEC and the nondispersive delay).

\begin{figure}[h]
  \centering
  \hspace{-0.82cm}
  \begin{adjustbox}{trim={175pt} {10pt} {155pt} {60pt}, clip, width=1.03\textwidth}
    \includegraphics{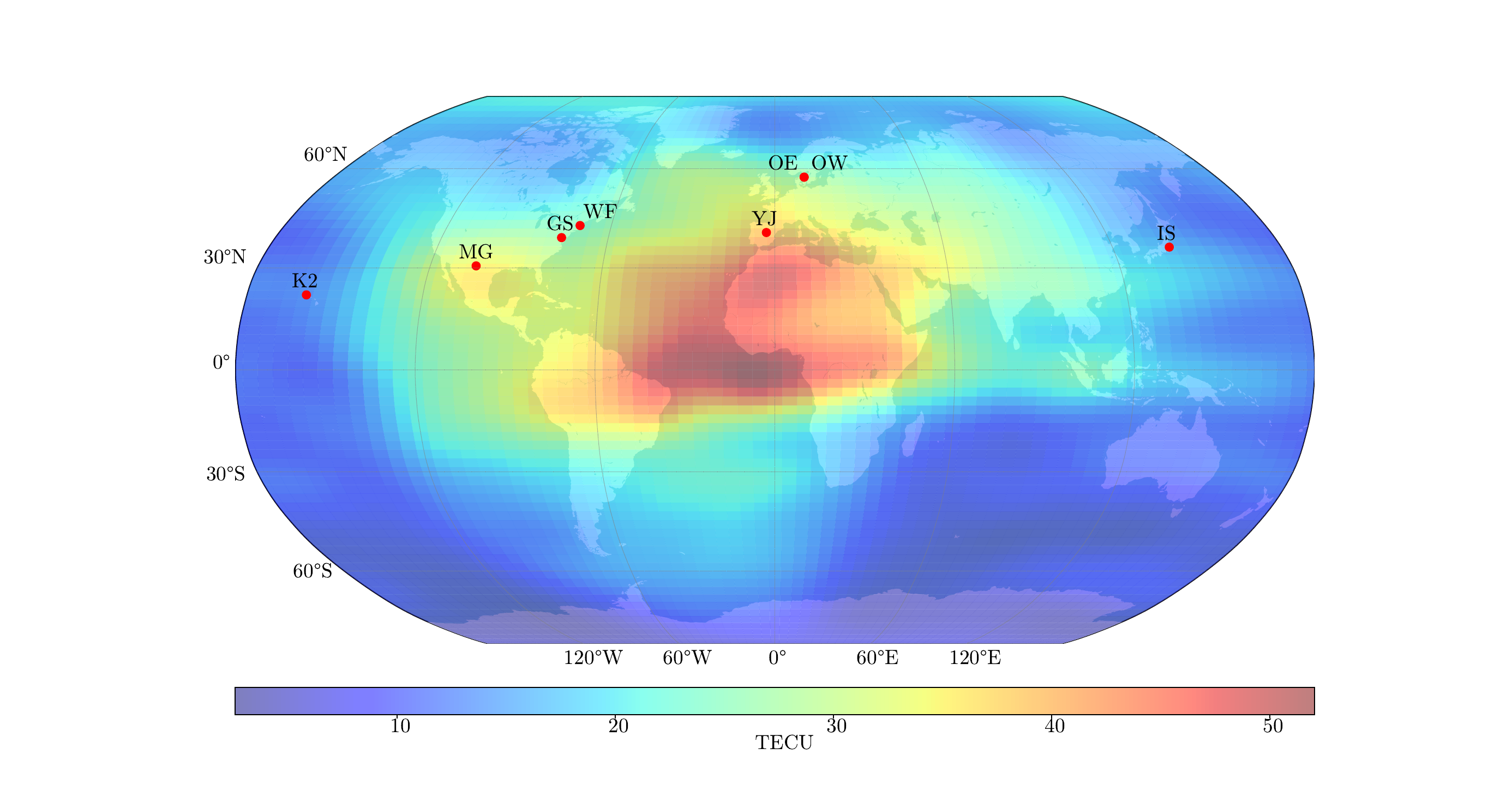}
  \end{adjustbox}
  \caption[Interpolated TEC map for scan 1749.]{Interpolated TEC map for scan 1749 (observed at 15.30\,UT), obtained using the global ionosphere map (IONEX) from the NASA Jet Propulsion Laboratory. The interpolation algorithm is similar to the one implemented in the AIPS task TECOR.}
  \label{fig:IONEX}
\end{figure}

In Fig. \ref{fig:IONEX}, we show the IONEX TEC model for the scan observed at 15:30\,UT, together with the geographical distribution of the antennas. In this scan, the ionosphere looks clearly more active for the antennas in Europe and in the continental US, where values of a few dozen TEC units (TECU) may affect their observations.

After the bulk of the ionosphere effects was removed from the data, we performed an ordinary global fringe fitting to each individual spw of the scan selected for the manual phase-cal with the IONEX model. From this GFF, we obtained the phase corrections that according to the manual phase-cal approach, definitely align the phases among all the spw in all the scans of the experiment, allowing us to execute the final wide-band global fringe fitting, as we explain in detail in the following section.

It is worth noting that any small difference (in the manual-phase-cal calibration scan) between the true ionosphere and the prediction from the IONEX model is inevitably absorbed in the estimated phase corrections. In any case, this effect is propagated to the remaining experiment as a constant (and small) residual ionosphere offset, and it is expected to affect the geodetic observables and not the dTEC.

In the official VGOS approach, a manual phase-cal is also performed, where many scans that have high S/N fringes on all four polarization products and the dTEC solution for each polarization product is estimated to be lower than 1 TEC unit \citep{Barrett2021}, are selected. However, in addition to not using any a priori information to subtract the dispersive delay, the main difference is that they use a nonglobal approach for the fringe fitting, while in our procedure, we performed a global fringe fitting, as we explain in the next section.

As a final remark, in our manual phase-cal, we impose a multiband delay (MBD) at each antenna that equals the median of its single-band delays (SBD) for all the spw. In this way, we can combine MBD and SBD in our WBGFF into one single delay quantity (i.e., a delay that describes the frequency dependence of phases within the spw, as well as across the whole VGOS band). This strategy is discussed in more detail in the next section.

\subsection{Wide Band Global Fringe Fitting (WBGFF)}
\label{sec:GFF}

A global fringe-fitting algorithm solves for the instrumental gains using all the data available for each scan, but parameterizing the gain information in an antenna-based manner. By fitting the data globally, we are making an optimum use of all the information available in a scan, while minimizing the parameter space (which is proportional to the number of antennas, whereas the data size increases quadratically with that number). Besides, the global gain solutions are more robust against source-intrinsic effects, compared to the baseline-based solutions. By globalizing the solutions, the closure phases \citep[quantities that encode information related to the brightness distribution of the observed sources, e.g.][]{TMS1986} remain unchanged in the data (i.e., not in the antenna gains) during the calibration, which allows us to perform image deconvolution. Actually, the effect of minimizing all the baseline phases, while keeping the phase closures unchanged, effectively concentrates the source brightness toward the image center. Baseline-based solutions, on the other hand, absorb all the effects related to the source structures (including closure phases) into their fitted quantities. 

To perform a global fringe fitting and be able to maintain the closure phases, all the visibilities at each scan must refer to the same time point, in this case, the moment in which the wavefront reaches the geocenter. However, in geodesy each visibility refers to the moment in which the wavefront reaches the reference antenna at each baseline. In this way, the position of the antennas can be known without performing other calculations, which is why a non-global approach is used in geodetic fringe fitting.

In VLBI observations, the antenna gains in an ordinary fringe fitting are modelled using three quantities: instrumental phases (discussed in the previous subsections), non-dispersive group delays, mainly due to the array geometry and atmosphere, and fringe rates. The group delays can be measured either within the frequency channels of a single spw (these are the single-band delays, SBDs) or from the phases among different spws (these are the multi-band delays, MBDs). In the official approach followed for VGOS, SBDs and MBDs are fitted independently, being the former taken from the stacking of the delay fringes across all the spw. In our case, though (and since SBD and MBD are indeed two ways of measuring the same physical quantity: the delay between signals arriving at the two antennas), we perform the manual phase-cal (see previous section) in such a way that SBDs and MBDs are well aligned (i.e., the median phase slope within the spw is aligned to the phase slope across the whole VGOS band). This way, we remove one of the fitting parameters in our WBGFF and only solve for the delay (i.e., MBD and SBD are combined). This method has some advantages, as compared to the approach followed by the official IVS calibration pipeline. On the one hand, it uses all the spectral channels of each spw in the fitting, whereas HOPS\footnote{\url{https://www.haystack.mit.edu/Haystack-observatory-postprocessing-system-hops}} collapses all the frequency information of each spw into one single value when it estimates the MBD \citep{Barrett2021}. In our case, by using all the spectral information in the fit of the MBD, we are effectively multiplying the MBD fringes by the Fourier transform of a 32 \,MHz boxcar function (i.e., the width of each spw, by means of the Convolution Theorem), hence decreasing the amplitudes of the MBD sidelobes and ambiguities. On the other hand, the WBGFF has only one delay parameter to model both, the SBDs and the MBDs, hence decreasing the number of degrees of freedom in the fitting.

In addition to this, if the fractional bandwidth is wide enough to detect dispersive effects due to the ionosphere, another parameter has to be introduced in the fringe fitting for each antenna: a dispersive delay, which depends on frequency as $\propto \nu^{-2}$. 

When fitting for dispersive and non-dispersive delays simultaneously, using Eq. \ref{eq:tec}, both parameters are usually highly coupled (i.e., the post-fit covariance matrix has large non-diagonal values), which may introduce biases in the estimate of the multiband non-dispersive delay (i.e., the most important geodetic observable in VGOS).

Our developed WBGFF algorithm performs simultaneous fitting of dispersive and non-dispersive terms, trying to minimize the problems related to the parameter coupling. Unlike the official approach \citep{Cappallo2014}, where fringe fitting is done separately for each baseline, we choose to perform the fit globally, using antenna-based quantities as parameters.

In this fitting process, we determine not only the phase and group delay but also the residual ionosphere. After subtracting the IONEX priors (using the same approach as the one described in Sect. \ref{sec:ManualPhasecal}), the bulk of the ionosphere effects are removed from the data, so we just adjust for the residual dTEC (i.e., the difference between the true ionosphere and that estimated from the IONEX model). 

The fringe search is initially performed on a discrete delay grid obtained from the Fourier transform of the visibility spectrum. A common approach to refining the location of the maximum is to increase the sampling of the delay spectrum by zero-padding the frequency-domain data. Although zero-padding does not add information to the data, it provides a more densely sampled representation of the fringe peak, at the cost of evaluating a larger Fourier transform.

In this work, the sub-bin location of the fringe peak is instead estimated using the interpolation method introduced by \citet{Quinn1994}. The Quinn estimator uses the complex Fourier coefficient at the discrete maximum and those in the two adjacent bins
to derive an analytical correction to the position of the peak. It therefore provides a continuous delay estimate without requiring a densely zero-padded Fourier transform. This is computationally advantageous in the WBGFF, where the fringe search must be repeated for multiple baselines and iterations.

The statistical performance of the estimator depends on the fringe signal-to-noise ratio and, critically, on the correct identification of the Fourier bin containing the main fringe lobe. When the correct bin is selected, thermal noise perturbs the three Fourier coefficients and increases the variance of the interpolated sub-bin displacement. For an isolated sinusoidal component in additive noise, Quinn-type estimators have favourable asymptotic properties, with a variance of the same order as that of statistically efficient frequency estimators \citep{Quinn1994, Quinn1997}. Their uncertainty nevertheless increases as the S/N decreases.

A more important limitation occurs in the low-S/N regime, where a noise fluctuation, a sidelobe, or a delay ambiguity may be selected instead of the main fringe peak. Since the Quinn estimator is local, it only refines the location of the previously selected bin and cannot recover the true delay when the wrong peak has been identified. Its performance can also degrade when the fringe spectrum contains several comparable peaks or departs substantially from the isolated-peak model.

Our WBGFF is implemented in the PyPhases script, which is part of the current distribution of PolConvert\footnote{https://github.com/marti-vidal-i/PolConvert}. Besides globalizing the solutions, it also determines possible TEC ambiguities (i.e., large dTEC values that deform the fringes so that a sidelobe becomes higher than the main peak, see Fig. \ref{fig:TEC-AMBG}). We summarize the procedure followed by PyPhases in the following lines. For each scan:

\begin{figure}
    \centering
    \includegraphics[width=\textwidth, trim={13pt 0pt 10pt 3pt}, clip]{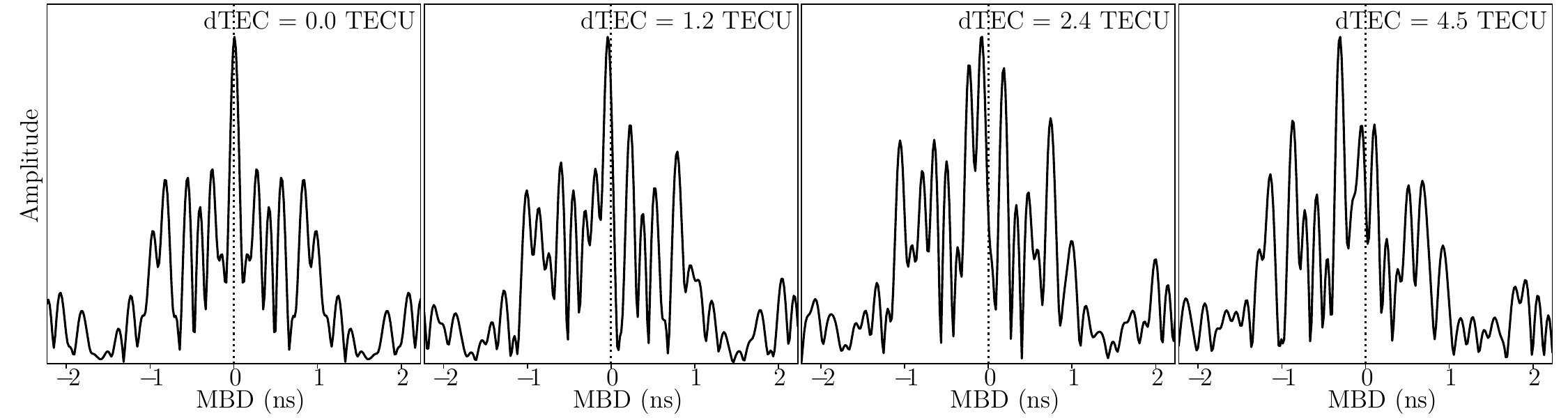}
    \caption[Simulated multi-band delay fringes for the VGOS frequency coverage.]{Simulated multi-band delay fringes for the VGOS frequency coverage, assuming a point source without noise. Different dTEC values have been added to each panel. The fringe peak with null dTEC is marked with a dotted line. Notice the ambiguity introduced for dTEC = 4.5\,TECU (i.e., the first sidelobe at the left becomes higher than the main fringe lobe).}
    \label{fig:TEC-AMBG}
\end{figure}

\begin{enumerate}

\item Perform an ordinary Global Fringe Fitting (GFF) using a fringe model with no dispersive terms (i.e., solving only for $\phi_0$, $\tau_{gr}$ and $\dot{\tau}$ in Eq. \ref{eq:tec}). From this fit, we obtain the so-called ``effective delay'', $\tau_{ef} = \tau_{gr}$, which corresponds to the fringe peaks in the data. We notice that, at this point, the fringe peaks are contaminated by the residual ionosphere.
\vspace{2mm}

\item Carry out a coarse exploration of TEC ambiguities, by adding/subtracting TECU and comparing the S/N of the resulting GFF solutions. The one with the highest S/N determines the TEC ambiguity.
\vspace{2mm}

\item Solve for the dispersive component using a modified version of Eq. \ref{eq:tec}:

\begin{equation}
\label{eq:taum}
\phi = \phi'_0 + 2\pi\left(\tau_{ef} \cdot \nu + \dot{\tau}(t - t_0)\right) + \frac{\kappa \cdot \text{dTEC}}{\nu} \left(1 + \frac{\nu^2}{\nu_{ef}^2}\right)
\end{equation}

\noindent where $\nu_{ef} = \frac{1}{\sum_i \nu_i^{-1}}$. In this new equation, $\dot{\tau}$ is fixed to the value found in step 1 and $\tau_{ef}$ is highly constrained to a small window around the value found in that step.
\vspace{2mm}

\item Recover $\tau_{gr}$ by just replacing:

\begin{equation}
    \tau_{gr} = \tau_{ef} + \frac{\kappa \cdot \mathrm{dTEC}}{\nu_{ef}^2}.
\end{equation}

\end{enumerate}

Using simulations, we have verified that this method accurately estimates the residual dTEC as long as it remains below 5\,TECU (a condition that can be relaxed by using a wider dTEC window in step 2). With these steps, the phases would be completely calibrated, having decoupled the dispersive delay caused by the ionosphere, eliminating the instrumental phases and obtaining the necessary non-dispersive group delay for geodesy.

The philosophy behind Eq. \ref{eq:taum} is that $\tau_{ef}$ has been designed to be equal (or very close) to the fringe peak when there is a small contamination from dispersive (ionosphere) delays. Hence, an ordinary (non-dispersive) GFF is able to provide a good estimate of $\tau_{ef}$, which can then be tightly constrained in the process of solving for the dispersive components in step 3. In that step, the model would basically solve for the non-linear component of the phase spectrum (i.e., the ``curvature'' in the phase model, which is only due to the ionosphere), while keeping the effective delay constrained.

On the other hand, in the HOPS procedure, they create a dTEC grid within a window of possible values and perform the fringe fitting within this grid, estimating the dTEC value as the one whose fringe has a higher S/N. By repeating this for each baseline, this makes the convergence slower than in our procedure, where dTEC is another parameter of the minimization.

\begin{figure}[t]
  \centering
\includegraphics[width=\textwidth, trim={10pt 0pt 10pt 0pt}, clip]{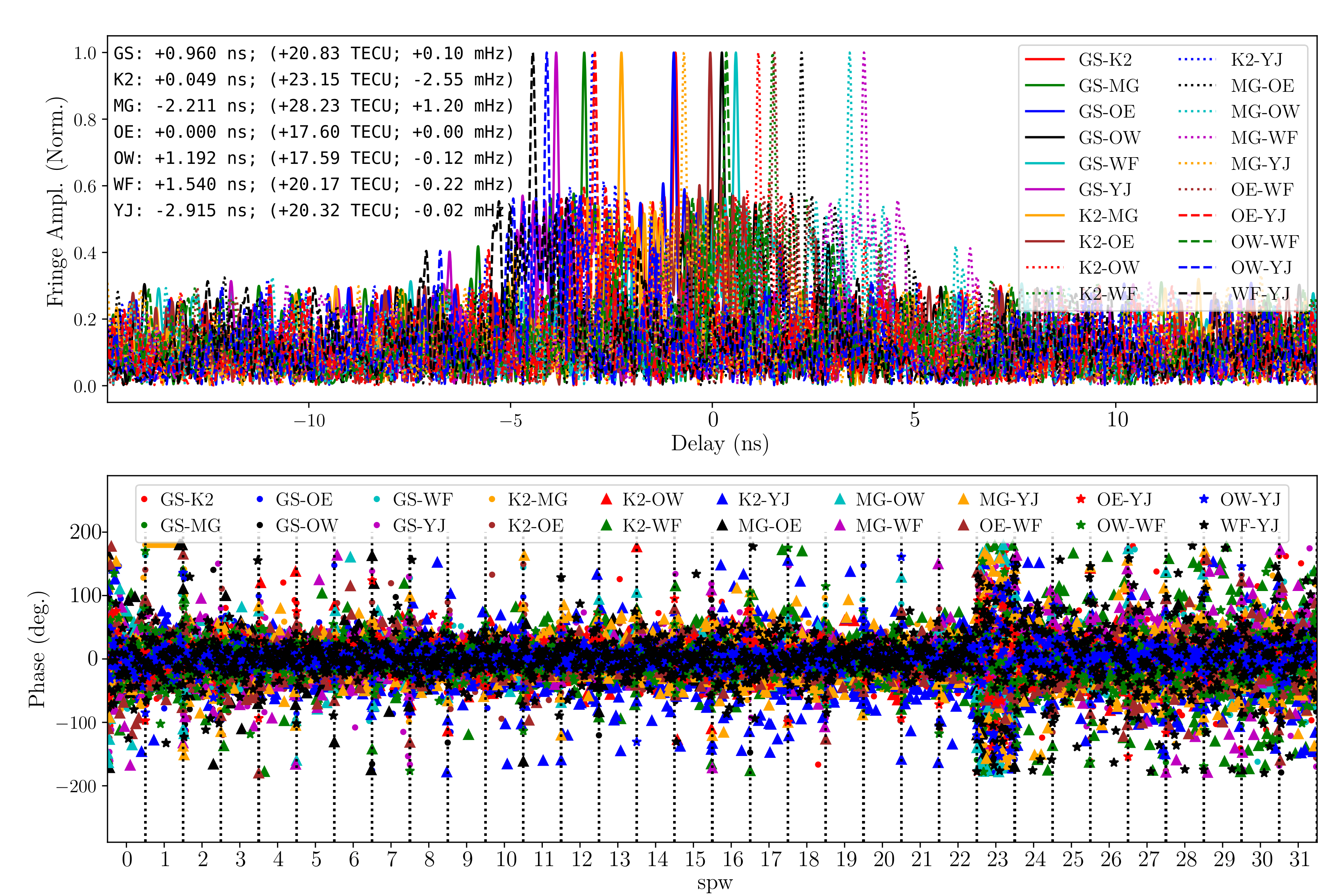}
  \caption[Wideband Global Fringe Fitting results for the first scan of the experiment (0059+581).]{Result of the Wideband Global Fringe Fitting for the first scan of our experiment. The observed source was 0059+581 and it was taken at 18:00\,UT. Top: Fringe amplitudes in delay space after removing the ionospheric dispersion. The values of the antenna-based gains from the GFF are shown at left (TECU is the full ionospheric contribution, including the prior from the IONEX map). Delay values are shown taking OE as the reference antenna. Bottom: visibility phases after the complete fringe calibration, shown as a function of frequency (ordered by spw). The IS antenna did not participate in this scan.}
  \label{fig:GFF}
\end{figure}

This modular WBGFF implementation (i.e., starting with an ordinary GFF, followed by a fit of the phase-curvature terms) is very efficient (a factor of several times faster than the HOPS implementation for VGOS, although the use of different programming languages could also affect the benchmarks). In Fig. \ref{fig:GFF}, we show the WBGFF results on the first scan of the experiment, which we find representative of the typical fringe solutions. All fringes in the delay space shown in the top panel (different colors for different baselines) are fitted using the antenna-based quantities listed in the left part of the panel (multiband delay, TEC, and fringe rate). The phase spectra for all the spw after applying the WBGFF calibration are shown in the bottom panel. No evidence of either significant residual phases, or multi-band or single-band delays, is seen, which indicates a successful calibration.

For clarity, the TEC values displayed in Fig.~\ref{fig:GFF} correspond to the absolute vertical TEC estimates obtained after adding the residual $d\mathrm{TEC}$ fitted by the WBGFF to the vertical TEC predicted by the IONEX model. Table~\ref{tab:tec_first_scan} separates these two contributions for each antenna in the first scan. There, it can be seen that the residual corrections derived by the WBGFF for this scan are much smaller than the absolute IONEX values. These small dispersive delay values can be decoupled from the non-dispersive delay, as described earlier in this section. The IONEX values listed in the table are the vertical TEC values; internally, these are mapped to the corresponding line-of-sight TEC when applying the ionospheric phase correction.

\begin{table}
    \centering
    \caption[Decomposition of the TEC values for the first scan of the experiment (0059+581).]{Decomposition of the TEC values shown in Fig.~\ref{fig:GFF} for the first scan of the experiment, corresponding to 0059+581. The second column gives the vertical TEC predicted by the IONEX model, the third column gives the residual $d\mathrm{TEC}$ fitted by the WBGFF, and the fourth column gives the total vertical TEC displayed in the figure.}
    \label{tab:tec_first_scan}
    \begin{tabular}{lrrr}
        \hline
        Antenna &
        IONEX TEC &
        WBGFF $d\mathrm{TEC}$ &
        Total TEC \\
        & (TECU) & (TECU) & (TECU) \\
        \hline
        GS & 20.85 & $-0.02$ & 20.83 \\
        K2 & 23.11 & $+0.04$ & 23.15 \\
        MG & 28.21 & $+0.02$ & 28.23 \\
        OE & 17.60 & $0.00$  & 17.60 \\
        OW & 17.60 & $-0.01$ & 17.59 \\
        WF & 20.15 & $+0.02$ & 20.17 \\
        YJ & 20.35 & $-0.03$ & 20.32 \\
        \hline
    \end{tabular}
\end{table}

\begin{figure}[t]
  \centering
  \includegraphics[width=0.85\textwidth]{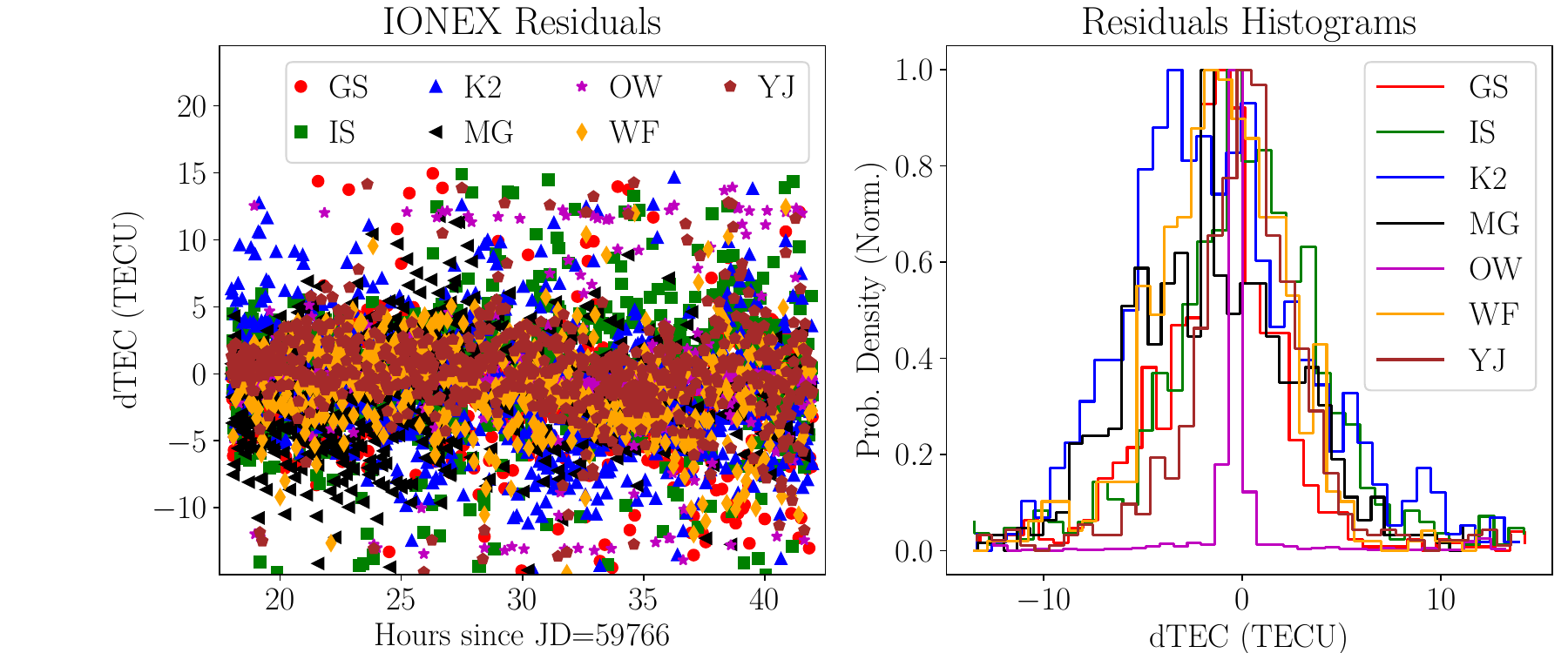}
  \caption[Residual dTEC values and their distributions, referenced to antenna OE.]{Left, residual dTEC (i.e., GFF estimates minus IONEX priors), referenced to antenna OE. Right, histogram distributions of the dTEC residuals, referenced to OE and normalized to the peak.}
  \label{fig:TECResid}
\end{figure}

Regarding the estimates of the dTEC from our WBGFF, the residual quantities found for the whole experiment, referenced to antenna OE, are shown in Fig. \ref{fig:TECResid}. In the left panel, the residuals are shown as a function of time, with different symbols for each antenna. The right panel shows the histogram distributions of the antenna-based dTEC residuals for the whole experiment. Most of the residual dTEC falls within a window of -10$-$10\,TECU, depending on the baseline length. A remarkable case is the dTEC of OW, which takes typical values below 1\,TECU. This is an expected result, since the very short baseline between the twin Onsala telescopes forces a very similar ionosphere for both. In any case, this result also gives a good argument in favor of a global approach for the fringe fitting: the very well constrained ionosphere over OW (when OE is taken as reference) is used to constrain the fit of all the other antennas, which also share baselines to the two Onsala twin stations.

Another point worth noticing in Fig. \ref{fig:TECResid} (right panel) is that the residual dTEC of K2 has a distribution peaking at a value clearly different from zero (around $-2$ to $-4$ TECU). This offset might be indicative of possible systematics in the IONEX model for a region above the Pacific Ocean.

\subsubsection{Comparison with the Haystack Pipeline}
The antenna-based GFF solutions from PyPhases can be re-written as baseline-based quantities easily, by forming the differences between the quantities of the antennas of each baseline. The delays (both dispersive and non-dispersive) can then be compared to the baseline-based solutions officially released for this VGOS epoch by the IVS, which were generated with the HOPS \texttt{fourfit} program following the control file generated by the Haystack VGOS pipeline \footnote{https://www.haystack.mit.edu} \texttt{vgoscf\_generate.py}, in its version on 8 September 2022. 

\begin{figure}[t]
  \centering
  \includegraphics[width=0.99\textwidth]{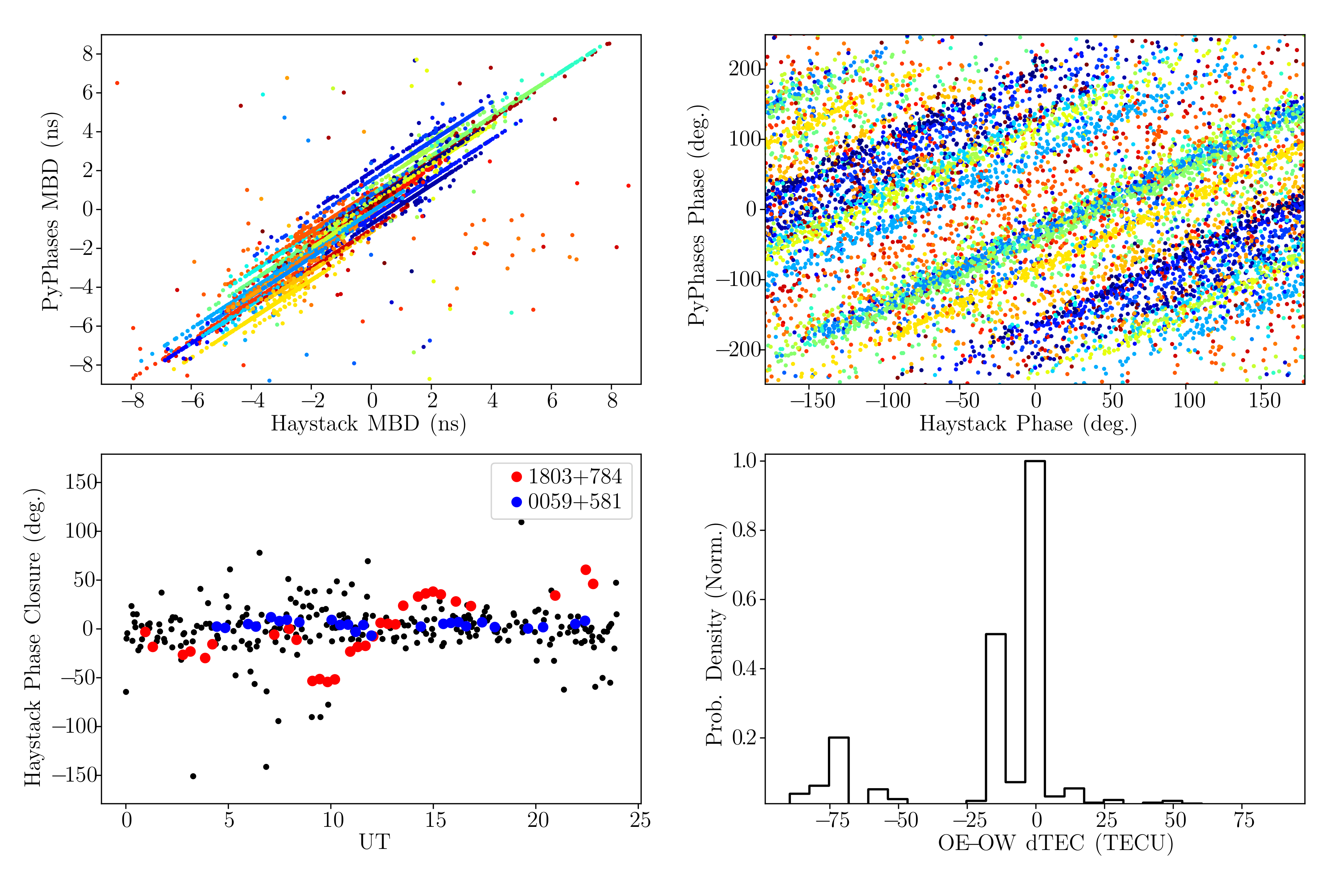}
  \caption[Comparison of fringe-fitting results between \texttt{fourfit} and PyPhases.]{Comparison of fringe-fitted quantities between \texttt{fourfit} (following the Haystack pipeline) and PyPhases. Top: Correlation plots for the multiband delays (left) and phases (right). Baselines are shown with different colors. Bottom: Closure phases from \texttt{fourfit}, avoiding the OE-OW baseline (left) and histogram of dTEC differences between OE and OW (also from \texttt{fourfit}).}
  \label{fig:HaystackVsGFF}
\end{figure}

We expect a correlation between PyPhases and Haystack results in the form of a tight linear trend with baseline-dependent offsets, directly related to the different methodologies followed for the manual phase calibration. This is actually what is seen for the multiband delays and fringe phases (Fig. \ref{fig:HaystackVsGFF}, top panels). However, the baseline-based phase solutions from \texttt{fourfit} contain non-zero closure phases (same figure, bottom-left panel), which are exactly zero (by construction) for the antenna-based PyPhases solutions.

Regarding the differential dispersive delay between the twin Onsala telescopes, OE and OW, in Fig. \ref{fig:HaystackVsGFF} (bottom right panel) we show the histogram from the \texttt{fourfit} Haystack pipeline. Even though the two stations should have very similar TEC corrections (since they are basically co-spatial, at scales of the Ionosphere structure function) we see clear deviations from zero, which may be related to a degraded fringe quality of the OE-OW baseline, likely due to common RFI, cross-talk between the two stations and correlation of their phase-cal tones (which can be partially mitigated by using notch filters). In our pipeline, we have not taken into account the OE-OW baseline data to avoid these drawbacks. However, because we use a global fringe fitting, we can obtain the dTEC independently, getting better results with a dTEC value close to 0. For comparison, the equivalent histogram from PyPhases is shown in Fig. \ref{fig:TECResid} (magenta line in the right panel).

\subsubsection{Comparison with CASA Fringe Fitting}
\label{sec:CASAComparison}
\begin{figure}[t]
  \centering
  \includegraphics[width=\textwidth]{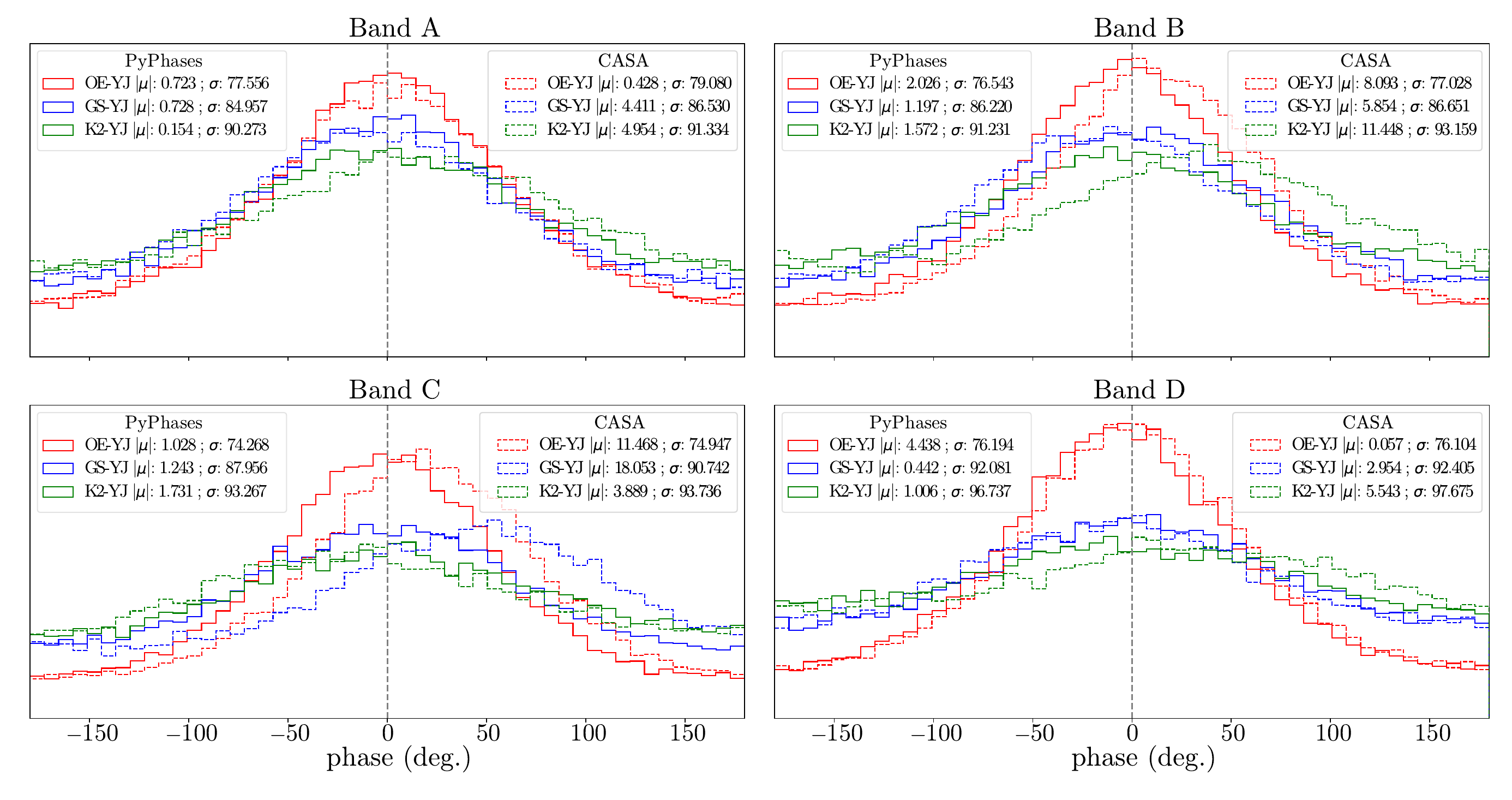}
  \caption[Phase distributions after fringe fitting with PyPhases and CASA.]{Phase histograms of the visibilities of scan 1362 (observed at 10:50\,UT), of the source 0059+581, for a selection of baselines, after the Global Fringe Fitting calibration. Solid lines, using our WBGFF algorithm implemented in PyPhases; dotted lines, using the implementation in CASA (see text). In the boxes, we show the absolute value of the mean (\(|\mu|\)) in deg. and the standard deviation (\(\sigma\)) in deg. for each distribution.}
    \label{fig:GFFvsCASA}
\end{figure}

The performance of our WBGFF is also comparable to (and, for the data reported here, slightly better than) other WBGFF implementations, like the one included in the latest versions of the CASA software \citep{CASA2022}. As an example, Fig. \ref{fig:GFFvsCASA} shows phase histograms (after the WBGFF calibration) of the scan observed at 10:50\,UT (source 0059$+$581, which is almost point-like), for a selection of baselines. In principle, all phase distributions should peak at values as close to zero as possible (as long as the least-squares minimization has converged successfully) and have standard deviations related to the S/N of each baseline.

For the case of our WBGFF, implemented in PyPhases, we find that the phase distributions are usually more centered around zero than the corresponding distributions obtained with the CASA implementation. This can be seen quantitatively by looking at the mean and standard deviation of the phase distributions (shown in the legends of Fig. \ref{fig:GFFvsCASA}). Some extreme cases are seen, for instance, in Band C, where the GS-YJ has an average phase of 18\,deg. (and a peak around 50\,deg.) for the CASA case, whereas it takes a value around only 1\,deg. for the PyPhases case.

The usual method in CASA also consists of carrying out a manual phase-cal, in which they calculate the single-band delay for each spw. However, their fringe fitting works differently, starting with a 2D Fast Fourier Transform that does not include dispersion and subsequently performing a least squares minimization stage that optionally includes a dispersion term. Furthermore, no a priori information is included in CASA. After applying the corrections from the manual phase-cal, the delay is calculated, either single-band delay (each spw separately) or combining every spw for calculating a multiband delay, without combining both types of delays \citep{vanBemmel2022}.

The improved phase centring obtained with PyPhases most likely results from the combined effect of several methodological differences. First, the dispersive and non-dispersive contributions are treated jointly over the full VGOS frequency range. If the ionospheric term is not included in the initial wideband search, part of its frequency-dependent curvature may be absorbed into an incorrect non-dispersive delay solution or may lead to the selection of a delay ambiguity. This effect is reduced in PyPhases by subtracting the IONEX prediction before the fit, so that only a much smaller residual $d\mathrm{TEC}$ has to be estimated.

In addition, the instrumental phase offsets and single-band delay information from the individual spectral windows are incorporated into a common antenna-based wideband calibration. This provides a coherent description of the phase slopes within each spectral window and across the full frequency range. The global antenna-based fit also uses all baselines in the scan simultaneously, increasing the redundancy of the solution and reducing the influence of individual noisy baselines while preserving the source-dependent closure information. The drawback is that data affected by a poorly calibrated spectral window may contaminate the global solution and must be removed, as the spw~23 in this experiment.

\subsection{Polarization Calibration}
\label{sec:polcal_obs}

After the phase calibration, we performed the polarization calibration independently for each VGOS band by solving for the instrumental polarization leakage terms (D-terms). This step was carried out with the \texttt{PolSolve} software \citep{PolSolve}, which implements a global fit of the full Radio Interferometer Measurement Equation (RIME) in order to solve simultaneously for the antenna polarization leakage and the source polarimetric structure. The method is based on the matrix formalism introduced by \citet{Hamaker1996}, where instrumental polarization appears as a non-diagonal Jones term coupling the nominally orthogonal polarization channels (as seen in Section \ref{sec:rime}).

In our case, the input data had already been converted to a circular polarization basis with \texttt{PolConvert} (Sect. \ref{sec:paperpolconvert}) and corrected for residual phase, delay, and rate terms through the wide-band fringe-fitting (Sect. \ref{sec:GFF}). At this stage, the dominant remaining non-diagonal instrumental corruption was the residual leakage between the right- and left-circular polarization channels. Even when these leakage terms are only at the level of a few percent, they can still bias the reconstruction of Stokes $Q$ and $U$, and, then, the polarized intensity and electric vector position angle (EVPA) obtained, particularly in sources with weak linear polarization \citep{Leppanen1995,Hales2017,PolSolve}.

The calibration was performed in CASA using a script that processes each band separately. First, all UVFITS files corresponding to each band were imported and concatenated into a single measurement set, so that the polarization solution could be derived globally using all sources together. This improves the robustness of the fit and helps disentangle source polarization from antenna-based instrumental effects.

Then, an initial full-Stokes deconvolution was produced for each source in order to define CLEAN masks and generate a first approximation to the source polarization structure. These initial masks were then kept fixed during the subsequent iterations. Starting from these source models, we alternated between imaging and instrumental-polarization calibration. In each iteration, all sources were re-imaged in Stokes $IQUV$ using the current masks. The updated models were then used for solving a new set of antenna D-terms for the whole band. In practice, we adopted ten iterations of this CLEAN/\texttt{PolSolve} loop, which was sufficient to stabilize the solutions.

Once the iterative process converged, the final D-term table was extracted and applied to the original UVFITS files of that band. In this way, we generated a new set of calibrated visibility files in which the residual instrumental polarization leakage had been removed.

\subsection{Hybrid Imaging}
\label{sec:HybridImaging}

As discussed in Sect.~\ref{sec:imaging}, imaging in VLBI is intrinsically coupled to calibration through the hybrid imaging procedure. In the present work, the primary role of hybrid imaging is the refinement of antenna-based amplitude gains after the Wide Band Global Fringe Fitting (WBGFF; Sect.~\ref{sec:GFF}).

In standard astronomical VLBI experiments, amplitude calibration is performed using measured system temperatures ($T_{\mathrm{sys}}$) together with well-characterized gain curves for each antenna. This allows a direct a priori conversion from correlator units to physical flux densities.

In the VGOS experiments analyzed in this work, this is not possible. Only a subset of stations provided reliable gain curves, and even fewer delivered calibrated $T_{\mathrm{sys}}$ measurements across the full wideband frequency range. As a result, a fully conventional amplitude calibration based on $T_{\mathrm{sys}}$ and gain curves could not be applied uniformly to the array.

After applying the WBGFF solutions, the visibilities are fully corrected for instrumental phases, non-dispersive and dispersive delays, and fringe rates. The dominant remaining calibration uncertainty is therefore associated with amplitude gain errors.

To mitigate these effects, we performed hybrid mapping independently in each of the four VGOS bands (A, B, C, and D). Imaging was initiated using the fully phase-calibrated visibilities. After a few iterations of CLEAN and phase self-calibration, converging rapidly due to the preservation of closure quantities, amplitude self-calibration was introduced.

Amplitude self-calibration was applied conservatively. Solution intervals were initially chosen to be long compared to the scan duration in order to avoid distorting the relative flux-density scale across the array. Only once stable phase solutions and a reliable source model were obtained, the timescale of the amplitude gains was refined.

\subsection{Model Fitting}
\label{sec:model_fitting}

To characterize the source structure, we performed model fitting directly to the visibility data following a procedure based on \citet{uvmultifit}. In this approach, the sky brightness distribution is represented as a superposition of a small number of Gaussian components, which provide a compact and physically meaningful description of the source morphology in the visibility domain.

For each source, the model consists of a core component, typically represented by an elliptical Gaussian, and a set of additional Gaussian components describing the extended jet emission. The number of jet components is determined on each source, depending on its structure.

The fitting procedure is initially performed in total intensity (Stokes $I$) using VGOS C band, that provides a good compromise between angular resolution and signal-to-noise ratio. The parameters of the Gaussian components include flux density, position offsets, size, axial ratio, and position angle. These parameters are optimized by minimizing the residuals between the observed and model visibilities.

To ensure a consistent description of the source structure across frequencies, we impose physical constraints on the model when fitting the multi-frequency data. In particular, the parameters of the optically thin jet components (positions and sizes) are assumed to be frequency-independent, as expected for synchrotron emission in the optically thin regime. Only their flux densities are allowed to vary with frequency. In contrast, the position of the core component is allowed to vary with frequency in order to capture the core-shift effect, which arises from synchrotron self-absorption in the innermost regions of the jet (as explained in Sect. \ref{sec:jetemissionmodel}).

The model fitting is performed independently for each intermediate frequency (IF), allowing us to exploit the full spectral resolution of the VGOS data. This strategy is essential to accurately trace the frequency dependence of the core position and to derive reliable core-shift measurements.

Finally, the polarization structure (Stokes $Q$, $U$, and $V$) is modeled using the same Gaussian components derived from Stokes $I$. In this case, only the flux densities of each component are fitted, while their positions and shapes are kept fixed. This ensures a consistent decomposition of the source into physical components across all Stokes parameters and frequencies, and allows for a robust determination of the polarized emission associated with each structural feature of the jet.

\subsection{WBGFF with Source Structure}
\label{sec:WBGFF_structure}

Once the source model has been derived from the visibility fitting described in Sect.~\ref{sec:model_fitting}, we used it as an input model for a new run of the Wide Band Global Fringe Fitting (WBGFF). In this way, the fringe fitting was no longer performed under the implicit assumption of a point-like source, but taking into account the resolved structure of each target.

In VLBI, source structure introduces baseline-dependent phase and delay terms that are not of instrumental origin. If they are not properly modeled, these contributions can bias the fringe solutions and propagate into the subsequent calibration and analysis. This effect is especially relevant in broadband geodetic observations, where the precision of the delay observables is high enough for source structure to become a non-negligible limitation.

In practice, our procedure consists of repeating the fringe fitting after replacing the point-source assumption by the source model obtained from the previous imaging and model-fitting stage. The model is incorporated only in the fringe-fitting step, that is, as a description of the expected source visibilities during the estimation of phases, delays, dispersive terms, and rates. Therefore, the aim of this stage is to improve the calibration by removing the contribution associated with resolved emission from the fringe solutions, obtaining more accurate gains.

Since the quality of the fringe solutions depends on the source model, and the source model itself improves once better-calibrated visibilities are available, the procedure is iterative. We found that the solutions converge very rapidly. In practice, after the second iteration, no further appreciable improvement is observed, so we adopted the solutions from this second WBGFF run as those defining the phase calibration. For the 441 scans in which both fringe-fitting runs converged to compatible solutions, subtracting the source models produced only minor changes in the parameters. The effect was larger for the more structured sources, reaching approximately 0.45\% for 3C418 and 0.53\% for DA426 in the phase relative change, while values of 0.07\% for 0059+581 and 0.09\% for 1803+784 were obtained.

\subsection{Calibrated Visibilities}
\label{CalibratedVisibilites}
\begin{figure}[H]
    \centering
    
    \includegraphics[width=\textwidth, trim={9pt 0pt 3pt 0pt}, clip]
    {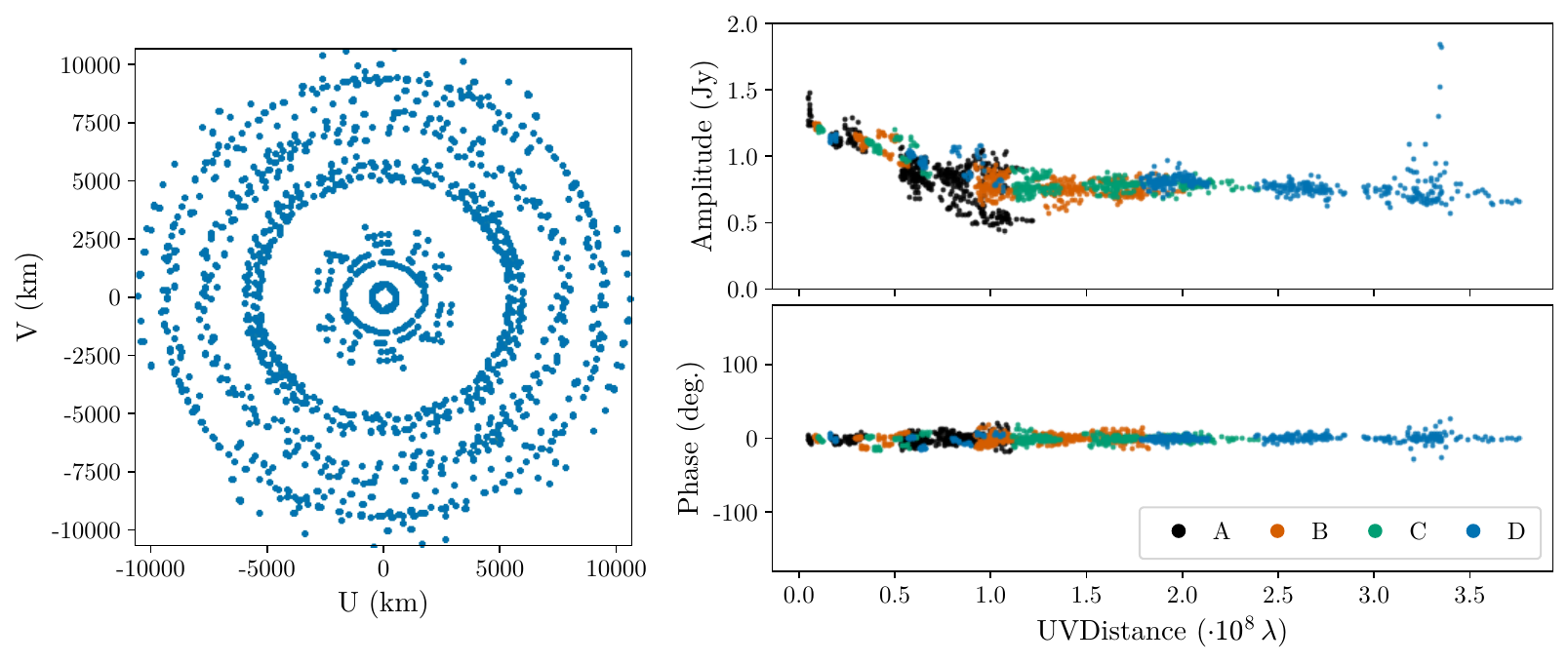}
    
    \vspace{0.5cm}
    
    \includegraphics[width=\textwidth, trim={9pt 0pt 3pt 0pt}, clip]
    {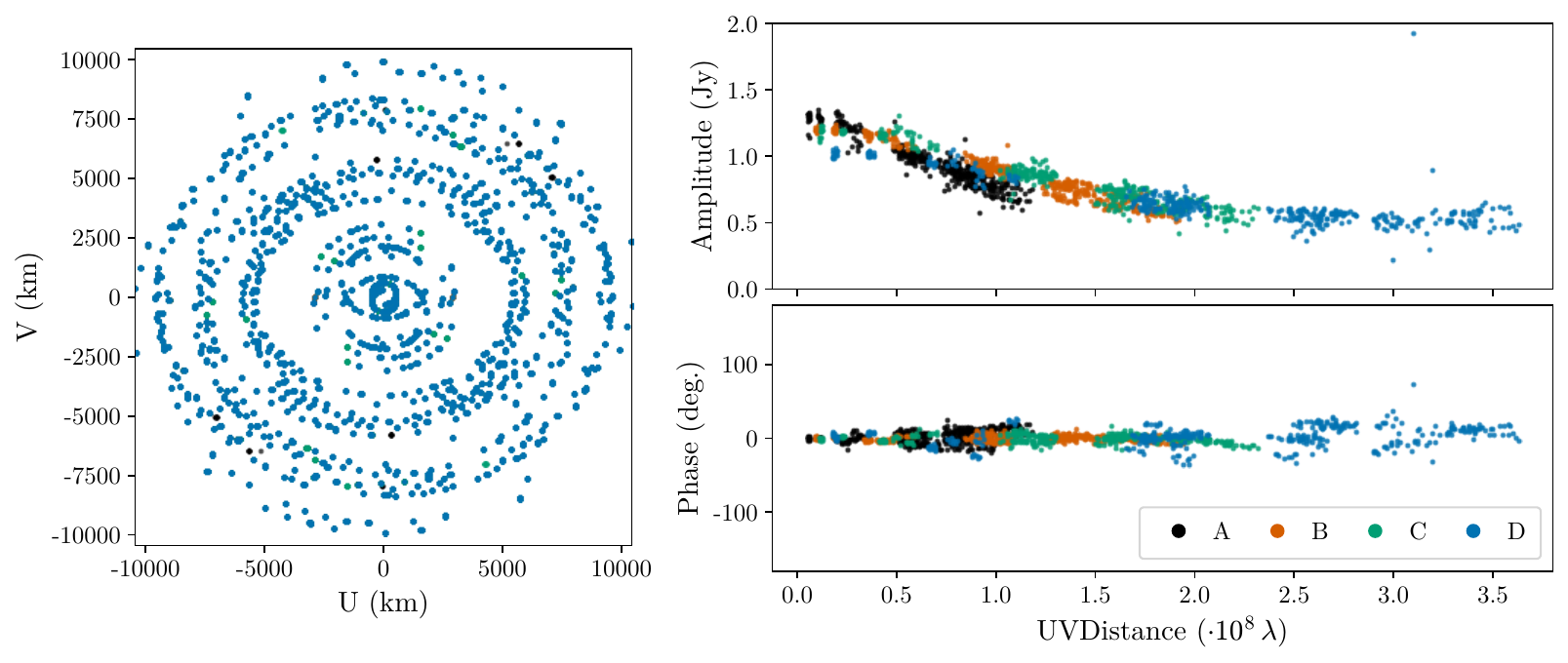}
    
    \caption[UV coverage and calibrated visibilities for sources 1803+784 and 0059+581 in experiment VO2187.]{UV coverage and calibrated visibilities for all the VGOS frequency bands in experiment VO2187. Top: source 1803+784. Bottom: source 0059+581. Left panels show the UV-plane distribution (in km) including conjugate points for bands A–D. Right panels show the visibility amplitude (top) and phase (bottom) as a function of UV distance (in units of $10^8 \lambda$).}
    
    \label{fig:CalVisibCombined}
\end{figure}

The data calibration procedure described in this work yields the observed visibilities properly calibrated, both in phase and amplitude. In Fig. \ref{fig:CalVisibCombined}, we show the final results after the phase calibration with dispersive fringe-fitting and the amplitude calibration, for two selected sources. We plot the phases and amplitudes of the RR and LL visibilities as a function of the distance in Fourier plane, showing the expanded UV coverage (thanks to all VGOS bands) in different colors. We obtain, for all bands, phases close to $0$ and amplitudes reflecting the structure of our sources, as we will see after imaging in the following section. Qualitatively similar results are obtained for the other sources.

\pagebreak
\section{Multi-frequency Imaging}
\label{sec:mfimaging}

After the calibration procedure described in the previous sections, the final VGOS visibilities were used to reconstruct the source brightness distribution across the full frequency coverage of the experiment. For this purpose, we used a regularized maximum likelihood (RML) imaging approach using the \texttt{eht-imaging} software \citep{Chael2016,Chael2018}. As discussed in Sect.~\ref{sec:imaging}, RML imaging is particularly useful for VLBI data because it allows fitting interferometric observables directly, while including regularization terms that introduce prior information about the image structure.

In the present analysis, the imaging was performed using closure phases (Eq. \ref{eq:closurephases}), log-closure amplitudes (Eq. \ref{eq:logclosureamps}), and visibilities (Eq. \ref{eq:visibility}) as data products. Closure quantities are insensitive to station calibration errors, as discussed in Sect. \ref{sec:VLBIobservables}. This is especially important for VGOS observations, which were not originally designed for astronomical imaging and for which residual calibration uncertainties (especially in amplitudes) may remain even after a proper calibration. The use of closure quantities therefore reduces the dependence of the final images on antenna-based errors, while the inclusion of visibility amplitudes helps to preserve part of the absolute flux information.

The total-intensity images were reconstructed using the multi-frequency mode implemented in \texttt{eht-imaging} \citep{Chael2023}, as discussed in Sect.~\ref{sec:imaging}. The image at each observing frequency is then obtained as
\begin{equation}
I_\nu =
I_0
\exp\left[
\alpha \log\!\left(\frac{\nu}{\nu_0}\right)
+
\beta \log^2\!\left(\frac{\nu}{\nu_0}\right)
\right].
\label{eq:mf_intensity_imaging}
\end{equation}
In our implementation, the three terms of this expansion (i.e., $I_0$, $\alpha$, and $\beta$) were allowed to vary. The reconstruction therefore solved simultaneously for the reference Stokes I image, the spectral index map, and the curvature map, minimizing the difference between the quantities of the model and the actual data in the four bands simultaneously.

\begin{figure}[H]
    \centering
    \begin{subfigure}{0.495\textwidth}
        \centering
        \includegraphics[width=\linewidth]{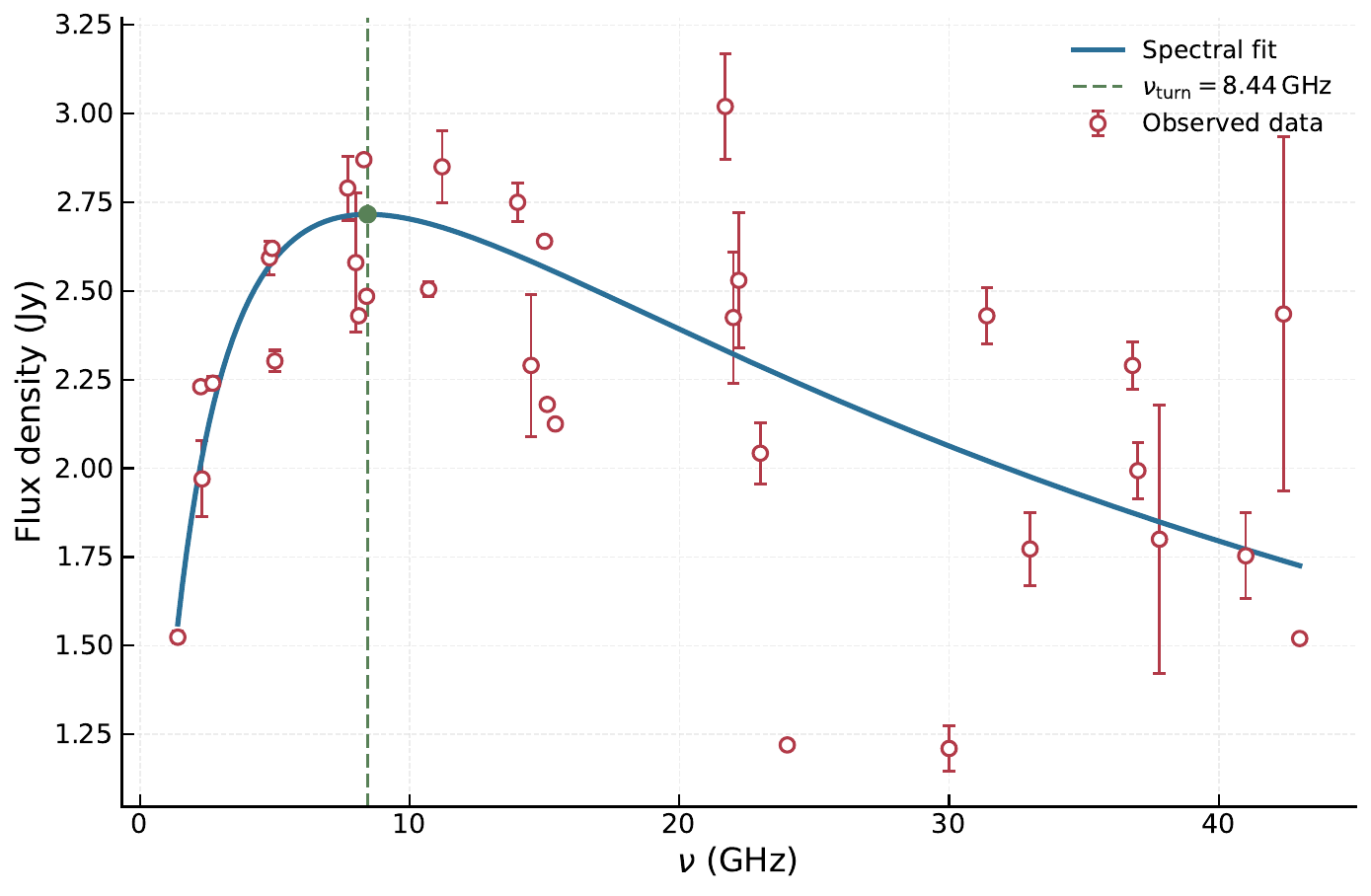}
        \caption{Linear scale}
        \label{fig:1803_spectrum_linear}
    \end{subfigure}
    \hfill
    \begin{subfigure}{0.495\textwidth}
        \centering
        \includegraphics[width=\linewidth]{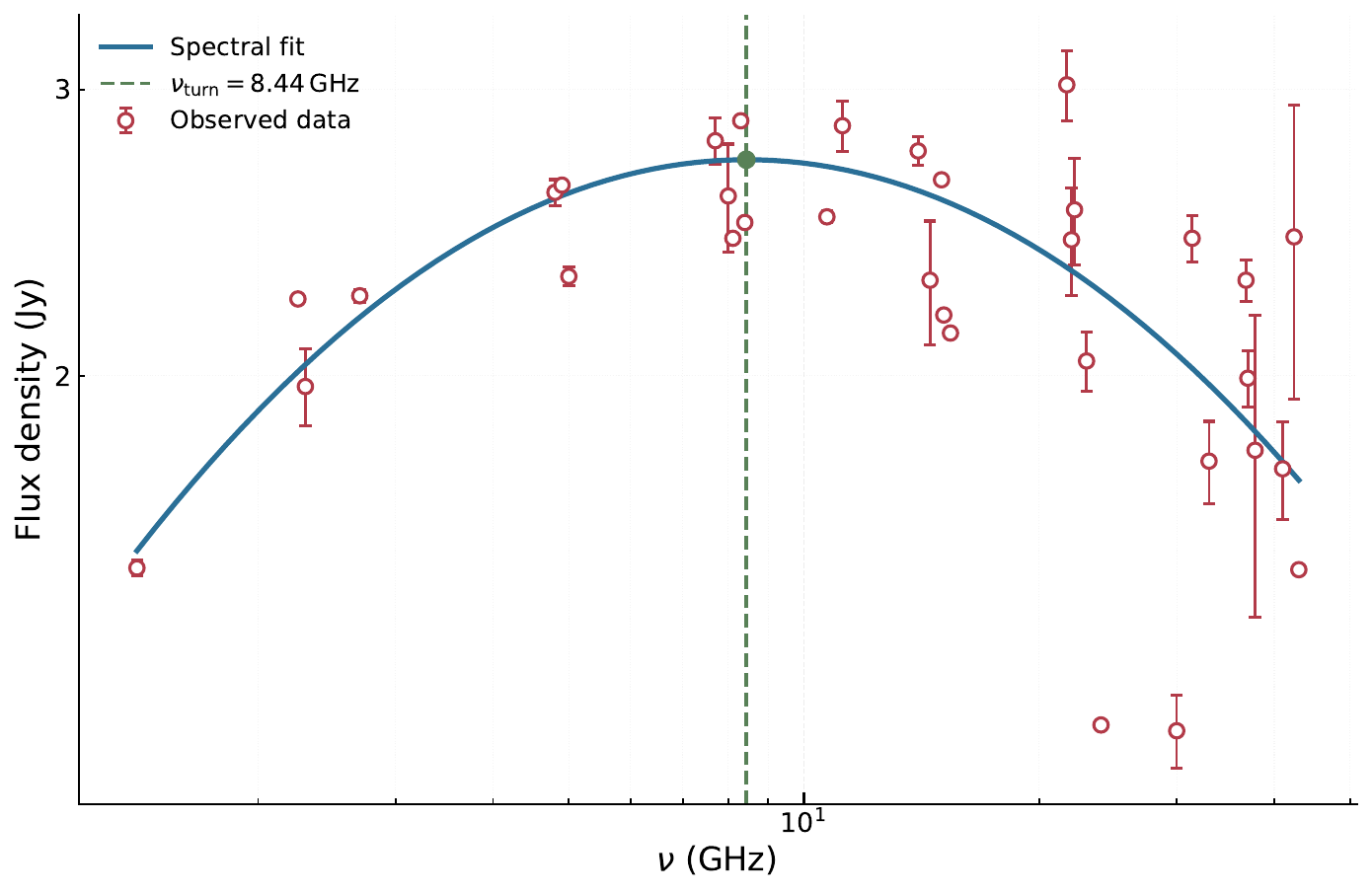}
        \caption{Logarithmic scale}
        \label{fig:1803_spectrum_loglog}
    \end{subfigure}
    \caption[Total-intensity spectrum of 1803+784 across the VGOS frequency range.]{Total-intensity spectrum of 1803+784 across the VGOS frequency range. Observations taken from 1981 to 2015. Given the highly variable nature of this blazar and the large range of years available, significant data dispersion is expected. Nevertheless, this interpretation is sufficient for estimating the turnover frequency, that lies within the VGOS observing frequencies.}
    \label{fig:1803_spectrum}
\end{figure}

Nevertheless, the official release of \texttt{eht-imaging} only includes the first two terms of the equation ($I_0$ and $\alpha$). This is because in the EHT data there are only 2 different bands, so curvature cannot be obtained. Furthermore, the EHT collaboration works at higher frequencies \citep{Issaoun2023}, where the turnover frequency is far from the observing range and the spectrum can be approximated with a constant slope. In VGOS, we have four frequency bands, so the curvature of the spectrum can be extracted from the data. Furthermore, the turnover frequency for the synchrotron radiation in this type of AGN lies close to our working frequencies, so it must be taken into account. This can be seen, for instance, in Fig. \ref{fig:1803_spectrum}, where we show the flux density measurements used for the radio spectrum of 1803+784 were retrieved from the NASA/IPAC Extragalactic Database (NED)\footnote{\url{https://ned.ipac.caltech.edu/}} \citep{Helou1991}, using data from \citet{Lister2015} and \citet{Lee2017}.

To include the spectral curvature effect in the imaging process, the multi-frequency algorithms of eht-imaging were modified. In the original implementation, the model vector solved by the optimizer contained the reference image and the spectral index map. This vector was extended to also include the curvature map, so that the optimized quantities became $(I_0, \alpha, \beta)$. In our modified version, the image compared with the data at each frequency is computed from Eq.~\ref{eq:mf_intensity_imaging}, rather than from a pure power-law dependence.

The code was extended to allow regularization of the curvature map itself, using a total variation term, in order to avoid small-scale fluctuations in $\beta$ that could otherwise be driven by noise or residual calibration errors. This allowed the information from the four VGOS bands to constrain the three maps simultaneously during the same optimization.

This approach is particularly suitable for VGOS data. The four VGOS bands cover a wide frequency range, from approximately 3 to 11\,GHz, but each individual band provides only a limited sampling of the Fourier plane. Imaging each band independently would discard part of the information available in the experiment. By reconstructing all bands simultaneously, the algorithm uses the combined multi-frequency uv-coverage to constrain a single model of the source brightness distribution as a function of frequency. In practice, at each step of the optimization the current multi-frequency model is evaluated at the frequency of each observation and compared with the corresponding closure phases, closure amplitudes, and visibilities. This same principle motivates multi-frequency imaging developments for the EHT and the ngEHT \citep{Chael2023}.

A further advantage of this method is that the frequency-dependent images are reconstructed on a common spatial grid. When only closure
quantities are used, the absolute source position is not preserved, because closure phases are invariant under a translation of the entire
image. Therefore, the reconstruction does not provide bona-fide absolute astrometry.

The simultaneous multi-frequency parameterization nevertheless provides an internally consistent relative registration between bands. The spectral-index and spectral-curvature regularizers favour solutions without arbitrary frequency-dependent offsets, and this approach has been shown to avoid evident image-misalignment artefacts in a range of simulated and real data sets \citep{Chael2023}. However, this relative registration is not equivalent to phase-referenced astrometry. Residual offsets may remain, particularly for poorly resolved sources or when few common structural features are present between bands. Consequently, the method can be used to study frequency-dependent structural variations and apparent relative shifts, but it does not by itself provide a bona-fide astrometric measurement of the core shift. The imaging was carried out iteratively. We started from a smooth Gaussian prior image, with a size large enough to cover the expected compact and extended emission of the sources. The initial Stokes I image was then obtained using closure data terms. The solution was progressively refined by including visibilities in the final stages of the reconstruction.

\subsection{Optimization of the RML Hyperparameters}

The RML objective function used in the imaging process included both data fidelity terms and regularizers. Since the final image can depend on the relative balance between these terms, we performed a set of controlled imaging simulations to determine a suitable combination of data products, regularizers, and weights before applying the method to the real VGOS data.

For this purpose, we generated synthetic Stokes I source models. Three different morphologies were considered: a point source, used as a reference case for an unresolved compact emitter; a triple component source, consisting of a compact core and two downstream jet components; and an extended source, designed to test the ability of the imaging procedure to recover resolved emission. In the triple component and extended models, the jet emission was kept fixed across frequency, while the apparent core position was allowed to vary with frequency in order to mimic the presence of a real core shift. One synthetic image was generated for each of the four VGOS bands. The field of view and pixel scale were chosen to match those used in the real imaging analysis.

Synthetic observations were then generated with eht-imaging from these models using the sampling of the real VGOS observations in each band, including the same array, observing times, baseline coordinates, and $u$-$v$ coverage. Thermal noise was added according to the uncertainties of the template observations, while no additional station-based gain or phase corruptions were introduced. The resulting synthetic visibilities could therefore be processed with the same imaging procedure as the real data.

The simulated data sets were imaged using the multi-frequency RML algorithm. Several combinations of data terms were explored, including closure phases, closure amplitudes, and visibilities. The use of closure quantities is particularly important in this context, since they are robust against residual antenna calibration errors, whereas the visibility amplitudes help to preserve the overall flux scale. Different image regularizers were also tested, including smoothness, compactness, sparsity, flux conservation, and spectral regularization terms. The optimization was performed over three successive imaging rounds, allowing the relative contribution of the data and regularization terms to vary from one round to the next.

The search over this parameter space was carried out using a genetic algorithm. The weights were sampled from a discrete logarithmic grid, allowing the algorithm to explore a broad range of relative contributions without requiring an excessively expensive continuous search. Each individual in the genetic population represented a complete imaging prescription, including the data terms, regularizers, and weights used in each of the three imaging rounds. For each candidate solution, the full multi-frequency reconstruction was performed for the three synthetic source morphologies and for all four VGOS bands.

The quality of each reconstruction was evaluated by comparing the recovered image with the corresponding input model. Before the comparison, both images were convolved with the same restoring beam, corresponding to 40\% of the band~C beam. This ensured that the reconstructed image was not penalized for unresolved structure that cannot be recovered by the interferometric data. The comparison was performed using a combined score based on the structural dissimilarity index, DSSIM, derived from the structural similarity index, SSIM, introduced by \citet{Wang2004}. This metric is useful in this context because it does not only compare images pixel by pixel, but also accounts for structural similarity in a way that is closer to human visual perception. Additional terms were included to account for the normalized image error, the recovered total flux density, and the linear correlation between the reconstructed and true brightness distributions.

The metrics were computed over the full field of view, without applying an emission mask. Consequently, in the sparsest models, the large number of near-zero background pixels may increase the apparent similarity, especially for SSIM. This limitation was partly mitigated by combining SSIM with morphology-normalized NRMSE, image correlation, and a separate total-flux penalty, and by assigning the largest weight to the extended core--jet model. Nevertheless, the possible contribution of the empty background should be kept in mind when interpreting the absolute metric values.

\begin{figure}[htpb]
    \centering
    \includegraphics[width=0.77\textwidth, trim={0pt 0pt 0pt 1pt}, clip]
    {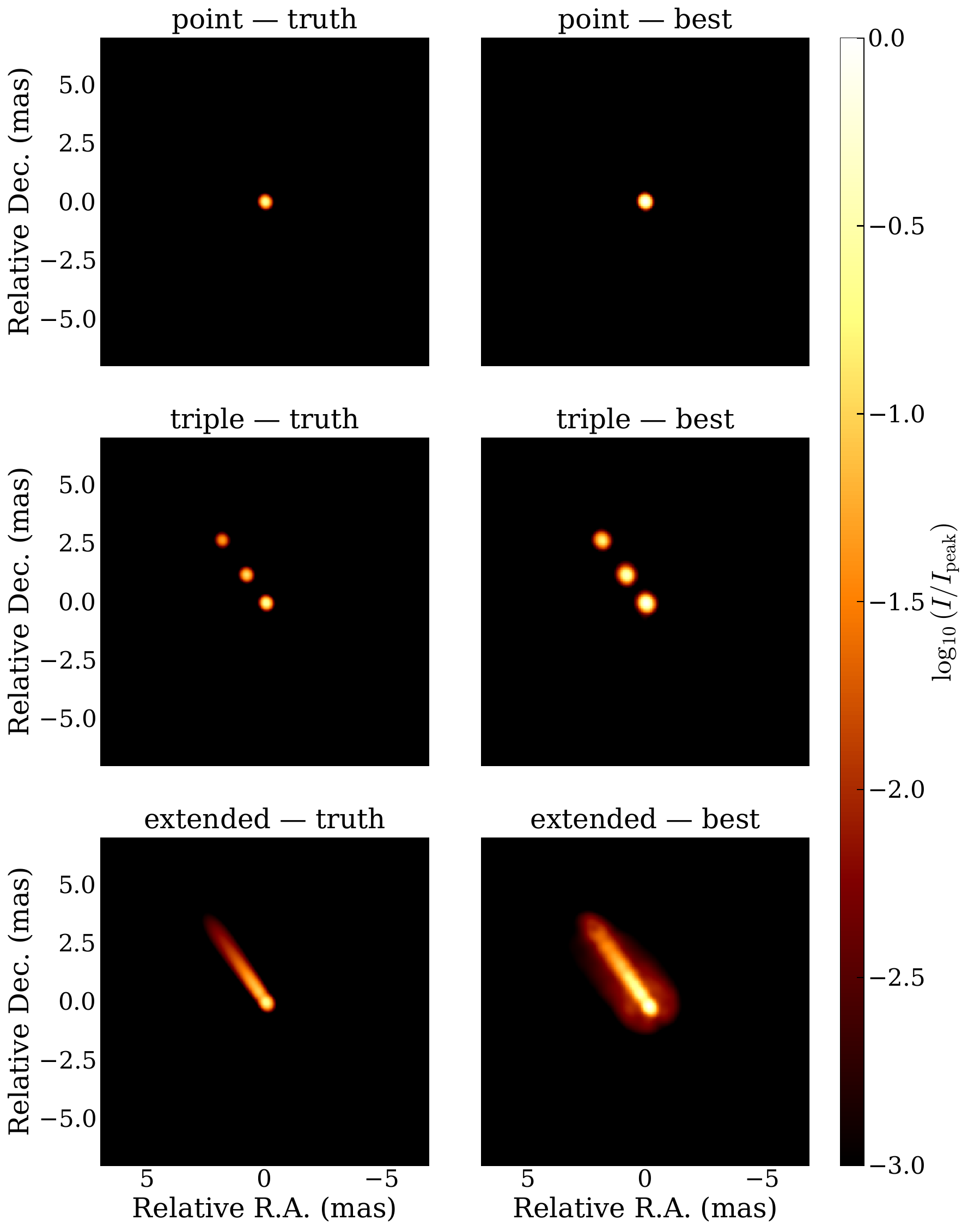}
    \caption[Comparison between synthetic images and their reconstructed counterparts for the simulated source morphologies.]{Comparison between the synthetic images and the corresponding reconstructions obtained with the best imaging setup found by the genetic algorithm. The panels show the band C images for the three simulated source morphologies: a point source, a triple source, and an extended source. For each morphology, the left panel shows the input model and the right panel shows the reconstructed image. All images were convolved with the same restoring beam, corresponding to 40\% of the band~C beam.}
    \label{fig:simul_comparison}
\end{figure}

\begin{table}[htpb]
\centering
\caption[Best imaging setup selected by the genetic algorithm.]{Best imaging setup selected by the genetic algorithm. The table lists the data terms and regularizers used in each imaging round, together with their corresponding weights. The notation follows the eht-imaging convention: \texttt{cphase} denotes closure phases, \texttt{logcamp} logarithmic closure amplitudes, \texttt{amp} visibility amplitudes, \texttt{tv2} second order total variation, \texttt{cm} the center of mass regularizer, \texttt{l1} sparsity, \texttt{flux} total-flux regularization, and \texttt{tv\_beta} total variation on the spectral curvature map.}
\label{tab:best_ga_imaging_setup}
\resizebox{\textwidth}{!}{
\begin{tabular}{cclcl}
\hline
Round & Data terms & Regularizers \\
\hline
1
& \texttt{cphase}: 100; \texttt{logcamp}: 5; \texttt{amp}: 2
& \texttt{tv2}: 10; \texttt{cm}: 100; \texttt{l1}: 50; \texttt{flux}: 50; \texttt{tv\_beta}: 100
\\

2
& \texttt{cphase}: 100; \texttt{logcamp}: 5
& \texttt{tv2}: 10; \texttt{l1}: 50; \texttt{flux}: 5; \texttt{tv\_beta}: 50
\\

3
& \texttt{cphase}: 50; \texttt{logcamp}: 10; \texttt{amp}: 1
& \texttt{tv2}: 50; \texttt{cm}: 5; \texttt{l1}: 5; \texttt{flux}: 10; \texttt{tv\_beta}: 10
\\
\hline
\end{tabular}
}
\end{table}

The best solution found by the genetic algorithm achieved an average score of 0.405 over the full set of simulated models and frequency bands. It is shown in Fig.~\ref{fig:simul_comparison}. The reconstruction quality depended on the complexity of the source structure. The point-source model was recovered very accurately, with an average score of 0.178, SSIM values of approximately 0.94, and flux-density errors below 1\% in all bands. The triple-component model was also well recovered, with an average score of 0.315 and SSIM values close to 0.98. In this case, the main differences were associated with small residual offsets and flux-density differences between the compact components, especially in band~A, where the angular resolution is poorer. The extended source represented the most challenging case, with an average score of 0.724, although the overall core-jet morphology was still recovered by the algorithm.

The final set of data terms and regularizers selected by the genetic algorithm is summarized in Table~\ref{tab:best_ga_imaging_setup}. This imaging prescription was then adopted for the reconstruction of the real VGOS data. The selected setup is dominated by closure phases in all imaging rounds, while logarithmic closure amplitudes are included with moderate weights and visibility amplitudes only with weak weights. This indicates that the source morphology is primarily constrained by phase information. This behaviour is expected, since closure phases are robust against residual antenna calibration errors and provide strong constraints on the relative structure of the source. In contrast, the visibility amplitudes are more sensitive to residual calibration errors, but they provide useful additional information on the total flux density scale. The genetic algorithm therefore favoured a configuration in which amplitudes contribute to the reconstruction without dominating the fit.

The best solution selected by the genetic algorithm was therefore characterized by a strong weight on closure phases in all imaging rounds, moderate weights on logarithmic closure amplitudes, and only a weak contribution from visibility amplitudes. The closure amplitudes help to stabilize the relative brightness distribution, while the visibility amplitudes provide an additional constraint on the total flux. The regularization was dominated by a combination of second-order total variation, sparsity, flux conservation, and spectral-curvature regularization. In the first imaging rounds, relatively strong regularization was favoured, which helps guide the reconstruction toward a stable solution. In the final round, the algorithm selected a more balanced setup, with a stronger smoothness term and weaker sparsity and flux constraints, allowing the image to adjust more closely to the data while still suppressing spurious small-scale structure.

The final imaging prescription obtained from these simulations was then adopted for the real VGOS data. This procedure provides a controlled and reproducible way to select the RML hyperparameters, rather than relying only on visual inspection of the final images. The simulation results show that the adopted setup reliably recovers compact and moderately resolved structures, while also providing a reasonable reconstruction of more extended jet emission.

\subsection{Imaging of Real VGOS Data}

After the first multi-frequency reconstruction, the resulting model was used to perform amplitude self-calibration. This self-calibration was applied independently to each VGOS band, using the corresponding band image extracted from the multi-frequency model. Only Stokes I amplitude self-calibration was performed, for three rounds. This conservative strategy was adopted to correct residual antenna gain errors without allowing the self-calibration to absorb real source-structure effects. After each self-calibration round, the multi-frequency imaging was repeated using the updated data.

For each source, the final Stokes I multi-frequency reconstruction provides the reference image, the spectral index map, and the curvature map. Images at the representative frequency of each VGOS band were then obtained by evaluating Eq.~\ref{eq:mf_intensity_imaging} at the corresponding frequency. The final images were convolved with a common restoring beam. This ensures that differences between bands are not dominated by the different angular resolutions and reflect intrinsic spectral changes in the source structure.

The other Stokes parameters were treated separately. The multi-frequency expansion described above was only used for Stokes I. For the polarized emission, Stokes $Q$ and $U$ were reconstructed independently in each VGOS band using the polarimetric imaging mode of eht-imaging. It does not include deconvolution for Stokes $V$ yet, but it is negligible in these sources, as explained in Sect. \ref{sec:polarization_theory}. The Stokes I image of each band was used as the reference intensity structure for the polarization reconstruction. From the resulting $Q$ and $U$ images, we derived the linearly polarized intensity, the fractional linear polarization, and the electric vector position angle, as given in Equations \ref{eq:linearlypolarizedintensity} and \ref{eq:evpa}. These quantities were masked in regions of low total intensity in order to avoid interpreting polarization fractions or EVPAs dominated by noise.

The final imaging products therefore consisted of a multi-frequency Stokes I model, one image for each VGOS band, the spectral index and curvature maps, and polarimetric images for each band.

\chapter{Application of the pipeline to the prototypical blazar 1803+784}
\label{ch:1803+784}

\vspace{-0.5cm}

\noindent
\hfill
\begin{minipage}{0.8\textwidth}
\small
\raggedleft
\textit{This chapter is mainly based on:}\\[0.3em]
\textit{``Evidence of a non-equipartition energy regime in 1803+784.\\
Core-shift and Faraday rotation measurements from simultaneous multi-frequency polarimetric VGOS observations''}\\[0.3em]
\textit{\textbf{V. Pérez-Díez}, I. Martí-Vidal, E. Albentosa-Ruiz and R. Bachiller.}\\[0.3em]
\textit{Astronomy \& Astrophysics, 700, L22 (2025).}\\
\end{minipage}

\vspace{0.8cm}

\lettrine{V}{  GOS} was designed to obtain geodetic results. However, this thesis attempts to exploit the large amount of AGN data to obtain astrophysical results from this type of source.

To verify the validity of our pipeline and confirm whether this is possible, we used data from 1803+784, a prototypical blazar that has been largely studied in astronomical VLBI. This allows us to compare our results with those obtained by other researchers to validate or refute our procedure.

To this end, our astrophysical objective is to study the core shift and Faraday rotation of 1803+784, compare it with the existing literature and extract some new information, as the validity of the energy equipartition hypothesis in this source.

\pagebreak
\section{Introduction}
\label{sec:introduction1803}

Compact jets from active galactic nuclei (AGN), particularly those associated with blazars and radio-loud quasars, are widely used in geodetic very long baseline interferometry (VLBI). However, the physical conditions and structure of these jets remain under active study.

In this chapter, we present an analysis of high-sensitivity VGOS observations of 1803+784. These multi-frequency data allowed us to derive a precise measurement of the core shift as a function of frequency. Additionally, we exploited the broad frequency coverage of VGOS to measure both internal and external Faraday rotation in the jet. 

Finally, we produced high-resolution multi-frequency images of the jet, and created detailed maps of the spectral index, spectral curvature and rotation measure. Furthermore, accurately characterizing the source structure in each band and its alignment through the core shift makes it possible to correct the structure effects described by \citet{Xu2021}. A direct application of this pipeline to VGOS data is shown in Chapter \ref{ch:geodesy}.

For this work, we used the calibrated data described in Chapter \ref{ch:vgos_pipeline}. To characterize the source structure, we performed model fitting to the visibilities following a procedure based on \citet{uvmultifit}. The source brightness model is a combination of several Gaussian components: a central elliptical Gaussian for the optically thick core, and two Gaussians to represent the extended optically thin jet emission. As already explained in Sect.~\ref{sec:model_fitting}, the fitting was initially performed in total intensity (Stokes I) at C band (6.75\,GHz), which provides the best compromise between resolution and signal-to-noise ratio for all observations.

To model the frequency-dependent core shift, the positions and sizes (but not the flux densities) of every optically thin jet component were kept fixed across bands and the position of the core component was left free. The Gaussian shape parameters of the core were also held fixed to minimize systematics related to jet blending at the lower frequencies.

Lastly, the polarization structure (Stokes Q, U, and V) was modelled using the same Gaussian components fitted to Stokes I, for the reasons already discussed in Sect. \ref{sec:model_fitting}.

\pagebreak
\section{Source Information}
\label{sec:1803+784}

1803+784 is located at 
$\alpha_{\mathrm{J2000}} = 18^{\mathrm h}00^{\mathrm m}45.6839^{\mathrm s}$, 
$\delta_{\mathrm{J2000}} = +78^\circ28'04.018''$, 
according to the ICRF3 S/X catalog \citep{Charlot2020}. 

It is a compact, flat-spectrum radio source identified with a BL Lac object at redshift $z=0.6797$ \citep{Lawrence1996, Stickel1993}. As a member of the complete S5 sample of high-declination flat-spectrum sources \citep{Kuehr1981}, it has been systematically observed for more than four decades across a wide range of angular resolutions and frequencies. The source exhibits the defining characteristics of BL Lac objects (see Sect. \ref{sec:AGN}): strong flux-density variability, high optical and radio polarization, and a core-dominated radio morphology indicative of a relativistically beamed jet viewed at a small angle to the line of sight, as seen in \citet{Wagner1995, Heidt1996}.

\begin{figure}[H]
    \centering
    \includegraphics[width=0.67\textwidth, trim={0pt 0pt 0pt 0pt}, clip]{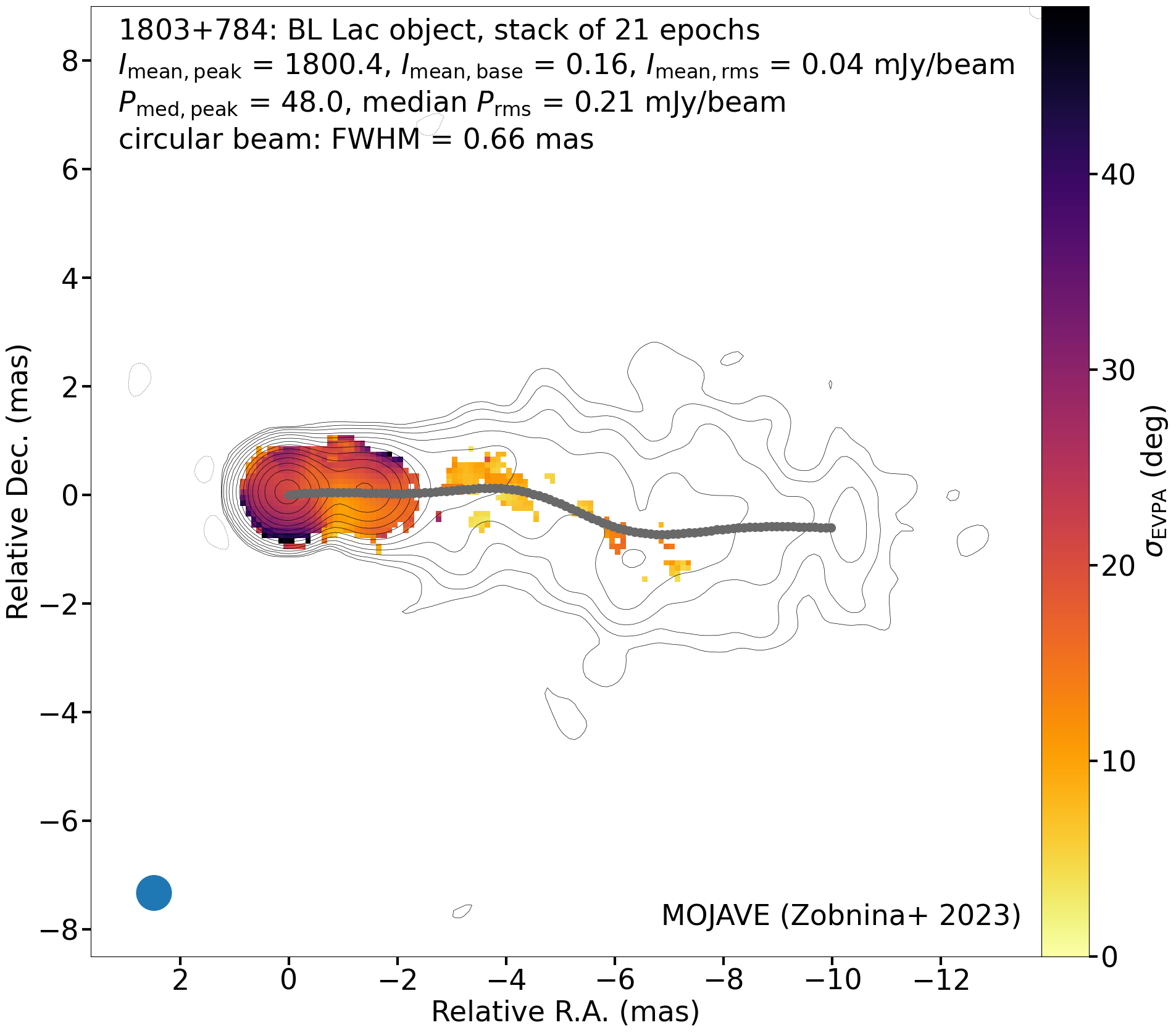}
    \caption[Stacked 15 GHz polarization VLBA image of 1803+784.]{Stacked 15 GHz polarization VLBA image of 1803+784 by the MOJAVE program \citep{Zobnina2023}.}
    \label{fig:1803+784_mojave}
\end{figure}

Geodetic S/X-band VLBI imaging from BVID \citep{Collioud2009} provides an X-band flux density of 1.45\,Jy, structure index 2.43, and compactness 0.64 \citep{Collioud2009}. 

Rapid variability has been detected in both the optical and radio bands. Optical monitoring has revealed fluctuations on timescales as short as tens of minutes \citep{Wagner1990}, while long-term radio variability has been documented over years to decades \citep{Nesci2002, Fan2007}. Such behavior strongly suggests Doppler-boosted synchrotron emission arising from a highly relativistic flow.

On kiloparsec scales, 1803+784 shows a pronounced misalignment between the parsec- and kiloparsec-scale jet directions. A secondary component is located $\sim45$ arcsec south-southwest of the core \citep{Antonucci1986, Strom1991}, while VLBI observations reveal that the jet bends by nearly $90^\circ$ at $\sim 0.5$ arcsec from the nucleus \citep{Britzen2005a}. The total misalignment between pc and kpc scales is approximately $76^\circ$ \citep{Britzen2005b}.

On parsec scales, the jet is predominantly oriented east–west and displays prominent components at core separations of approximately 1.4, 5, and 12~mas \citep{Eckart1986}. The inner jet structure is notably complex, with both stationary and superluminal features coexisting.

Within the MOJAVE program \citep{Lister2009, Lister2018}, 1803+784 has been regularly monitored at 15 GHz with the VLBA. MOJAVE images (see Fig. \ref{fig:1803+784_mojave}) confirm the highly curved parsec-scale morphology and the presence of multiple bright features within the inner 10 mas. The jet geometry of 1803+784 has also been quantified in the MOJAVE XIV analysis of AGN jet shapes and opening angles by \citet{Pushkarev2017}. From their Table 4, the apparent jet opening angle at 15~GHz is $\alpha_{\rm app}=21.8\pm0.5^\circ$, while the intrinsic opening angle is $\alpha_{\rm int}=0.3\pm0.2^\circ$. The work reports an apparent jet speed of $\beta_{\rm app}=1.16\pm0.31$, a Doppler factor $\delta=12.1$, a bulk Lorentz factor $\Gamma=6.2\pm1.5$, and a viewing angle $\theta=0.9\pm0.5^\circ$.

\pagebreak
\section{Core Shift}
\label{sec:coreshift1803}
\begin{figure}[H]
   \centering
   \includegraphics[width=0.67\textwidth]{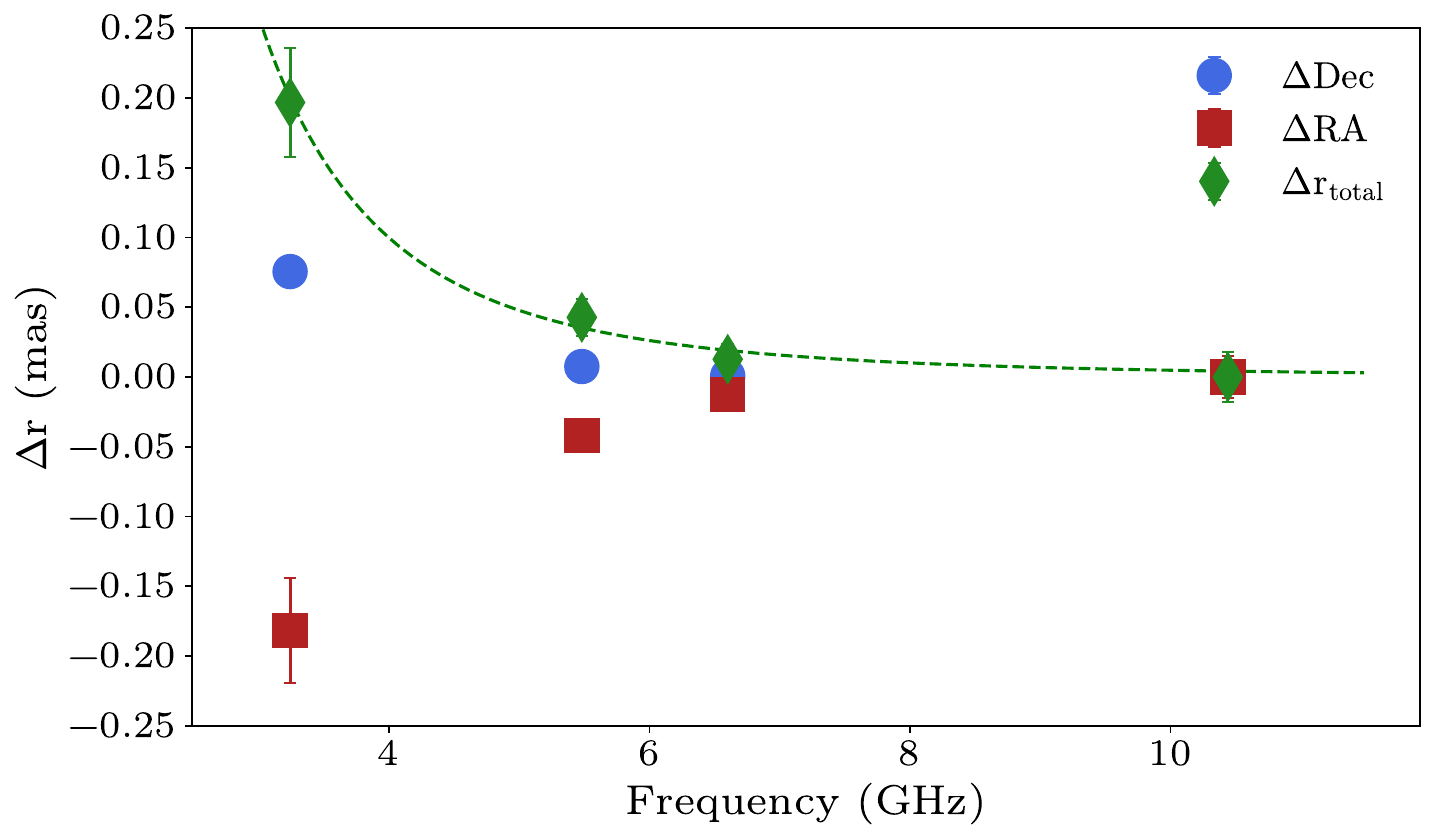}
   \caption[Measured relative core shift as a function of frequency for 1803+784.]{Measured relative core shift as a function of frequency for 1803+784. The total shift vector, along with its components in right ascension and declination, are shown. The error bars represent 3$\sigma$ uncertainties. The fit model is shown as a dashed green line.}
   \label{fig:core_shift_1803+784}
\end{figure}

Using the model fitting, we have measured a core shift of 0.24 milliarcseconds (mas) between 3 and 11\,GHz. In Fig.~\ref{fig:core_shift_1803+784} we show the core positions relative to the one at the highest frequency (10.66\,GHz). The shift is decomposed into its right ascension (RA) and declination (Dec) components, both exhibiting consistent trends with frequency. The total core-shift vector, $\Delta r$, follows a clear power-law behaviour with frequency. The wide bandwidth of VGOS, with eight  spws per band, allows for relatively small uncertainties in the measurements. The uncertainty in band A is larger due to greater scatter among spws, possibly caused by its bigger beam.

By fitting the frequency dependence of the core shift described in Eq.~\ref{eq:coreshift}, we obtain a fit power-law index of $k_r = 0.73^{+0.12}_{-0.19}$. This result represents a significant deviation from the standard equipartition condition ($k_r \sim 1$), which assumes a balance between the energy densities of relativistic particles and magnetic fields in a conical, freely expanding jet \citep{Lobanov1998}. A value of $k_r < 1$ indicates that the equipartition condition, the conical jet geometry, or both may not hold at the region we are observing at these frequencies. These findings underline the necessity of carefully evaluating the core-shift behaviour in blazars when deriving physical jet parameters, such as magnetic field strength, and when conducting geodetic studies.

\pagebreak
\section{Rotation Measure}

The wide simultaneous frequency coverage provided by VGOS allows us, for the first time, to separate the internal and external contributions of the Faraday rotation observed in 1803+784. While a distant plasma screen introduces an external rotation measure, $RM_E$, that rotates the electric vector position angle (EVPA), $\chi$, as $\Delta \chi = RM_E \, \Delta \lambda^2$\,, an internal (and/or surrounding) screen around the optically thick core produces a rotation measure $RM_I$ that depends on frequency as $RM_I \propto \nu^{a}$, as derived in \citet{Jorstad2007}. The exponent $a$ describes the radial dependence of the electron density, assuming a power-law decrease with distance from the black hole: $n_e \propto r^{-a}$ \citep{Hovatta2019}. Reported values of the exponent $a$ in the literature typically range from 0.4 to 4, with average values of around 2 \citep{Jorstad2007, OSullivan2009, Kravchenko2017, Hovatta2019}. This average value of $a=2$ suggests that the Faraday rotation takes place in a sheath surrounding a conically expanding jet. Lower values could be attributed to a more collimated jet structure \citep{Hovatta2019}, different geometry of the jet, flaring activity, or a possible filamentary structure of the Faraday screen \citep{Kravchenko2017}.

\begin{figure}[H]
   \centering
   \includegraphics[width=0.66\textwidth, trim=0.0cm 0.3cm 0.0cm 0]{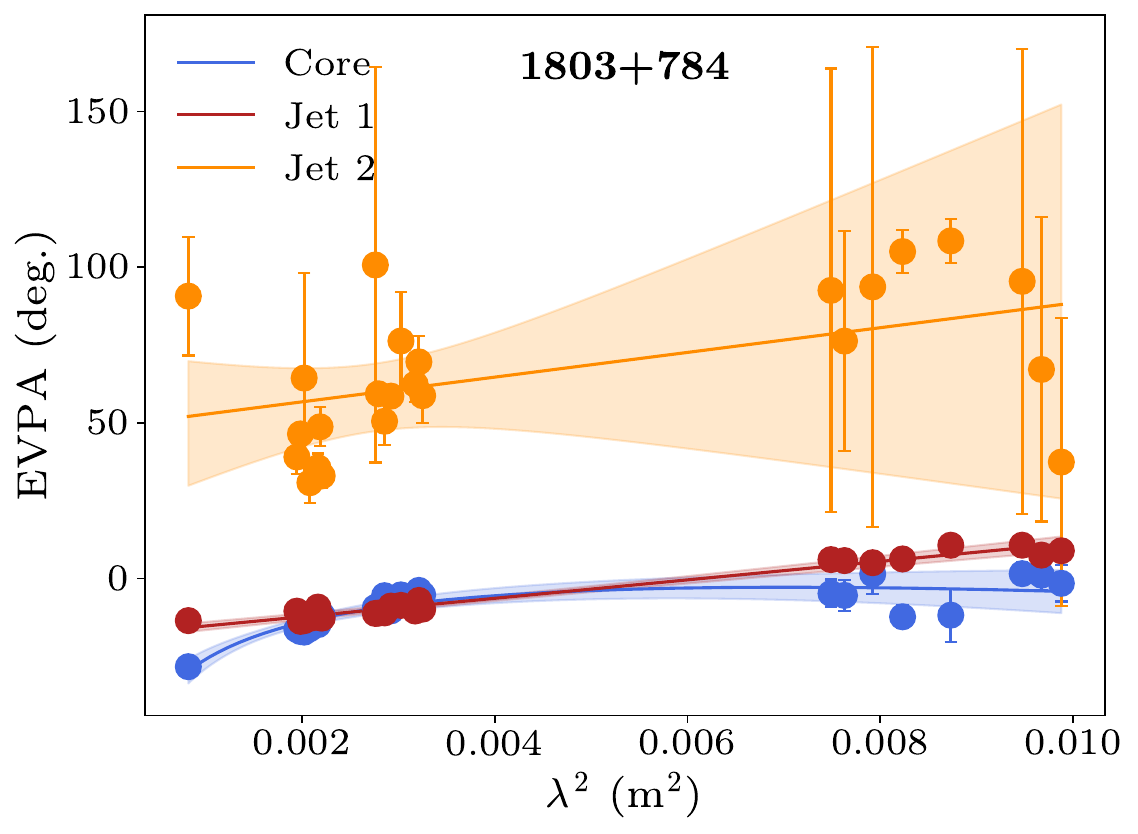}
   \caption[Fit of the polarization angle as a function of $\lambda^2$ for 1803+784.]{Fit of the polarization angle as a function of $\lambda^2$ for the core and jet regions of 1803+784. The shaded regions represent the 3$\sigma$ confidence intervals of the model fits. Band D data have been averaged because of the low S/N in this band.}
   \label{fig:RM_fit}
\end{figure}

By fitting the polarization angle, $\chi$, of the VLBI core as a function of $RM_E$ and $RM_I$ (see Fig.~\ref{fig:RM_fit}), we measure an external Faraday rotation of $RM^{\rm core}_E = -44 \pm 9$ rad m$^{-2}$ and an internal rotation of $RM^{\rm core}_I = 121 \pm 8$ rad m$^{-2}$ at the reference frequency of 5.5\,GHz.  When fitting for the exponent of $RM^{\rm core}_I$ with $\nu$, the result is $a = 2.18 \pm 0.95$. Due to large uncertainties, we cannot confirm this value, but it seems to be consistent with the theoretical expectation for a common expanding jet.

On the other hand, the plasma in the jet components is optically thin, so there is no core shift and the effects of $RM^{\rm jet}_E$ and $RM^{\rm jet}_I$ both follow a $\lambda^2$ relation. In this case, the fitted rotation measure is $RM^{\rm jet}_{\rm tot} = RM^{\rm jet}_E + RM^{\rm jet}_I = 49 \pm 2$ rad m$^{-2}$, and $\chi$ follows a clean $\lambda^2$ dependence, consistent with the presence of a total optically thin Faraday screen (Fig.~\ref{fig:RM_fit}). 

We also fitted the EVPA behaviour of a second, fainter jet component. For this component we obtain $RM^{\rm jet2}_{\rm tot} = 60 \pm 50$ rad m$^{-2}$. The uncertainty is much larger than for the first jet component, mainly because of the lower S/N of this region. Nevertheless, the fitted value is compatible, within the errors, with the rotation measure obtained for the brighter component. This suggests that both jet regions may be affected by a similar screen, although the second component does not provide a strong independent constraint because of its large uncertainty.

Considering that the effect of the distant plasma screen does not change between the core and the inner jet, we can further infer the $RM^{\rm jet}_I$ contribution by taking $RM^{\rm jet}_E \sim RM^{\rm core}_E$. With this assumption, the internal rotation measure in the optically thin jet is approximately $RM^{\rm jet}_I \sim 93 \pm 10$ rad m$^{-2}$, which is lower than $RM^{\rm core}_I$. This is in agreement with a jet magnetic field that decreases with distance to the black hole, as is implied in the model, $RM \propto \nu^p$, used to fit the $\chi$ at the core. Hence, the Faraday rotation at the core and the jet are both self-consistent with the picture of having two Faraday screens, one distant and extended and a second one located in the immediate jet surroundings (and/or internal to the jet itself).

These results are summarized in Table \ref{tab:astro_results_summary}, along with results from other sources. 

\pagebreak
\section{Multi-frequency Imaging}
\label{sec:mf_imaging_1803+784}

\begin{figure}[htpb]
   \centering
   \includegraphics[
      width=\textwidth,
      trim=20 10 10 10,
      clip
   ]{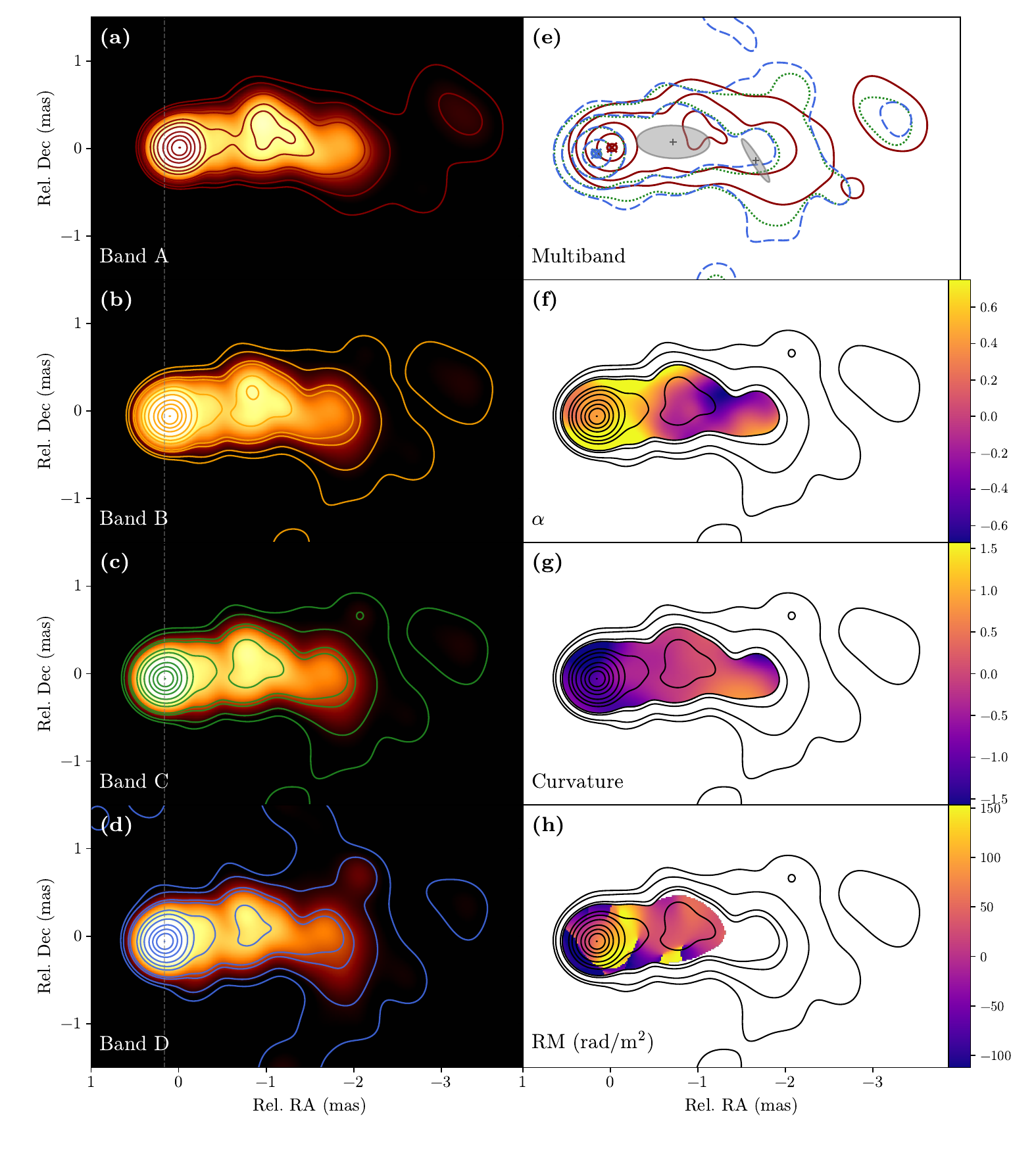}
   \caption[Multi-frequency polarimetric image of 1803+784.]{Multi-frequency polarimetric image of 1803+784. Panels (a)-(d) show the total-intensity maps at bands A, B, C, and D, respectively, displayed in logarithmic scale and normalized to the peak intensity of each band. The dashed vertical line marks the reference core position, making the frequency-dependent core shift visible across the VGOS bands. Panel (e) displays the total-intensity contours for bands A (solid red), C (dotted green) and D (dashed blue) in a common reference frame, together with the core and jet Gaussian components obtained from the visibility-domain model fitting. Panel (f) shows the spectral index map, $\alpha$, panel (g) the spectral curvature map, and panel (h) the rotation-measure map. Only regions above 1\% of the peak intensity are shown.}
   \label{fig:polarimetric_1803+784}
\end{figure}

Figure~\ref{fig:polarimetric_1803+784} shows the multi-frequency image of 1803+784 obtained with ehtim from the calibrated data. The total-intensity images in panels (a)-(d) correspond to bands A, B, C, and D, respectively. They are displayed in logarithmic scale and normalized to the peak intensity of each band. All four bands recover the same general morphology: a compact, bright core and a jet extending predominantly towards the east. The jet is more extended at the lower frequencies, as expected from the steeper spectrum of the optically thin emission. At the highest frequency, the image is more compact and dominated by the innermost region of the source.

The dashed vertical line marks the reference core position used to visualize the frequency-dependent displacement of the core. The relative shift between the low- and high-frequency peaks is consistent with the core-shift measurement obtained from the visibility-domain model fitting in Sect.~\ref{sec:coreshift1803}. This agreement is important because the two approaches are largely independent: the core-shift fit is based on Gaussian model components fitted directly to the visibilities, whereas Fig.~\ref{fig:polarimetric_1803+784} is obtained from regularized maximum-likelihood imaging. The fact that both methods recover the same qualitative displacement supports the robustness of the measured frequency-dependent core position.

Panel (e) overlays the total intensity contours from bands A, C and D (band B is omitted for ease of visualization of the results due to its similarity to band C) on a common reference frame. This representation provides a direct visual comparison of the source morphology across the VGOS frequency range. The core and jet Gaussian components obtained from the model fitting analysis are also shown. The optically thin jet components remain approximately fixed with frequency, while the apparent core position changes, as expected if the observed VLBI core corresponds to the surface where the optical depth is close to unity.

Panel (f) shows the spectral index map, $\alpha$, obtained from the multi-frequency RML reconstruction. The core region shows an inverted spectrum, with $\alpha > 0$, consistent with synchrotron self-absorption in the compact, optically thick base of the jet. In contrast, the downstream jet shows a negative spectral index, as expected for optically thin synchrotron emission. The transition from positive to negative $\alpha$ therefore provides an independent confirmation of the physical separation between the opaque core region and the optically thin jet.

Panel (g) shows the spectral curvature map. It demonstrates the importance of taking into account a second order term while deconvolving images from AGN at these frequencies.

Finally, panel (h) presents the rotation measure map. The RM distribution is not uniform across the source: larger values are found close to the core, while the jet shows a smoother and lower RM structure. This spatial behaviour is consistent with the model fitting, with a RM of $\sim80~{\rm rad~m^{-2}}$ for the core and $\sim50~{\rm rad~m^{-2}}$ for the jet.

Overall, the multi-frequency image confirms that VGOS observations can recover not only the total intensity structure of a bright AGN, but also spatially resolved spectral and Faraday rotation information. In the case of 1803+784, the images are fully consistent with the results obtained from model fitting and the previous images of the MOJAVE program.

\pagebreak
\section{Discussion}

The results presented in this chapter demonstrate that VGOS observations can be used not only for geodetic applications, but also to extract physically meaningful information of AGN jets. In the case of 1803+784, the simultaneous frequency coverage between 3 and 11\,GHz allows us to measure the core shift, reconstruct the spectral structure of the source, and study the Faraday rotation in the core and jet within a single observing epoch.

\subsection*{Core Shift}

The most relevant result is the measured core shift index, $k_r = 0.73^{+0.12}_{-0.19}$, which deviates from the standard equipartition prediction $k_r \simeq 1$ \citep{Lobanov1998}. In the framework of the conical jet model, $k_r = 1$ is expected when the magnetic field and particle energy densities are in equipartition and the jet follows the standard radial scalings $B \propto r^{-1}$ and $N \propto r^{-2}$. Therefore, the value obtained here suggests that at least one of these assumptions is not fully satisfied at centimeter wavelengths for 1803+784 during this epoch. This may indicate a departure from equipartition, a deviation from a purely conical geometry, or a combination of both effects.

This result is especially relevant because 1803+784 is a bright and compact source commonly observed in geodetic VLBI experiments. If the core shift followed exactly a $\nu^{-1}$ dependence, the frequency dependent position of the core would not introduce a group delay offset \citep{Porcas2009}. However, when $k_r \neq 1$, the group delay position becomes frequency dependent. This implies that the same physical effect used here as an astrophysical diagnostic can also introduce systematic errors in high precision astrometric and geodetic analyses. Therefore, measuring $k_r$ directly in VGOS sources is important both for jet physics and for the interpretation of geodetic observables.

A comparison with previous measurements supports the consistency of our result. \citet{Kovalev2008A} reported a core shift of 0.72\,mas between 2.3 and 8.6\,GHz for 1803+784. By extrapolating our fit model to 2.3\,GHz, we obtain a core-shift value of approximately 0.61\,mas. The difference between both values is moderate and may be related to the fact that the observations were separated by about 17 years. Given the known variability and structural evolution of 1803+784, such a difference is not unexpected. It may also reflect changes in the opacity structure of the core, the activity state of the source, or the position of the dominant emitting region at different epochs. Another comparison comes from \citet{OSullivan2008}, where they report a core shift of $1.27 \pm0.35$ mas between 1.6 and 5 GHz. Extrapolating our results to these frequencies, we obtain a core shift of $1.8 ^{+1.2}_{-0.8}$ mas, which agrees within the uncertainties. \citet{Plavin2019} studied the core shift of 1803+784 from 1994 to 2014, obtaining a mean value of $\Delta r_{8-2} = 0.62 \pm 0.23~{\rm mas}$, between 8 and 2 GHz, with a minimum of 0.290 and a maximum of 1.503 mas. The extrapolation of our fit to this pair of frequencies estimates: $\Delta r_{8-2} = 0.9^{+0.4}_{-0.3}~{\rm mas}$, which lies within their range of values.

Interestingly, similar values of $k_r$ have been reported in other AGN jets. For example, \citet{Boccardi2021} analysed the radio galaxy NGC~315 using six frequencies over a range comparable to ours and obtained $k_r = 0.84 \pm 0.06$. A later reanalysis by \citet{Ricci2022}, including an additional point at 43\,GHz, found two different regimes: $k_r = 0.93 \pm 0.01$ below 10\,GHz and $k_r = 0.57 \pm 0.17$ at higher frequencies. They interpreted this behaviour as possible evidence for a transition between different jet regions, associated with the change from a parabolic, magnetically dominated zone to a more conical regime, as explained in Fig. \ref{fig:JetModel}.

Motivated by this interpretation, we also tested the effect of separating the VGOS frequency range around 10\,GHz. When the D-band data are excluded and only the three frequencies below 10\,GHz are fitted, we obtain $k_r = 0.85^{+0.03}_{-0.06}$. This value is closer to unity than the full-band result, although it still does not fully reach the equipartition expectation. This may suggest that the transition proposed by \citet{Ricci2022} for NGC~315 could also be relevant for 1803+784, but occurring at a lower frequency or over a different spatial scale. Observations below band A would be needed to confirm whether the low frequency core-shift behaviour approaches $k_r = 1$ more clearly. The advantage of the present VGOS measurement is that all bands are observed simultaneously, avoiding ambiguities caused by intrinsic variability between observing epochs.

\subsection*{Faraday Rotation}

The Faraday rotation analysis provides an additional view of the physical conditions in the source. The broad instantaneous frequency coverage of VGOS allows us to separate, for the first time, the Faraday rotation affecting the optically thick core into an external and an internal contribution. We obtain a negative external component, $RM^{\rm core}_E = -44 \pm 9$\,rad\,m$^{-2}$, and a positive internal component, $RM^{\rm core}_I = 121 \pm 8$\,rad\,m$^{-2}$, at the reference frequency of 5.5\,GHz.

The fitted frequency dependence of the internal core RM gives $a = 2.18 \pm 0.95$, where $RM_I \propto \nu^a$. Although the uncertainty is large, this value is compatible with the expectation for a Faraday rotating sheath surrounding a conically expanding jet. In this interpretation, the decrease of the particle density and magnetic field strength with distance from the black hole naturally produces a stronger Faraday rotation closer to the core. The lower internal RM inferred for the jet, together with the smoother RM distribution seen in the image plane, is consistent with this explanation.

The jet components show a cleaner $\lambda^2$ behaviour, as expected for optically thin synchrotron regions. For the brightest jet component we obtain $RM^{\rm jet1}_{\rm tot} = 49 \pm 2$\,rad\,m$^{-2}$, while for the second, fainter component we obtain $RM^{\rm jet2}_{\rm tot} = 60 \pm 50$\,rad\,m$^{-2}$. The uncertainty of the second value is much larger because of the lower S/N, but it is still compatible with the RM measured in the brighter jet component. The multi-frequency image reinforces this interpretation. The spectral index map shows an inverted spectrum in the core and a steep spectrum in the jet, as expected from the transition between an optically thick compact core and optically thin downstream emission.

\citet{Gabuzda2003} measured RM for different components of 1803+784. Their observed RM in the farthest component should not be affected by the internal Faraday screen because of the large distance to the core, so it should be similar to our estimated external component. They obtain $RM = -20 \pm 15$\,rad\,m$^{-2}$, with the same sign and order of magnitude as our estimated $RM^{\rm core}_E$. Our results are therefore in agreement with their measurements.

\subsection*{Physical Parameters}

Using the core-shift formalism described in Sect.~\ref{sec:physicaljet}, and adopting the jet parameters listed in Sect.~\ref{sec:1803+784}, we obtain $B_1 = 30 \pm 10$~G and $N_1 = (10 \pm 4) \times10^6~{\rm cm^{-3}}$ at a deprojected distance of 1 pc from the black hole. At the observed VGOS core positions, the inferred magnetic field ranges from $B_{\rm core}=0.17 \pm 0.05$~G at 3.24~GHz to $B_{\rm core}=0.54 \pm 0.11$~G at 10.44~GHz, while the corresponding particle density increases from $N_{\rm core}=(2.9 \pm 1.5)\times10^2~{\rm cm^{-3}}$ to $N_{\rm core}=(3.0 \pm 1.5) \times10^3~{\rm cm^{-3}}$. A useful comparison is provided by \citet{OSullivan2008}, who used the core shift between 1.6 and 4.8 GHz in 1803+784 to estimate $B_{\rm core}(4.8\,{\rm GHz})=0.11\pm0.02$~G at $r_{\rm core}(4.8\,{\rm GHz})=20\pm5$~pc. These values are of the same order of magnitude as ours.

In this discussion we are assuming the standard energy equipartition between particles and fields, as well as a conical jet shape. However, our measured value $k_r=0.73^{+0.12}_{-0.19}$ indicates a departure from the ideal equipartition regime and/or conical jet geometry. Even so, our estimates from the standard jet model may not be regarded as precise estimates, but used for an order-of-magnitude discussion.

Using the internal rotation measure of the core at the reference frequency ($5.5\,{\rm GHz}$), $RM^{\rm core}_I=121\pm8~{\rm rad~m^{-2}}$, we can obtain a rough estimate of the line-of-sight component of the magnetic field, using Eq.~\ref{eq:rm_definition}. For this calculation, we assume that the internal Faraday rotation is produced close to (if not at) the emitting region. Under this assumption:
\[
B_{\parallel}^i =
\frac{RM}{812\,N_{\rm core}\,dl}
=
1.8\times10^{-7}~{\rm G},
\]
where $N_{core} = 8.39 \times 10^2 {\rm cm}^{-3}$, and we use $dl = 1\, {\rm pc}$.

This value is much smaller than the total magnetic field strength inferred from the core-shift analysis at a similar frequency, $B_{\rm core}(5.5\,{\rm GHz})\simeq0.28$~G. The ratio between both quantities is
\[
\frac{B_{\parallel}^i}{B_{\rm core}}
\simeq 6.4\times10^{-7}.
\]

Another possibility to explain the internal RM would be that it is produced at the immediate surroundings of the jet (i.e., at or around the jet sheath, see Sect. \ref{sec:jetlaunching}). In this case, we can assume \citep{Gabuzda2003} energy equipartition with the particles being at a temperature of $T \sim 10^4 K$ \citep[typical of the narrow-line region, which should be surrounding the jet at these scales][]{Koski1978}. By assuming a temperature, we can decouple particle density and magnetic field for a given RM, hence allowing to estimate the magnetic field directly \citep{Gabuzda2003}:
\[
B_{\parallel}^s  \sim \left[ \frac{RM(1+z)^2\pi k T}{10^5 dl} \right]^{1/3},
\]
where $k$ is the Boltzmann constant and $T$, the plasma temperature. From this equation we estimate $B_{\parallel}^s  \sim 2.5 \times 10^{-5} \rm{G}$. This value is about two orders of magnitude higher than the one corresponding to the assumption of an internal jet origin ($1.8 \times 10^{-7} \rm{ G}$). In any case, both estimates are much lower than the magnetic field derived from the core shift. 

The much lower values of $B_{\parallel}$ compared to the magnetic field estimated from the core shift indicate that only a tiny effective component of the magnetic field is projected along the line of sight. This is the expected result for a blazar such as 1803+784, whose jet is observed at a very small viewing angle \citep[$0.9^\circ$, ][]{Pushkarev2017}. If the helical magnetic field \citep[e.g., ][]{Blandford1982, Marscher2014, Zamaninasab2014} has a predominantly azimuthal component, most of the field will be projected transversely to the line of sight, hence explaining our low estimated ratio $B_\parallel^i/B_{\rm core}$. Even accounting for the RM suppression due to the relativistic electron energies (in case the RM would be internal to the emitting region), the ratio $B_\parallel^i/B_{\rm core}$ would only be increased by a factor $\sim 20$ \citep[$\Gamma \sim 6.2$, ][]{Pushkarev2017}.

All this discussion is based on qualitative order-of-magnitude estimates. On the one hand, the result depends directly on the assumed depth of the Faraday screen, here taken to be $dl=1$~pc as an example. Besides, we also make several assumptions about the particle density and physical conditions of the plasma. Therefore, the small values of $B_\parallel$ are physically plausible, but they should not be regarded as a precise measurement of the magnetic field.

\subsection*{Additional Validations and Conclusions}

Another test for the validity of this pipeline is the comparison with the analysis developed in \citet{AlbentosaRuiz2025}. They use the same VGOS VO2187 data, but perform the analysis with closure traces. Their results are highly consistent with those obtained here. They obtain an external Faraday rotation of $RM^{\rm core}_E = -30 \pm 20$ rad m$^{-2}$ and an internal rotation of $RM^{\rm core}_I = 122 \pm 55$ rad m$^{-2}$. For the jet, $RM^{\rm jet}_{\rm tot} = 40 \pm 35$ for the first component and $RM^{\rm jet2}_{\rm tot} = 30 \pm 50$ rad m$^{-2}$ for the second one. The results from both methods are in good agreement. Nevertheless, the estimate of the magnetic field at 1 pc is a bit different. They obtain $B_1 = 0.45 \pm 0.05$~G. This disagreement is produced by the different geometry parameters used by \citet{AlbentosaRuiz2025}, which are based on optical observations \citep{Impey1991}, whereas our geometry comes from VLBI observations \citep{Pushkarev2017}.

Overall, 1803+784 provides a clear demonstration of the astrophysical potential of VGOS. The same dataset allows us to measure core shift, constrain the opacity structure of the jet, separate different Faraday-rotation contributions, and produce spatially resolved spectral and RM maps. The results are summarized in Table \ref{tab:1803+784_results}. These results are also directly relevant for geodesy: if compact reference sources do not always satisfy the $k_r = 1$ condition, their frequency-dependent structure may introduce effects in broadband group delay measurements. This reinforces the need to characterize the source structure of geodetic AGN and motivates the application of the same methodology to a larger sample of VGOS sources.

\begin{table}[t]
\centering
\small
\setlength{\tabcolsep}{8pt}
\renewcommand{\arraystretch}{1.08}
\caption{Summary of the main parameters estimated for 1803+784.}
\label{tab:1803+784_results}

\begin{tabular}{lc}
\hline
Parameter & Estimated value \\
\hline
$\Delta r_{\rm 3-11\,GHz}$ 
& $0.24 \pm 0.05~{\rm mas}$ \\

$k_r$ 
& $0.73^{+0.12}_{-0.19}$ \\

$a$ 
& $2.2 \pm 1.0$ \\

$RM_{\rm E}^{\rm core}$ 
& $-44 \pm 9~{\rm rad~m^{-2}}$ \\

$RM_{\rm I}^{\rm core}$ 
& $121 \pm 8~{\rm rad~m^{-2}}$ \\

$RM_{\rm tot}^{\rm jet\,1}$ 
& $49 \pm 2~{\rm rad~m^{-2}}$ \\

$RM_{\rm tot}^{\rm jet\,2}$ 
& $60 \pm 50~{\rm rad~m^{-2}}$ \\

$RM_{\rm I}^{\rm jet}$ 
& $93 \pm 10~{\rm rad~m^{-2}}$ \\

$\alpha$ 
& $-0.6$ \\

$B_1$ 
& $30 \pm 10~{\rm G}$ \\

$N_1$ 
& $(10 \pm 4)\times10^6~{\rm cm^{-3}}$ \\

$B_{\rm core}(3.24\,{\rm GHz})$ 
& $0.17 \pm 0.05~{\rm G}$ \\

$B_{\rm core}(10.44\,{\rm GHz})$ 
& $0.54 \pm 0.11~{\rm G}$ \\

$N_{\rm core}(3.24\,{\rm GHz})$ 
& $(2.9 \pm 1.5)\times10^2~{\rm cm^{-3}}$ \\

$N_{\rm core}(10.44\,{\rm GHz})$ 
& $(3.0 \pm 1.5)\times10^3~{\rm cm^{-3}}$ \\

$B_{\parallel}^i$ 
& $1.8\times10^{-7}~{\rm G}$ \\

$B_{\parallel}^s$ 
& $2.5\times10^{-5}~{\rm G}$ \\

\hline
\end{tabular}
\end{table}

Finally, the agreement between our results and previous studies based on other VLBI networks provides an additional validation of the calibration and imaging pipeline developed in this thesis. This consistency confirms that VGOS data, when processed with this pipeline, can deliver reliable astrophysical information on AGN jets. In Chapter \ref{ch:astro}, we apply this analysis to all the sources of experiment VO2187 for which it was possible to measure core shift and rotation measure with enough S/N.

\chapter{Astrophysical Analysis}
\label{ch:astro}

\vspace{-0.5cm}

\noindent
\hfill
\begin{minipage}{0.8\textwidth}
\small
\raggedleft
\textit{This chapter is mainly based on:}\\[0.3em]
\textit{``Towards an astronomical use of new-generation geodetic observations.\\
II. Core-shift and multi-band polarization for a selection of sources''}\\[0.3em]
\textit{\textbf{V. Pérez-Díez}, I. Martí-Vidal, E. Albentosa-Ruiz, J. González-García,\\
F. Jaron, T. Savolainen, M. H. Xu and R. Bachiller.}\\[0.3em]
\textit{In prep.}\\
\end{minipage}

\vspace{0.8cm}

\lettrine{H}{aving} validated the procedure in the previous chapter, we now apply it to the remaining sources in the sample in order to extract astrophysical information from the VGOS observations.

In this chapter, we present the results for core shift, equipartition hypothesis validity, Faraday rotation and multi-frequency images for a sample of AGN observed with VGOS.

\pagebreak
\section{3C418}

\begin{figure}[H]
    \centering
    \includegraphics[width=0.67\textwidth]{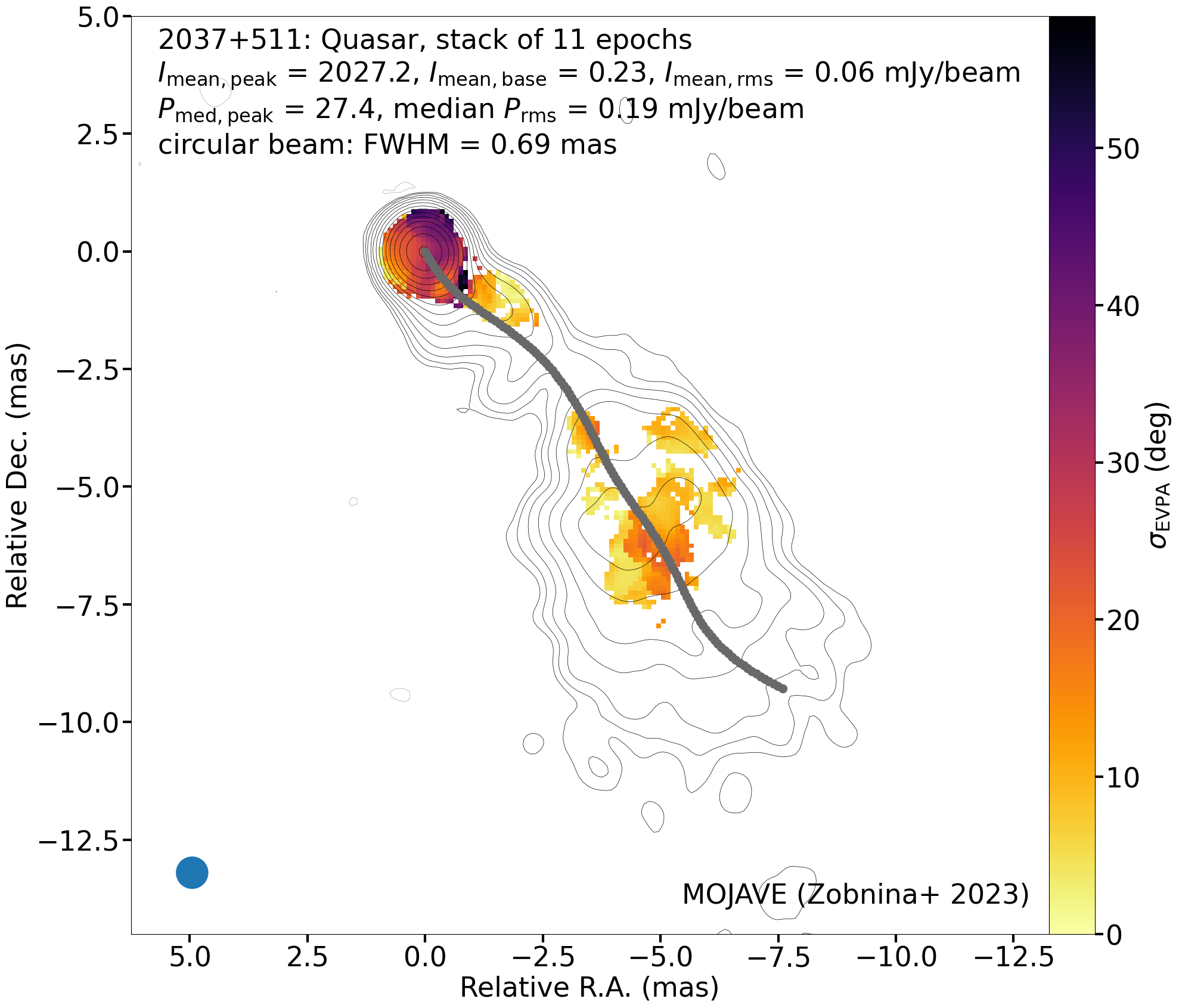}
    \caption[Stacked 15 GHz polarization VLBA image of 3C418.]{Stacked 15 GHz polarization VLBA image of 3C418 from the MOJAVE program \citep{Zobnina2023}.}
    \label{fig:3C418_mojave}
\end{figure}

3C418, also known as 2037+511 or J2038+5119, is a bright radio-loud quasar at redshift $z=1.686$ \citep{Spinrad1985}. The source is included in geodetic and astrometric VLBI catalogues and is classified as a non-defining ICRF3 source \citep{Charlot2020}. Despite its frequent use in geodetic VLBI observations, the source shows a significant structure, as shown in Fig. \ref{fig:3C418_mojave}. It has structure indices of 3.36 at X band and 2.25 at S band, with compactness values 0.4, indicating that a substantial fraction of the radio emission is resolved on VLBI scales \citep{Collioud2019}.

MOJAVE classifies 3C418 as a low-spectral peaked flat spectrum quasar \citep{Lister2018}. Its images at 15~GHz reveal a jet with apparent superluminal motions reaching $6.6 \pm 1.3\,c$ based on five moving features. \citet{Punsly2017} showed that 3C418 is one of the most kinetically dominated quasars known.

The combination of high compact flux density, pronounced source structure, and strong frequency-dependent opacity effects makes 3C418 a particularly relevant target for multi-frequency VLBI studies.

\subsection{Core Shift}

\begin{figure}[H]
   \centering
   \includegraphics[width=0.67\textwidth]{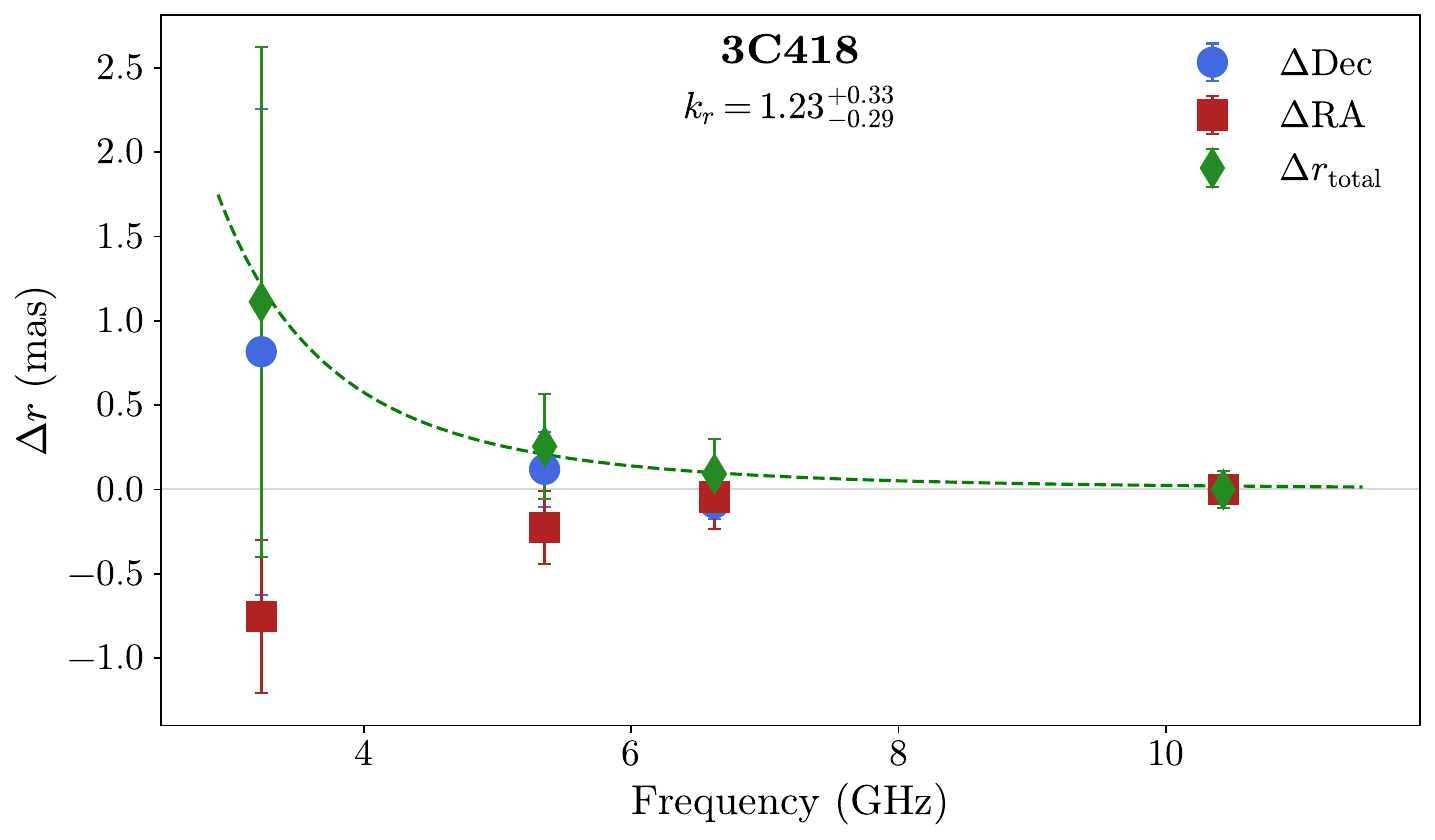}
   \caption[Measured relative core shift as a function of frequency for 3C418.]{Measured relative core shift as a function of frequency for 3C418. The total shift vector, along with its components in right ascension and declination, are shown. The error bars represent 3$\sigma$ uncertainties. The fit model is shown as a dashed green line.}
   \label{fig:core_shift_3C418}
\end{figure}

Applying the same procedure described in Chapter~\ref{ch:1803+784}, we obtained the core-shift measurements for 3C418, shown in Fig.~\ref{fig:core_shift_3C418}.

We measure a total core shift of approximately $1.0$ mas between 3 and 11 GHz. The largest uncertainty is found in band A, where the error is larger than the measured displacement. This is likely caused by a combination of the poorer angular resolution at the lowest frequencies and blending between the core and the closest jet emission. The remaining bands also show relatively large uncertainties, which is expected given the limited number of observations available for this source in VO2187 (shown in Table \ref{tab:sources}).

Despite these limitations, the measured displacements follow a curve that can be described by the standard core shift relation (Sect. \ref{sec:jetemissionmodel}) with a value of
\[
  k_r = 1.23^{+0.33}_{-0.29}.
\]

Although the uncertainty is large, this result is consistent with the equipartition expectation, $k_r \simeq 1$, within the errors. Therefore, while the individual core shift measurements should be interpreted with caution, the global frequency dependence does not show evidence for a strong departure from equipartition.

The only published core shift measurements for this source are those reported by \citet{Pushkarev2012}. For 3C418, listed as 2037+511, they measured shifts of 0.024, 0.051, and 0.052 mas for the 15.4--8.1, 15.4--8.4, and 15.4--12.1 GHz frequency pairs, respectively. Extrapolating our fitted core-shift model to the same frequency pairs, we obtain corresponding values of 0.044, 0.038, and 0.007 mas. Given the small magnitude of the shifts involved, the different observing epochs, the different frequency coverage, and the large uncertainties in our low-frequency measurements, our extrapolated values are broadly consistent with those of \citet{Pushkarev2012}. The agreement is better for the 15.4--8.1 and 15.4--8.4 GHz pairs than for the 15.4--12.1 GHz pair, where the expected displacement is very small and therefore particularly sensitive to the fitted value of $k_r$.

\subsection{Rotation Measure}

\begin{figure}[H]
   \centering
   \includegraphics[width=0.66\textwidth, trim=0.0cm 0.3cm 0.0cm 0]{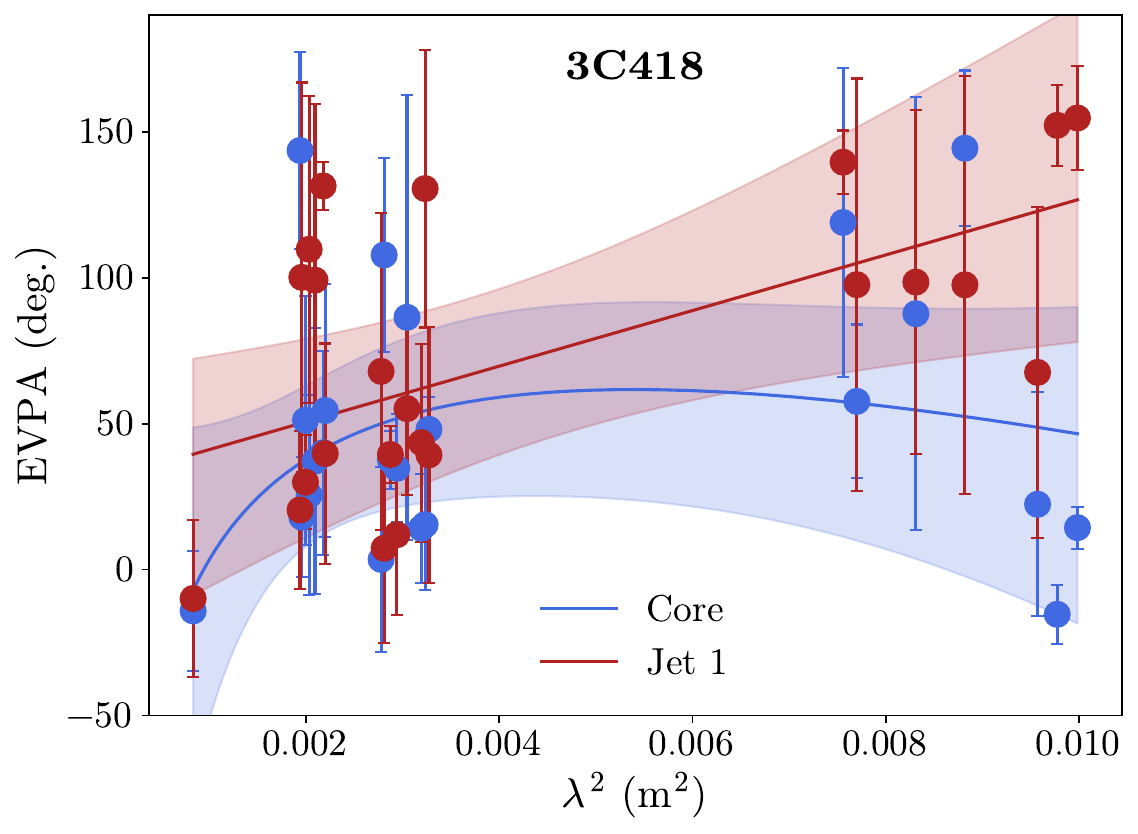}
   \caption[Fit of the polarization angle as a function of $\lambda^2$ for 3C418.]{Fit of the polarization angle as a function of $\lambda^2$ for the core and jet regions of 3C418. The shaded regions represent the 3$\sigma$ confidence intervals of the model fits. Band D data have been averaged because of the low S/N in this band.}
   \label{fig:RM_fit_3C418}
\end{figure}

Figure~\ref{fig:RM_fit_3C418} shows the fit of the electric vector position angle (EVPA) as a function of $\lambda^2$ for the core and jet regions of 3C418.

The uncertainties in the EVPA measurements do not allow us to reliably fit the exponent of the internal Faraday-rotation model as a free parameter. For the core, we obtain an external Faraday rotation of
\[
    RM^{\rm core}_{\rm E} = -200 \pm 100~{\rm rad~m^{-2}},
\]
and an internal contribution of
\[
    RM^{\rm core}_{\rm I} = 378 \pm 150~{\rm rad~m^{-2}}.
\]

For the jet region, a linear fit gives
\[
    RM^{\rm jet}_{\rm tot} = 166 \pm 50~{\rm rad~m^{-2}}.
\]
Following the same argument used in Chapter~\ref{ch:1803+784}, and assuming that the external contribution affecting the jet is comparable to that inferred for the core region, we estimate an internal jet contribution of
\[
    RM^{\rm jet}_{\rm I} = 366 \pm 150~{\rm rad~m^{-2}}.
\]

\subsection{Multi-frequency Imaging}

\begin{figure}[htpb]
   \centering
   \includegraphics[
      width=\textwidth,
      trim=35 10 30 10,
      clip
   ]{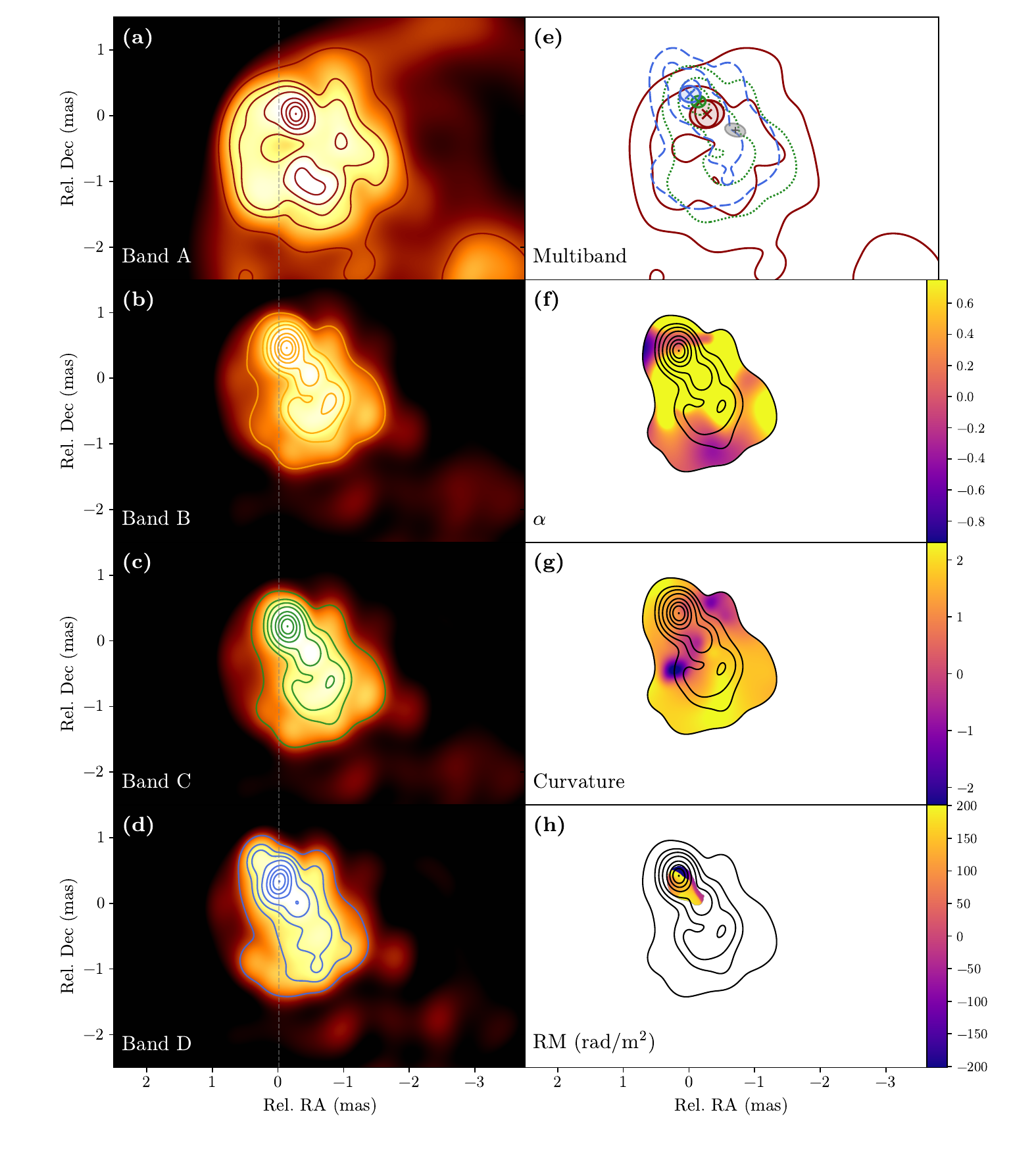}
   \caption[Multi-frequency polarimetric image of 3C418.]{Multi-frequency polarimetric image of 3C418. Panels (a)--(d) show the total intensity maps at bands A, B, C, and D, respectively, displayed in logarithmic scale. The color scale in each panel is normalized to its peak intensity. The dashed vertical line indicates the core position at band A, marking the core shift across bands. Panel (e) displays the total-intensity contours for bands A (solid red), C (dotted green) and D (dashed blue) in a common reference frame, together with the core and jet Gaussian components obtained from the visibility-domain model fitting. Panel (f) shows the spectral index map ($\alpha$), and panel (g), the curvature map. Panel (h) presents the rotation measure (RM) map. Only regions above 1\% of the peak intensity are included.}
   \label{fig:polarimetric_3C418}
\end{figure}

The multi-frequency image of 3C418 is shown in Fig.~\ref{fig:polarimetric_3C418}. The total intensity maps reveal a compact bright core and a jet extending away from it, to the south-west, in agreement with the morphology seen in MOJAVE 15 GHz images. The source is clearly resolved by the VGOS observations, especially at the lower frequency bands, where the poorer angular resolution and the stronger contribution from optically thin jet emission increase the apparent extent of the source.

The superposition of the contours from the four VGOS bands shows a clear frequency dependent displacement of the brightness peak. The lowest frequency emission is shifted downstream with respect to the high-frequency core position, as expected for a typical core shift (as seen in the model presented in Sect. \ref{sec:jetemissionmodel}). This behaviour is consistent with the quantitative core shift fit shown in Fig.~\ref{fig:core_shift_3C418}. The spectral index map shows the expected transition between an optically thick spectrum core to an optically thin jet.

The RM map provides spatial information complementary to the integrated EVPA fits. Although the limited S/N prevents a detailed interpretation, the map indicates that the Faraday rotation is not uniform across the source and the order of magnitudes are the same as in the model fitting.

\subsection{Discussion}

3C418 is a particularly useful source for assessing the astrophysical potential of VGOS observations because it combines high compact flux density with significant structure. The presence of moving jet components \citep{Lister2019} provides a framework in which frequency-dependent core positions are important.

The morphology recovered in our VGOS images is consistent with the 15~GHz VLBA images from MOJAVE. Both datasets show a compact core and a jet extending downstream towards the south-west. The spectral-index map also agrees qualitatively with the MOJAVE 8--15~GHz map \citep{Hovatta2014}: the core is associated with an optically thick spectrum ($\alpha \approx 0.6$), while the jet shows an optically thin spectrum ($\alpha < -0.4$).

The core shift measurement obtained from the VGOS bands gives a displacement of about 1 mas between 3 and 11 GHz. This is larger than the shifts reported by \citet{Pushkarev2012} at 8--15 GHz, as expected from the lower frequencies sampled by VGOS. When our fitted model is extrapolated to the frequency pairs used by \citet{Pushkarev2012}, the predicted shifts are of the same order as their measurements.

Regarding the rotation measure, \citet{Hovatta2012} reported observed-frame RM values of approximately $RM^{\rm core}_{\rm obs} = -450 \pm 76~{\rm rad~m^{-2}}$ for the core and $RM^{\rm jet}_{\rm obs} = -248 \pm 133~{\rm rad~m^{-2}}$ for the jet of 3C418. Their estimation of the RM of the jet differs greatly from ours, so it should be similar to our value for the external screen, as it is: $-200$ vs. $-248\,\mathrm{rad\,m^{-2}}$. Nevertheless, our estimated RM for the core has the opposite sign to that of \citet{Hovatta2012}, which can suggest changes between epochs, since there are 16 years of difference between observations. 

The relevance of 3C418 for geodetic VLBI has been emphasized in several recent studies. \citet{Xu2021} describe 3C418 as an extremely extended source in geodetic VLBI and show that its structure produces significant closure delay signatures in VGOS data. More recently, \citet{Xu2025} interpreted the astrometric position variations of 3C418 as the result of source structure and jet kinematics, with the brightness centroid shifting downstream at lower frequencies.

Thus, 3C418 illustrates the dual role of source structure in VGOS observations. From an astrophysical point of view, the broadband data provide information on opacity, spectral gradients, Faraday rotation, and jet physics. From a geodetic point of view, the same structure introduces systematic effects in closure delays, post-fit residuals, and source positions. The fact that our measured core shift, spectral index structure, and RM behaviour are broadly consistent with previous VLBA results demonstrates that VGOS data can be used not only to identify and correct source structure effects, but also to extract physically meaningful information about the innermost regions of radio loud AGN.

\pagebreak
\section{0059+581}
\label{sec:0059+581}

0059+581, also known as J0102+5824, is a bright radio-loud quasar at redshift $z=0.644$ \citep{SowardsEmmerd2005}. The source is included in the ICRF3 S/X catalogue, where it is classified as a defining source \citep{Charlot2020}. Its ICRF3 coordinates are
$\alpha_{\mathrm{J2000}} = 01^{\mathrm h}02^{\mathrm m}45.76237^{\mathrm s}$ and
$\delta_{\mathrm{J2000}} = +58^\circ24'11.1366''$.

\begin{figure}[htpb]
    \centering
    \includegraphics[width=0.67\textwidth]{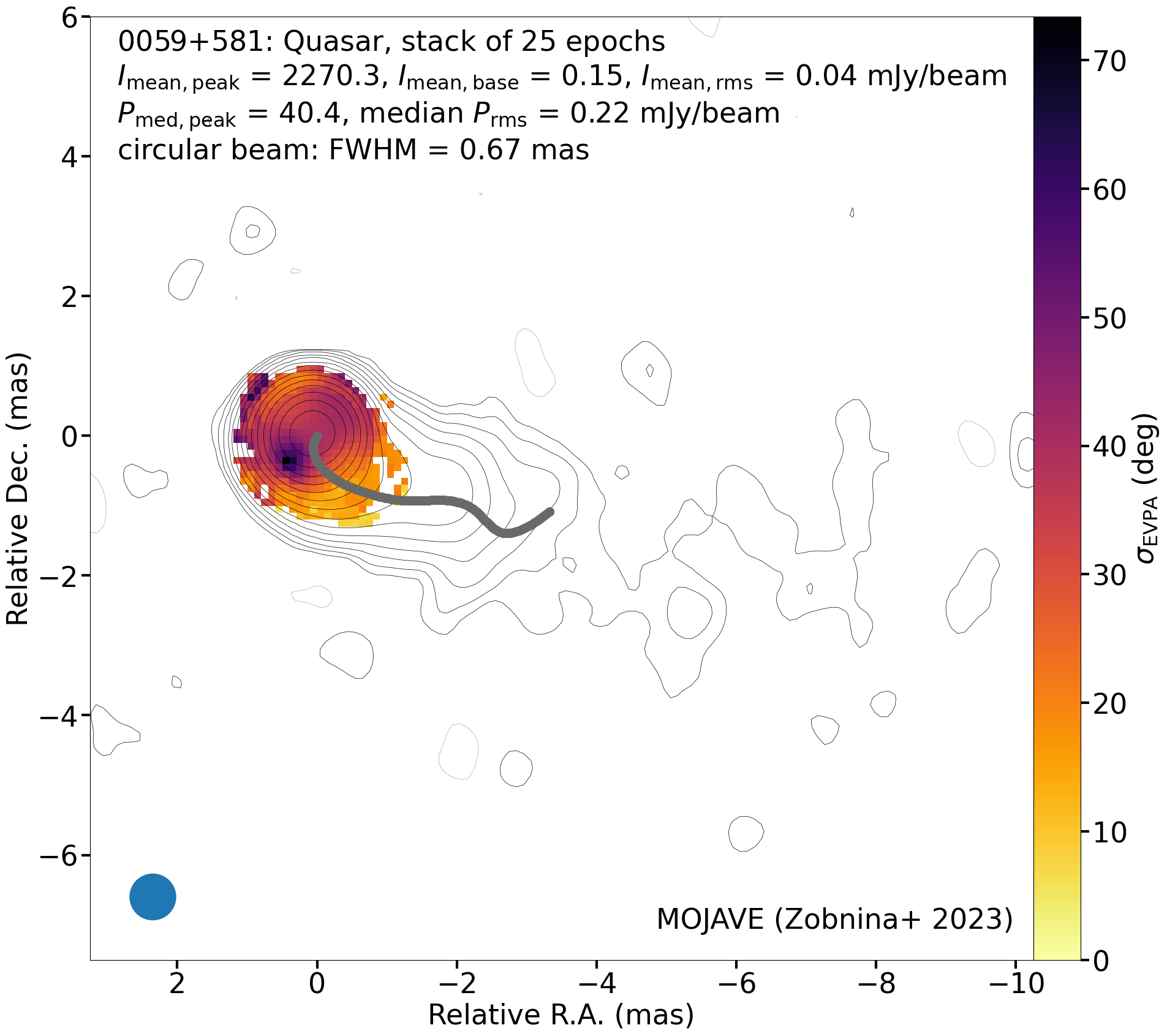}
    \caption[Stacked 15~GHz polarization VLBA image of 0059+581.]{Stacked 15~GHz polarization VLBA image of 0059+581 from the MOJAVE program \citep{Zobnina2023}.}
    \label{fig:0059+581_mojave}
\end{figure}

In the MOJAVE classification, 0059+581 is categorized as a low-spectral peaked quasar \citep{Lister2018}. Long-term 15~GHz VLBA monitoring shows a compact core and a parsec-scale jet, as shown in Fig.~\ref{fig:0059+581_mojave}. The fastest reported angular speed is $224\pm11~\mu$as~yr$^{-1}$, corresponding to an apparent speed of $8.28\pm0.41\,c$, based on four moving features \citep{Lister2021}. This confirms that 0059+581 hosts a relativistic jet with significant structural evolution on parsec scales.

The jet geometry and relativistic parameters of 0059+581 have also been studied in the context of the MOJAVE analysis of AGN jet opening angles. \citet{Pushkarev2017} report an apparent opening angle at 15~GHz of $\alpha_{\rm app}=26.9^\circ\pm2.0^\circ$, a Doppler factor $\delta=21.9\pm3.6$, a bulk Lorentz factor $\Gamma=12.0\pm1.6$, and a viewing angle $\theta=1.5^\circ\pm0.5^\circ$. They also derive an intrinsic full opening angle of $\alpha_{\rm int}=0.7^\circ\pm0.2^\circ$. These parameters provide the geometrical and relativistic quantities required to estimate the magnetic field strength from the measured core shift under the equipartition assumption.

0059+581 is particularly interesting because it combines high compact flux density with measurable extended structure. For the epoch RV156 (2022-11-28), the BVID database reports an X-band flux density of 1.73~Jy, a structure index of 3.08, and a compactness of 0.39 \citep{Collioud2009}. These values indicate that, despite being a bright source commonly used in geodetic VLBI, a significant fraction of its radio emission is resolved on VLBI scales.

\subsection{Core Shift}

\begin{figure}[htpb]
   \centering
   \includegraphics[width=0.67\textwidth]{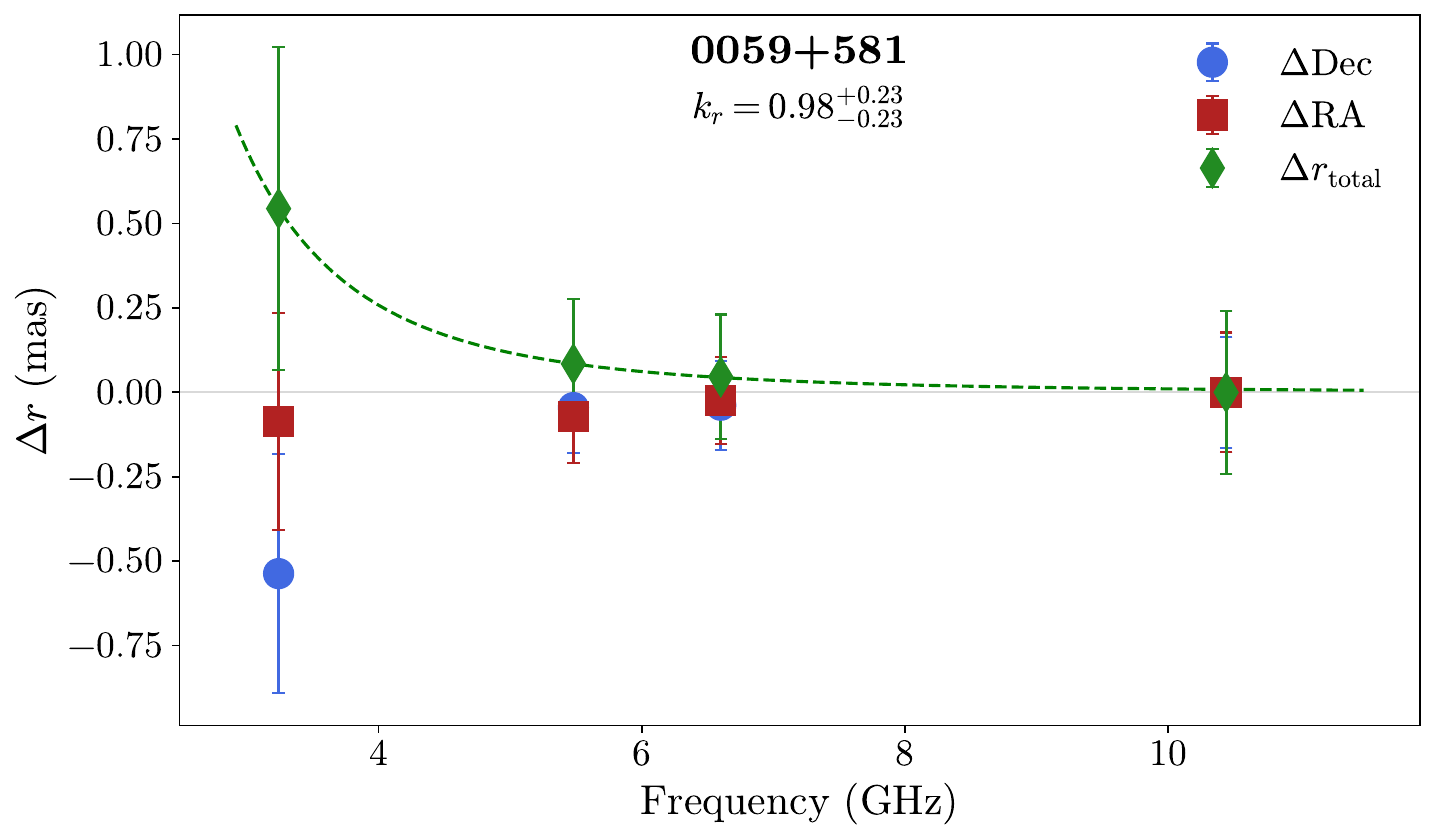}
   \caption[Measured relative core shift as a function of frequency for 0059+581.]{Measured relative core shift as a function of frequency for 0059+581. The total shift vector, along with its components in right ascension and declination, are shown. The error bars represent 3$\sigma$ uncertainties. The fit model is shown as a dashed green line.}
   \label{fig:core_shift_0059+581}
\end{figure}

Applying the same procedure described in Chapter~\ref{ch:1803+784}, we obtained the core-shift measurements for 0059+581, shown in Fig.~\ref{fig:core_shift_0059+581}.

We measure a total core shift of approximately 0.55~mas between 3 and 11~GHz. The uncertainties are large, especially at the lowest frequencies, where the poorer angular resolution and the blending between the compact core and the innermost jet emission make the position of the brightness peak more difficult to determine. As in the case of 3C418, the limited number of observations available for this source in VO2187 also contributes to the size of the uncertainties.

Despite these limitations, the measured displacements follow the expected frequency-dependent behaviour of the VLBI core position. Fitting the standard core-shift relation described in Sect.~\ref{sec:jetemissionmodel}, we obtain
\[
    k_r = 0.98^{+0.23}_{-0.23}.
\]
This value is fully consistent with the equipartition expectation, $k_r \simeq 1$. Therefore, although the individual measurements should be interpreted with caution, the global frequency dependence indicates that the opacity gradient traced by the VGOS bands is compatible with the standard model of a conical synchrotron jet.

A previous measurement for this source was reported by \citet{Plavin2019}, who found a core shift of $0.42\pm0.19$~mas between 2.3 and 8.6~GHz. Extrapolating our fitted core-shift model to the same frequency pair, we obtain a value of approximately $0.81\pm0.70$~mas. The uncertainty is large, but the result is compatible with the measurement of \citet{Plavin2019}. The difference between the central values is not unexpected, given the different observing epochs, the different frequency setup, and the intrinsic variability of core shifts in compact relativistic jets. In fact, \citet{Plavin2019} showed that core shifts measured between S and X bands can vary significantly with time, especially in sources with variable core flux density. In this context, our measurement should be regarded as an estimate of the core shift at this epoch, not as a fixed property of the source.

Since the fitted value of $k_r$ is consistent with unity, we used the equipartition formalism described in Sect.~\ref{sec:physicaljet} to estimate the physical parameters of the jet. From the measured core shifts, we obtain a core-position offset of
\[
    \Omega_{r\nu} = 12.99^{+3.77}_{-3.37}~{\rm pc\,GHz}.
\]

For the geometrical and relativistic jet parameters, we adopted the values reported by \citet{Pushkarev2017}: $\delta_j=21.9\pm3.6$, $\theta=1.5^\circ\pm0.5^\circ$, and an intrinsic full opening angle $\alpha_{\rm int}=0.7^\circ\pm0.2^\circ$. Since the formalism of \citet{Lobanov1998} uses the jet half opening angle, we take $\phi=\alpha_{\rm int}/2=0.35^\circ\pm0.10^\circ$.

Using Eq.~\ref{eq:magnetic_field}, we obtain a magnetic field strength at 1~pc of
\[
    B_1 = 13.6^{+4.9}_{-6.7}~{\rm G}.
\]

Assuming $\gamma_{\max}/\gamma_{\min}=100$ (as in \citet{Lobanov1998} and \citet{Lisakov2017}), Eq.~\ref{eq:N1_B} then gives a particle density at 1~pc of
\[
    N_1 = \left(2.0^{+2.1}_{-1.5}\right)\times10^6~{\rm cm^{-3}}.
\]

The large fractional uncertainty mainly reflects the uncertainty in the viewing angle, since the estimate of $B_1$ depends strongly on $\sin\theta$.

The corresponding deprojected distances of the VGOS cores from the jet origin range from
\[
    r_{\rm core} \simeq 185~{\rm pc}
\]
at band A to
\[
    r_{\rm core} \simeq 57~{\rm pc}
\]
at band D. These large deprojected distances are a direct consequence of the small viewing angle of this source. Applying the radial scalings of Eqs.~\ref{eq:magnetic_field_r} and \ref{eq:N1_r}, we obtain magnetic fields at the observed VGOS core positions of
\[
    B_{\rm core} \simeq 0.08 - 0.25~{\rm G},
\]
and particle densities of
\[
    N_{\rm core} \simeq 60 - 650~{\rm cm^{-3}},
\]
from band A to band D.

\subsection{Rotation Measure}

\begin{figure}[htpb]
   \centering
   \includegraphics[width=0.66\textwidth, trim=0.0cm 0.3cm 0.0cm 0]{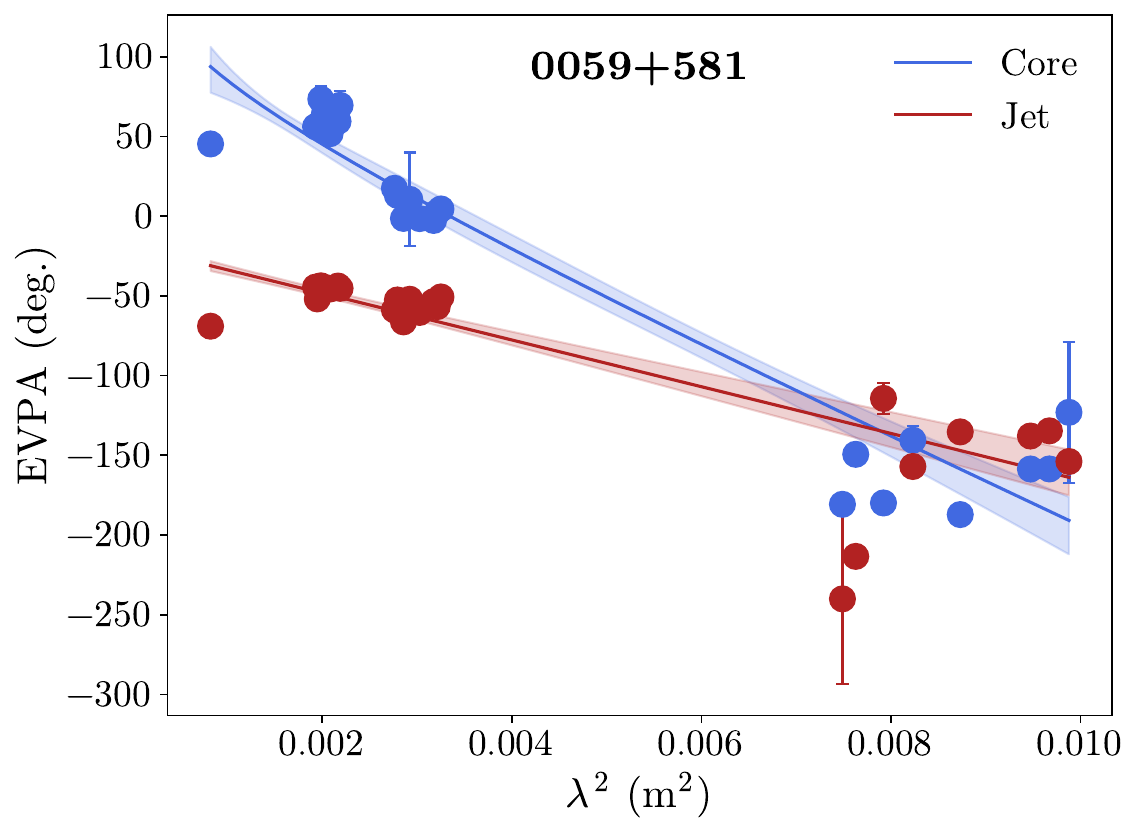}
   \caption[Fit of the polarization angle as a function of $\lambda^2$ for 0059+581.]{Fit of the polarization angle as a function of $\lambda^2$ for the core and jet regions of 0059+581. The shaded regions represent the 3$\sigma$ confidence intervals of the model fits. Band D data have been averaged because of the low S/N in this band.}
   \label{fig:RM_fit_0059+581}
\end{figure}

Figure~\ref{fig:RM_fit_0059+581} shows the fit of the electric vector position angle (EVPA) as a function of $\lambda^2$ for the core and jet regions of 0059+581.

As for the other sources analysed in this chapter, the EVPA uncertainties do not allow us to reliably fit the exponent of the internal Faraday rotation model as a free parameter. We therefore fix it to $a=2.1$, following the same criterion used in the previous cases. For the core, this fit gives an external Faraday rotation of
\[
    RM^{\rm core}_{\rm E} = -456 \pm 40~{\rm rad~m^{-2}},
\]
and an internal contribution of
\[
    RM^{\rm core}_{\rm I} = -113 \pm 50 ~{\rm rad~m^{-2}}.
\]
Both components have the same sign, indicating that the Faraday rotation affecting the compact core region is dominated by a negative RM contribution along the line of sight.

For the jet region, the EVPA behaviour is described by a linear dependence with $\lambda^2$. We obtain
\[
    RM^{\rm jet}_{\rm tot} = -256 \pm 10~{\rm rad~m^{-2}},
\]
\[
    RM^{\rm jet}_{\rm I} = 200 \pm 60~{\rm rad~m^{-2}},
\]

The RM inferred for the core can be compared with previous VLBA measurements of this source. In particular, \citet{Hovatta2012} reported an RM of $-147 \pm 72~{\rm rad~m^{-2}}$ for the core of 0059+581. This value has the same sign as the RM components obtained here.

\subsection{Multi-frequency Imaging}

\begin{figure}[htpb]
   \centering
   \includegraphics[
      width=\textwidth,
      trim=35 10 35 10,
      clip
   ]{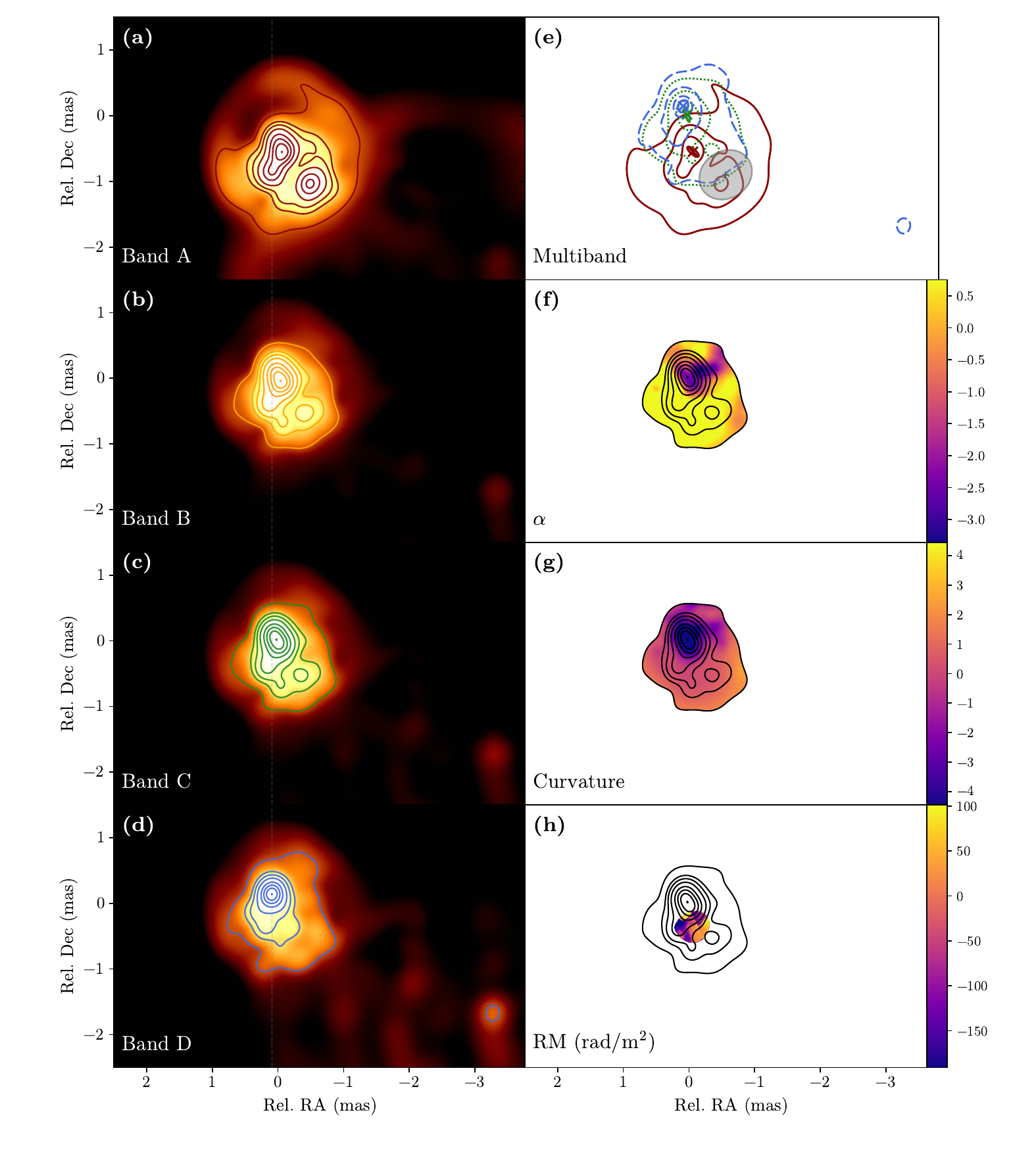}
   \caption[Multi-frequency polarimetric image of 0059+581.]{Multi-frequency polarimetric image of 0059+581. Panels (a)--(d) show the total intensity maps at bands A, B, C, and D, respectively, displayed in logarithmic scale. The color scale in each panel is normalized to its peak intensity. The dashed vertical line indicates the core position at band A, marking the core shift across bands. Panel (e) displays the total-intensity contours for bands A (solid red), C (dotted green) and D (dashed blue) in a common reference frame, together with the core and jet Gaussian components obtained from the visibility-domain model fitting. Panel (f) shows the spectral index map ($\alpha$), and panel (g), the curvature map. Panel (h) presents the rotation measure (RM) map. Only regions above 1\% of the peak intensity are included.}
   \label{fig:polarimetric_0059+581}
\end{figure}

The multi-frequency image of 0059+581 is shown in Fig.~\ref{fig:polarimetric_0059+581}. The total intensity maps reveal a compact bright core and a jet extending downstream towards the south west, in agreement with the morphology observed in previous VLBI images of this source.

A remarkable property of 0059+581 is that the jet does not follow a straight trajectory. In our VGOS images, the emission initially emerges from the core towards the south, but then progressively bends in a counter-clockwise direction. This behaviour is also seen in previous VLBI images of the source, including the MOJAVE images presented by \citet{Cooper2007}, \citet{Hovatta2014}, and \citet{Zobnina2023}, as well as in the images available from the Bordeaux VLBI Image Database \citep{Collioud2019}. In the images with a larger field of view, the downstream emission can be traced farther from the core, where the jet bends towards the north-west. Therefore, the morphology recovered from the VGOS observations is consistent with the twisted structure previously observed in this source.

The superposition of the contours from the four VGOS bands shows a clear displacement of the brightness peak. The low-frequency emission is shifted downstream with respect to the high-frequency core position, as expected for a typical core shift. This qualitative behaviour is consistent with the quantitative fit shown in Fig.~\ref{fig:core_shift_0059+581}. The direction of the core shift is also consistent with the local direction of the inner jet, supporting the interpretation that the measured displacement is associated with opacity effects along the jet axis.

The spectral index map shows a compact region with a flat spectrum associated with the core, while the downstream emission is characterized by an optically thin spectrum. In the jet, we obtain greater values of the spectral index than those shown in \citet{Hovatta2014}. This difference may be related to the use of a curved spectral model in our multi-frequency reconstruction, since part of the frequency dependence can be absorbed by the curvature term when fitting the wide VGOS frequency coverage.

The RM map shows strong spatial gradients, suggesting that the S/N of the polarized emission in this source is not sufficient to obtain a robust pixel-by-pixel RM distribution over the whole jet. Nevertheless, the map qualitatively shows different RM signs between the core and the jet, in agreement with the behaviour obtained from the model fitting.

\subsection{Discussion}

0059+581 provides a useful example of a source that is sufficiently compact to be used regularly in geodetic VLBI, but sufficiently structured to produce measurable astrophysical and astrometric effects. The VGOS images recover the same general morphology seen in 15 GHz VLBA observations: a compact core and a one-sided jet extending towards the south-west. The spectral index structure also follows the expected pattern for a self-absorbed VLBI core and an optically thin jet. In addition, the images show that the jet is not straight, but progressively bends downstream. This curved morphology is also seen in previous VLBI images of the source, where the jet initially emerges towards the south and then rotates counter-clockwise, eventually pointing towards the north-west on larger angular scales.

The core shift measured from the VGOS bands is about 0.55~mas between 3 and 11~GHz. This value is consistent with the lower-frequency nature of the VGOS observations: the opacity-driven displacement is expected to be larger at lower frequencies than in the 8--15~GHz range. When our fitted model is extrapolated to 2.3 and 8.6~GHz, the result is compatible with the S/X-band value reported by \citet{Plavin2019}, although with a large uncertainty.

The fitted value $k_r = 0.98^{+0.23}_{-0.23}$ is consistent with the equipartition expectation. This suggests that, at least within the uncertainties of the present data, the frequency dependence of the apparent core position can be described by the standard conical jet model. Nevertheless, 0059+581 is known to be variable. \citet{Varenius2022} reported a bright multi-frequency flare in this source using the Onsala twin telescopes in the VGOS frequency range. Since changes in the particle density and opacity of the inner jet can modify the apparent core position, the core shift measured here should not be interpreted as a permanent source property. Instead, it represents the state of the opacity structure at the epoch sampled by VO2187.

The physical parameters derived from the core-shift analysis are also consistent with the range of values reported for compact relativistic jets, although the inferred value of $B_1$ is relatively high. We obtain $B_1=13.6^{+4.9}_{-6.7}$~G and $N_1=(2.0^{+2.1}_{-1.5})\times10^6~{\rm cm^{-3}}$ under the equipartition assumption. The value of $B_1$ is larger than the median values found in large core-shift samples, the typical $B_1\sim0.9$~G reported for quasars by \citet{Pushkarev2012}. Regarding the particle density, we notice that a value of order $10^6\,{\rm cm}^{-3}$ may be abnormally large. In any case, these estimates are strongly affected by the geometrical parameters adopted for the jet; a small change in the viewing angle will result in large changes in both magnetic field and particle density (Eq. \ref{eq:magnetic_field}).

The RM results show that both the core and the jet are affected by negative Faraday rotation. For the core, the model with fixed $a=2.1$ gives $RM^{\rm core}_{\rm E}=-456~{\rm rad~m^{-2}}$ and $RM^{\rm core}_{\rm I}=-113~{\rm rad~m^{-2}}$, while the linear fit to the jet gives $RM^{\rm jet}_{\rm tot}=-256\pm10~{\rm rad~m^{-2}}$. The sign of these values is consistent with the core RM reported by \citet{Hovatta2012}, $RM=-147\pm72~{\rm rad~m^{-2}}$.

An important point of this analysis is that the EVPA points in both the jet and the core are not perfectly described by a simple linear dependence with $\lambda^2$. In particular, the highest-frequency data, corresponding to band D, shows a small offset with respect to the trend defined by the lower frequency bands. A similar behaviour can also be seen in the data of \citet{Hovatta2012}, where the EVPAs do not follow an exact $\lambda^2$ law and there are offsets compared to their fit. This suggests that the assumption of a single, homogeneous, external Faraday screen is probably too simple for this source. As discussed by \citet{Hovatta2019}, deviations from a linear $\lambda^2$ relation may appear when the observed polarization is produced by several unresolved components that undergo different amounts of Faraday rotation. This interpretation is plausible for 0059+581, given its curved jet morphology (Fig. \ref{fig:polarimetric_0059+581}): different polarized regions within the beam may be seen through different magnetized plasma screens, or through different parts of the same inhomogeneous screen.

A more complete treatment of the Faraday structure of 0059+581 would therefore require Faraday rotation measure synthesis \citep{Brentjens2005}, possibly followed by RM-CLEAN deconvolution \citep{Heald2009}. This approach would make it possible to search for polarized emission as a function of Faraday depth. It would also help to identify possible multiple Faraday components and to diagnose $n\pi$ ambiguities in the EVPAs. Such an analysis lies outside the scope of this thesis.

From the geodetic point of view, 0059+581 is especially relevant because it is a defining ICRF3 source and is very frequently observed in VLBI sessions. Recent quad-band astrometric analyses have shown that 0059+581 exhibits systematic position variations, mainly in declination, which are likely related to source structure and jet evolution \citep{Xu2025}. The combination of a measurable core shift, a curved parsec-scale jet, and a complex Faraday-rotation behaviour shows that even sources suitable for geodetic use can contain astrophysical structure that needs to be taken into account in high-precision astrometric and geodetic applications.

\pagebreak
\section{0613+570}
\label{sec:0613+570}

0613+570 (J0617+5701), is located at $\alpha_{\mathrm{J2000}} = 06^{\mathrm h}17^{\mathrm m}16.9226^{\mathrm s}$, $\delta_{\mathrm{J2000}} = +57^\circ01'16.423''$, as listed in the ICRF3 S/X catalogue \citep{Charlot2020}. It is classified as a defining source in ICRF3.

The source is a compact radio-loud AGN extensively monitored in geodetic VLBI programs. A recent statistical analysis of astrometric variability metrics by \citet{Cigan2024} places this source among the ``Best Defining Sources'', based on thresholds in mean positional offset, combined scatter, weighted rms, and structure parameters. The source has been observed in 418 sessions with more than 66,000 delays, demonstrating long-term monitoring within the IVS network \citep{Cigan2024}.

\begin{figure}[hptb]
    \centering
    \includegraphics[width=0.67\textwidth]{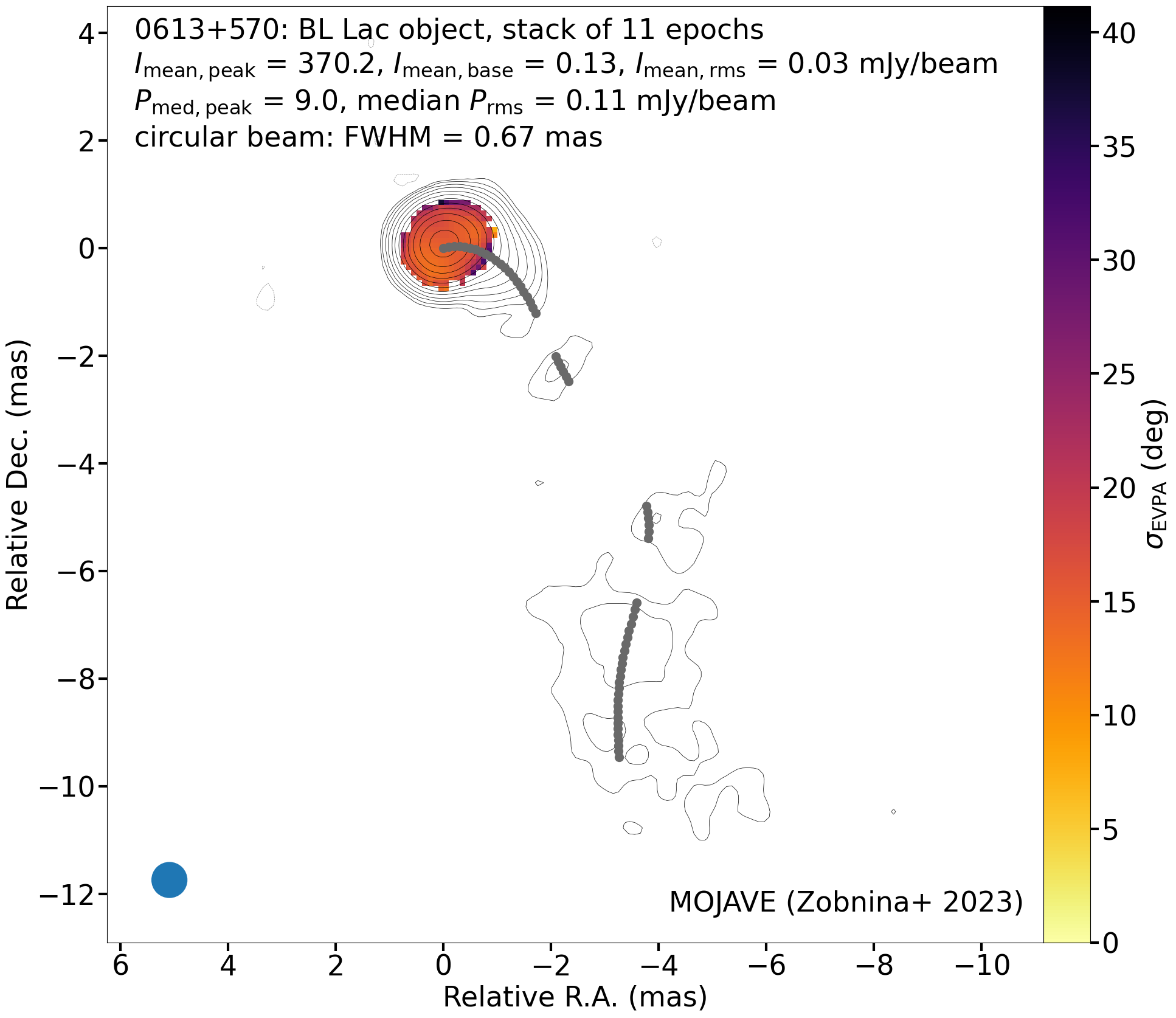}
    \caption[Stacked 15~GHz polarization VLBA image of 0613+570.]{Stacked 15~GHz polarization VLBA image of 0613+570 from the MOJAVE program \citep{Zobnina2023}.}
    \label{fig:0613+570_mojave}
\end{figure}

The source is part of the MOJAVE 15~GHz VLBA monitoring program \citep{Lister2018}. It is classified as a low-synchrotron-peaked BL Lac object with a flat radio spectrum and detected gamma-ray emission by Fermi-LAT. For 0613+570, we adopt the redshift estimate $z=0.592$ listed by \citet{Foschini2022} and \citet{Shaw2013}. Multi-epoch VLBA imaging reveals a compact, core-dominated morphology and measurable jet kinematics, with a maximum apparent angular speed of $10 \pm 3~\mu$as~yr$^{-1}$ based on four moving components \citep{Lister2021}.

Geodetic S/X imaging products from the Bordeaux VLBI Image Database (BVID) provide additional structure diagnostics \citep{Collioud2009}. For epoch RV128 (2018-02-20), the source shows an X band flux density of 0.55~Jy, a structure index of 2.36, and a compactness of 0.82 \citep{Collioud2009}. These values confirm that 0613+570 is considerably more compact than sources such as 3C418 and 0059+581, and explain its suitability for reference frames.

\subsection{Core Shift}

\begin{figure}[H]
   \centering
   \includegraphics[width=0.7\textwidth]{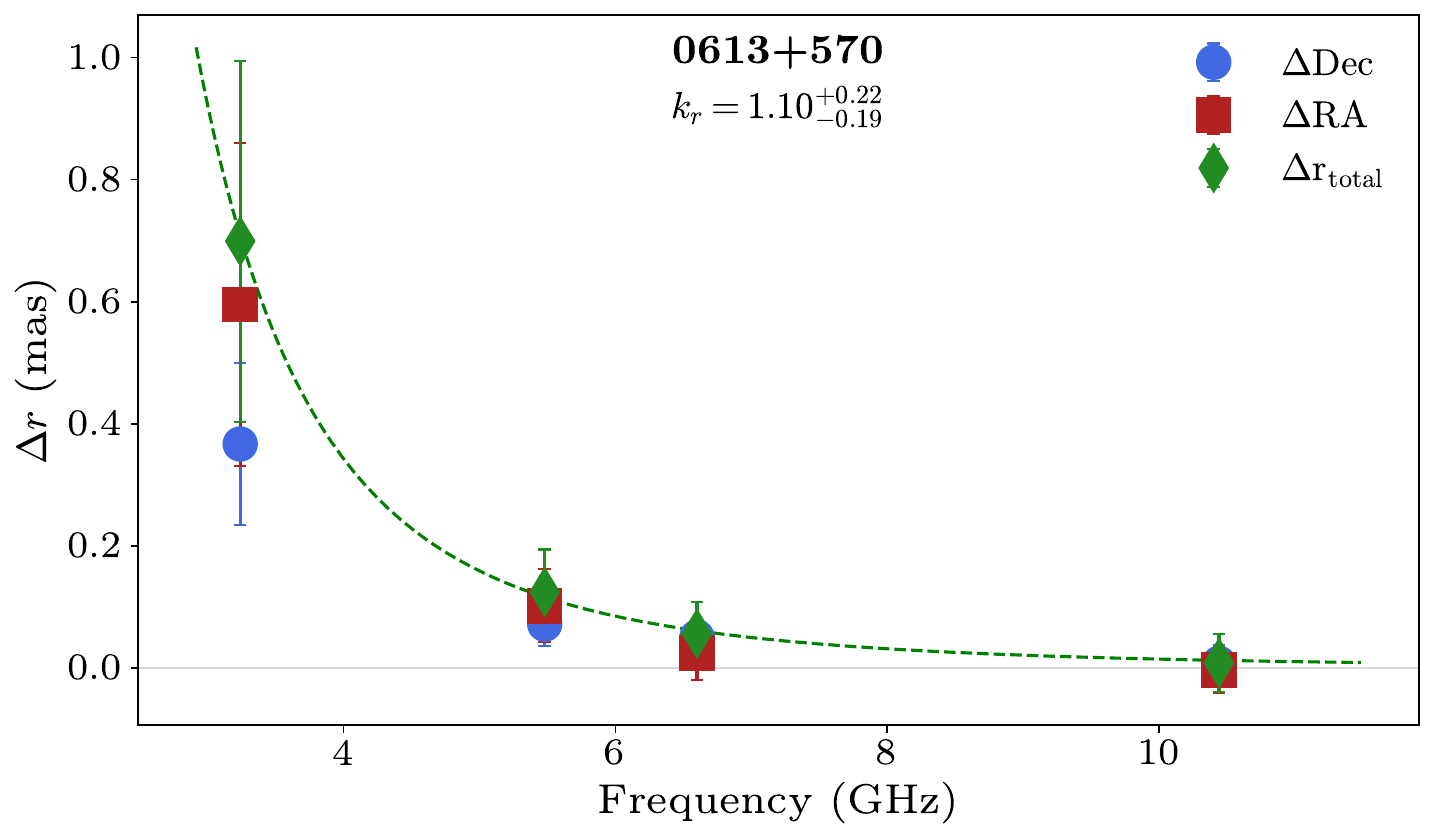}
   \caption[Measured relative core shift as a function of frequency for 0613+570.]{Measured relative core shift as a function of frequency for 0613+570. The total shift vector, along with its components in right ascension and declination, are shown. The error bars represent 3$\sigma$ uncertainties. The fit model is shown as a dashed green line.}
   \label{fig:core_shift_0613+570}
\end{figure}

Applying the same procedure described in Chapter~\ref{ch:1803+784}, we obtained the core-shift measurements for 0613+570, shown in Fig.~\ref{fig:core_shift_0613+570}.

We measure a total core shift of approximately $0.7\pm0.3$~mas between 3 and 11~GHz. The uncertainty is significant, but the displacement is detected across the VGOS frequency range and follows the expected trend of a downstream shift at lower frequencies. This behaviour is consistent with the standard interpretation of the VLBI core of the AGN.

Fitting the standard core-shift relation described in Sect.~\ref{sec:jetemissionmodel}, we obtain
\[
    k_r = 1.10^{+0.22}_{-0.19}.
\]
This value is compatible with the equipartition expectation, $k_r \simeq 1$, within the uncertainties. Therefore, despite the compactness of the source and the limited angular resolution at the lowest VGOS frequencies, the global frequency dependence of the core position is consistent with the standard conical jet model.

The only published core-shift measurement for this source is listed in \citet{Pushkarev2012}, who report a shift of 0.128~mas between 15.4 and 8.1~GHz. Extrapolating our fitted model to the same frequency pair, we obtain a smaller value of
\[
    \Delta r_{15.4-8.1} = 0.034 \pm 0.013~{\rm mas}.
\]

The two estimates are therefore not in close agreement. This difference may be related to the different observing epochs and the difficulty of extrapolating a fit constrained mainly by lower frequency to the 8--15~GHz range. Since the expected displacement at these higher frequencies is small, even modest changes in the opacity structure of the inner jet can produce a significant fractional difference. For this reason, the comparison should be interpreted as showing the same order of magnitude, but not as a strict agreement.

\subsection{Rotation Measure}

\begin{figure}[H]
   \centering
   \includegraphics[width=0.66\textwidth, trim=0.0cm 0.3cm 0.0cm 0]{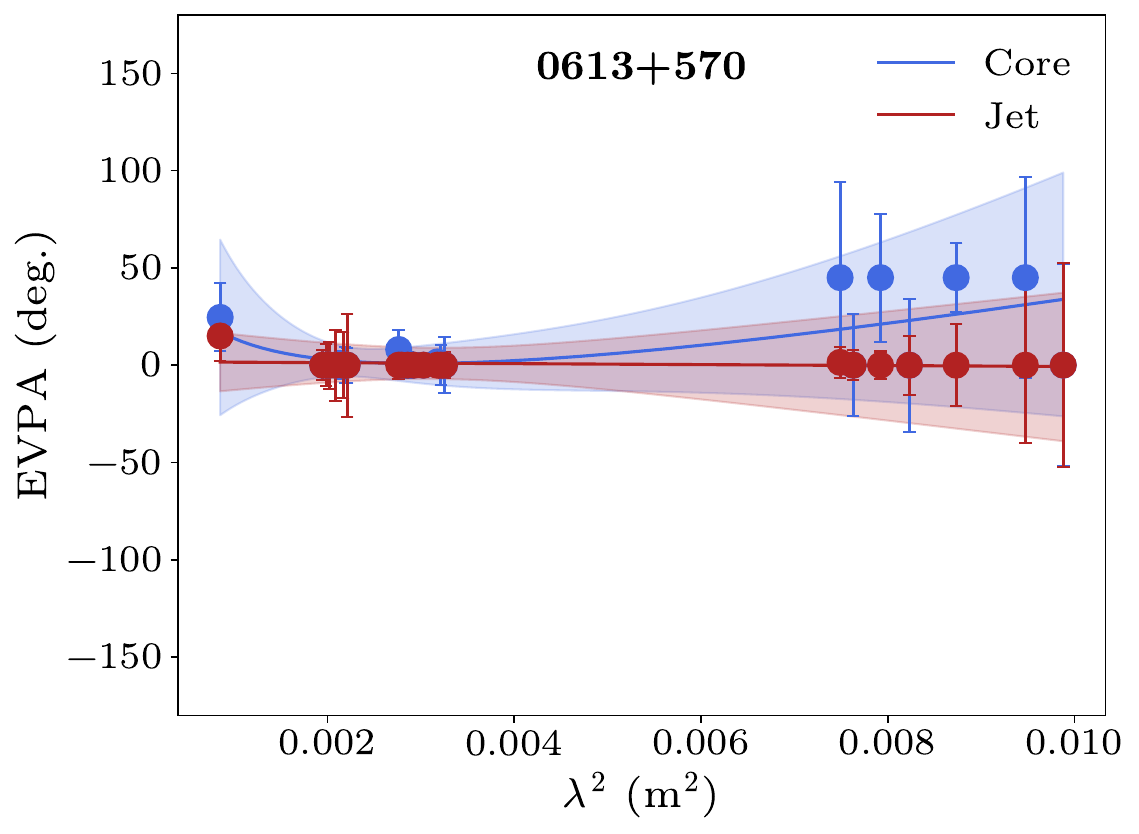}
   \caption[Fit of the polarization angle as a function of $\lambda^2$ for 0613+570.]{Fit of the polarization angle as a function of $\lambda^2$ for the core and jet regions of 0613+570. The shaded regions represent the 3$\sigma$ confidence intervals of the model fits. Band D data have been averaged because of the low S/N in this band.}
   \label{fig:RM_fit_0613+570}
\end{figure}

Figure~\ref{fig:RM_fit_0613+570} shows the fit of the EVPA as a function of $\lambda^2$ for the core and jet regions of 0613+570.

As for the other sources analysed in this chapter, the EVPA uncertainties do not allow us to reliably fit the exponent of the internal Faraday rotation model as a free parameter. We therefore fix it to $a=2.1$, following the same criterion used in the previous cases. For the core, we obtain an external Faraday rotation of
\[
    RM^{\rm core}_{\rm E} = 170 \pm 60~{\rm rad~m^{-2}},
\]
and an internal contribution of
\[
    RM^{\rm core}_{\rm I} = -170 \pm 70~{\rm rad~m^{-2}}.
\]

For the jet region, a linear fit gives
\[
    RM^{\rm jet}_{\rm tot} = -4 \pm 15~{\rm rad~m^{-2}}.
\]

If the external contribution affecting the jet is assumed to be comparable to that inferred for the core region, this implies an internal jet contribution of
\[
    RM^{\rm jet}_{\rm I} = 165 \pm 80~{\rm rad~m^{-2}}.
\]

\subsection{Multi-frequency Imaging}

\begin{figure}[htpb]
   \centering
   \includegraphics[
      width=\textwidth,
      trim=55 10 50 10,
      clip
   ]{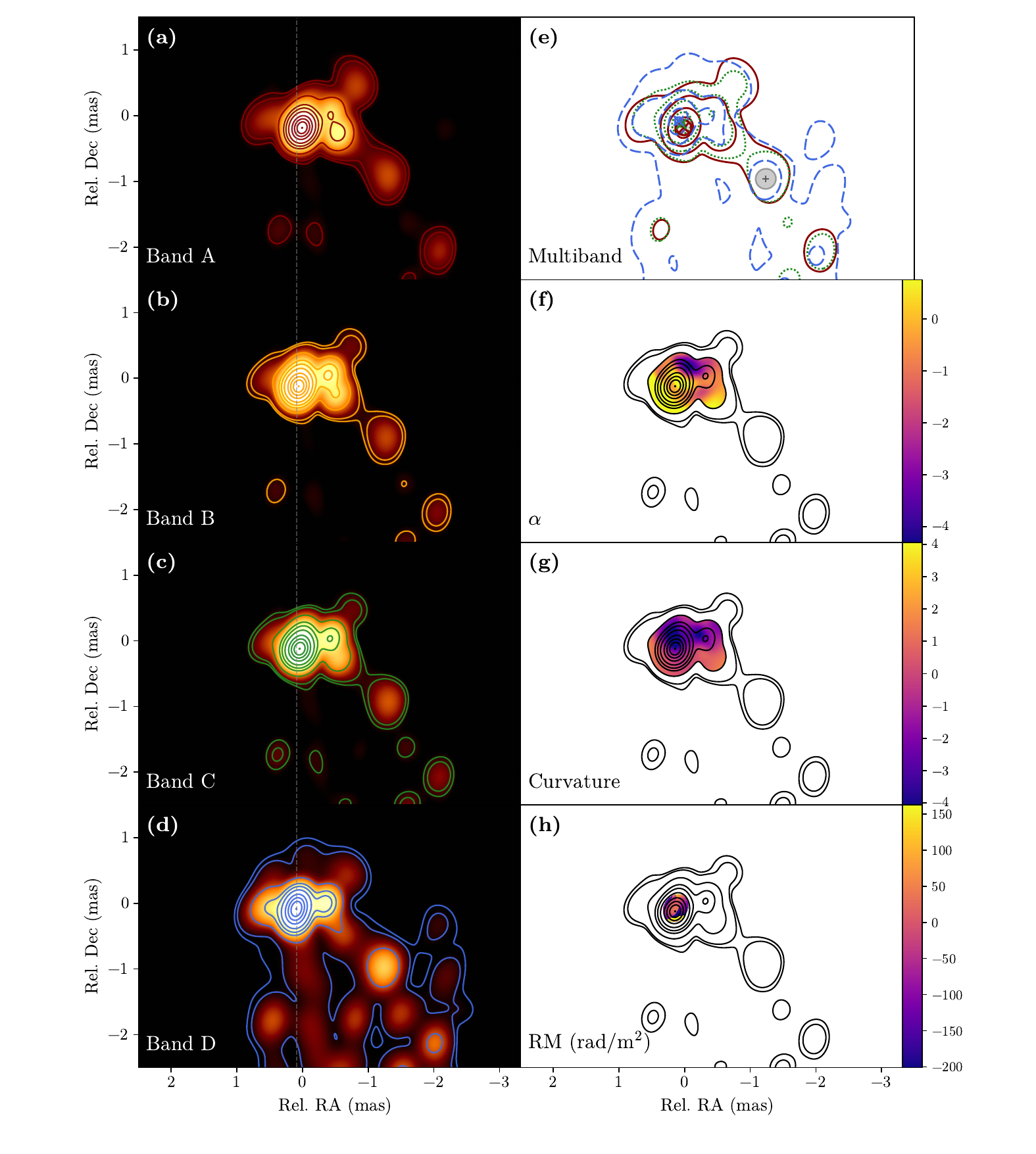}
   \caption[Multi-frequency polarimetric image of 0613+570.]{Multi-frequency polarimetric image of 0613+570. Panels (a)--(d) show the total intensity maps at bands A, B, C, and D, respectively, displayed in logarithmic scale. The color scale in each panel is normalized to its peak intensity. The dashed vertical line indicates the core position at band A, marking the core shift across bands. Panel (e) displays the total-intensity contours for bands A (solid red), C (dotted green) and D (dashed blue) in a common reference frame, together with the core and jet Gaussian components obtained from the visibility-domain model fitting. Panel (f) shows the spectral index map ($\alpha$), and panel (g), the curvature map. Panel (h) presents the rotation measure (RM) map. Only regions above 0.1\% of the peak intensity are included.}
   \label{fig:polarimetric_0613_570}
\end{figure}

The multi-frequency image of 0613+570 is shown in Fig.~\ref{fig:polarimetric_0613_570}. In contrast to 3C418 and 0059+581, the source appears more compact and more strongly core dominated. The total intensity maps show a bright compact core and weak downstream emission, consistent with the high compactness reported by geodetic S/X band imaging products.

Even in this compact source, the superposition of the contours from the four VGOS bands reveals a measurable displacement of the brightness peak. The low-frequency emission is shifted with respect to the high-frequency core position, in agreement with the core-shift fit shown in Fig.~\ref{fig:core_shift_0613+570}. This result shows that frequency-dependent opacity effects are detectable even in sources that are among the most compact and stable defining sources in the celestial reference frame.

The spectral index map shows a flat spectrum core and a steeper spectrum in the downstream emission, as expected for a synchrotron self absorbed core and optically thin jet emission. No published spectral index map could be found for 0613+570 to compare, but in the image it can be seen that the core region is flat and the jet has typical values of $-0.5 - -1$.

The curvature map suggests that a single power law does not fully describe the broadband spectrum in all regions, since negative values close to -1 appear in some regions.

Regarding the RM map, the low intensity of the jet and the relatively low polarization S/N make a detailed interpretation difficult. Nevertheless, the RM distribution is consistent with the model fitting in the sense that the combined internal and external RM is close to 0, producing soft gradients in the EVPA as a function of the frequency.

\subsection{Discussion}

0613+570 is an especially useful source for this analysis because it represents a different regime from sources such as 3C418. It is compact, core dominated, and astrometrically stable enough to be classified as one of the best defining sources in recent variability studies \citep{Cigan2024}. Nevertheless, the VGOS data reveal measurable astrophysical structure, including a frequency-dependent core shift, a non-uniform spectral index distribution, and some Faraday rotation.

The measured core shift of approximately $0.7\pm0.3$~mas between 3 and 11~GHz shows that opacity effects can be relevant even in compact sources. The fitted value $k_r = 1.10^{+0.22}_{-0.19}$ is consistent with the equipartition expectation, suggesting that the apparent core position follows the standard frequency dependence expected for a conical synchrotron jet. This result supports the use of VGOS broadband data to constrain physical jet parameters, even when the source is not strongly extended.

The comparison with the 15.4--8.1~GHz measurement reported by \citet{Pushkarev2012} is less straightforward. Our extrapolated value is smaller than their reported shift, although both values are small in absolute terms. Since the 8--15~GHz displacement probes a more compact region closer to the true jet base, it is particularly sensitive to changes in the inner jet opacity and to the identification of the core position. The discrepancy may reflect real time variability, differences in angular resolution and frequency coverage, or limitations of extrapolating the low frequency VGOS fit to higher frequencies.

The RM behaviour is also different from that found in more strongly polarized or more extended sources. The jet RM obtained from the linear fit is consistent with zero, while the core fit suggests external and internal contributions of comparable amplitude but opposite sign. 

\pagebreak
\section{1849+670}
\label{sec:1849+670}

1849+670 is a compact flat-spectrum radio quasar at redshift $z=0.657$ \citep{Stickel1993}. Its ICRF3 position is
$\alpha_{\mathrm{J2000}} = 18^{\mathrm h}49^{\mathrm m}16.0723^{\mathrm s}$ and
$\delta_{\mathrm{J2000}} = +67^\circ05'41.6803''$ \citep{Charlot2020}. The source is classified as defining in ICRF3 \citep{Charlot2020}.

\begin{figure}[H]
    \centering
    \includegraphics[width=0.67\textwidth]{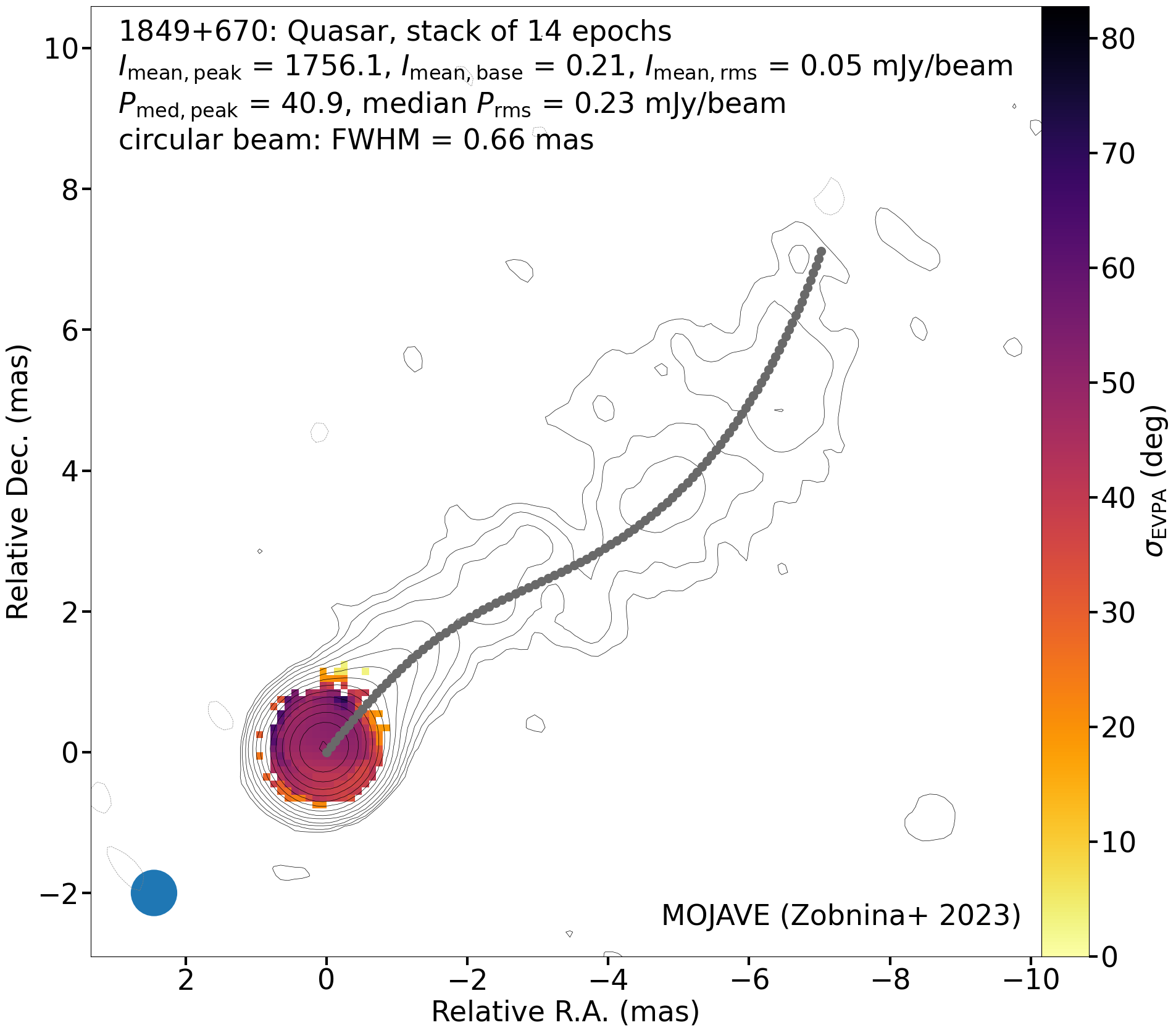}
    \caption[Stacked 15~GHz polarization VLBA image of 1849+670.]{Stacked 15~GHz polarization VLBA image of 1849+670 from the MOJAVE program \citep{Zobnina2023}.}
    \label{fig:1849+670_mojave}
\end{figure}

The source is included in the MOJAVE 15~GHz VLBA monitoring program, where it is classified as a low-spectral peaked quasar with a flat radio spectrum and detected gamma-ray emission by {\it Fermi}-LAT \citep{Lister2018}. The MOJAVE kinematic analysis reports multiple moving features in the parsec-scale jet, with the fastest component reaching $589 \pm 33~\mu$as~yr$^{-1}$, corresponding to an apparent speed of $22.1 \pm 1.2\,c$ \citep{Lister2019}. This indicates that 1849+670 hosts a highly relativistic jet viewed at a small angle to the line of sight.

The jet geometry and relativistic parameters of 1849+670 have also been characterized in the MOJAVE analysis of AGN jet opening angles. \citet{Pushkarev2017} report an apparent opening angle at 15~GHz of $\alpha_{\rm app}=18.9^\circ\pm0.4^\circ$, an apparent speed of $\beta_{\rm app}=21.43\pm1.39\,c$, a Doppler factor $\delta=8.1\pm1.4$, a bulk Lorentz factor $\Gamma=32.5\pm5.6$, and a viewing angle $\theta=4.7^\circ\pm0.3^\circ$. They also derive an intrinsic full opening angle of $\alpha_{\rm int}=1.6^\circ\pm0.1^\circ$. These parameters provide the geometrical and relativistic quantities required to estimate the magnetic field strength from the measured core shift under the equipartition assumption.

On larger scales, 1849+670 has also been studied as part of the MOJAVE-Chandra sample of hybrid radio sources. \citet{Stanley2015} analysed archival VLA data together with {\it Hubble} and {\it Chandra} observations. They identify two large-scale knots in 8C~1849+670 and discuss the broadband emission mechanisms of the extended jet.

Geodetic S/X-band imaging products from the Bordeaux VLBI Image Database provide additional structure diagnostics \citep{Collioud2009}. For the BVID epoch RV130 (2018-08-21), the source has an X-band flux density of 1.17~Jy, a structure index of 2.04, and a compactness of 0.84. These values indicate that the source is relatively compact on geodetic VLBI baselines, consistent with its use as a defining ICRF3 source. The MOJAVE images clearly show a parsec-scale jet, but most of the flux is concentrated in the core.

\subsection{Core Shift}

\begin{figure}[H]
   \centering
   \includegraphics[width=0.7\textwidth]{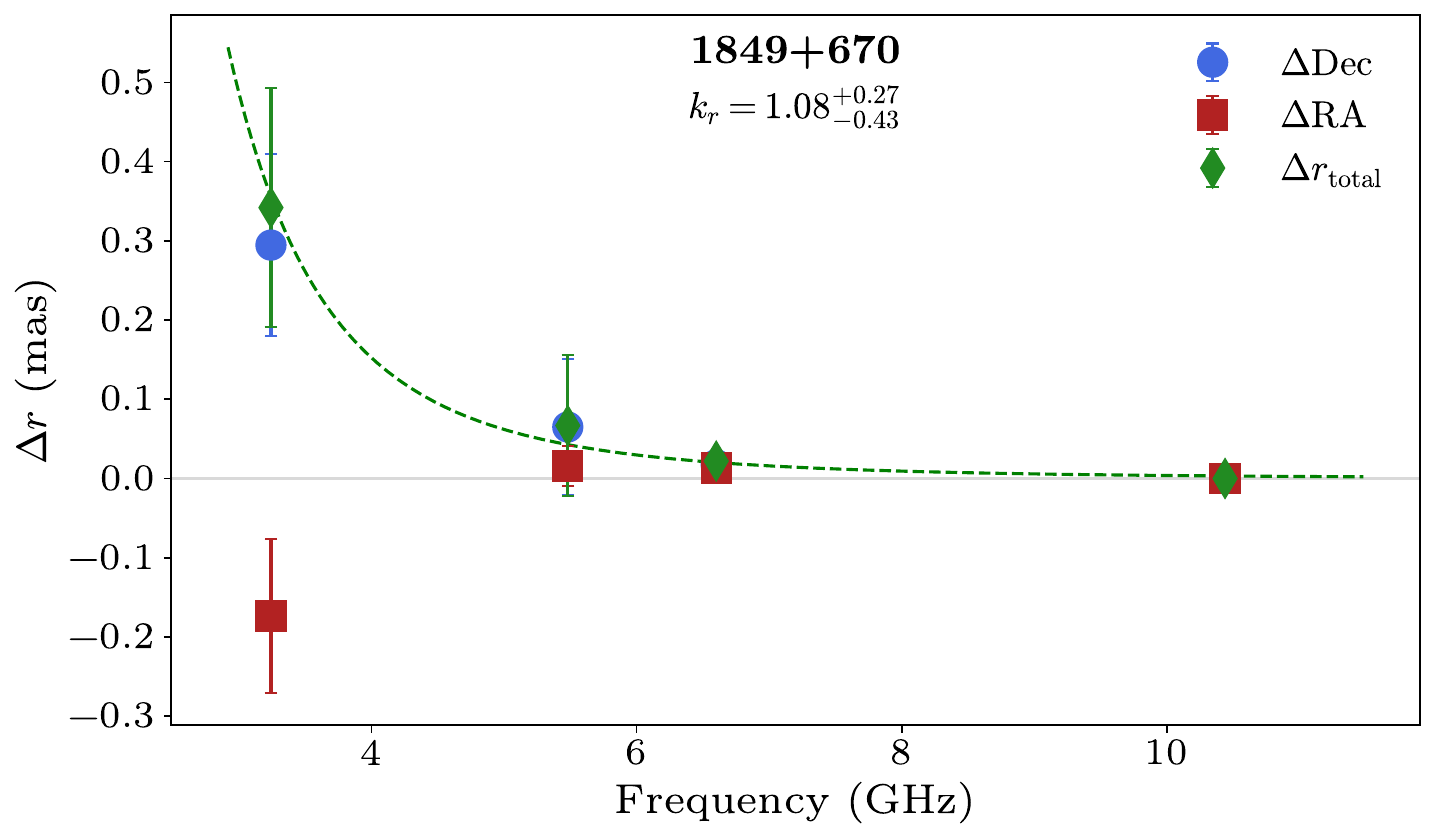}
   \caption[Measured relative core shift as a function of frequency for 1849+670.]{Measured relative core shift as a function of frequency for 1849+670. The total shift vector, along with its components in right ascension and declination, are shown. The error bars represent 3$\sigma$ uncertainties. The fit model is shown as a dashed green line.}
   \label{fig:core_shift_1849+670}
\end{figure}

Applying the same procedure described in Chapter~\ref{ch:1803+784}, we obtained the core-shift measurements for 1849+670, shown in Fig.~\ref{fig:core_shift_1849+670}.

We measure a total core shift of approximately 0.7~mas between 3 and 11~GHz. The displacement is detected across the VGOS frequency range and follows the expected behaviour of a downstream shift at lower frequencies.

Fitting the standard core-shift relation described in Eq.~\ref{eq:coreshift}, we obtain
\[
    k_r = 1.08^{+0.27}_{-0.43}.
\]

Although the uncertainty is asymmetric and relatively large, the fitted value is compatible with the equipartition expectation, $k_r \simeq 1$. Thus, the frequency dependence of the core position is broadly consistent with the standard conical jet model.

The only direct published core-shift measurements that we found for this source are those from the MOJAVE opacity study by \citet{Pushkarev2012}. They report shifts of 0.024, 0.010, and 0.014~mas for the 15.4--8.1, 15.4--8.4, and 15.4--12.1~GHz frequency pairs, respectively. Extrapolating our fitted model to the same frequency pairs, we obtain
\[
    \Delta r_{15.4-8.1} = 0.010 \pm 0.003~{\rm mas},
\]
\[
    \Delta r_{15.4-8.4} = 0.009 \pm 0.003~{\rm mas},
\]
and
\[
    \Delta r_{15.4-12.1} = 0.0014 \pm 0.0008~{\rm mas}.
\]

These extrapolated values are smaller than those reported by \citet{Pushkarev2012}, with the exception of the pair 15.4-8.4 GHz.

Since the fitted value of $k_r$ is compatible with unity, we used the equipartition formalism described in Sect.~\ref{sec:physicaljet} to estimate the physical parameters of the jet. From the measured core shifts, we obtain a core-position offset of
\[
    \Omega_{r\nu} = 10.8^{+3}_{-1.3}~{\rm pc\,GHz}.
\]

For the geometrical and relativistic jet parameters, we adopted the values reported by \citet{Pushkarev2017}: $\delta_j=8.1\pm1.4$, $\theta=4.7^\circ\pm0.3^\circ$, and an intrinsic full opening angle $\alpha_{\rm int}=1.6^\circ\pm0.1^\circ$, which gives a jet half-opening angle of $\phi=\alpha_{\rm int}/2=0.8^\circ\pm0.05^\circ$.

Using Eq.~\ref{eq:magnetic_field}, we obtain a magnetic field strength at 1~pc of
\[
    B_1 = 3.52^{+0.9}_{-0.8}~{\rm G}.
\]

Assuming $\gamma_{\max}/\gamma_{\min}=100$, Eq.~\ref{eq:N1_B} gives a particle density at 1~pc of
\[
    N_1 = \left(1.3^{+0.8}_{-0.6}\right)\times10^5~{\rm cm^{-3}}.
\]

The corresponding deprojected distances of the VGOS cores from the jet origin range from
\[
    r_{\rm core} \simeq 43~{\rm pc}
\]
at band A to
\[
    r_{\rm core} \simeq 13~{\rm pc}
\]
at band D. Applying the radial scalings of Eqs.~\ref{eq:magnetic_field_r} and \ref{eq:N1_r}, we obtain magnetic fields at the observed VGOS core positions of
\[
    B_{\rm core} \simeq 0.08\text{--}0.26~{\rm G},
\]
and particle densities of
\[
    N_{\rm core} \simeq 70\text{--}720~{\rm cm^{-3}},
\]
from band A to band D.

\subsection{Rotation Measure}

\begin{figure}[htpb]
   \centering
   \includegraphics[width=0.66\textwidth, trim=0.0cm 0.3cm 0.0cm 0]{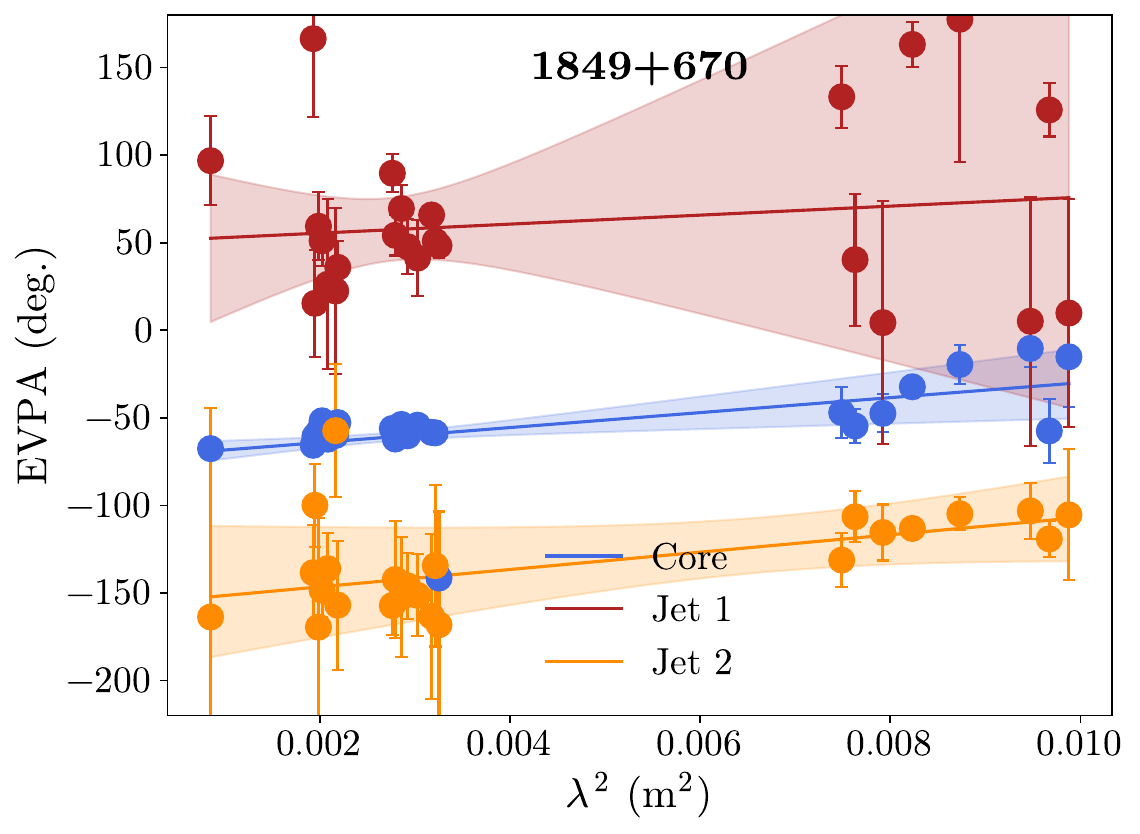}
   \caption[Fit of the polarization angle as a function of $\lambda^2$ for 1849+670.]{Fit of the polarization angle as a function of $\lambda^2$ for the core and jet regions of 1849+670. The shaded regions represent the 3$\sigma$ confidence intervals of the model fits. Band D data have been averaged because of the low S/N in this band.}
   \label{fig:RM_fit_1849+670}
\end{figure}

Figure~\ref{fig:RM_fit_1849+670} shows the EVPA fits as a function of $\lambda^2$ for the core and two jet components of 1849+670.

In contrast to the behaviour found in some of the other sources, the EVPA of the core is well described by a linear dependence on $\lambda^2$. A purely optically-thin Faraday rotation law is therefore sufficient, and no additional internal (optically-thick) Faraday rotation component is required within the present uncertainties. For the core, we obtain
\[
    RM^{\rm core} = 74 \pm 15~{\rm rad~m^{-2}}.
\]

The jet emission was modelled with two separate Gaussian components. The first jet component gives
\[
    RM^{\rm jet1} = 45 \pm 100~{\rm rad~m^{-2}},
\]
whereas the second jet component gives
\[
    RM^{\rm jet2} = 100 \pm 30~{\rm rad~m^{-2}}.
\]

Although the uncertainty in the first jet component is large, the three regions show positive RM values that are mutually consistent within the errors. In particular, the core and the second jet component both indicate a modest Faraday rotation of order $80~{\rm rad~m^{-2}}$.

The MOJAVE 8-15~GHz RM map of 1849+670 from 2006-05-24 shows a positive RM in the core region, with a fitted value of approximately $520 \pm 91~{\rm rad~m^{-2}}$ \citep{Hovatta2012}. Our VGOS RM is therefore significantly lower, but has the same sign. This discrepancy may be due to variability and/or the different angular resolution (the fact that MOJAVE samples the compact 8--15~GHz core region, while our fit averages a broader core component). The comparison nevertheless supports the presence of a positive Faraday rotation contribution towards the compact core.

The similarity between the RM values measured in our core and jet regions suggests that, at the angular resolution of the VGOS data, the observed Faraday rotation may be dominated by an external screen affecting the compact emission on milliarcsecond scales, rather than an internal component that would depend on distance to the central engine.

\subsection{Multi-frequency Imaging}

\begin{figure}[hptb]
   \centering
   \includegraphics[
      width=\textwidth,
      trim=45 10 35 10,
      clip
   ]{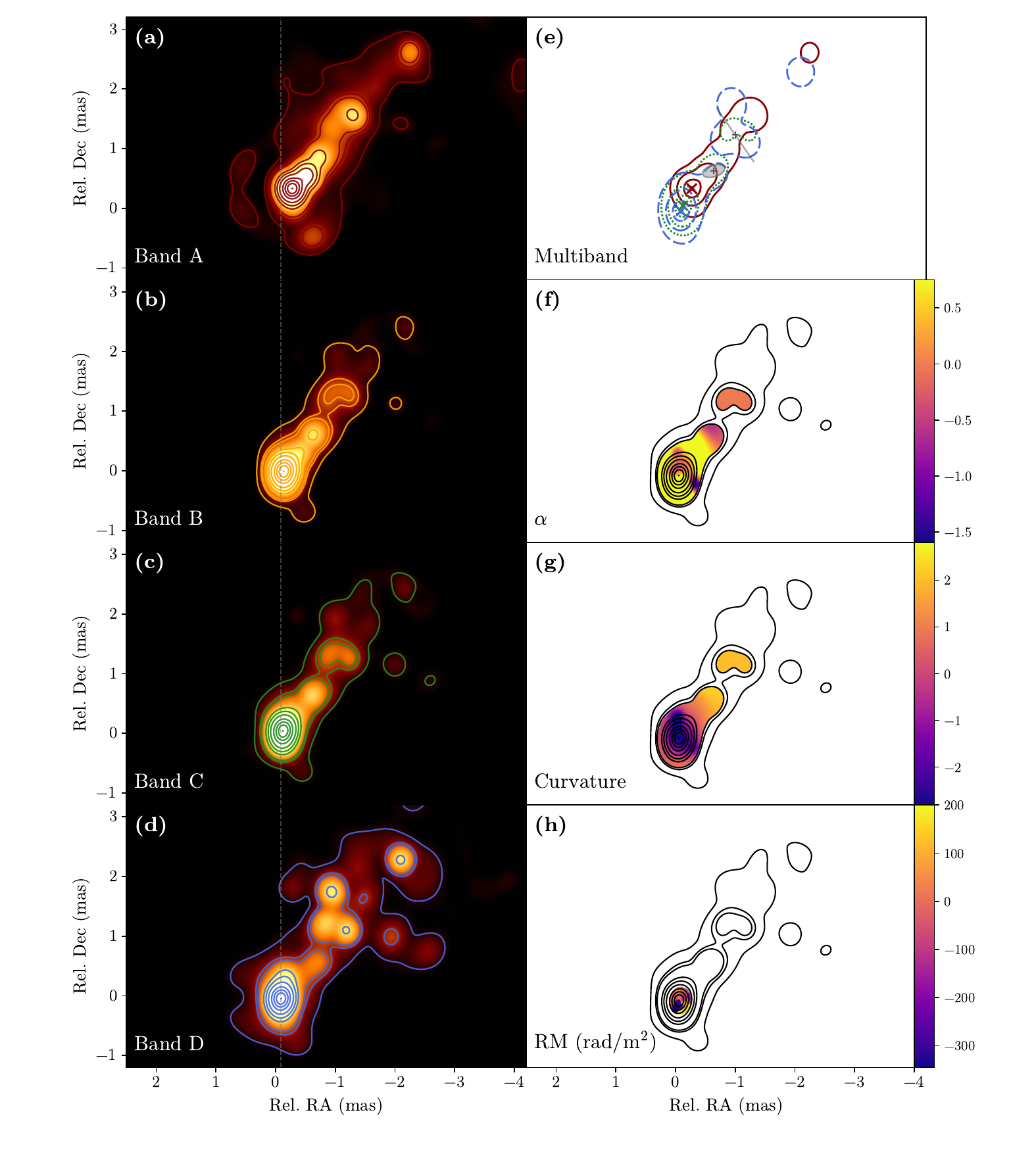}
   \caption[Multi-frequency polarimetric image of 1849+670.]{Multi-frequency polarimetric image of 1849+670. Panels (a)--(d) show the total intensity maps at bands A, B, C, and D, respectively, displayed in logarithmic scale. The color scale in each panel is normalized to its peak intensity. The dashed vertical line indicates the core position at band A, marking the core shift across bands. Panel (e) displays the total-intensity contours for bands A (solid red), C (dotted green) and D (dashed blue) in a common reference frame, together with the core and jet Gaussian components obtained from the visibility-domain model fitting. Panel (f) shows the spectral index map ($\alpha$), and panel (g), the curvature map. Panel (h) presents the rotation measure (RM) map. Only regions above 0.15\% of the peak intensity are included.}
   \label{fig:polarimetric_1849+670}
\end{figure}

The multi-frequency image of 1849+670 is shown in Fig.~\ref{fig:polarimetric_1849+670}. The total intensity maps reveal a compact bright core and a jet extending to the north-east, consistent with the morphology seen in the MOJAVE 15~GHz image. The source remains relatively compact on VGOS baselines, but the jet is clearly detected and contributes to the resolved structure.

The superposition of the contours from the four VGOS bands shows a frequency-dependent displacement of the brightness peak. The low-frequency emission is shifted downstream with respect to the high-frequency core position. This qualitative behaviour agrees with the quantitative fit shown in Fig.~\ref{fig:core_shift_1849+670}.

The spectral index map shows a spectral index of $\sim 0.3$ close in the core region and a spectral index of $\sim -0.7$ in the downstream jet. This strongly agrees with the public MOJAVE 8--15~GHz spectral-index map of 1849+670 from 2006-05-24 \citep{Hovatta2014}.

\subsection{Discussion}

1849+670 occupies an intermediate regime among the sources analysed in this chapter. It is compact enough to be used as a defining source in ICRF3, and the BVID structure parameters indicate relatively high compactness at X band. At the same time, the images reveal a clear jet with highly relativistic apparent motions, and multiwavelength studies show that the source also has extended emission on kiloparsec scales \citep{Stanley2015}, but most of the intensity comes from the core region. It is therefore a useful target for connecting the geodetic and astrophysical interpretations of source structure.

The VGOS core shift measurement gives a displacement of about 0.5~mas between 3 and 11~GHz. When our fitted model is extrapolated to the MOJAVE frequency pairs, the predicted shifts are smaller than the published values, although the shifts are, in any case, close to the astrometric resolution of our method (Fig.~\ref{fig:core_shift_1849+670}). The difference may reflect intrinsic variability of the inner jet opacity, differences in angular resolution and frequency coverage, or systematic effects in the image registration. Despite this, the fitted value of $k_r$ remains compatible with the equipartition expectation.

The physical parameters inferred from the core-shift analysis are also consistent with the expected range for compact relativistic jets. Under the equipartition assumption, and using the geometrical and relativistic parameters reported by \citet{Pushkarev2017}, we obtain
\[
    B_1 = 3.52^{+0.98}_{-0.84}~{\rm G}
\]
and
\[
    N_1 = \left(1.31^{+0.83}_{-0.55}\right)\times10^5~{\rm cm^{-3}}.
\]

The value of $B_1$ is higher than the median value of $B_1\sim0.9$~G found for quasars in the large core-shift sample of \citet{Pushkarev2012}, but it remains within the range found for strongly beamed compact jets. A more direct comparison can be made at the observed VGOS core positions, where we obtain
\[
    B_{\rm core}\simeq0.08-0.26~{\rm G}.
\]

These values are comparable to the magnetic fields of a few tenths of a Gauss inferred for VLBI cores in previous core-shift studies, such as 3C~345 \citep{Lobanov1998} and the multifrequency AGN sample of \citet{OSullivan2009}. Therefore, although the magnetic field extrapolated to 1~pc is a bit high, the magnetic field at the actual observed core positions is physically reasonable.

The particle density follows directly from the equipartition relation between $B_1$ and $N_1$, and therefore inherits the same model dependence. At the VGOS core positions, the inferred densities decrease to
\[
    N_{\rm core}\simeq70\text{--}720~{\rm cm^{-3}},
\]
from band A to band D.

It is also possible to compare the derived magnetic field with model estimates of the central black hole mass. \citet{Lobanov1998} discussed this connection in the context of 3C~345, using models for magnetically driven accretion disks and compact objects embedded in strong magnetic fields \citep{Field1993,Kardashev1995}. In particular, the magnetic field at 1~pc can be related to the black hole mass through the approximate scaling
\[
    M_{\rm BH} \simeq 2.7\times10^9
    \left(\frac{B_1}{1~{\rm G}}\right) M_\odot.
\]

Using our value of $B_1$, this gives
\[
    M_{\rm BH}
    =
    \left(9.5^{+2.6}_{-2.3}\right)\times10^9~M_\odot.
\]

This estimate is higher than previous values reported for this source, though still of the same order of magnitude. For example, \citet{Xiong2014} gives $\log(M_{\rm BH}/M_\odot)=9.14$, corresponding to $M_{\rm BH}=1.38^{+1.37}_{-0.69}\times10^9\,M_\odot$. This provides another consistency check showing that the magnetic field inferred from the VGOS core shift is reasonable.

The RM behaviour of 1849+670 is simpler than in the sources where both internal (optically-thick) and external Faraday rotation terms were required. The EVPA--$\lambda^2$ relation is well described by a linear law, and the core and jet regions all show positive RM values that are consistent within the uncertainties. This suggests that the dominant Faraday rotation can be approximated by an external screen rather than a non-uniform internal component associated with the jet. The comparison with the MOJAVE RM map supports the same sign of RM in the core, although the MOJAVE value is substantially larger. Furthermore, the spectral-index structure recovered with VGOS is also consistent with previous VLBA results.

\pagebreak
\section{2229+695}
\label{sec:2229+695}

2229+695 (also known as J2230+6946) is a compact radio-loud AGN located at $\alpha_{\mathrm{J2000}} = 22^{\mathrm h}30^{\mathrm m}36.4697^{\mathrm s}$, $\delta_{\mathrm{J2000}} = +69^\circ46'28.077''$, as listed in the ICRF3 S/X catalogue \citep{Charlot2020}. The source is included in ICRF3 as a defining source and is frequently observed in geodetic VLBI sessions. It is classified as a flat-spectrum radio quasar \citep{Wenger2000}. We adopt a spectroscopic redshift of $z=1.426$, following the value reported by \citet{Healey2008}. This redshift was later confirmed by \citet{Pursimo2013}.

\begin{figure}[H]
    \centering
    \includegraphics[width=0.67\textwidth, trim=0.0cm 2.7cm 0.0cm 4.7cm]{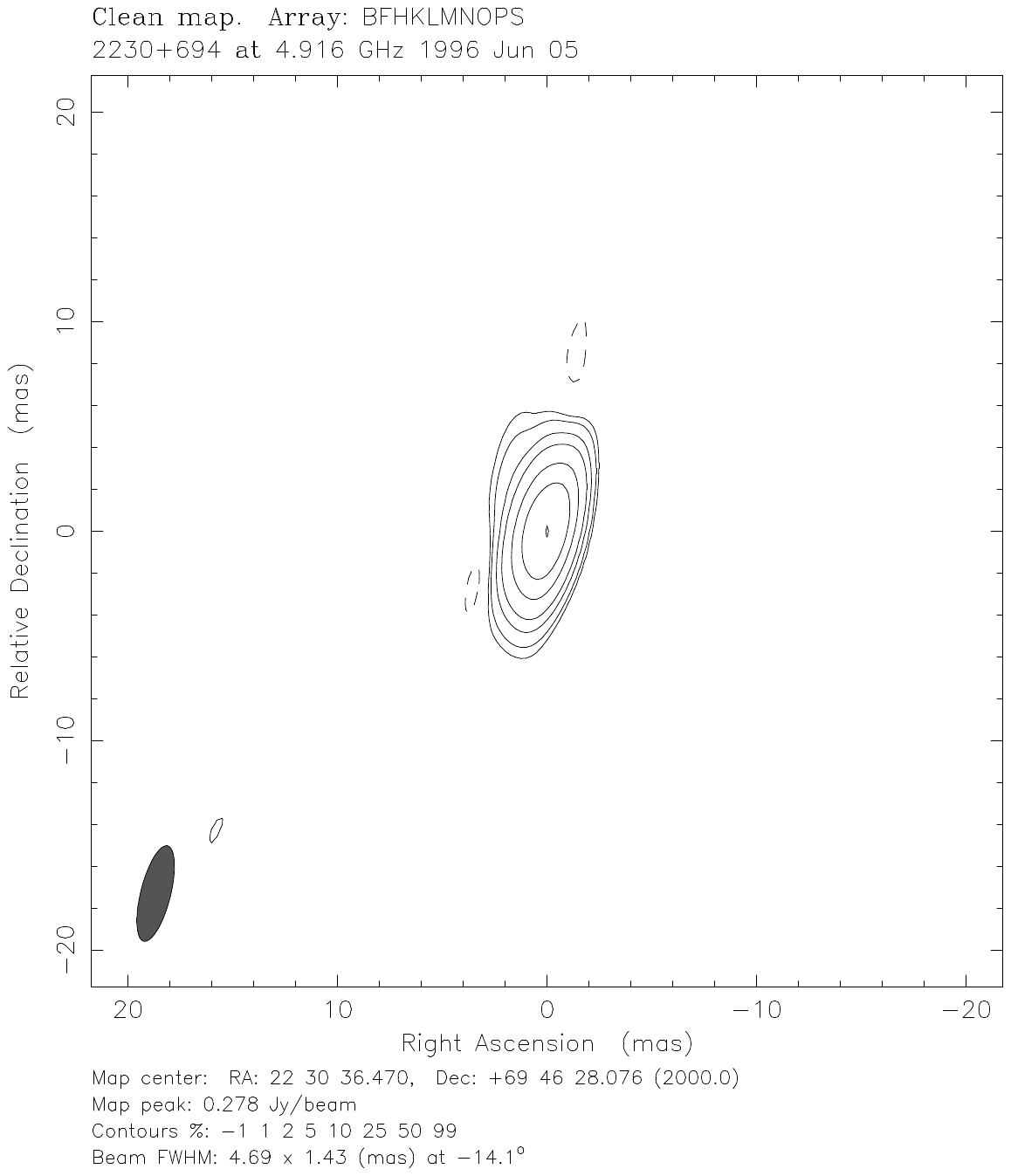}
    \caption[VLBI image of 2229+695 (J2230+6946) at 4.916~GHz from the Caltech--Jodrell Bank flat-spectrum (CJF) survey.]{VLBI image of 2229+695 (J2230+6946) at 4.916~GHz from the Caltech--Jodrell Bank flat-spectrum (CJF) survey \citep{Taylor1996}.}
    \label{fig:2229+695_cjf}
\end{figure}

Unlike several of the other sources analysed in this chapter, 2229+695 is not part of the MOJAVE 15~GHz monitoring sample. Nevertheless, it is a very relevant target for geodetic and astrometric VLBI. It appears in the IVS and ICRF catalogues as a compact source and has been observed in a large number of geodetic sessions.

Geodetic S/X-band images from the Bordeaux VLBI Image Database (BVID) provide additional information on its radio structure \citep{Collioud2009}. For epoch RV156 (2022-11-28), the source has an X band flux density of 0.65~Jy, a structure index of 3.05, and a compactness of 0.55 \citep{Collioud2009}. These values indicate that, although 2229+695 is sufficiently compact to be used as a geodetic reference source, some fraction of its emission is resolved on VLBI scales.

The structured nature of this source has also been emphasized at higher radio frequencies. \citet{deWitt2023} presented K band VLBA images of J2230+6946 as part of their celestial reference frame imaging programme. They described the source as a close double system, with two compact components whose relative brightness changes with time. This behaviour is particularly relevant for astrometry, because changes in the relative brightness of nearby components can shift the apparent reference point of the source, even when the total structure remains compact.

\subsection{Core Shift}

\begin{figure}[t]
   \centering
   \includegraphics[width=0.67\textwidth]{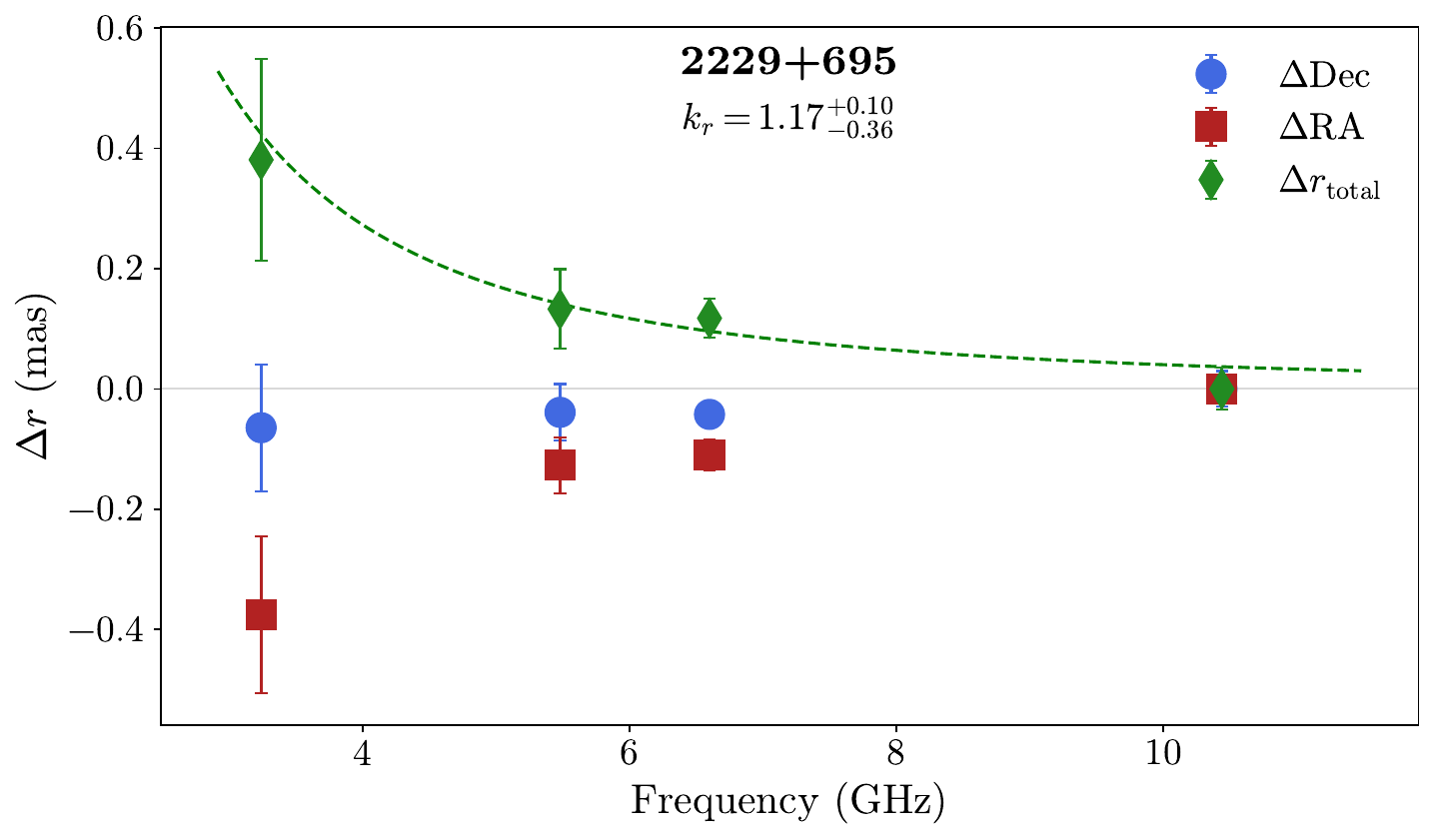}
   \caption[Measured relative core shift as a function of frequency for 2229+695.]{Measured relative core shift as a function of frequency for 2229+695. The total shift vector, along with its components in right ascension and declination, are shown. The error bars represent 3$\sigma$ uncertainties. The fit model is shown as a dashed green line.}
   \label{fig:core_shift_2229+695}
\end{figure}

Applying the same procedure described in Chapter~\ref{ch:1803+784}, we obtained the core-shift measurements for 2229+695, shown in Fig.~\ref{fig:core_shift_2229+695}.

We measure a total core shift of approximately $0.39\pm0.15$~mas between 3 and 11~GHz. The displacement is smaller than in other sources, but it is still detectable across the VGOS frequency range.

Fitting the standard core-shift relation described in Sect.~\ref{sec:jetemissionmodel}, we obtain
\[
    k_r = 1.17^{+0.10}_{-0.36}.
\]

This value is compatible with the equipartition expectation, $k_r\simeq1$, within the uncertainties. Therefore, although the individual measurements should be interpreted with caution, the global frequency dependence of the core position is consistent with the standard picture of a conical synchrotron jet. The result suggests that, even in this compact and geodetically useful source, the observed position of the VLBI core is affected by opacity in a physically plausible way.

We are not aware of any published core-shift measurement for 2229+695. Previous studies have mainly discussed this source in the context of geodetic and astrometric VLBI, where its structure can produce significant position variations. In particular, recent VLBI analyses identify 2229+695 as a jet dominated source for which the measured astrometric position can be strongly affected by visible source structure \citep{Xu2025}. Our VGOS measurements therefore provide a first estimate of the frequency-dependent displacement of the compact emission in this source over the 3--11~GHz range.

\subsection{Rotation Measure}

\begin{figure}[t]
   \centering
   \includegraphics[width=0.66\textwidth, trim=0.0cm 0.3cm 0.0cm 0]{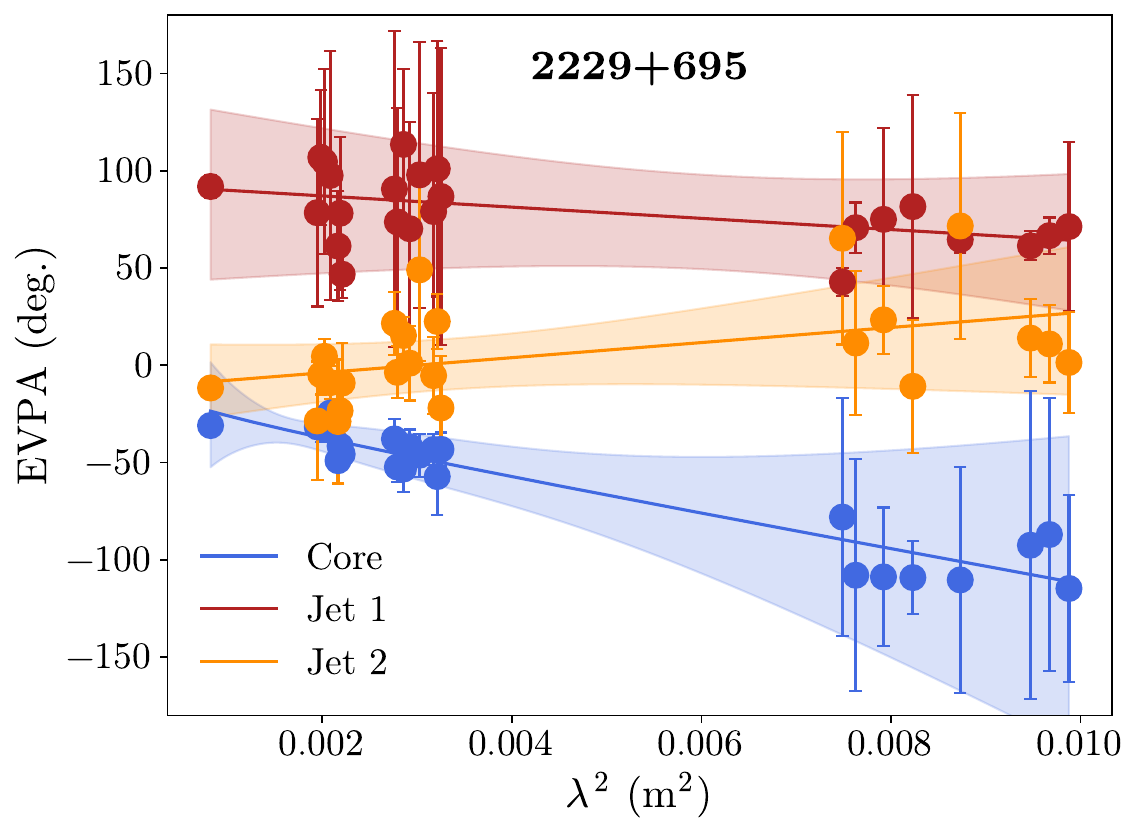}
   \caption[Fit of the polarization angle as a function of $\lambda^2$ for 2229+695.]{Fit of the polarization angle as a function of $\lambda^2$ for the core and jet regions of 2229+695. The shaded regions represent the 3$\sigma$ confidence intervals of the model fits. Band D data have been averaged because of the low S/N in this band.}
   \label{fig:RM_fit_2229+695}
\end{figure}

Figure~\ref{fig:RM_fit_2229+695} shows the EVPA fits as a function of $\lambda^2$ for the core and two jet components of 2229+695.

As in some of the other sources analysed in this chapter, the EVPA uncertainties do not allow us to reliably fit the exponent of the internal Faraday rotation model as a free parameter. We therefore fix it to $a=2.1$, following the same criterion used in the previous cases. For the core, we obtain an external Faraday rotation of
\[
    RM^{\rm core}_{\rm E} = -150 \pm 120~{\rm rad~m^{-2}},
\]
and an internal contribution of
\[
    RM^{\rm core}_{\rm I} = -23 \pm 110~{\rm rad~m^{-2}}.
\]

The uncertainties are large, and the internal component is fully compatible with zero. The observed EVPA behaviour is mainly constrained by the external component, whose amplitude is itself only marginally determined.

The jet emission was modelled with two separate Gaussian components. For the first jet component, we obtain
\[
    RM^{\rm jet1}_{\rm tot} = -50 \pm 40~{\rm rad~m^{-2}},
\]
whereas the second jet component gives
\[
    RM^{\rm jet2}_{\rm tot} = 68 \pm 30~{\rm rad~m^{-2}}.
\]

The two jet components therefore show RM values with opposite signs. This can be explained by the closeness of the first component of the jet to the core, where the internal Faraday still dominates, while in the second component, the external screen is the one that dominates.

No published RM map or resolved RM measurement could be found for 2229+695 with which to compare our results.

\subsection{Multi-frequency Imaging}

\begin{figure}[htpb]
   \centering
   \includegraphics[width=\hsize, trim=0 0 0 0, clip]{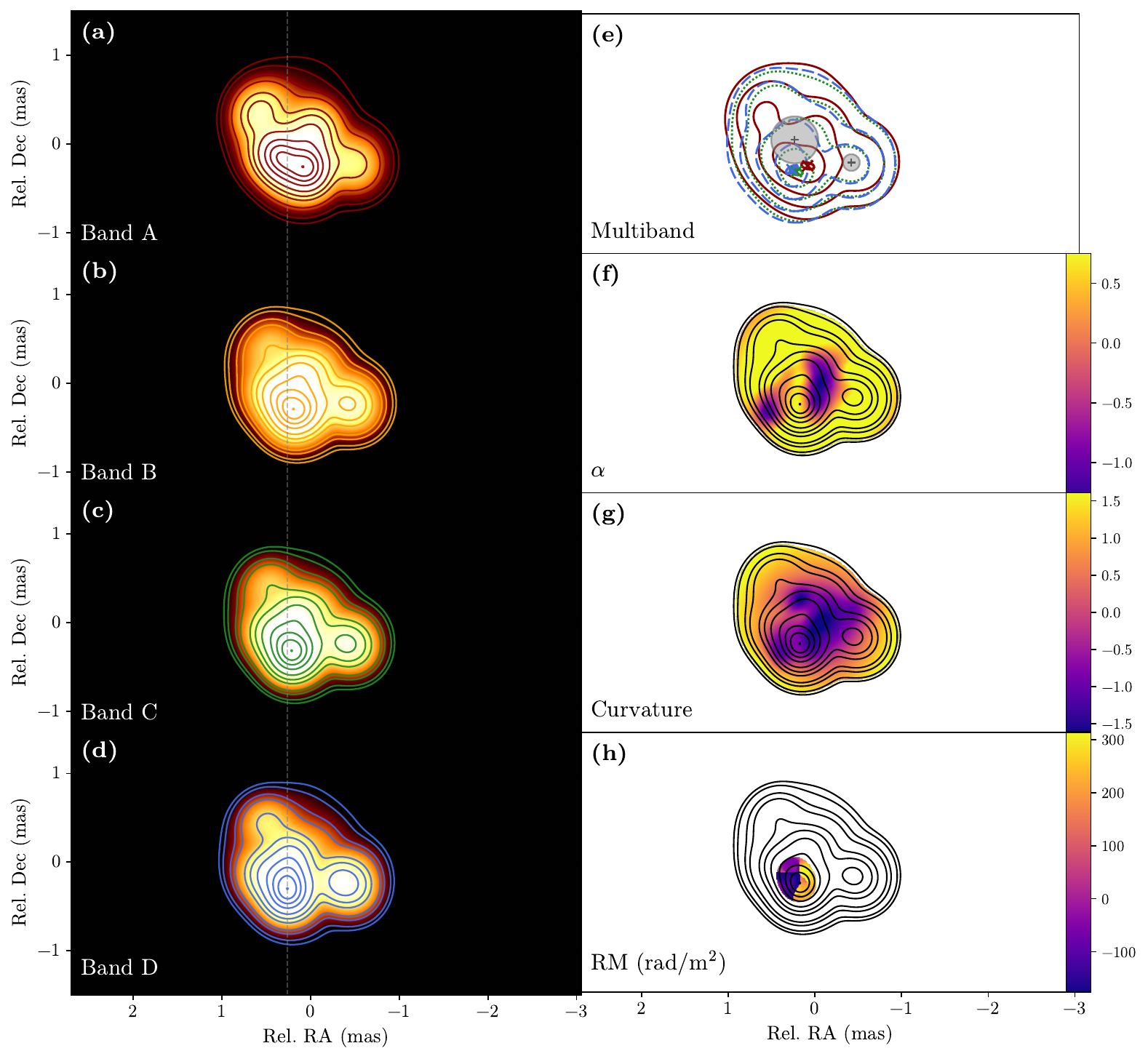}
   \caption[Multi-frequency polarimetric image of 2229+695.]{Multi-frequency polarimetric image of 2229+695. Panels (a)--(d) show the total intensity maps at bands A, B, C, and D, respectively, displayed in logarithmic scale. The color scale in each panel is normalized to its peak intensity. The dashed vertical line indicates the core position at band A, marking the core shift across bands. Panel (e) displays the total-intensity contours for bands A (solid red), C (dotted green) and D (dashed blue) in a common reference frame, together with the core and jet Gaussian components obtained from the visibility-domain model fitting. Panel (f) shows the spectral index map ($\alpha$), and panel (g), the curvature map. Panel (h) presents the rotation measure (RM) map. Only regions above 0.1\% of the peak intensity are included.}
   \label{fig:polarimetric_2229_695}
\end{figure}

The multi-frequency image of 2229+695 is shown in Fig.~\ref{fig:polarimetric_2229_695}. The total intensity maps reveal a compact but structured source. The emission is not dominated by a single unresolved component; instead, the contours indicate an elongated morphology to the west, consistent with the structured appearance reported in \citet{Xu2025}.

The superposition of the contours from the four VGOS bands shows a measurable core shift of the brightness peak. This displacement is smaller than in some of the other sources analysed in this chapter, but it is consistent with the core-shift measurement shown in Fig.~\ref{fig:core_shift_2229+695}.

The spectral index map shows a flat spectrum in the compact upstream region and an optically thin region ($\alpha \approx -0.7$) downstream. This behaviour is consistent with the classification of 2229+695 as a flat-spectrum radio quasar and with the standard interpretation of the VLBI core as a synchrotron self-absorbed region. Since no published resolved spectral index map could be found for this source, the VGOS map provides new spatial information on the distribution of the radio spectrum across the compact emission.

The RM map provides a qualitative view of the Faraday rotation structure. The limited polarization S/N and the compactness of the source prevent a detailed interpretation of small scale RM gradients, but the map is consistent with the model fitting presented before, where huge gradients are shown.

\subsection{Discussion}

2229+695 is a particularly interesting source because it is both a defining ICRF3 source and a source with clear evidence for compact structure. The BVID structure parameters show that the source is not completely point-like at S/X band, and K-band VLBA imaging has revealed a close double morphology with time-dependent relative component brightness \citep{deWitt2023}. This makes the source especially relevant for understanding the connection between compact jet structure and astrometric position stability.

Our VGOS images recover this structured nature. Although the source is compact, the emission is elongated and the brightness peak changes with frequency. We measure a core shift of approximately $0.39\pm0.15$~mas between 3 and 11~GHz. This confirms that the apparent position of the compact emission is frequency dependent. The fitted value $k_r = 1.17^{+0.10}_{-0.36}$ is compatible with the equipartition expectation, suggesting that the frequency dependence of the core position can be interpreted within the standard core-shift framework.

This behaviour is consistent with recent astrometric results. \citet{Xu2025} identify 2229+695 as a jet-dominated source and discuss it as an example where source structure can strongly affect the measured position. In such a case, the position variations need not be caused only by a smooth core shift. They can also arise from changes in the relative brightness of nearby compact components, which shift the centroid or the dominant brightness peak. This interpretation is supported by the K-band imaging results of \citet{deWitt2023}, where the source appears as a close double source.

The RM results are less conclusive. The internal RM of the core is compatible with zero, and the external component has a large uncertainty. Nevertheless, the second component of the jet has a positive RM. This may indicate that there is also an internal optically-thin Faraday screen near the core, that also affects the first component of the jet, because of its closeness, and it dominates in this region. In the more external component, the external positive RM screen dominates.

Overall, 2229+695 illustrates the importance of imaging even for defining sources. Its use for reference frames does not imply that it is intrinsically point-like. Instead, the VGOS observations show that compact structure, core shift, spectral gradients, and Faraday rotation are all present at the milliarcsecond level. 

\pagebreak
\section{Conclusions}

The results presented in this chapter show that VGOS observations can be used to extract astrophysical information about the sample of observed radio-loud AGN. In particular, the broad frequency coverage of VGOS allows us to measure core shifts, to constrain the frequency dependence of the VLBI core position, and to obtain estimations of the Faraday rotation.

Table~\ref{tab:astro_results_summary} summarizes the main quantitative results obtained for the sources analysed in this chapter, including 1803+784 from Chapter~\ref{ch:1803+784}. The table lists the fitted value of $k_r$ of the core-shift fit, together with the Faraday rotation measures inferred for the core and jet regions.

\begin{table}[htpb]
\centering
\scriptsize
\renewcommand{\arraystretch}{1.18}
\setlength{\tabcolsep}{3.5pt}
\caption[Summary of the main astrophysical parameters obtained for the analysed sources.]{Summary of the main astrophysical parameters obtained for the sources analysed in this work. RM values are given in ${\rm rad~m^{-2}}$ for the reference frequency of 5.4~GHz. Quantities marked with an asterisk were adopted from the literature.}
\label{tab:astro_results_summary}
\resizebox{\textwidth}{!}{
\rowcolors{2}{gray!7}{white}
\begin{tabular}{lcccccc}
\hline
\rowcolor{white}
Parameter
& 1803+784
& 3C418
& 0059+581
& 0613+570
& 1849+670
& 2229+695 \\
\hline

$z^\ast$
& $0.6797$
& $1.686$
& $0.644$
& $0.592$
& $0.657$
& $1.426$ \\

$\beta_{\rm app}^\ast$
& $1.16\pm0.31$
& --
& $6.84\pm0.48$
& --
& $21.43\pm1.39$
& -- \\

$\delta^\ast$
& $12.1\pm3.0$
& --
& $21.9\pm3.6$
& --
& $8.1\pm1.4$
& -- \\

$\Gamma^\ast$
& $6.2\pm1.5$
& --
& $12.0\pm1.6$
& --
& $32.5\pm5.6$
& -- \\

$\theta^\ast$
& $0.9^\circ\pm0.5^\circ$
& --
& $1.5^\circ\pm0.5^\circ$
& --
& $4.7^\circ\pm0.3^\circ$
& -- \\

$\alpha_{\rm int}^\ast$
& $0.3^\circ\pm0.2^\circ$
& --
& $0.7^\circ\pm0.2^\circ$
& --
& $1.6^\circ\pm0.1^\circ$
& -- \\

$\phi^\ast=\alpha_{\rm int}/2$
& $0.15^\circ\pm0.10^\circ$
& --
& $0.35^\circ\pm0.10^\circ$
& --
& $0.8^\circ\pm0.05^\circ$
& -- \\

$\Omega_{r\nu}$
& $9.40^{+0.11}_{-0.13}~{\rm pc\,GHz}$
& $40^{+8}_{-10}~{\rm pc\,GHz}$
& $13^{+4}_{-3}~{\rm pc\,GHz}$
& $18^{+4}_{-5}~{\rm pc\,GHz}$
& $10.8^{+3}_{-1.3}~{\rm pc\,GHz}$
& $11.7^{+0.6}_{-1.6}~{\rm pc\,GHz}$ \\

$\Delta r_{\rm 3-11\,GHz}$
& $0.24\pm0.05~{\rm mas}$
& $1.1\pm0.9~{\rm mas}$
& $0.54\pm0.17~{\rm mas}$
& $0.7\pm0.3~{\rm mas}$
& $0.5\pm0.2~{\rm mas}$
& $0.39\pm0.15~{\rm mas}$ \\

$k_r$
& $0.73^{+0.12}_{-0.19}$
& $1.23^{+0.33}_{-0.29}$
& $0.98^{+0.23}_{-0.23}$
& $1.10^{+0.22}_{-0.19}$
& $1.08^{+0.27}_{-0.43}$
& $1.17^{+0.10}_{-0.36}$ \\

$a$
& $2.2\pm1.0$
& $2.1^\dagger$
& $2.1^\dagger$
& $2.1^\dagger$
& $2.1^\dagger$
& $2.1^\dagger$ \\

$RM^{\rm core}_{\rm E}$
& $-44\pm9$
& $-200\pm100$
& $-456\pm40$
& $170\pm60$
& $74\pm15$
& $-150\pm120$ \\

$RM^{\rm core}_{\rm I}$
& $121\pm8$
& $378\pm150$
& $-113\pm50$
& $-170\pm70$
& --
& $-23\pm110$ \\

$RM^{\rm jet1}_{\rm tot}$
& $49\pm2$
& $166\pm50$
& $-256\pm10$
& $-4\pm15$
& $45\pm100$
& $-50\pm40$ \\

$RM^{\rm jet2}_{\rm tot}$
& $60\pm50$
& --
& --
& --
& $100\pm30$
& $68\pm30$ \\

$RM^{\rm jet}_{\rm I}$
& $93\pm10$
& $366\pm150$
& $200\pm60$
& $165\pm80$
& --
& -- \\

$\alpha_{\rm jet}$
& $-0.6$
& $-0.4$
& $-0.4$
& $-0.7$
& $-0.4$
& $-0.8$ \\

$B_1$
& $30\pm10~{\rm G}$
& --
& $13.6\pm5~{\rm G}$
& --
& $3.5\pm0.9~{\rm G}$
& -- \\

$N_1$
& $(10\pm4)\times10^6~{\rm cm^{-3}}$
& --
& $(2.0\pm2)\times10^6~{\rm cm^{-3}}$
& --
& $(1.3\pm0.8)\times10^5~{\rm cm^{-3}}$
& -- \\

$B_{\rm core}(3\,{\rm GHz})$
& $0.17\pm0.05~{\rm G}$
& --
& $0.08\pm0.02~{\rm G}$
& --
& $0.08 \pm 0.01~{\rm G}$
& -- \\

$B_{\rm core}(10\,{\rm GHz})$
& $0.54\pm0.11~{\rm G}$
& --
& $0.25\pm0.05~{\rm G}$
& --
& $0.26 \pm 0.03~{\rm G}$
& -- \\

$N_{\rm core}(3\,{\rm GHz})$
& $290\pm150~{\rm cm^{-3}}$
& --
& $60\pm30~{\rm cm^{-3}}$
& --
& $70 \pm 20 ~{\rm cm^{-3}}$
& -- \\

$N_{\rm core}(10\,{\rm GHz})$
& $3000\pm1500~{\rm cm^{-3}}$
& --
& $700\pm200~{\rm cm^{-3}}$
& --
& $700 \pm 200~{\rm cm^{-3}}$
& -- \\

$B_{\parallel}^{i}$
& $1.8\times10^{-7}~{\rm G}$
& --
& $7.7\times10^{-7}~{\rm G}$
& --
& $4.5\times10^{-4}~{\rm G}$
& -- \\

$B_{\parallel}^{s}$
& $2.5\times10^{-5}~{\rm G}$
& --
& $2.4\times10^{-5}~{\rm G}$
& --
& $2.1\times10^{-5}~{\rm G}$
& -- \\

\hline
\end{tabular}
}
\begin{flushleft}
\footnotesize
$^\ast$ Adopted from the literature and used as input parameters in the physical interpretation. $^\dagger$ Fixed to 2.1 when the EVPA uncertainties did not allow a reliable free fit. The values of $B$ and $N$ were derived only for the sources for which suitable geometrical and relativistic parameters were available. The exponent $a$ was fixed to 2.1 when the EVPA uncertainties did not allow a reliable free fit. For 1849+670, the EVPA--$\lambda^2$ relation was modelled with a purely external Faraday screen, so no internal RM component is listed.
\end{flushleft}
\end{table}

A common result across the sample is that all fitted values of $k_r$ are compatible with the equipartition expectation, except for 1803+784. This is an important outcome because the sources span different regimes of compactness and structure. Some, such as 3C418 and 0059+581, show clearly resolved jet emission and relatively large structure effects, whereas others, such as 0613+570 and 2229+695, are compact enough to be used as defining sources in ICRF3. The fact that all of them show a physically plausible frequency dependence of the core position indicates that VGOS can recover the opacity structure of the inner jet even in sources primarily observed for geodetic purposes.

The measured core shifts are also relevant from the astrometric point of view. The apparent position of the VLBI core changes across the VGOS frequency range, typically by several tenths of a milliarcsecond and up to about one milliarcsecond in the most structured cases. Such displacements are not negligible for high precision astrometry. 

The Faraday rotation results are more source-dependent. In several cases, the EVPA behaviour cannot be described by a linear trend over the full VGOS frequency range, and a decomposition between external and internal Faraday rotation is required. However, these decompositions should be interpreted with caution, because the limited polarization S/N prevents a robust fit of all model parameters. 

Overall, these results demonstrate that VGOS data contain significant astrophysical information. The broadband frequency coverage, combined with polarimetric imaging, makes it possible to measure core shifts, spectral index, and Faraday rotation in compact AGN. At the same time, the same source structure produces systematic effects that are relevant for geodesy and astrometry.

\pagebreak
\cleardoublepage

\chapter{Geodetic Analysis}
\label{ch:geodesy}

\vspace{-0.5cm}

\noindent
\hfill
\begin{minipage}{0.8\textwidth}
\small
\raggedleft
\textit{This chapter is based on the author’s contribution to:}\\[0.3em]
\textit{``Removing the Effect of Source Structure from Correlated VLBI Data''}\\[0.3em]
\textit{F. Jaron, \textbf{V. Pérez-Díez}, H. Krásná, J. Böhm, P. Charlot, A. Collioud,\\
J. Gruber, I. Martí-Vidal and A. Nothnagel}\\[0.3em]
\textit{(manuscript submitted to Radio Science).}
\end{minipage}

\vspace{0.8cm}

\lettrine{T}{  raditionally}, active galactic nuclei (AGN) have been used in geodetic VLBI as ideal targets due to their high brightness, which allows them to be detected within very short integration times, and their compact structure. Under this assumption, these objects are commonly treated as point-like sources in geodetic analyses. However, this approximation breaks down at the angular resolutions achieved by VLBI, where the contribution of the jet becomes significant and the intrinsic source structure can no longer be neglected.

In this chapter, we describe how geodetic VLBI is progressively incorporating source structure effects in order to improve the accuracy of its derived products. This is achieved by making use of the high-fidelity images obtained through the calibration and imaging pipeline presented in this thesis, which allow the contribution of the source structure to the interferometric delays to be quantified and removed.

As a result, the inclusion of source structure corrections leads to a significant improvement in geodetic observables, reducing post-fit delay residuals and enhancing the overall precision and stability of the estimated parameters.

\pagebreak
\section{Introduction}
Very long baseline interferometry (VLBI) is a fundamental technique in geodesy and astrometry, playing a central role in the realization of global reference frames. In particular, VLBI provides the celestial reference frame through the International Celestial Reference Frame (ICRF; \citet{Charlot2020}), contributes to the terrestrial reference frame (ITRF \citep{Altamimi2023}), and is uniquely sensitive to the full set of Earth orientation parameters. These capabilities rely on the precise measurement of signal arrival times from distant extragalactic radio sources.

The primary targets of geodetic VLBI are radio-loud AGN, which have traditionally been treated as ideal point-like sources due to their high brightness and compact appearance. However, as discussed in previous chapters of this thesis, this approximation is not strictly valid at the angular resolutions achieved by VLBI. The radio emission in AGN originates from relativistic jets, which often exhibit extended and complex structures that can significantly affect the interferometric observables.

The fundamental observable in geodetic VLBI is the group delay (Eq. \ref{eq:groupdelay}). Under the assumption of a point-like source, the phase is expected to follow a simple behaviour across frequency. However, an extended brightness distribution introduces additional phase contributions, commonly referred to as source structure effects. These contributions propagate into the group delays, introducing systematic errors in the geodetic analysis.

Historically, source structure has been recognized as a potential limitation in geodetic VLBI since its early development \citep{Coates1975, Thomas1980, Charlot1990}. While its impact was initially considered secondary compared to atmospheric and instrumental effects, improvements in observational precision have made source structure one of the dominant remaining sources of systematic error \citep{Fey1997, Fey2000, Anderson2018}. Traditional mitigation strategies have relied on selecting sources with minimal structure, quantified through the structure index, and avoiding sources with complex morphologies. However, this approach does not fully exploit the available information and becomes increasingly insufficient as the accuracy requirements of geodetic VLBI continue to improve.

The transition from legacy S/X-band observations to broadband systems such as VGOS emphasizes the need to properly account for source structure. The increased bandwidth of VGOS enables a more precise determination of group delays, but also makes the observations more sensitive to frequency-dependent structure effects. Meeting the ambitious accuracy goals of VGOS requires the modelling and correction of source structure contributions.

In this chapter, we present an approach to mitigate source structure effects by directly removing their contribution from the interferometric phases. This is achieved by making use of images of the observed sources, obtained through the calibration and imaging pipeline described in Chapter~\ref{ch:vgos_pipeline}. From these images, we derive source models that allow us to compute and subtract the structure phase terms from the correlated VLBI data. This correction can be applied at the level of the correlator output, and can be used both for the session from which the images are derived and for subsequent observations of the same sources.

The software developed and used for the source-structure modelling and correction presented in this chapter is publicly available as the \textsc{VieSOFT} package \citep{Jaron2026soft}.

\pagebreak
\section{Observations}

The performance of the proposed source structure mitigation strategy was tested using three operational VGOS sessions of the IVS. The first session, VO2187 (July 6-7, 2022), has already been introduced in Chapter~\ref{ch:vgos_pipeline}, where it was used to develop and validate the full calibration and imaging pipeline. Two further sessions, VO2244 (September 1-2, 2022) and VO3124 (May 4-5, 2023), were included in the analysis. These sessions were observed at later epochs and involve slightly different station networks, resulting in variations in uv-coverage and sampling of the source structure. The inclusion of these additional datasets allows us to assess the robustness of the proposed method under realistic conditions, where both the observing geometry and the intrinsic source structure may vary with time. The participating VGOS stations and their reference geocentric coordinates are listed in Table~\ref{tab:antennas2}.

\begin{table}[ht]
\centering
\caption[Antennas used in experiments VO2187, VO2244 and VO3124.]{Antennas used in experiments VO2187, VO2244 and VO3124, together with their reference Geocenter coordinates used in the DiFX correlation.}
\resizebox{\textwidth}{!}{%
\begin{tabular}{l c| r r r}
\multicolumn{2}{c|}{Antenna} & \multicolumn{3}{c}{GC Coords (m)} \\ 
\hline
IVS Component Name & Code & X & Y & Z \\
\hline 
Goddard Geophysical and Astronomical Observatory & GS & 1130729.877 & -4831245.972 & 3994228.300 \\
Hobart 12m, Mt. Pleasant Radio Observatory & HB & -3950237.160 & 2522347.572 & -4311562.320 \\
Ishioka VLBI Station & IS & -3959636.203 & 3296825.448 & 3747042.571 \\
Kokee Park Geophysical Observatory & K2 & -5543831.745 & -2054585.590 & 2387828.974 \\
McDonald Geodetic Observatory & MG & -1330788.462 & -5328106.593 & 3236427.492 \\
Ny-Alesund Geodetic Observatory & NN & 1202462.513 & 252734.517 & 6237766.223 \\
Onsala Space Observatory East & OE & 3370889.298 & 711571.199 & 5349692.048 \\
Onsala Space Observatory West & OW & 3370946.779 & 711534.507 & 5349660.925 \\
Westford Antenna, Haystack Observatory & WF & 1492206.223 & -4458130.552 & 4296015.629 \\
Geodetic Observatory Wettzell & WS & 4075539.548 & 931735.649 & 4801629.602 \\
IGN Yebes Observatory & YJ & 4848831.021 & -261629.388 & 4122976.576 \\
\hline
\end{tabular}
}
\label{tab:antennas2}
\end{table}
The correlated visibility data were obtained in the form of DiFX SWIN files from the Crustal Dynamics Data Information System (CDDIS; \citet{Noll2010}). For the analysis, we focus on three representative radio sources: 1849+670, 3C418, and OJ287.

This dataset allows for two complementary tests. First, the structure correction can be applied to the same session from which the images were derived (VO2187), ensuring full consistency between the source model and the observed data. Second, the same structure information can be applied to later sessions (VO2244 and VO3124), enabling an assessment of the temporal stability of the correction. Since AGN jets evolve on timescales of days to weeks, structural changes are expected between sessions, which may limit the effectiveness of corrections based on non-contemporaneous images.

\subsection{Image Model Fitting}
\label{sec:imagemodelfitting}
In this work, we describe the source structure as a superposition of elliptical Gaussian components, which can be directly used within the formalism of \citet{Charlot1990} to compute structure-related observables. In contrast to the approach followed in Sect.~\ref{sec:model_fitting}, where the model fitting was performed directly in the visibility (uv) domain, in this study we adopt an image-plane modeling strategy. This choice is motivated by the availability of high-fidelity images obtained through the multi-frequency regularized maximum likelihood (RML) imaging framework described in Section \ref{sec:mfimaging}. These images incorporate prior information and constraints across frequency, resulting in a physically consistent reconstruction of the source structure. Modeling in the image plane allows us to take advantage of these priors, which are not trivially incorporated in traditional uv-plane Gaussian fitting.

We express the extended radio brightness distribution as a superposition of $n$ elliptical Gaussian components,
\begin{equation}
S(x, y) \approx \sum_{k=1}^{n} S_k \exp \left[
-\frac{1}{2} \left( \frac{x_k'^2}{a_k^2} + \frac{y_k'^2}{b_k^2} \right)
\right],
\end{equation}
where $S_k$ is the peak flux density of the $k$-th component, and
\begin{equation}
x_k' = (x - x_k)\cos\psi_k + (y - y_k)\sin\psi_k,
\end{equation}
\begin{equation}
y_k' = -(x - x_k)\sin\psi_k + (y - y_k)\cos\psi_k,
\end{equation}
with $(x_k, y_k)$ the position of the component in the image plane and $\psi_k$ its position angle, measured from North to East.

The fitting of the Gaussian model to the image data is performed using a custom Python implementation based on the \texttt{Astropy} and \texttt{SciPy} libraries. Each VGOS band is treated independently. The fitting is carried out in a semi-automated way. We initialize the model with a given number of Gaussian components and initial parameter estimates, and then optimize all parameters through a least-squares minimization. The number of components is not treated as a free parameter, but is instead increased iteratively in a controlled way, starting from a single component and adding new components in successive iterations until a satisfactory representation of the source structure is achieved. This strategy ensures that the model remains as simple as possible while still capturing the relevant structural features. The resulting Gaussian representations provide a compact and physically meaningful description of the source brightness distribution, suitable for further use in geodetic analysis through the computation of structure delays and phases.

\begin{figure}[htbp]
    \centering
    \includegraphics[width=1.05\textwidth]{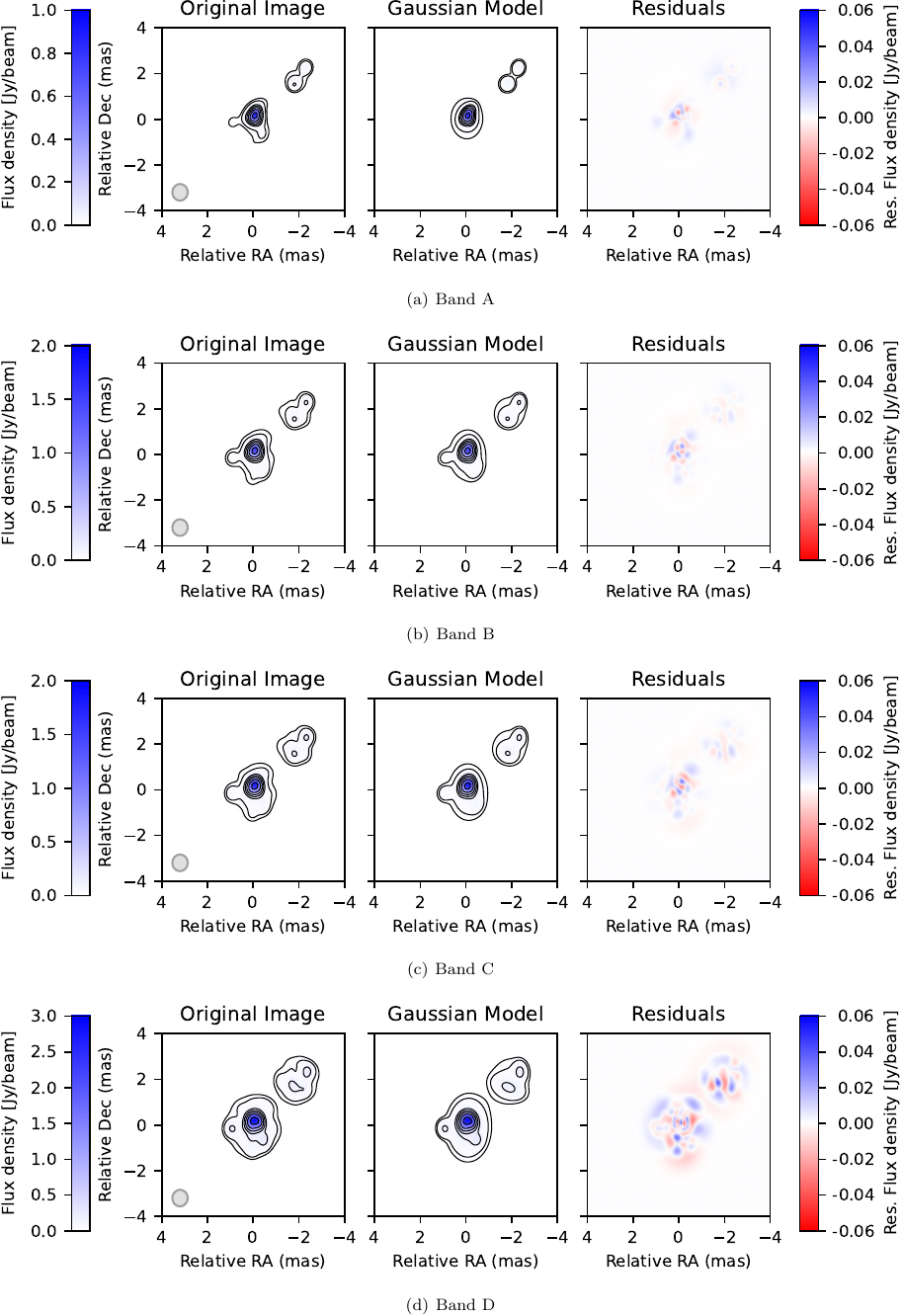}
    \caption[Model fitting of 1849+670.]{Model fitting of 1849+670. Each panel shows the residual map for a VGOS band, from A to D. The contour levels are 0.5, 1, 5, 10, 20, 40, and 80\,\%.}
    \label{fig:model1849_670}
\end{figure}

\begin{figure}[htbp]
    \centering
    \includegraphics[width=1.05\textwidth]{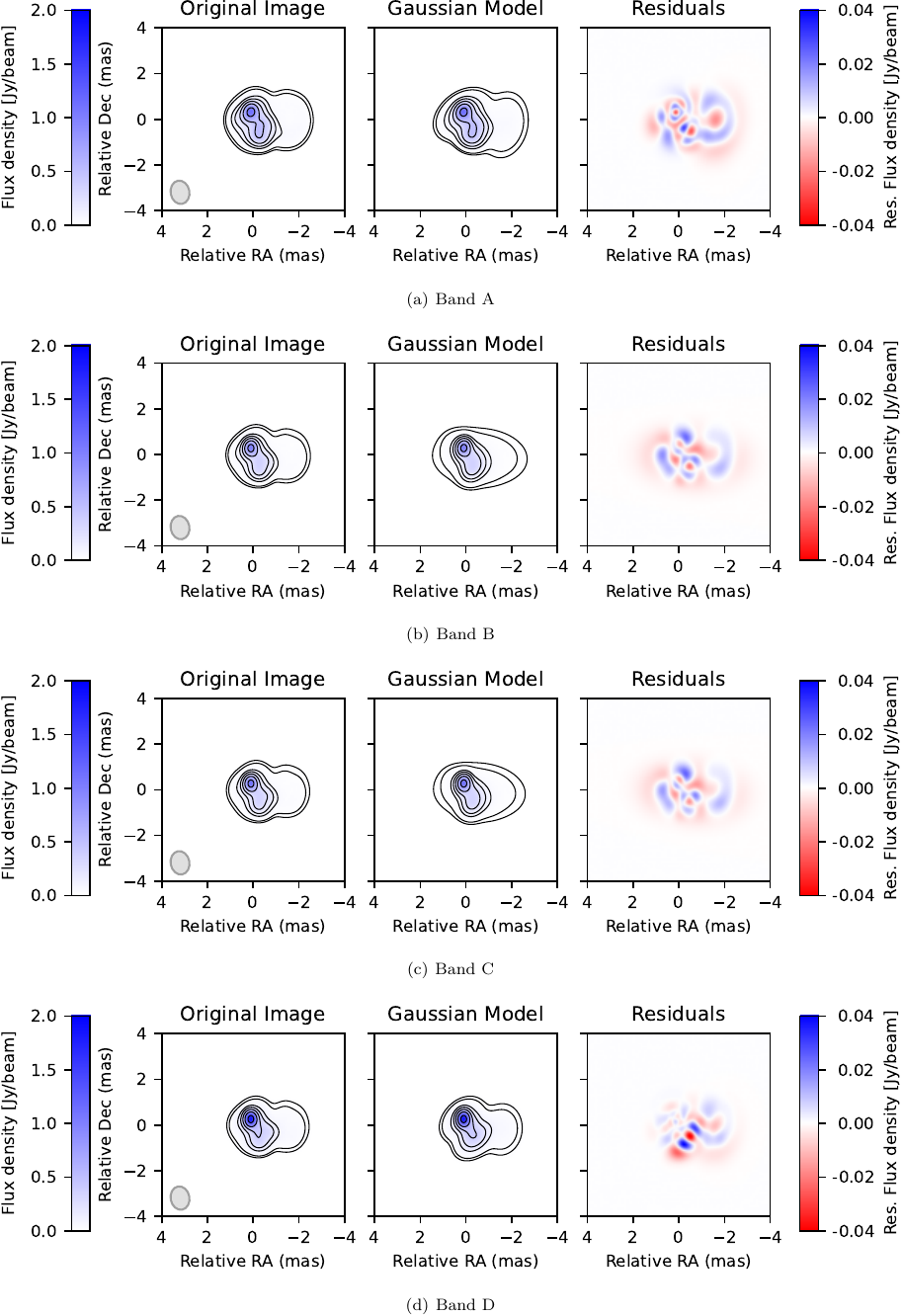}
    \caption[Model fitting of 3C418.]{Model fitting of 3C418. Each panel shows the residual map for a VGOS band, from A to D. The contour levels are 5, 10, 20, 40, and 80\,\%.}
    \label{fig:model3C418}
\end{figure}

\begin{figure}[htbp]
    \centering
    \includegraphics[width=1.05\textwidth]{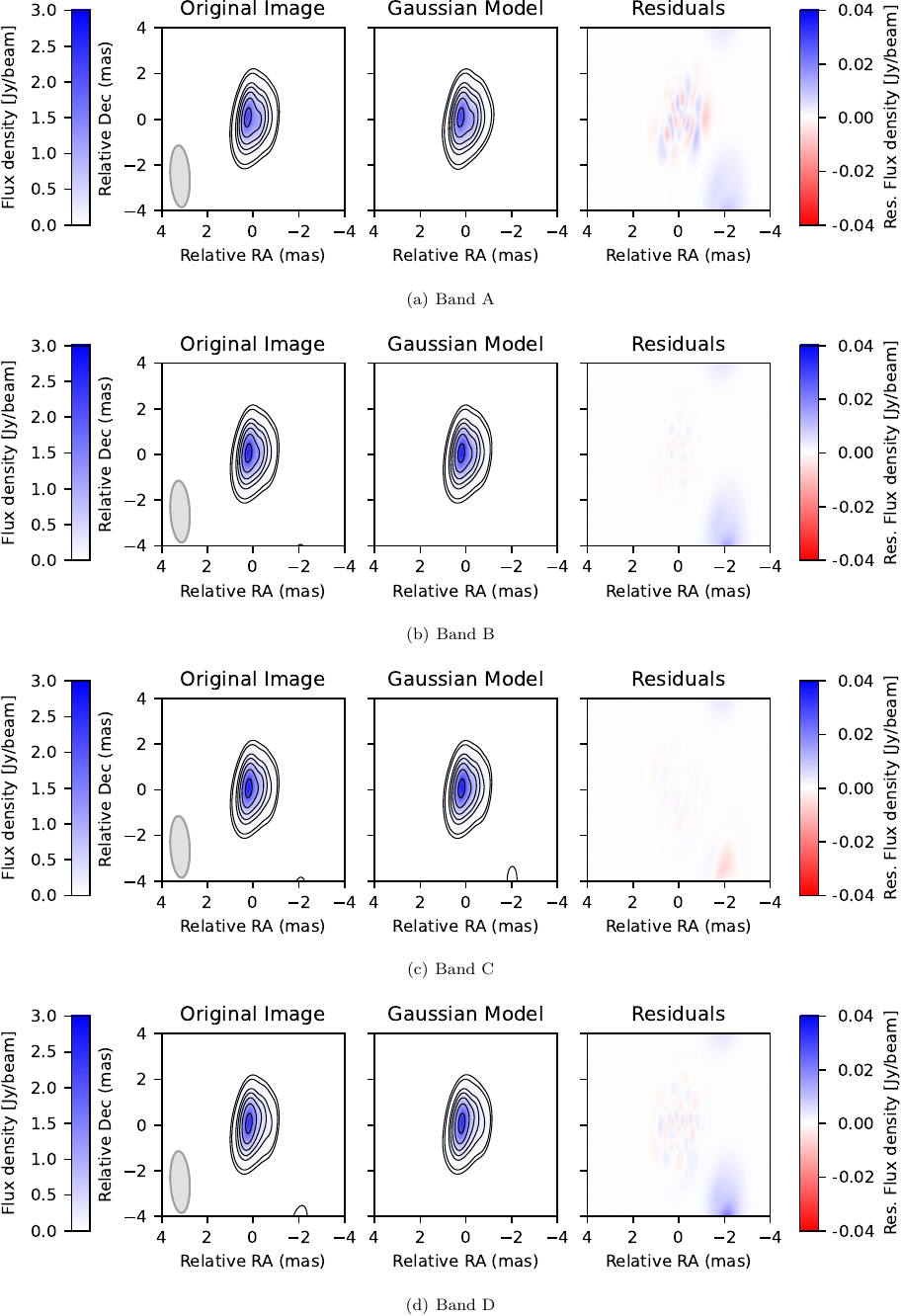}
    \caption[Model fitting of OJ287.]{Model fitting of OJ287. Each panel shows the residual map for a VGOS band, from A to D. The contour levels are 5, 10, 20, 40, and 80\,\%.}
    \label{fig:modelOJ287}
\end{figure}

Figures \ref{fig:model1849_670}, \ref{fig:model3C418}, and \ref{fig:modelOJ287} summarize the Gaussian modeling for the sources 1849+670, 3C418, and OJ287, respectively. For each source, the four panels correspond to the VGOS A, B, C, and D bands. The displayed maps show the residuals obtained after subtracting the Gaussian model from the corresponding image. In this way, they provide a direct visual assessment of how well the fitted components reproduce the main structural features of the source at each frequency.

The residual maps show that the adopted Gaussian parameterization captures the dominant source morphology in all three objects, while leaving only low-level extended residual emission. The source 1849+670 is dominated by a compact bright feature, plausibly associated with the core, together with weaker secondary emission extending mainly towards the north-west. Its overall structure is therefore relatively compact, and the fitted Gaussian components beyond the brightest one contribute only a modest fraction of the total flux density. In contrast, 3C418 shows a clearly more extended morphology, dominated by a bright primary component and a second prominent feature displaced to the south-west. In this case, the secondary components are comparatively strong, which reflects the larger structural contribution of the jet emission.

OJ287 represents an intermediate case in terms of morphological complexity, with several secondary components contributing a non-negligible fraction of the total flux density. Although the number of fitted components is comparable to that of 1849+670, the relative importance of the non-core emission is significantly greater. This makes OJ287, together with 3C418, a suitable target to test the impact of source structure corrections on geodetic observables.

The three selected sources span different degrees of structural complexity, from the comparatively compact morphology of 1849+670 to the more extended cases of 3C418 and OJ287. This makes them a useful set for evaluating the performance of the proposed source-structure mitigation strategy under different source morphologies. The present analysis is limited to these three sources. Although they were selected to sample different degrees of structural complexity, they cannot be considered representative of the full VGOS source population.

\pagebreak
\section{Structure Correction to VO2187}

In this section, we evaluate the performance of the proposed structure correction when applied to the same dataset (VGOS session VO2187) used to derive the source models. This represents the most favourable scenario, as the source structure information is fully consistent with the observed visibilities.

The source structure correction is explained in \citet{Jaron2026}. It is based on the formalism of \citet{Charlot1990}. In this approach, the extended brightness distribution of the source is represented by a set of Gaussian components (explained in section \ref{sec:imagemodelfitting}), from which the corresponding structure phase contribution can be computed for each baseline and channel. This structure phase is an additive term in the interferometric phase and, if not accounted for, propagates into the group delay as a source-dependent systematic error.

The correction is applied directly to the Level-1 VLBI data, i.e. to the complex visibilities produced by the DiFX correlator. In practice, the structure phase predicted from the Gaussian source model is subtracted from the visibility phase by rotating each complex visibility in the complex plane. The resulting SWIN files therefore contain corrected visibilities, while the original correlator output is preserved. These corrected data can then be processed with the standard geodetic pipeline, including fringe fitting and the generation of geodetic files, in exactly the same way as the uncorrected data.

An important aspect of this strategy is that the correction is introduced as early as possible in the VLBI processing chain, before the estimation of group delays and geodetic parameters. In the ideal case, the corrected visibilities should behave, for the subsequent analysis, as if they had been obtained from a point source. The implementation used in this chapter only corrects the phase contribution of the source structure, without modifying the amplitudes. Therefore, the corrected images are not expected to be perfectly point-like, but rather to show a significant reduction of the extended structure.

After applying the source structure correction to the visibility data, we generated images following exactly the same calibration and imaging procedure described in Chapter~\ref{ch:vgos_pipeline}. Using an identical pipeline for both the original and corrected datasets is essential in order to ensure that any differences in the resulting images are solely due to the structure correction, and not to differences in the imaging process itself.

\begin{figure}
  \centering
  \includegraphics[width=.94\linewidth]{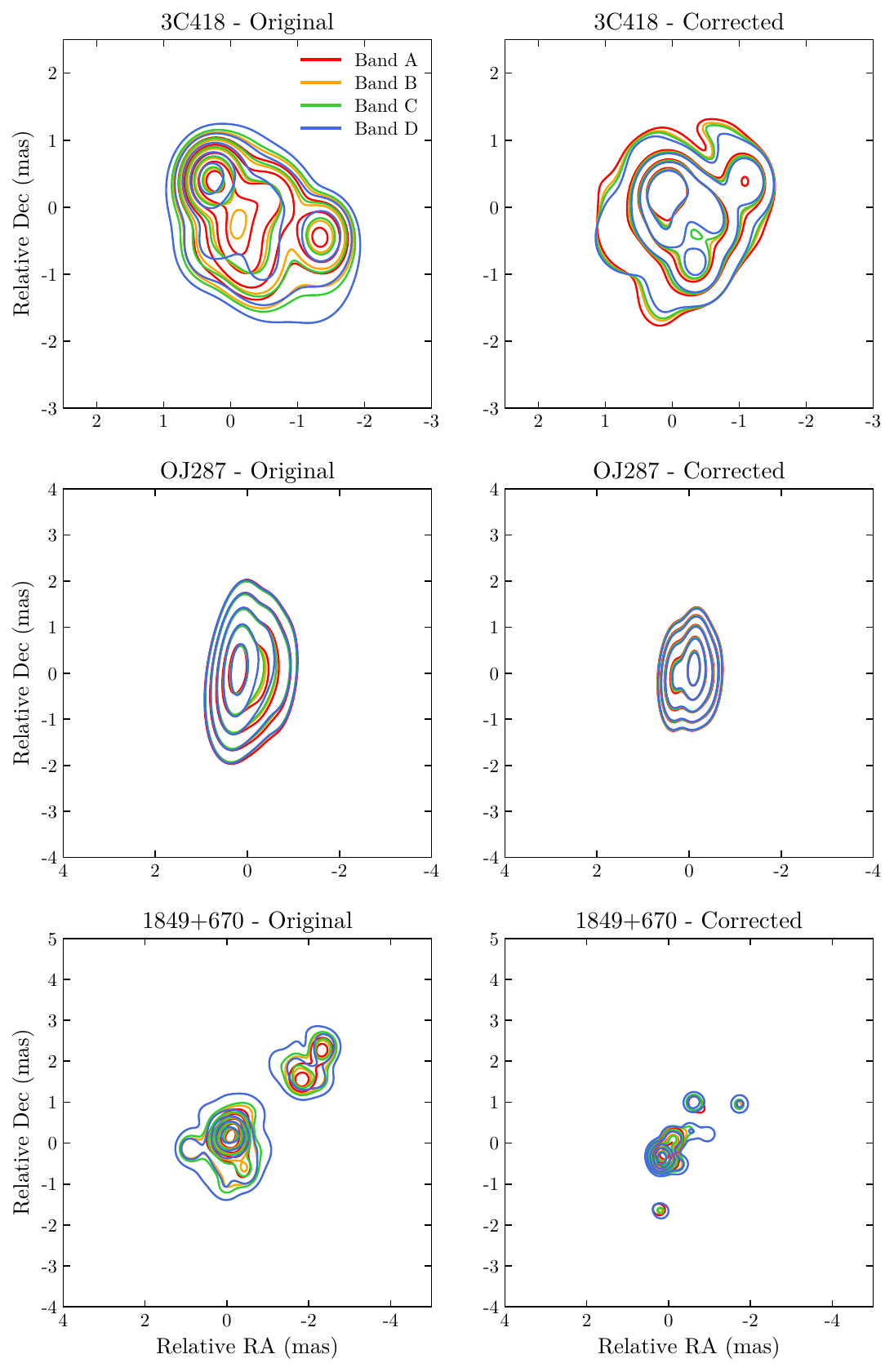}\\
  \caption[Images of the three sources before and after the correction.]{Images of the three sources. In each row, the image on the left side has been generated from the original visibility data. The right panels show the images of the corrected data, applying the same procedure as for the original image.}
  \label{fig:image-correction}
\end{figure}

The comparison between original and corrected images is shown in Fig. \ref{fig:image-correction}. For each source, the left panel shows the original image, while the right panel corresponds to the corrected visibilities. In all cases, the images have been convolved with 50\% of the C band beam, providing a common resolution reference across frequencies.

From a qualitative point of view, the impact of the structure correction is clearly visible. For the source 1849+670, the extended emission present in the original image, particularly towards the north-west direction, is strongly reduced after the correction. The remaining emission becomes significantly more compact and concentrated around the core region. 

A similar behaviour is observed for 3C418, where the extended structure is largely suppressed, leaving a dominant compact component together with weaker residual features. In contrast, the case of OJ287 shows a more subtle visual difference between the original and corrected images. However, as discussed below, quantitative metrics reveal that a reduction of the source structure is still achieved.

It is important to note that, in all cases, some residual structure remains after the correction. This is expected, since the method only accounts for phase contributions of the source structure, while leaving the visibility amplitudes unchanged.

To quantify the reduction of source structure suggested by the images, we define the compactness parameter $c$ as

\begin{equation}
c = \frac{\sum_{i=1}^{M} \sum_{j=1}^{N} d_{ij} S_{ij}}{\sum_{i=1}^{M} \sum_{j=1}^{N} S_{ij}},
\end{equation}
where $S_{ij}$ is the flux density at pixel $(i,j)$, and $d_{ij}$ is its distance to a reference point $(x_{\mathrm{ref}}, y_{\mathrm{ref}})$, defined as the position of the brightest pixel in the image. This quantity can be interpreted as the flux-weighted average distance of the emission from the source center, such that lower values of $c$ correspond to more compact structures.

The compactness values before and after applying the correction are listed in Table~\ref{tab:compactness}. In almost all cases, the corrected images show lower compactness values than the original ones, indicating a clear reduction of the source structure. The only exception is 3C418 in band A, where a slight increase is observed, likely reflecting the limitations of the model or the sensitivity of the metric in this particular case.

\begin{table}[]
    \centering
    \caption[Quantitative comparison of the compactness of the source structure before and after the correction.]{Quantitative comparison of the compactness of the source structure before and after the correction. The compactness~$c$ is defined in the text. A smaller value of $c$ corresponds to a more compact structure.}
    \label{tab:compactness}
    \begin{tabular}{l|rr|rr|rr|rr}
                & \multicolumn{2}{c}{Band A} & \multicolumn{2}{c}{Band B} & \multicolumn{2}{c}{Band C} & \multicolumn{2}{c}{Band D} \\
         Source & $c_{\rm orig}$ & $c_{\rm corr}$ & $c_{\rm orig}$ & $c_{\rm corr}$ & $c_{\rm orig}$ & $c_{\rm corr}$ & $c_{\rm orig}$ & $c_{\rm corr}$  \\
         \hline
        1849+670 & 16.2 & 9.6 & 19.3 & 10.3 & 20.7 & 11.1 & 24.3 & 15.3 \\
        3C418 & 22.4 & 22.7 & 24.0 & 22.1 & 25.0 & 21.9 & 27.4 & 21.5 \\
        OJ287 & 26.2 & 17.2 & 25.5 & 17.0 & 25.5 & 16.9 & 25.6 & 16.7 \\
        \hline
    \end{tabular}
\end{table}

The effect of the correction is particularly significant for 1849+670 and OJ287, where the compactness is reduced by a substantial fraction across all frequency bands. For 3C418, although the visual reduction of extended emission is evident, the compactness values show a more moderate improvement, consistent with the intrinsically more complex and extended morphology of this source.

Overall, both the qualitative inspection of the images and the quantitative analysis based on the compactness parameter consistently demonstrate that the proposed method effectively reduces the impact of source structure when applied to the same session used for model derivation.

\subsection{Assessment of the Phase-Only Structure Correction}
\label{sec:phase_only_validation}

The structure correction applied in the previous section modifies only the phase of the observed visibilities, while leaving their amplitudes unchanged. Although this procedure removes the modelled source contribution to the visibility phase, it does not constitute a complete removal of the source structure from the complex visibilities. In particular, the baseline-dependent amplitude modulation produced by the source brightness distribution remains present after the correction.

Consequently, the corrected images shown in Fig.~\ref{fig:image-correction} and the compactness values reported in Table~\ref{tab:compactness} should be interpreted as the result of a phase-only structure correction. They quantify the degree to which removing the structural phase contribution makes the reconstructed brightness distribution more centrally concentrated, but they do not represent the compactness that would be obtained after removing the full complex source structure. 

To quantify the magnitude of this effect, we performed a controlled simulation using the real $(u,v)$ coverage and visibility uncertainties of the VO2187 experiment. The synthetic source was chosen to have an asymmetric and clearly resolved morphology, consisting of three elliptical Gaussian components with different flux densities, sizes, and offsets from the phase centre. This choice provides a representative test case in which both the visibility phases and amplitudes contain a significant contribution from source structure, similar to the studied sources.

\begin{figure}[hptb]
    \centering
    \includegraphics[width=0.98\textwidth]
    {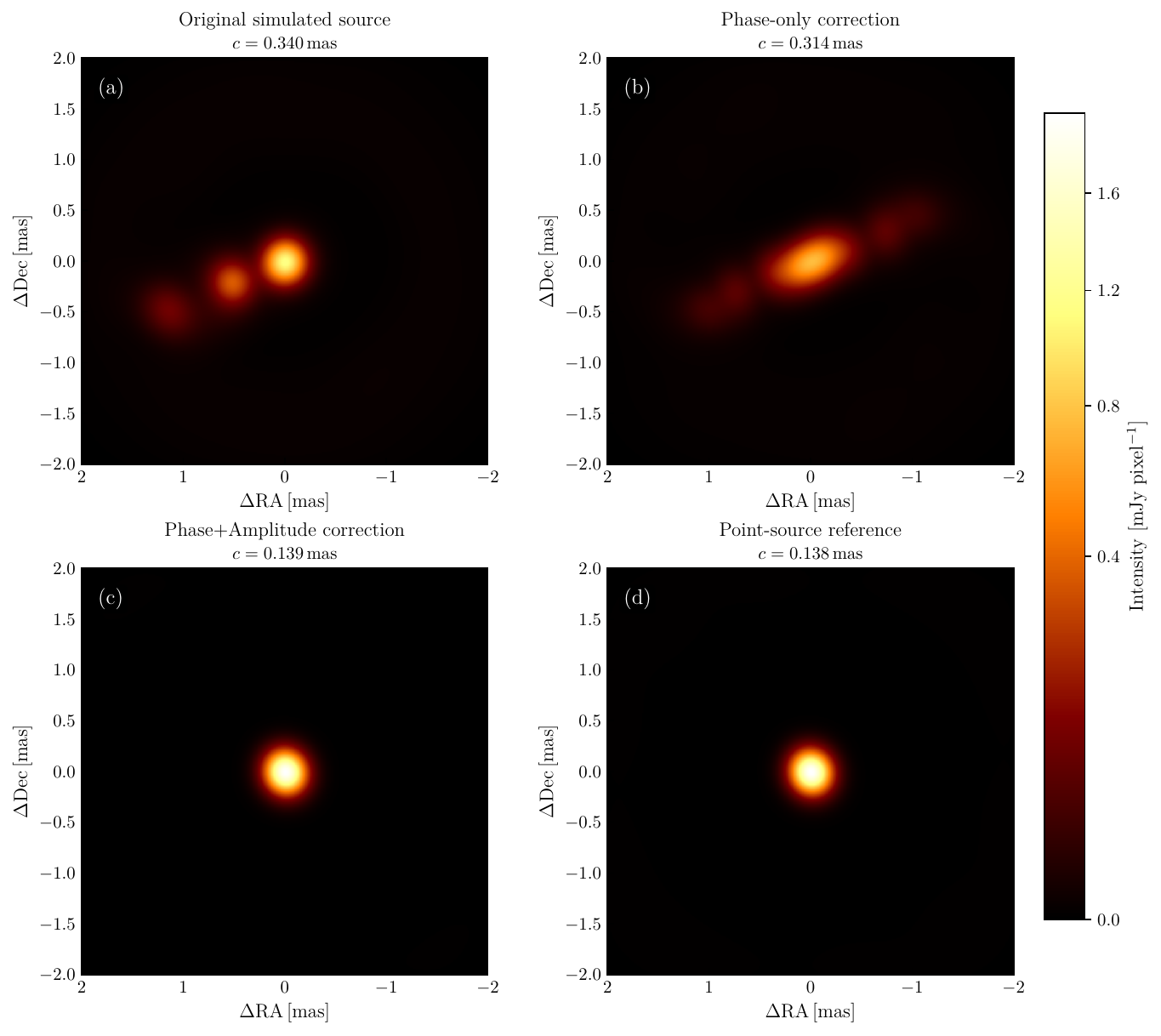}
    \caption[Comparison of reconstructed images from the source-structure simulation.]{Comparison of the reconstructed images obtained from the source-structure simulation.
    Panel (a) shows the original simulated source.
    Panel (b) shows the result after applying the phase-only correction.
    Panel (c) shows the result after correcting both the phase and amplitude.
    Panel (d) shows the corresponding point-source reference.
    The compactness parameter $c$ is indicated in each panel.}
    \label{fig:structure_correction_simulation}
\end{figure}

The model visibilities were sampled at the same $(u,v)$ points as the real observations, and thermal noise was added using the visibility uncertainties measured in the original data. From this common simulated dataset, four visibility sets were generated: the original structured source, the same data after applying the phase-only correction, the data after applying the full phase+amplitude correction, and a point-source reference with the same total flux density. All four datasets were reconstructed using the same RML imaging pipeline described in Sect.~\ref{sec:mfimaging}.

The results are shown in Fig.~\ref{fig:structure_correction_simulation}. The original simulated source has a compactness of $c=0.340$~mas. After applying only the structural phase correction, the compactness decreases to $c=0.314$~mas, corresponding to a reduction of approximately $7.5\%$. The source therefore becomes more centrally concentrated, but a significant elongated structure remains visible in the reconstructed image. In contrast, when both the visibility phase and amplitude are corrected, the reconstructed source becomes almost indistinguishable from the point-source reference. The corresponding compactness is $c=0.139$~mas, compared with $c=0.138$~mas for the point-source case. The difference between these two values is below $1\%$, demonstrating that, for an accurate source model, the full complex correction effectively removes the structural information from the visibilities.

This test shows that the residual structure visible in the phase-corrected images of Sect.~5.3 is not necessarily evidence of an inaccurate phase correction. Part of this residual morphology is expected because the observed visibility amplitudes, which still contain source-structure information, are left unchanged. Accordingly, the compactness values reported in Table~\ref{tab:compactness} should not be interpreted as the residual structure that would remain after a complete complex correction, but rather as a measure of the morphological change produced by removing the structural phase contribution alone.

Importantly, the geodetic results presented in Sect.~5.4 are not expected to change significantly when extending the correction from phase-only to phase+amplitude. The closure delays and post-fit delay residuals analysed in that section are derived from the frequency dependence of the visibility phases, and the additional amplitude correction does not modify these phases. Therefore, both correction approaches contain the same direct source-structure phase correction. Small numerical differences could nevertheless arise indirectly because correcting the amplitudes also modifies the effective signal-to-noise ratio and, consequently, the statistical weights assigned to the delay observables. These effects are expected to be secondary compared with the change observed in the reconstructed source morphology.
\pagebreak

\section{Geodetic Results}
\label{sec:geodetic_results}

The image comparison presented in the previous section provides a direct way to verify that the structure correction reduces the extended emission present in the reconstructed images. However, the ultimate goal of the correction is not only to obtain more compact images, but to improve the geodetic observables derived from the VLBI data. For this reason, the corrected visibilities were processed through the standard geodetic analysis pipeline in \citet{Jaron2026}. Here, we summarize its main results in order to assess whether the source structure correction derived from the images presented in this thesis leads to a measurable improvement in geodetic quantities. The original and corrected datasets were processed using the same pipeline, the one shown in this thesis, so the only difference between them was the phase correction applied to the Level-1 visibility data.

\subsection{Statistical Distribution of Closure Delays}
\label{sec:closure_delay_statistics}

Closure delays provide a useful diagnostic for evaluating the effect of source structure on the geodetic observables. For a closed triangle of stations, most station contributions cancel, and the closure delay is expected to be close to zero in the absence of baseline dependent effects. It is the same case as seen in Sect. \ref{sec:VLBIobservables} with the closure phases (Eq. \ref{eq:closurephases}). An extended source brightness distribution, however, introduces baseline dependent phase terms and can therefore produce non-zero closure delays. The closure delay is defined as:
\begin{equation}
\tau_{abc} = \tau_{ab} + \tau_{bc} + \tau_{ca}
\end{equation}
where $\tau_{ij}$ is the geocentric group delay between stations $i$ and $j$. The use of geocentric delays ensures that the three delays entering the closure relation correspond to the same incoming wavefront.

\begin{figure}[tp]
  \centering
  \includegraphics[width=\textwidth]{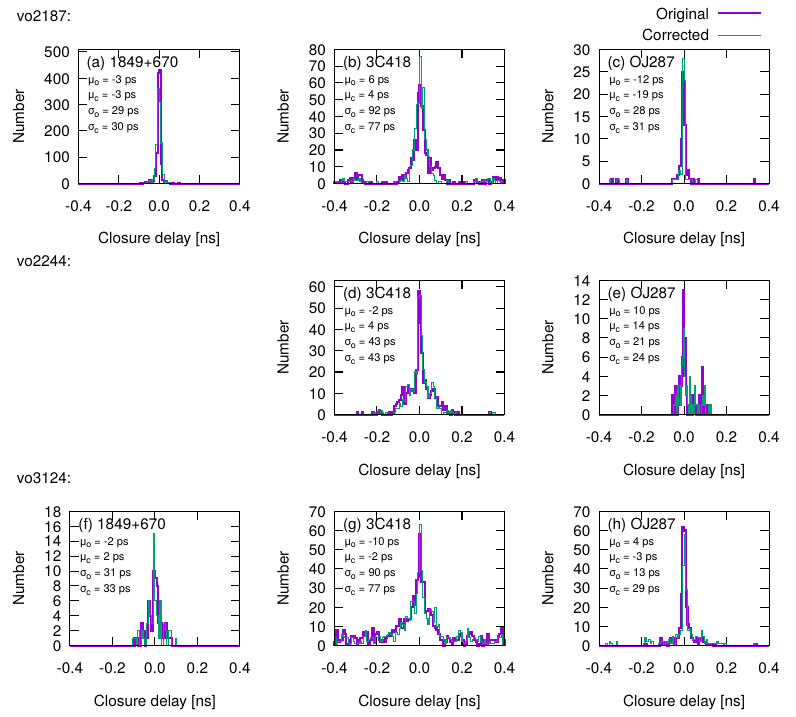}
  \caption[Closure-delay distributions before and after the structure correction.]{Histograms of closure delays for the three sources observed in the three VGOS sessions. The data derived from the original visibilities are plotted in purple, and the corrected data are plotted in green. Each panel shows the distribution of closure delays for one source. In each plot, the average value $\mu$ and the standard deviation $\sigma$ are shown for the original (index o) and corrected (index c) data.}
  \label{fig:closuredelays}
\end{figure}

The statistical distributions of closure delays, shown in Fig. \ref{fig:closuredelays}, before and after applying the structure correction show that the effect depends strongly on the intrinsic morphology of the source. For 1849+670, which is the most compact of the three sources considered here, the correction does not lead to a significant improvement. This is expected, since the source contains only weak extended components and therefore produces relatively small structure closure errors.

\begin{figure}[tp]
  \centering
  \includegraphics[width=\textwidth]{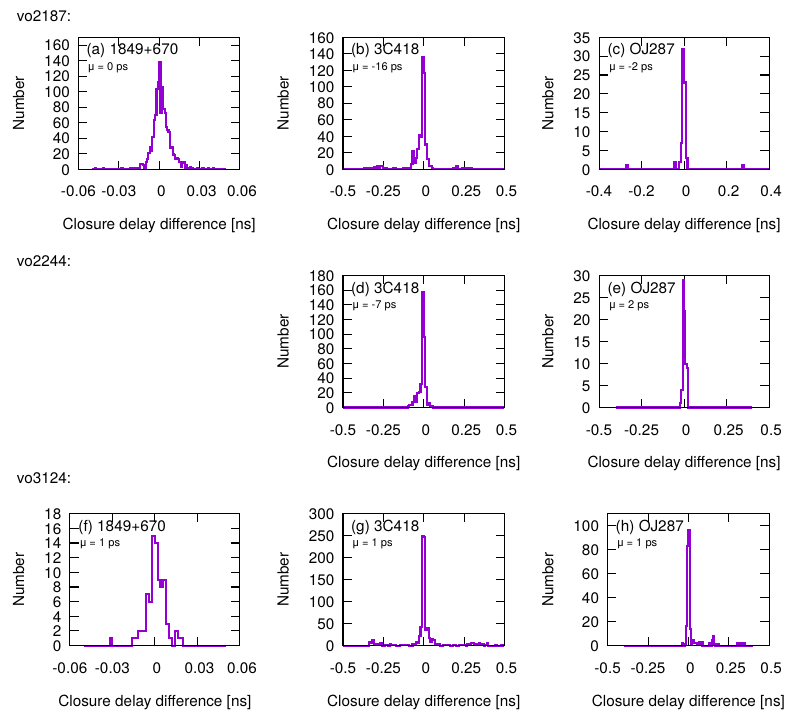}
  \caption[Differences between the absolute closure delays before and after the correction.]{Histograms of the differences between the absolute values of the closure delays. Negative values indicate that the corrected closure delay is closer to zero than the original one. The average values $\mu$ refer to the distributions shown in the panels. A negative value of $\mu$ indicates that, on average, the correction improves the closure delays.}
  \label{fig:closuredelays-diff}
\end{figure}

The clearest improvement is found for 3C418. This source has the most prominent extended structure, and the correction reduces the width of the closure delay distribution. In the VO2187 session, the standard deviation decreases from 92 ps in the original to 77 ps in the corrected data. This indicates that a significant fraction of the delay contribution was associated with source structure and was successfully mitigated by the correction. Also, the mean is a 66\% closer to 0.

For OJ287, the result is more moderate. Although the visual comparison of the histograms suggests a partial concentration of the corrected closure delays towards zero, the global statistical indicators do not show as clear an improvement as in the case of 3C418. This can be explained because it is the source with the fewest visibilities (as shown in Tab. \ref{tab:sources}), which is translated to a worse image and, so, to a worse image model fitting.

The lower rows of Figs.~\ref{fig:closuredelays} and \ref{fig:closuredelays-diff} show the same comparison for the later sessions VO2244 and VO3124, using also the source models derived from VO2187.

For VO2244, observed approximately two months after VO2187, the effect of the correction is still positive for 3C418, although weaker than in the same session. The standard deviation of the closure delay distribution remains almost unchanged in the global histogram, 43 ps, but the distribution of absolute closure delay differences has a negative mean value of $\mu=-7$~ps. This indicates that, on average, the corrected closure delays are slightly closer to zero than the original ones. In contrast, OJ287 does not show a significant improvement in this session. Its standard deviation increases from
21 ps to 24 ps after the correction, and the mean of the absolute difference distribution is positive, $\mu=2$~ps. This suggests that the VO2187 structure
model is not sufficiently representative of OJ287 in VO2244 (because of physical changes in the jet), or that the available observations do not sample the same structure in a favourable way (because of different uv coverages).

The results are less favourable for VO3124, observed ten months after VO2187. For 1849+670, the correction has essentially no positive effect: the standard deviation slightly increases from 31~ps to 33~ps, and the mean of the absolute difference distribution is $\mu=1$~ps. For 3C418, the global histogram shows a decrease in the standard deviation, from 90~ps to 77~ps, but the direct comparison of absolute closure delays does not confirm a systematic improvement, with $\mu=1$~ps. The worst case is found for OJ287, where the standard deviation increases from 13~ps to 29~ps after applying the correction. In this case, the outdated source model appears to introduce additional delay contributions rather than removing them.

\subsection{Post-Fit Residuals}
\label{sec:postfit_residuals}

\begin{figure}[tp]
  \includegraphics{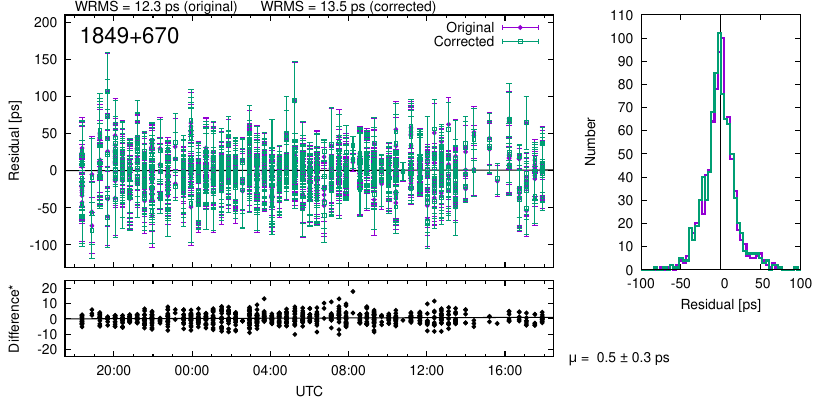}
  \includegraphics{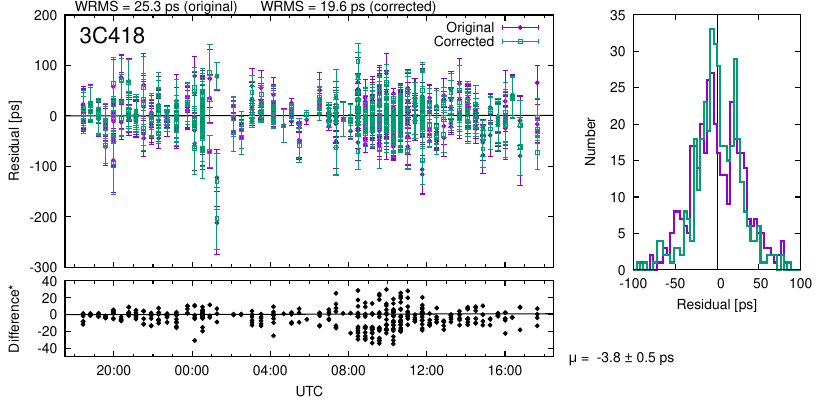}
  \includegraphics{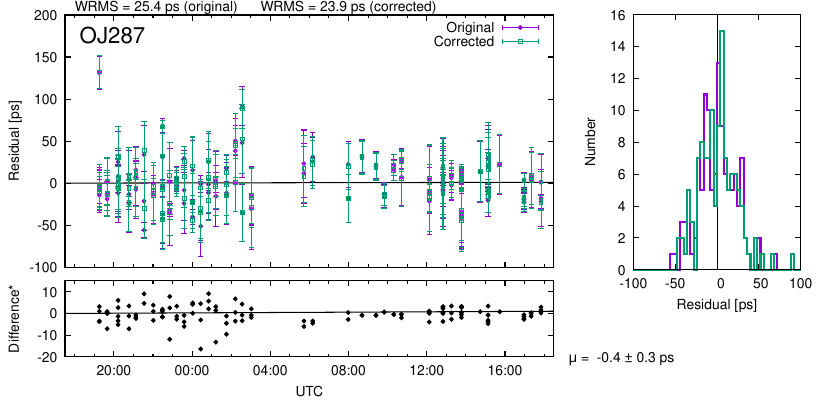}
  \caption[Post-fit residuals for session VO2187.]{Post-fit residuals for the session VO2187.
  Left: Time series of residuals, original data plotted as purple dots, and corrected data as green squares.
  Right: Histograms of the residuals.}
  \label{fig:postfit-vo2187}
\end{figure}

The most relevant test for geodetic applications is whether the correction improves the results of a full geodetic solution. For this purpose, \citet{Jaron2026} compared the post-fit residuals obtained from the original and corrected databases, using the same analysis in both cases, as shown in Fig. \ref{fig:postfit-vo2187}.

For 1849+670, the post-fit residuals remain essentially unchanged after the correction. This agrees with the closure delay analysis and reflects the relatively compact nature of the source. Since the structure contribution is small, only a limited improvement can be expected.

For 3C418, the impact of the correction is more evident. The corrected residuals are more concentrated around zero during part of the session, particularly in the time interval where the closure delays also show the strongest improvement. The weighted root mean square of the residuals decreases from 25.3 ps in the original solution to 19.6 ps after the correction. This provides evidence that the phase correction applied at the visibility level propagates into an improvement of the geodetic observables.

The case of OJ287 is less conclusive. The weighted root mean square of the residuals shows only a small reduction, from 25.4 ps to 23.9 ps, and the number of observations is not sufficient to draw a strong conclusion. Nevertheless, the result is compatible with a modest improvement, weaker than that obtained for 3C418.

\begin{figure}[tp]
    \centering
    \includegraphics[width=\textwidth]{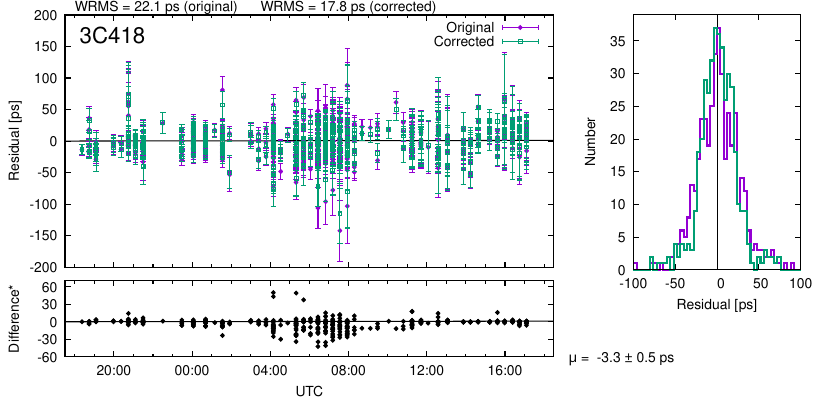}
    \includegraphics[width=\textwidth]{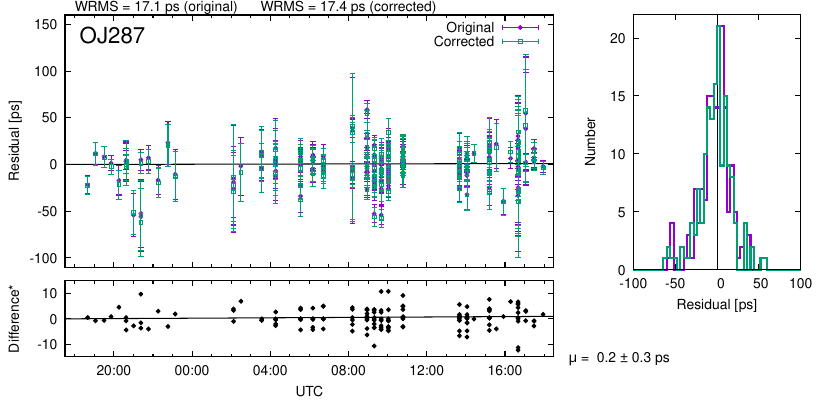}
    \caption{
        Post-fit residuals for the session VO2244. 
    }
    \label{fig:postfit-vo2244}
\end{figure}

\begin{figure}[tp]
    \centering
    \includegraphics[width=\textwidth]{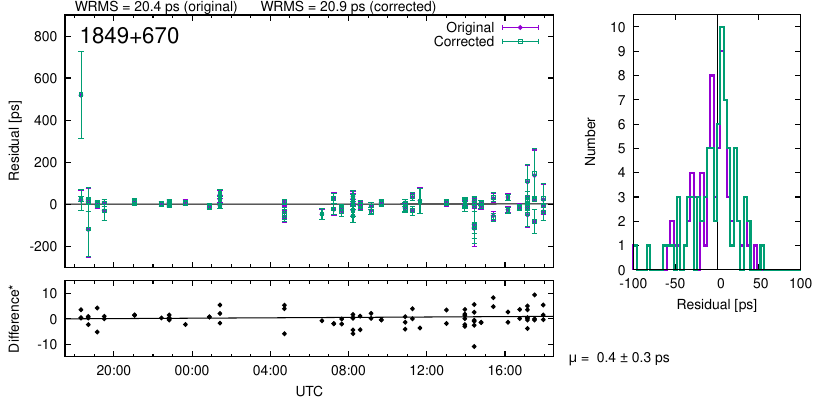}
    \includegraphics[width=\textwidth]{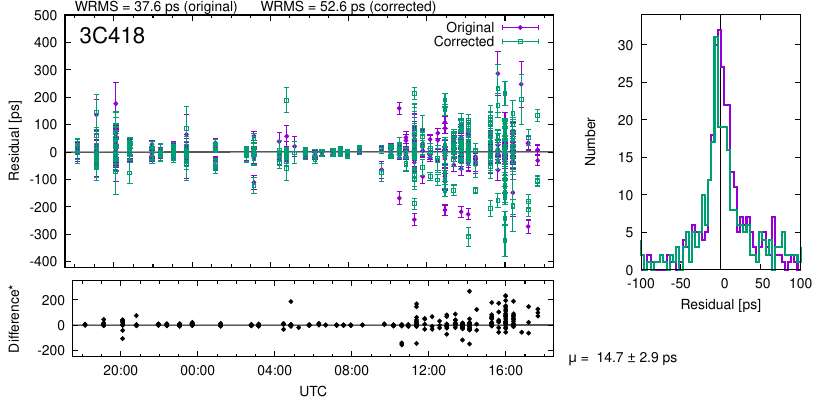}
    \includegraphics[width=\textwidth]{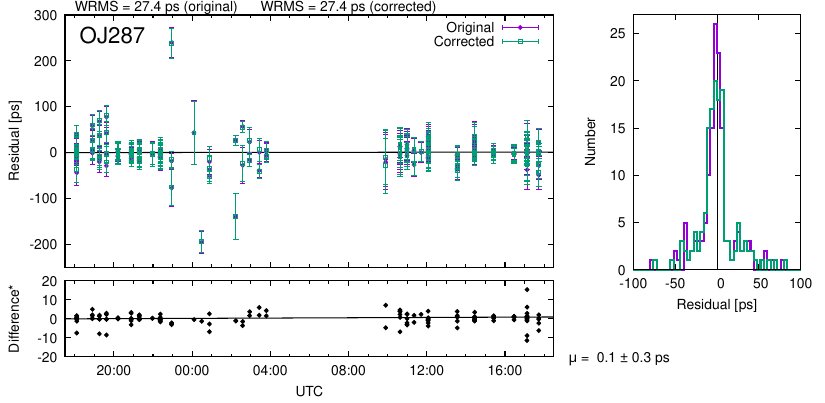}
    \caption{
        Post-fit residuals for the session VO3124.
    }
    \label{fig:postfit-vo3124}
\end{figure}

The same comparison was also carried out for the later sessions VO2244 and VO3124, shown in Figs.~\ref{fig:postfit-vo2244} and \ref{fig:postfit-vo3124}, respectively. 

For VO2244, observed two months after VO2187, the post-fit residuals still show a positive effect for 3C418. The corrected residuals are more concentrated around zero during part of the session, and the weighted root mean square decreases from 22.1 ps in the original solution to 17.8 ps after the correction. This is consistent with the closure delay analysis, which also showed that the VO2187 source model still provides a useful correction for 3C418 in this session. For OJ287, however, the correction has almost no effect on the post-fit residuals. The weighted root mean square remains essentially unchanged, indicating that the correction does not lead to a measurable improvement for this source in VO2244.

The results for VO3124, observed about ten months after VO2187, are less favourable. For 1849+670, the post-fit residuals are only weakly affected by the
correction, with the weighted root mean square not changing. This is consistent with the compact nature of the source and with the closure
delay results, which did not show a significant improvement. For 3C418, the effect of the correction is clearly negative in the full geodetic solution: the weighted root mean square increases from 37.6 ps to 52.6 ps. This indicates that the structure model derived from VO2187 is no longer representative of the source structure sampled in VO3124, and that applying an outdated model can degrade the geodetic solution. Finally, OJ287 again shows almost no global effect, with the weighted root mean square remaining unchanged at 27.4 ps.

\pagebreak
\section{Conclusions}
\label{sec:geodetic_conclusions}

In this chapter, we have explored the geodetic application of the calibration and imaging pipeline developed in this thesis. The main goal was to assess whether the VGOS images produced by this pipeline can be used to mitigate the impact of source structure on geodetic VLBI observables. For this purpose, the source models derived from the images were used in the structure correction method developed by \citet{Jaron2026}, which removes the structure phase contribution directly from the raw visibility data.

The results show that the correction is effective when the source model is close in time to the observations to which it is applied. This is most clearly seen in the case of 3C418 in session VO2187. This source has the most prominent extended structure among the three targets considered here, and therefore provides the strongest test of the method. After applying the correction, the closure-delay distribution becomes narrower and the post-fit residuals of the geodetic solution are reduced. This demonstrates that the phase correction applied at the visibility level propagates into the group delays and can lead to a measurable improvement in the final geodetic analysis.

The behaviour of the three sources is consistent with their different morphologies. For 1849+670, which is the most compact source in the sample, the correction produces little improvement. This is expected, since the contribution of source structure to the geodetic observables is already small. For OJ287, the results are less conclusive. Although some diagnostics suggest a modest improvement, the limited number of observations and the corresponding uncertainty in the image model make the effect weaker than for 3C418. These results indicate that the benefit of the correction depends on both the amount of resolved structure in the source and the quality of the image model used to represent it.

The application of the same VO2187 source models to later sessions provides an important test of the temporal stability of the correction. For VO2244, observed two months after VO2187, the correction still improves the post-fit residuals for 3C418, although the effect is weaker than in the same session. For VO3124, observed ten months after the imaging session, the positive effect is largely lost and can even become negative, as shown by the degradation of the post-fit residuals for 3C418. This indicates that source structure corrections based on outdated images must be applied with caution. The
brightness distribution of AGN jets can evolve with time, and different observing schedules or station networks can also sample different regions of the uv plane, reducing the validity of a model derived from a previous epoch.

VGOS imaging is not only useful for astronomical studies, but can also provide valuable input for geodetic analysis. The pipeline developed in this thesis makes it possible to obtain source images from VGOS data themselves, and these images can be transformed into compact source models for structure correction. At the same time, the results highlight the need for regular imaging of geodetic sources. A single image or source model cannot be assumed to remain valid indefinitely. For a robust operational implementation, the structure correction should ideally be based on images obtained from the same experiment, or from observations separated by no more than a short time interval.

\cleardoublepage

\chapter{Conclusions and Future Work}
\label{ch:conclusions}

\lettrine{T}{  his} thesis has explored the astronomical exploitation of next-generation geodetic VLBI observations. 
Although VGOS was designed to improve the precision of geodetic VLBI, the results presented in this thesis show that the wide frequency coverage and high time-cadence of observations of extragalactic radio sources also make it a powerful instrument for astrophysical studies.

The main goal of this work was to demonstrate that VGOS observations can be processed beyond their standard geodetic use and transformed into scientifically useful full-polarization, multi-frequency VLBI images. To achieve this, a complete calibration and imaging pipeline has been developed and tested. The pipeline converts the VGOS linear polarization data into a circular basis, calibrates the instrumental phase and polarization response, applies a new wide-band global-fringe-fitting algorithm for the phase calibration, performs amplitude calibration, and, finally, reconstructs total-intensity and polarimetric images of the observed sources.

The results presented here show that VGOS observations contain robust astronomical information. The reconstructed images reproduce the expected core-jet morphologies of the analysed sources, and the derived spectral index, core shift, and Faraday rotation results are broadly consistent with those obtained from other astronomical VLBI arrays. This agreement validates the calibration strategy developed in this work and confirms that geodetic VGOS observations can be used to study the structure of AGN jets.

Chapter~\ref{ch:vgos_pipeline} presented the calibration pipeline developed for VGOS data. The first key step of this pipeline is the conversion of the linear polarization visibilities into a circular polarization basis using PolConvert. This is necessary because, although the use of linear feeds is advantageous for wide instantaneous bandwidths and low instrumental polarization, it complicates the calibration of VLBI data. The quality of the polarization conversion was assessed through several independent tests. The comparison between the parallel and cross-hand fringe amplitudes showed that the converted data behave as expected for circular polarization VLBI observations. The estimated cross-polarization gains were also compared with those obtained in a later VGOS experiment, showing that the instrumental delays were generally stable at the level of a few picoseconds. It also introduced a new wide-band global-fringe-fitting. This algorithm subtracts an a priori ionospheric dispersive delay model using IONEX maps and estimates dispersive and non-dispersive delays. When applied to the polconverted VGOS data, the algorithm produced calibrated visibility phases and enabled the generation of full-polarization images. When comparing its performance with existing algorithms, we demonstrate that our WBGFF performs better, both in time and effectiveness. 

Chapter~\ref{ch:1803+784} used the calibrated data of the prototypical blazar 1803+784. This source was selected because it is bright, compact, frequently observed in geodetic VLBI sessions, and extensively studied in previous astronomical VLBI works. The VGOS images recovered the expected compact core and downstream jet structure. The simultaneous multi-frequency coverage between approximately 3 and 11~GHz allowed us to measure the frequency-dependent position of the VLBI core and to estimate a core shift below the canonical equipartition expectation of $k_r \simeq 1$, suggesting that the jet of 1803+784 does not fully satisfy the assumptions of a simple conical jet in energy equipartition between particle and magnetic field.

The wide instantaneous frequency coverage of VGOS also allowed us to study the Faraday rotation of 1803+784, separating for the first time, with a broad simultaneous frequency coverage, into an external Faraday screen and a frequency-dependent internal contribution.

The total intensity, spectral index and Faraday rotation maps derived from the multi-frequency imaging agree qualitatively with previous VLBI studies, including results from the MOJAVE programme. This agreement confirms the reliability of both the calibration pipeline and the physical interpretation of the VGOS images.

Chapter~\ref{ch:astro} extended the astrophysical analysis to the remaining sources in the sample. The sources analysed in this thesis include 0059+581, 0613+570, 3C418, 1849+670, and 2229+695, in addition to 1803+784. For these sources, VGOS data were used to produce multi-frequency images, measure core shifts, and estimate Faraday rotation properties. In several cases, these results provide new astrophysical information for sources that had not been studied before.

The core-shift measurements obtained in Chapter~\ref{ch:astro} show that the position of the VLBI core can be measured with VGOS data for a variety of AGN. Most of the fitted values of $k_r$ are broadly compatible with the equipartition expectation within the uncertainties, although 1803+784 remains the clearest case of a significant deviation. These measurements demonstrate that VGOS observations can be used to probe opacity effects and the physical conditions in the innermost regions of AGN jets.

For the two sources for which the required geometrical and relativistic jet parameters were available in the literature, 0059+581 and 1849+670, the core-shift measurements were also used to estimate physical jet parameters under the equipartition assumption. Using the Doppler factors, viewing angles, and intrinsic opening angles reported by \citet{Pushkarev2017}, we derived the magnetic field strength and relativistic particle density both at 1~pc from the jet apex and at the observed VGOS core positions.

The spectral index maps obtained for the source sample show physically meaningful structures. In general, the compact core regions are associated with optically thick spectra, while the downstream jet regions show optically thin emission. This is the expected behaviour for synchrotron AGN jets, where the observed VLBI core corresponds to the surface of optical depth close to unity and the downstream jet is dominated by optically thin plasma.

The Faraday rotation analysis of the extended source sample further illustrates the potential of VGOS. Although the EVPA uncertainties and the limited angular resolution of some sources make the interpretation of individual RM values more uncertain, the results show that VGOS data can recover Faraday rotation measures of the same order as those reported in previous VLBI studies. Furthermore, thanks to the four instantaneous bands used in the observations, we can separate, for the first time, the contributions of external and internal Faraday screens.  

Chapter~\ref{ch:geodesy} explored the geodetic application of the VGOS images produced with the pipeline developed in this thesis. The same source structures that provide information about AGN jet physics also introduce systematic errors in geodetic VLBI observables. Extended source structure affects visibility phases and group delays, producing closure delays different from zero and increasing post-fit residuals in geodetic solutions. Therefore, imaging geodetic sources is not only useful from an astrophysical point of view, but also necessary for improving the accuracy of geodetic products.

The multi-frequency model images obtained in this thesis have been used as part of a geodetic analysis, where the contribution of the source structures was removed from the visibilities prior to the fringe fitting. Comparison of this geodetic analysis with the case where no contribution from source structures was removed shows that there is a general improvement of the post-fit residuals when the source structures are accounted for.

The impact of the source-structure correction was also evaluated through triangular closure delays. For sources with sufficiently extended structure, the closure-delay distributions improved significantly after applying the correction. The clearest examples are 3C418 and OJ287, where the structure contribution is strong enough to produce a measurable effect. In contrast, little or no improvement was found for compact sources such as 1849+670, as expected. This confirms that the usefulness of source-structure corrections depends strongly on the intrinsic morphology of the source and on the quality of the image model used for the correction.

Chapter~\ref{ch:geodesy} also showed that the temporal validity of the source model is a critical limitation. When the structure correction was applied to a session close in time to the one used for imaging, the effect remained positive. However, when the model was applied to observations separated by a longer time interval, the positive effect disappeared and could even become negative. This suggests that the structure of compact AGN can evolve on timescales relevant for geodetic analysis. Based on the cases studied here, source images used for structure correction should ideally be obtained from the same session or from observations separated by no more than about one month.

Overall, the results of this thesis demonstrate the dual scientific value of VGOS data. On the one hand, VGOS observations can be used to study the physical properties of AGN jets, including core shift, spectral index structure, Faraday rotation, and deviations from equipartition. On the other hand, the same images can be used to mitigate one of the main systematic effects in geodetic VLBI. This connection between astrophysics and geodesy is one of the central conclusions of the thesis: VGOS should not be regarded only as a geodetic observing system, but also as a valuable astrophysical instrument whose products can improve both AGN science and geodetic precision.

Finally, the comparison between the results obtained in this work and previous studies based on other VLBI arrays confirms the validity of the calibration and imaging pipeline developed in this thesis. The agreement between VGOS and other astronomical VLBI arrays supports the robustness of the analysis. At the same time, the results presented here provide new astrophysical information for several sources, like the first resolved broadband VGOS characterization of 0613+570, including its multi-frequency structure, spectral-index and curvature maps, Faraday-rotation estimates, and the frequency dependence of the core position. Another example is 2229+695, for which we provide a first estimate of the core-shift index, $k_r=1.17^{+0.10}_{-0.36}$, together with resolved Faraday-rotation measurements in the core and in two jet regions. In addition, for sources such as 3C418 and 0059+581, the VGOS data provide new estimates of the core-shift, spectral and Faraday-rotation properties over the 3--11 GHz range, including the separation between external and internal Faraday-rotation contributions when allowed by the data.

\pagebreak
\section{Proposed Strategies to Improve VGOS Data Quality}
\label{sec:future_work_vgos_improvements}

VGOS observations are currently scheduled primarily to maximize geodetic precision. This strategy is optimal for the main goal of the VGOS network, but it is not necessarily optimal for astronomical purposes and for source structure correction. The results presented in this thesis show that relatively modest changes in the observing strategy and instrumental calibration could significantly improve the scientific return of VGOS data, while also benefiting the geodetic analysis by providing better information about frequency-dependent source structure and polarization.

A first improvement would be the inclusion of longer calibration scans on bright and compact sources during VGOS experiments. Calibration scans of about one or two minutes, repeated a few times during an observation, would allow the cross-polarization gains to be determined and tracked with higher precision. This would improve the quality of the polarization conversion from linear to circular basis and would reduce residual instrumental polarization in the final data products.

The phase-calibration system should also be optimized. In particular, the frequency separation of the phase-cal tones should be as small as possible in order to obtain robust and precise estimates of the instrumental delays within each spectral window. This is especially relevant for Yebes, where the 10~MHz separation between phase-cal tones can make the delay estimate vulnerable to missing or unreliable tone phases. If a small fraction of the phase-cal tones cannot be correctly determined, the relatively large tone spacing may lead to an unreliable estimate of the instrumental delay of the spectral window, degrading the phase calibration and, consequently, the polarization conversion and fringe-fitting performance.

Another instrumental improvement concerns twin or nearby stations. The phase-cal tones at twin stations should be offset with respect to each other in order to avoid spurious cross-correlations that can degrade the quality of the fringes. Including the baselines between twin stations in the analysis would add robustness to the global antenna-based calibration of gains and instrumental polarization. This would be particularly useful for the determination of cross-polarization delays and phases, which are essential for the processing of VGOS data.

Amplitude calibration is another key aspect that should be improved in future VGOS observations. Amplitude calibration should ideally be performed using gain curves and system temperature information, which is incomplete for some VGOS stations. A systematic characterization of gain curves and a routine monitoring of system temperatures would allow a more accurate amplitude calibration of VGOS visibilities. Some stations are incorporating this, so in the near future, a complete amplitude calibration will be possible (see Appendix \ref{sec:NyAlesund} for an example of a VGOS antenna characterization). This would directly improve image fidelity, spectral-index measurements, and polarimetric products. It would also enable the development of more complete source-structure corrections that account not only for visibility phases, but also for visibility amplitudes.

Future VGOS schedules could also be modified to improve their astronomical capability without compromising their geodetic goals. In addition to occasional longer calibration scans, phase-referencing observations could be included throughout VGOS experiments in order to better quantify and remove the influence of core shift on group-delay measurements. This would improve the astrometric accuracy, especially in sources where the jets are too compact for a reliable image alignment. Such changes would help transform VGOS into an observing system capable of producing both high-quality geodetic observables and scientifically useful astronomical images.

\pagebreak
\section{Future Work}
\label{sec:future_work}

The first natural continuation of this thesis is the systematic application of the calibration and imaging pipeline to a larger number of VGOS sessions and sources. This work has focused mainly on experiment VO2187, but VGOS weekly observes a large number of AGN. Applying the same calibration and imaging strategy to a larger sample would make it possible to build a homogeneous catalogue of VGOS source images, core shifts, spectral-index maps, and Faraday-rotation measurements. Such a catalogue would be valuable both for AGN physics and for identifying the sources that introduce the largest structure errors in geodetic solutions.

A larger VGOS imaging catalogue would also allow the results obtained in this thesis to be placed on a firmer statistical basis. In particular, it would make it possible to study how common deviations from the equipartition value of $k_r \simeq 1$ are among geodetic AGN, how frequently significant Faraday rotation gradients or internal RM components are detected, and whether the sources that are most problematic for geodesy are also those with the most complex spectral and polarimetric structures. This would provide a direct connection between the astrophysical properties of AGN jets and their impact on geodetic VLBI observables.

Another important line of future work is the use of time-dependent imaging algorithms. New dynamic imaging methods \citep[e.g., ][]{Mus2024} offer a promising way to model the temporal evolution of AGN structure instead of treating each epoch independently. Algorithms such as Kine \citep{Foschi2024} are designed to reconstruct time-variable source structures from sparse interferometric data. The application of these methods to VGOS could provide a continuous description of the structural evolution of AGN.

These dynamic imaging techniques are especially relevant for VGOS because the same sources are observed repeatedly in different experiments, but sometimes with a sparse uv coverage. Instead of producing independent static images for each session, future work could combine observations obtained at different epochs to reconstruct the temporal evolution of the brightness distribution, polarization structure, spectral index, and core position. This would be particularly useful for sources with significant structural variability, for which a source model derived from a single epoch may not remain valid for later geodetic experiments.

From the geodetic point of view, the next step is to extend the source-structure correction analysis to a larger number of sources and experiments. A systematic study over many VGOS sessions would make it possible to determine under which conditions the correction is beneficial, how long source models remain valid, and which source properties best predict the improvement in the geodetic solution.

The source structure correction method itself can also be extended. The implementation used in this thesis corrects the phase contribution of source structure, but does not correct the visibility amplitudes. This explains why part of the source structure remains in the corrected images. A future improvement would be to incorporate the amplitude response of the source model into the correction procedure and to analyse its contribution separately. This would provide a more complete transformation of the observed visibilities towards the response expected for a point source and could further reduce the residual structure contribution in both imaging and geodetic analysis.

Besides source structure, the use of wide-band global fringe fitting may help to enhance the quality of the geodetic products, since, by construction, the closure delays and phases of the solution are exactly zero, being most of the contribution from source structure left in the post-fit global-fringe-fitting residuals. Therefore, adapting our WBGFF for its use in geodetic analyses is a clear goal for future work.

A further extension of this thesis would be to obtain the additional jet parameters required to convert VGOS core-shift measurements into physical estimates of the magnetic-field strength and particle density for a larger number of sources. This calculation could only be performed for 1803+784, 0059+581 and 1849+670, because these were the only sources in the sample for which suitable values of the Doppler factor, viewing angle, and intrinsic opening angle were available in the literature. For the remaining sources, the lack of reliable estimates of $\delta_j$, $\theta$, and $\phi$ prevented a robust application of the full equipartition hypothesis.

Future observing proposals with astronomical VLBI arrays could be designed to obtain these missing parameters for geodetic AGN. Multi-epoch, high-resolution observations at centimetre wavelengths would allow the apparent motions of jet components to be measured, providing estimates of the apparent speed and, in combination with variability or brightness temperature information, constraints on the Doppler factor and viewing angle. 

Such observations should ideally be obtained close in time to the VGOS experiments used for the core-shift analysis. The results of this thesis show that source structure and core shifts can vary on timescales relevant for geodetic VLBI, so combining VGOS core shifts with jet parameters measured years apart may introduce additional systematic uncertainties. Coordinated or quasi-simultaneous VGOS and astronomical VLBI observations would therefore provide a much more reliable way to estimate $B_1$, $N_1$, $B_{\rm core}$, and $N_{\rm core}$ for a larger sample of geodetic sources. This would open the possibility of building a homogeneous catalogue of magnetic fields and particle densities in AGN jets, directly linking geodetic source monitoring with jet physics.

\cleardoublepage
\appendix
\cleardoublepage

\chapter{Appendices}
\section{Characterization of the Ny-Ålesund North (Nn) 13-m Telescope}
\label{sec:NyAlesund}

The Ny-\AA lesund Geodetic Observatory, located in Svalbard, Norway, operates two twin 13.2~m radio telescopes devoted to the IVS VGOS VLBI project. Yebes Observatory built and installed two VGOS receivers for this observatory. At the time of the characterization described here, only the northern antenna, Nn, was equipped with one of these VGOS receivers, while the other antenna was operating with a tri-band receiver, also developed at Yebes Observatory. The VGOS receivers are designed to observe over a broad frequency range, from 2 to 14~GHz. The Nn radio telescope has recently started successful operations within the VGOS network.

This appendix summarizes the characterization of the Nn antenna, including the pointing and focus models, gain curves, aperture efficiency, and system equivalent flux density (SEFD).

\subsection{Pointing Model}
\label{sec:app_nyalesund_pointing}

The pointing model of the antenna is described by a set of eight parameters with a physical meaning, relating the azimuth and elevation pointing errors to the antenna position. The procedure follows the method described by de Vicente and Barcia (2007).

\begin{figure}[htbp]
\centering
\includegraphics[width=0.7\textwidth, trim=0 0pt 0 20pt, clip]{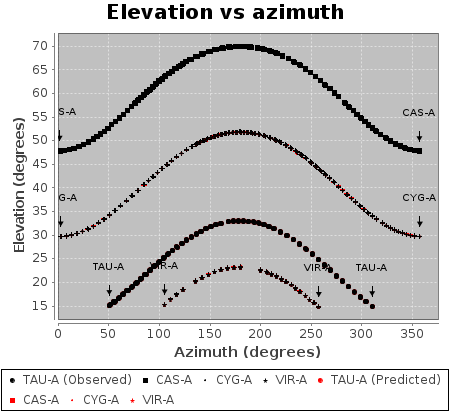}
\caption{Elevation and azimuth coverage of the pointing drifts used to derive the pointing model.}
\label{fig:app_nyalesund_coverage}
\end{figure}

To determine these parameters, several 24-hour observing sessions were carried out. These observations consisted of pointing drifts along the azimuth and elevation axes toward four bright calibrators: \emph{Cas-A}, \emph{Vir-A}, \emph{Cyg-A}, and \emph{Tau-A}. These sources are bright, have accurately known positions, and are smaller than the telescope beam at the observing frequencies. They also provide a wide coverage in elevation and azimuth, allowing a reliable pointing model to be derived over a large fraction of the sky. This coverage is shown in Fig.~\ref{fig:app_nyalesund_coverage}.

\begin{figure}[htbp]
\centering
\begin{minipage}{0.495\textwidth}
\centering
\includegraphics[width=0.975\textwidth, trim=0 0pt 0 20pt, clip]{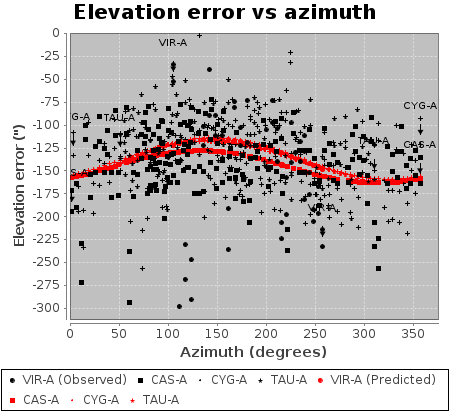}
\caption[Elevation error as a function of azimuth.]{Elevation error as a function of azimuth. The residual root mean square deviation is 39.16'', compared with 141.65'' before applying the model.}
\label{fig:app_nyalesund_elerraz}
\end{minipage}%
\hspace*{15pt}
\begin{minipage}{0.495\textwidth}
\centering
\includegraphics[width=0.975\textwidth, trim=0 0pt 0 20pt, clip]{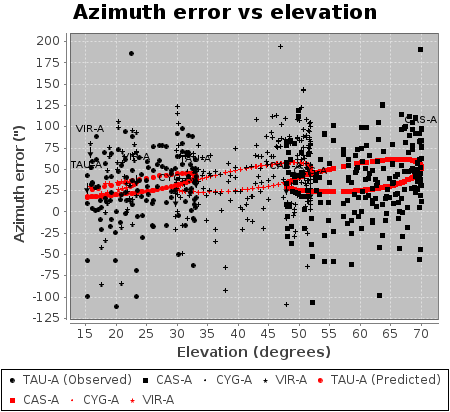}
\caption[Azimuth error as a function of elevation.]{Azimuth error as a function of elevation. The residual root mean square deviation is 38.97'', compared with 56.14'' before applying the model.}
\label{fig:app_nyalesund_azerrel}
\end{minipage}
\end{figure}

\begin{figure}[htbp]
\centering
\begin{minipage}{0.495\textwidth}
\centering
\includegraphics[width=0.975\textwidth, trim=0 0pt 0 20pt, clip]{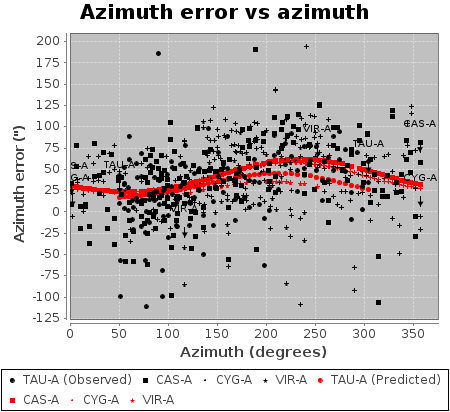}
\caption{Azimuth error as a function of azimuth.}
\label{fig:app_nyalesund_azerraz}
\end{minipage}%
\hspace*{15pt}
\begin{minipage}{0.495\textwidth}
\centering
\includegraphics[width=0.975\textwidth, trim=0 0pt 0 20pt, clip]{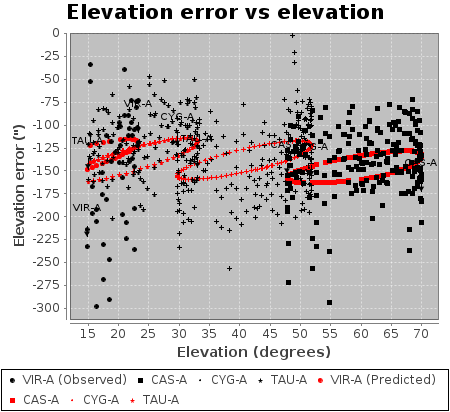}
\caption{Elevation error as a function of elevation.}
\label{fig:app_nyalesund_elerrel}
\end{minipage}
\end{figure}

A least-squares fit was performed to obtain the model parameters that best reproduce the observed pointing offsets. The observations were performed at 10~GHz, the highest-frequency band of the VGOS receiver, in order to obtain the best angular resolution and the most precise pointing model. The observed and fitted pointing errors are shown in Figs.~\ref{fig:app_nyalesund_elerraz}, \ref{fig:app_nyalesund_azerrel}, \ref{fig:app_nyalesund_azerraz}, and \ref{fig:app_nyalesund_elerrel}.


The pointing model parameters obtained from this fit are:

\begin{center}
$P1 = 44''\hspace{1cm}  P2 = -38''\hspace{1cm} P3 = 66''\hspace{1cm} P4 = 13'' $ \
$P5 = -18''\hspace{1cm} P7 = -210''\hspace{1cm} P8 = 39''\hspace{1cm} P9 = 62'' $
\end{center}

After implementing these values in the antenna control system, additional pointing observations were performed. These confirmed the validity of the model, with residual pointing errors much smaller than the antenna beam. No further iterations were required to refine the pointing model.

\subsection{Focus Model}
\label{sec:app_nyalesund_focus}

The focus model was determined from observations lasting approximately 30 hours for each focus axis. These observations consisted of pointing drifts, similar to those used for the pointing model, but with the focus position set to five different values.

\begin{figure}[htbp]
\centering
\includegraphics[width=.9\textwidth]{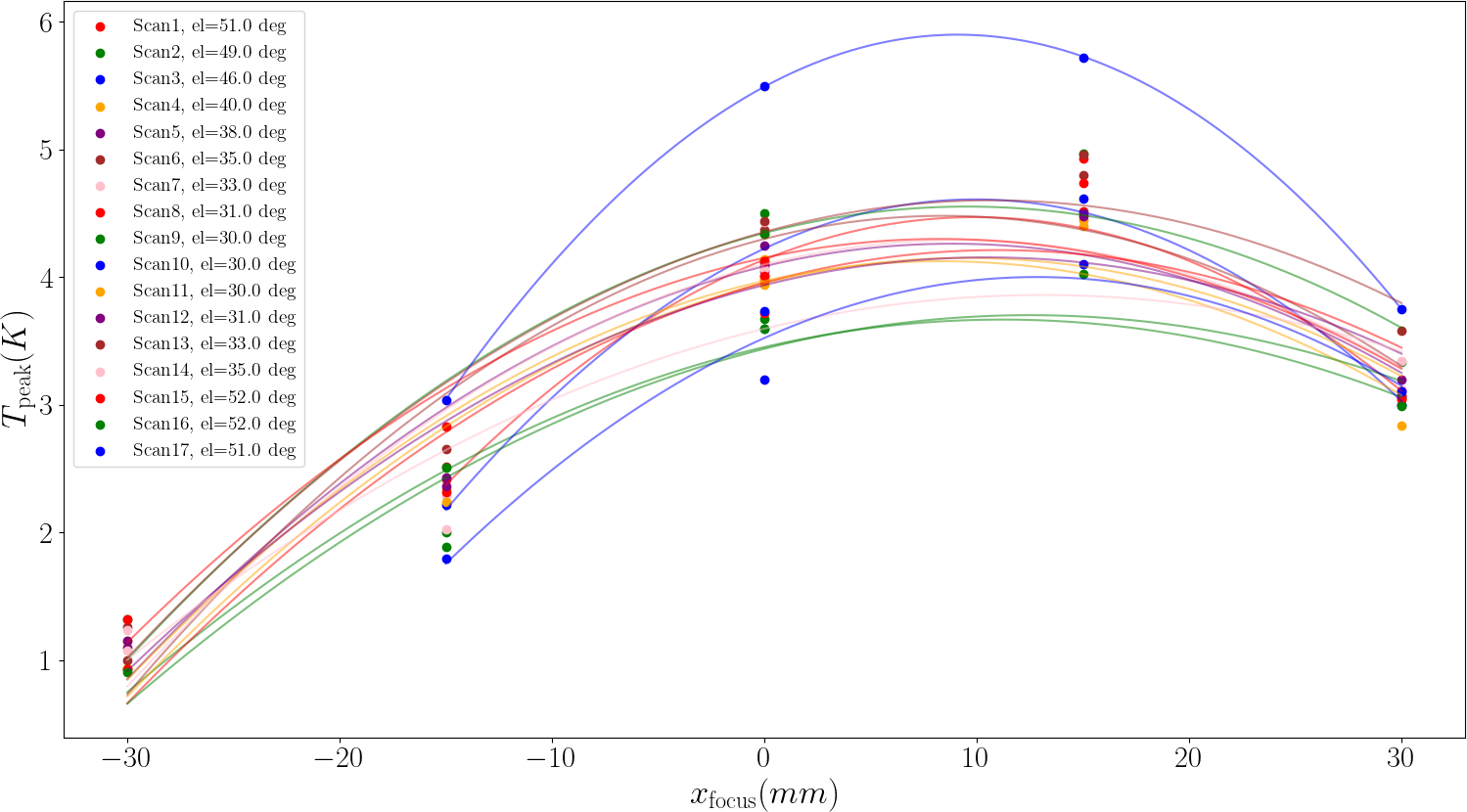}
\caption{Maximum antenna temperature observed in the pointing drifts for each focus shift in the X axis.}
\label{fig:app_nyalesund_tvsxfocus}
\end{figure}

\begin{figure}[htbp]
\centering
\includegraphics[width=.9\textwidth]{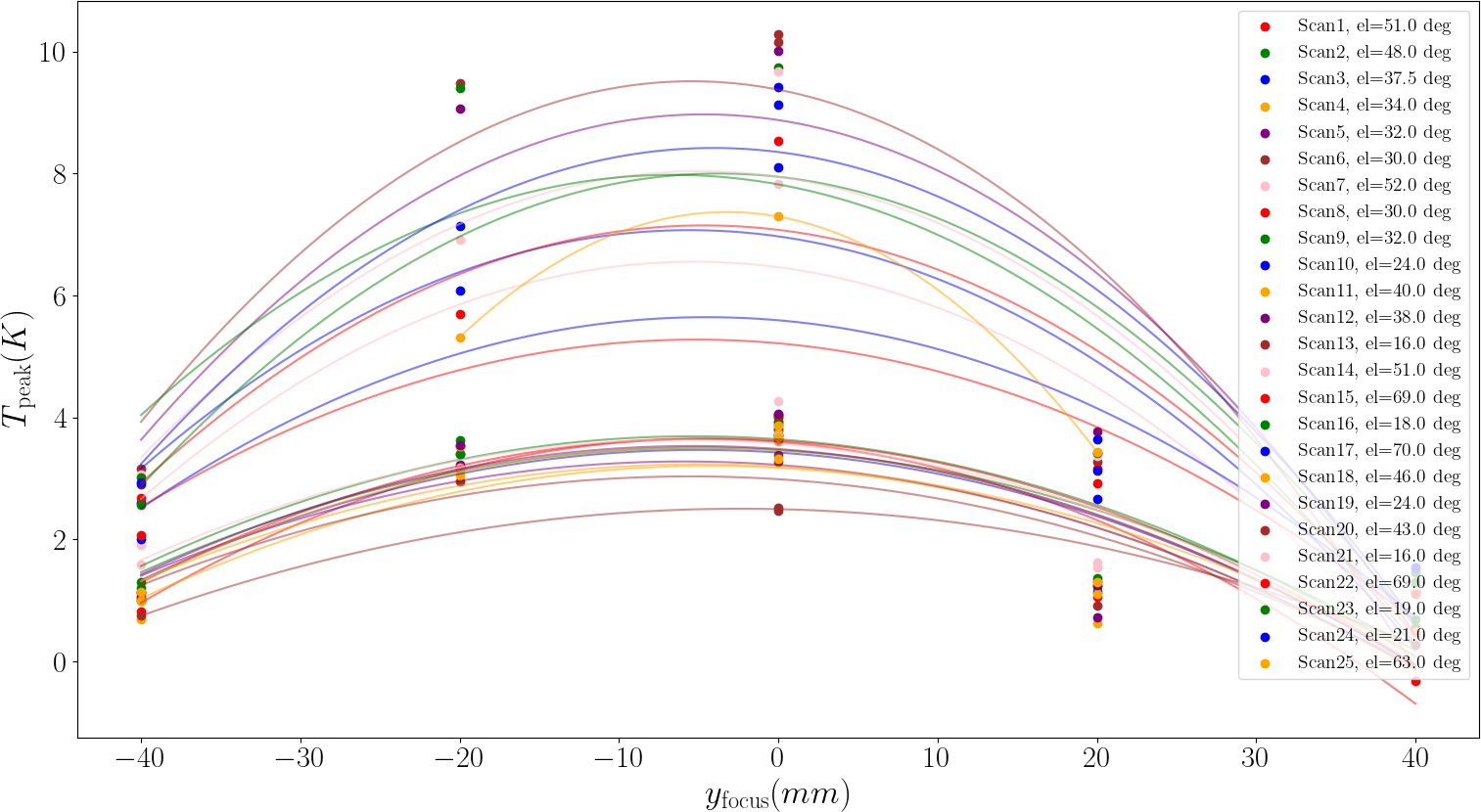}
\caption{Maximum antenna temperature observed in the pointing drifts for each focus shift in the Y axis.}
\label{fig:app_nyalesund_tvsyfocus}
\end{figure}

The maximum antenna temperature observed in each scan can be represented as a function of the focus position, as shown in Figs.~\ref{fig:app_nyalesund_tvsxfocus} and \ref{fig:app_nyalesund_tvsyfocus}.


\begin{figure}[htbp]
\centering
\includegraphics[width=.9\textwidth]{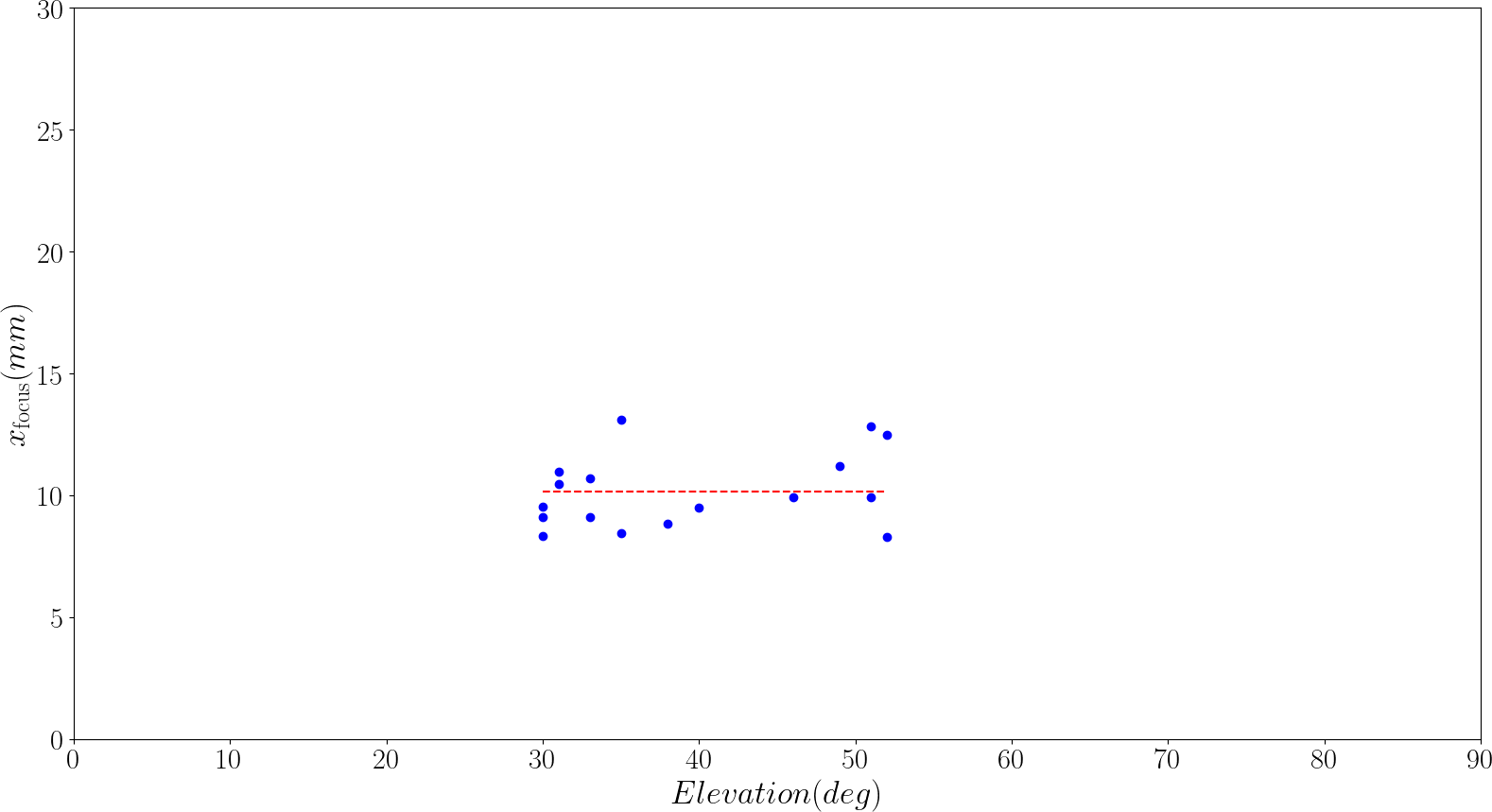}
\caption{Optimum X-axis focus position as a function of elevation.}
\label{fig:app_nyalesund_xfocusvsel}
\end{figure}

\begin{figure}[htbp]
\centering
\includegraphics[width=.9\textwidth]{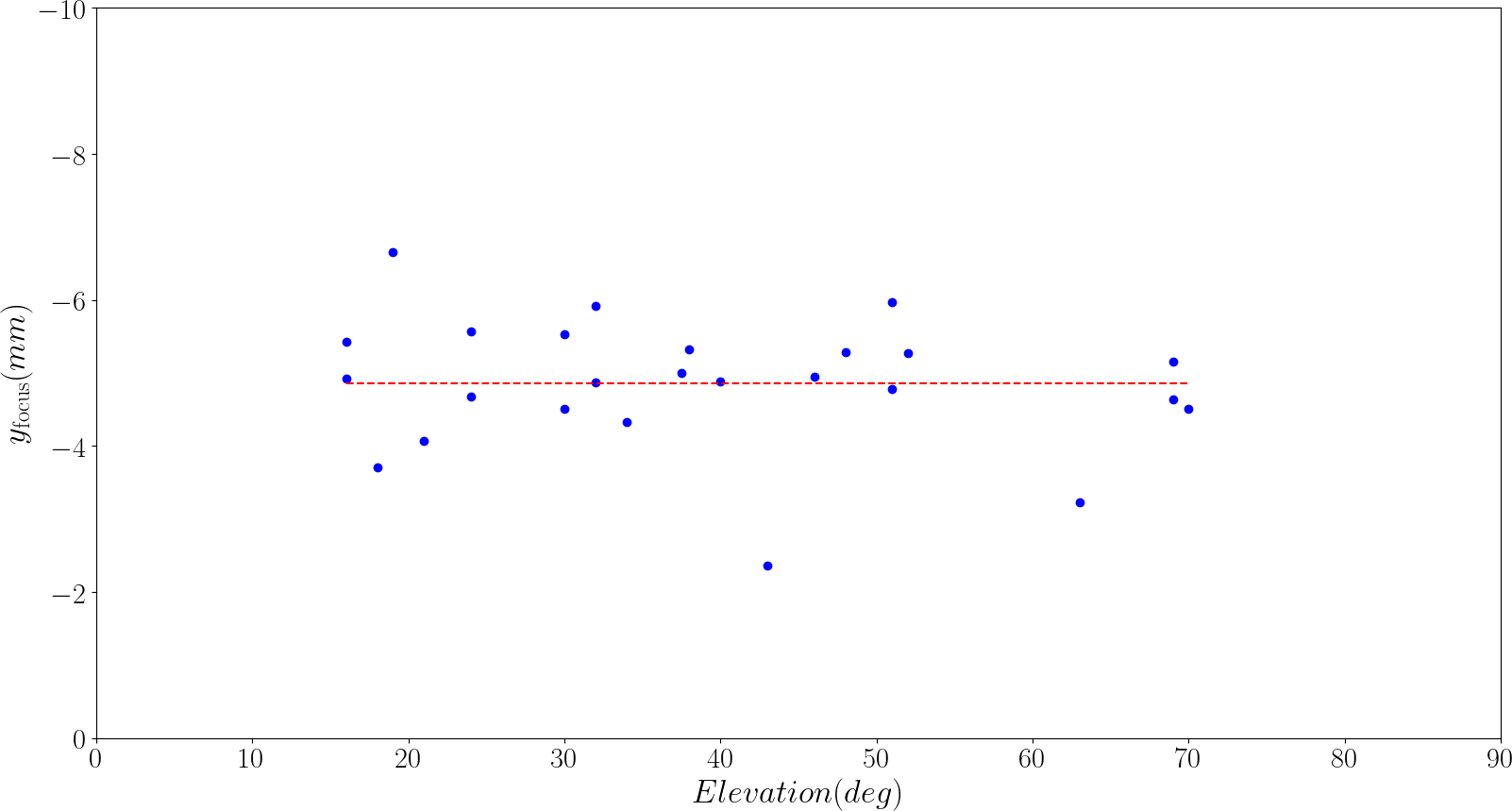}
\caption{Optimum Y-axis focus position as a function of elevation.}
\label{fig:app_nyalesund_yfocusvsel}
\end{figure}

After fitting these measurements with a polynomial function, the focus position that maximizes the observed antenna temperature, and therefore the antenna efficiency, was obtained. The resulting optimum focus positions as a function of elevation are shown in Figs.~\ref{fig:app_nyalesund_xfocusvsel} and \ref{fig:app_nyalesund_yfocusvsel}.

The results show that the optimum focus position does not depend significantly on elevation. Therefore, the adopted optimum focus positions are 10.04~mm for the X axis and $-4.86$~mm for the Y axis.

A change of the focus position in the X and Y axes introduces pointing changes. For this reason, a new pointing model session was performed after the focus model had been determined, following the procedure described in Sect.~\ref{sec:app_nyalesund_pointing}.

\subsection{Gain Curve and Aperture Efficiency}
\label{sec:app_nyalesund_gain}

The gain curve and aperture efficiency of the antenna were determined from sequences of \emph{fivept} and \emph{onoff} observations carried out with the VLBI Field System over three days. The \emph{fivept} procedure performs pointing drifts to locate the peak intensity of the source, while \emph{onoff} observations determine the antenna power associated with the source by subtracting the background contribution. By comparing this value with the known flux density of the calibrator, the antenna gain can be calculated.

The observations were carried out using three calibrators: \emph{Cas-A}, \emph{Cyg-A}, and \emph{Tau-A}. These sources allow a broad range of elevations to be sampled. They also have well-known flux densities and are smaller than the antenna beam at these frequencies, making them suitable flux calibrators at centimetre wavelengths.

The analysis showed that there may be an error in the calibration-temperature measurement, since the $T_{\rm cal}$ ratio values differ from unity. This may be due either to an incorrect $T_{\rm cal}$ measurement or to diode degradation.

\begin{figure}[htbp]
\centering
\begin{minipage}{0.495\textwidth}
\centering
\includegraphics[width=0.975\textwidth]{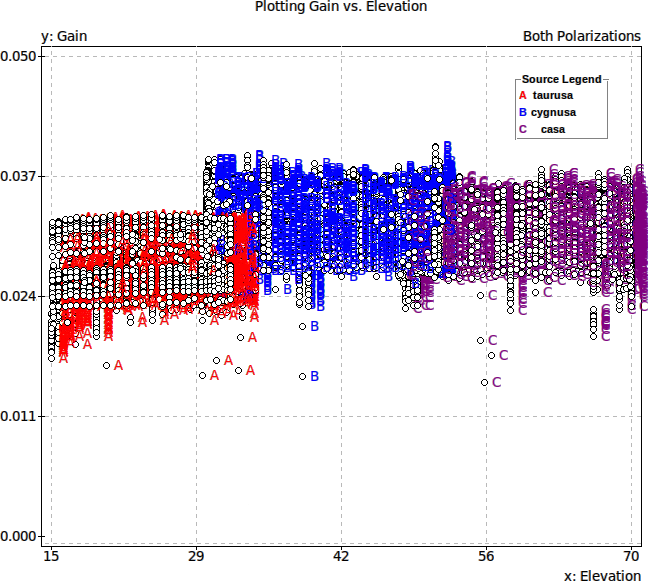}
\caption{Gain curve for VGOS Band A, from 3.016 to 3.464~GHz.}
\label{fig:app_nyalesund_gainA}
\end{minipage}%
\hspace*{15pt}
\begin{minipage}{0.495\textwidth}
\centering
\includegraphics[width=0.975\textwidth]{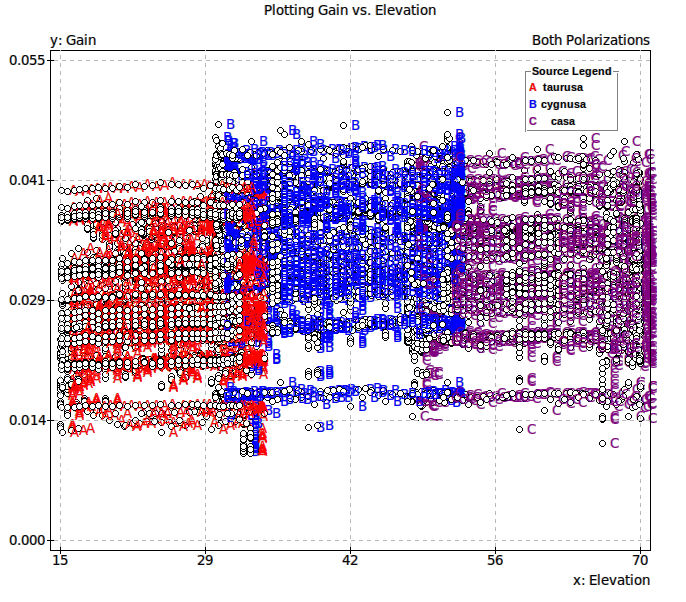}
\caption{Gain curve for VGOS Band B, from 5.288 to 5.736~GHz.}
\label{fig:app_nyalesund_gainB}
\end{minipage}
\end{figure}

\begin{figure}[htbp]
\centering
\begin{minipage}{0.495\textwidth}
\centering
\includegraphics[width=0.975\textwidth]{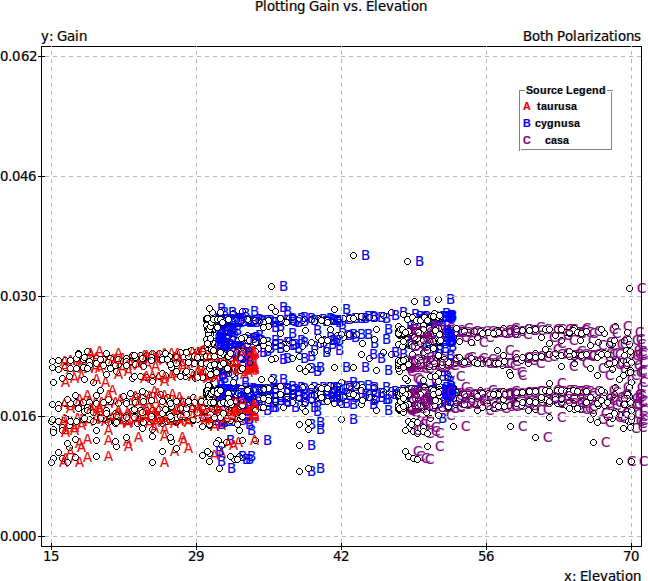}
\caption{Gain curve for VGOS Band C, from 6.408 to 6.856~GHz.}
\label{fig:app_nyalesund_gainC}
\end{minipage}%
\hspace*{15pt}
\begin{minipage}{0.495\textwidth}
\centering
\includegraphics[width=0.975\textwidth]{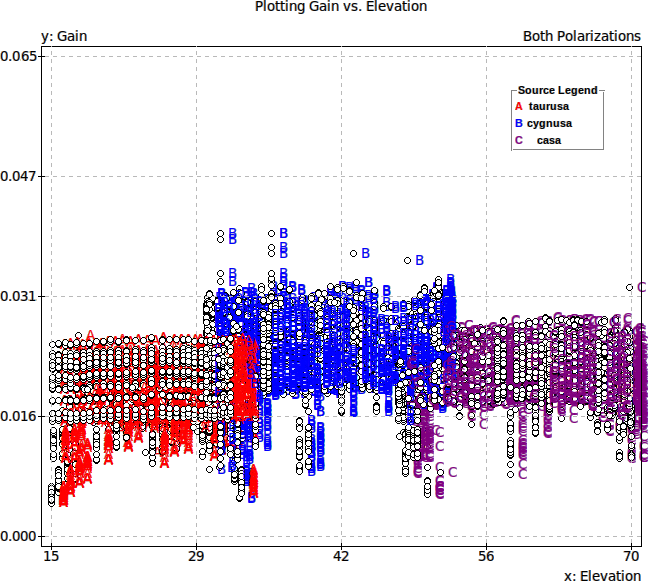}
\caption{Gain curve for VGOS Band D, from 10.248 to 10.696~GHz.}
\label{fig:app_nyalesund_gainD}
\end{minipage}
\end{figure}

The gain curves, expressed in K/Jy, were obtained using the VLBI Field System \emph{Gain Plot 2} software. They are shown in Figs.~\ref{fig:app_nyalesund_gainA}, \ref{fig:app_nyalesund_gainB}, \ref{fig:app_nyalesund_gainC}, and \ref{fig:app_nyalesund_gainD}.

The gain curves are fairly flat, in agreement with the behaviour expected for other ring-focus antennas in the VGOS network. Due to the geographical location of the antenna, it was not possible to obtain measurements at higher elevations, since no suitable flux calibrator reached those positions in the sky.

The aperture efficiency was calculated from the relation

\begin{equation}
G = \frac{A \eta_A}{2760},
\end{equation}
where $G$ is the antenna gain in K/Jy, $A$ is the geometrical collecting area in m$^2$, and $\eta_A$ is the aperture efficiency.

The aperture efficiency was found to be approximately independent of elevation, with the following values for the four VGOS bands:

\begin{center}
\begin{tabular}{l}
$\eta_{A,3,{\rm GHz}} = 0.60\pm0.13$ \\
$\eta_{A,5,{\rm GHz}} = 0.59\pm0.15$ \\
$\eta_{A,6,{\rm GHz}} = 0.42\pm0.11$ \\
$\eta_{A,10,{\rm GHz}} = 0.44\pm0.10$
\end{tabular}
\end{center}

\begin{figure}[htbp]
\centering
\includegraphics[width=.9\textwidth]{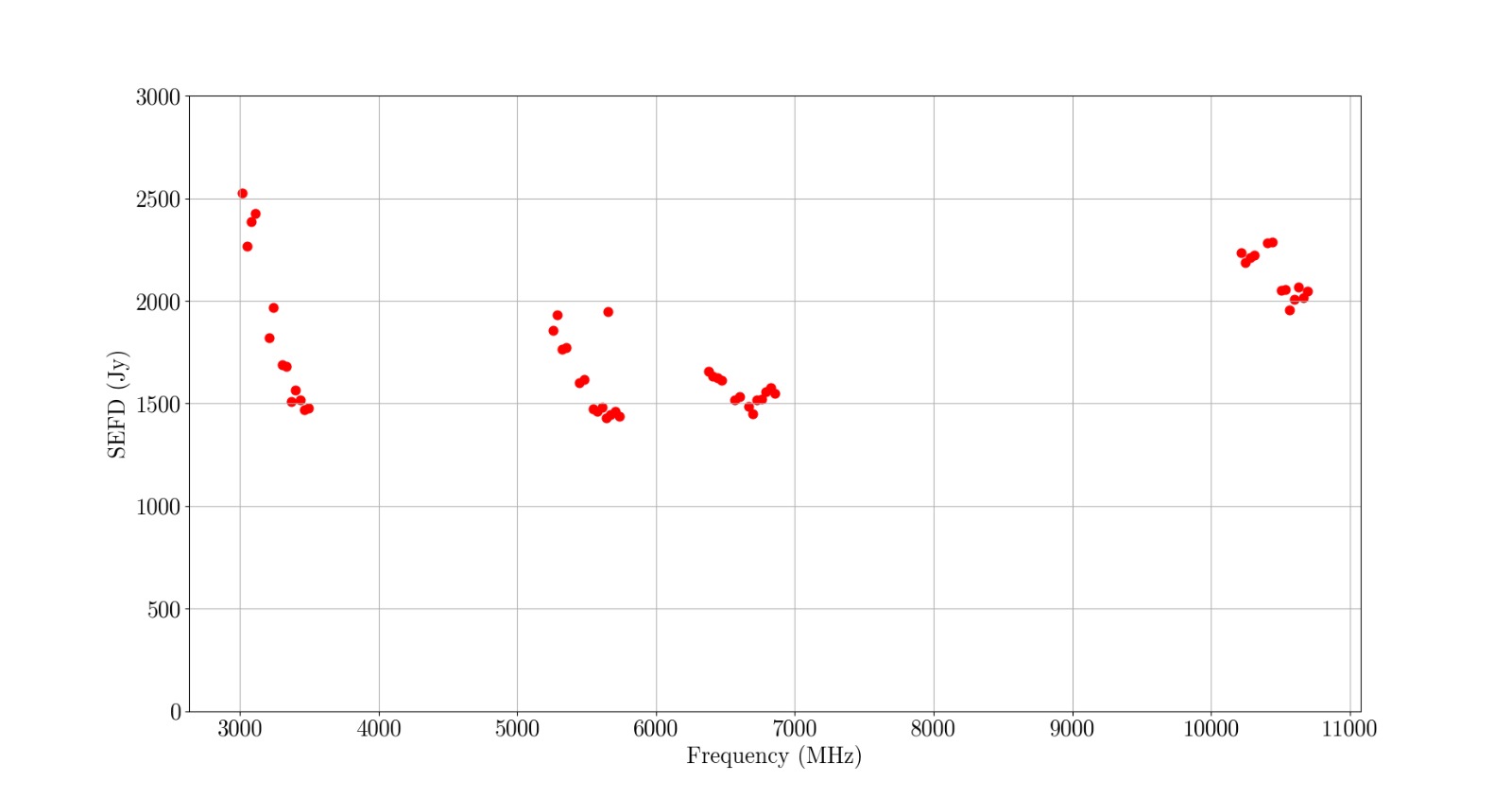}
\caption{SEFD as a function of frequency.}
\label{fig:app_nyalesund_SEFD}
\end{figure}

These observations also provide information about the system equivalent flux density (SEFD), defined as the flux density of a radio source that doubles the system temperature. Figure~\ref{fig:app_nyalesund_SEFD} shows the SEFD, calculated as the average of all observations for each frequency channel, as a function of frequency.

\pagebreak

\section{List of Publications and Contributions}

This appendix summarizes the main scientific contributions produced during the development of this thesis. The list includes refereed journal publications, refereed conference proceedings, software, technical reports, outreach publications, and scientific contributions presented at conferences.

\subsection{Refereed Publications}

\begin{enumerate}

\item Jaron, F., Martí-Vidal, I., Schartner, M., González-García, J.,
Albentosa-Ruiz, E., Bernhart, S., Böhm, J., Gruber, J., Modiri, S.,
Nothnagel, A., \textbf{Pérez-Díez, V.}, Savolainen, T., Soja, B.,
Varenius, E., \& Xu, M. H.
\textit{``Cross-Polarization Gain Calibration of Linearly Polarized VLBI Antennas by Observations of 4C 39.25''}.
Radio Science, 59, e2023RS007892, 2024.
doi: \href{https://doi.org/10.1029/2023RS007892}{10.1029/2023RS007892}.

\item \textbf{Pérez-Díez, V.}, Martí-Vidal, I., Albentosa-Ruiz, E., González-García, J.,
Jaron, F., Savolainen, T., Xu, M. H., \& Bachiller, R.
\textit{``Towards an astronomical use of new-generation geodetic observations. I. From the correlator to full-polarization images''}.
Astronomy \& Astrophysics, 688, A151, 16 pp., 2024.
doi: \href{https://doi.org/10.1051/0004-6361/202348633}{10.1051/0004-6361/202348633}.
arXiv: \href{https://arxiv.org/abs/2406.12509}{2406.12509}.

\item \textbf{Pérez-Díez, V.}, Martí-Vidal, I., Albentosa-Ruiz, E., \& Bachiller, R.
\textit{``Evidence of a non-equipartition energy regime in 1803+784: Core-shift and Faraday rotation measurements from simultaneous multi-frequency polarimetric VGOS observations''}.
Astronomy \& Astrophysics, 700, L22, 6 pp., 2025.
doi: \href{https://doi.org/10.1051/0004-6361/202555706}{10.1051/0004-6361/202555706}.
arXiv: \href{https://arxiv.org/abs/2602.20746}{2602.20746}.

\item Jaron, F., \textbf{Pérez-Díez, V.}, Krásná, H., Böhm, J.,
Charlot, P., Collioud, A., Gruber, J., Martí-Vidal, I., \& Nothnagel, A.
\textit{``Removing the Effect of Source Structure from Correlated VLBI Data''}.
Radio Science, accepted.

\item \textbf{Pérez-Díez, V.}, Martí-Vidal, I., Albentosa-Ruiz, E.,
González-García, J., Jaron, F., Savolainen, T., Xu, M. H., \& Bachiller, R.
\textit{``Towards an astronomical use of new-generation geodetic observations.
II. Core-shift and multi-band polarization for a selection of sources''}.
In preparation.

\end{enumerate}

\subsection{Refereed Conference Proceedings}

\begin{enumerate}

\item Albentosa, E., Alef, W., Bernhart, S., Böhm, J.,
Bolaño González, R., Choi, Y. K., Gansmoe, T.,
García Espada, S., García Miró, C., Girdiuk, A.,
González García, J., Grøslie Wennesland, S. A.,
Gruber, J. F., Haas, R., Hammargren, R., Heinkelmann, R.,
Jaron, F., Kareinen, N., Kirkvik, A.-S., Krásná, H.,
Martinez, E., Martí-Vidal, I., Meldahl, A., Melnikov, A.,
Modiri, S., Neidhardt, A., Nothnagel, A., Panzenböck, O.,
\textbf{Pérez, V.}, Petrov, L., Plötz, C., Rottmann, H.,
Savolainen, T., Schartner, M., Schüler, T., Schuh, H.,
Soja, B., Varenius, E., de Vicente, P., Wagner, J.,
Xu, M. H., Zubko, N., Kettenis, M., Matsumoto, S.,
Porcas, R., Small, D., \& Verkouter, M.
\textit{``Current Status of the EU-VGOS Project''}.
In \textit{International VLBI Service for Geodesy and Astrometry 2022 General Meeting Proceedings},
K. L. Armstrong, D. Behrend, \& K. D. Baver, eds.,
NASA/CP--20220018789, 2023, pp.~85--89.

\item Jaron, F., Baldreich, L., Böhm, J., Charlot, P., Collioud, A.,
Gruber, J., Krásná, H., Martí-Vidal, I., Nothnagel, A., \&
\textbf{Pérez-Díez, V.}
\textit{``Mitigating Source Structure in Geodetic VLBI on the Visibility Level''}.
In \textit{Proceedings of the 16th European VLBI Network Symposium},
E. Ros, P. Benke, S. Dzib, I. Rottmann, \& J. A. Zensus, eds.,
Bonn, Germany, 2--6 September 2024, pp.~145--148.

\item \textbf{Pérez-Díez, V.}, Martí-Vidal, I., \& Albentosa-Ruiz, E.
\textit{``Towards an astronomical use of new generation geodetic observations --
Imaging and astronomical results''}.
In \textit{Highlights of Spanish Astrophysics XII}, Proceedings of the XVI Scientific Meeting of the Spanish Astronomical Society, Granada, Spain, 15--19 July 2024,
M. Manteiga, F. González-Galindo, A. Labiano-Ortega, M. J. Martínez-González,
N. Rea, M. Romero-Gómez, A. Ulla-Miguel, G. Yepes, C. Rodríguez-López,
A. Gómez-García, \& C. Dafonte, eds., 2025, p.~141.

\item Martí-Vidal, I., \textbf{Pérez-Díez, V.}, \& Albentosa-Ruiz, E.
\textit{``Full-Polarization Imaging of PolConverted VLBI Observations''}.
In \textit{International VLBI Service for Geodesy and Astrometry 2024 General Meeting Proceedings},
D. Behrend, K. D. Baver, \& K. L. Armstrong, eds.,
NASA/CP--20250002586, 2025, p.~137.

\item Albentosa-Ruiz, E., Martí-Vidal, I., \& \textbf{Pérez-Díez, V.}
\textit{``Optimizing Global VLBI Calibration with PolConvert''}.
In \textit{International VLBI Service for Geodesy and Astrometry 2024 General Meeting Proceedings},
D. Behrend, K. D. Baver, \& K. L. Armstrong, eds.,
NASA/CP--20250002586, 2025, p.~144.

\end{enumerate}

\subsection{Conference Contributions}

\subsubsection*{First Author Contributions}

\begin{enumerate}

\item \textbf{Pérez-Díez, V.}
\textit{``Study of ultra-wideband VLBI data with both astronomical and geodetic applications''}.
Oral contribution, 51st Young European Radio Astronomers Conference (YERAC),
Aalto University Metsähovi Radio Observatory, Helsinki, Finland,
24--26 August 2022.

\item \textbf{Pérez-Díez, V.}
\textit{``Complete Calibration of the VLBI Global Observing System (VGOS)''}.
Contribution, PhDay Físicas 2023, Facultad de Ciencias Físicas,
Universidad Complutense de Madrid, Spain, 2023.

\item \textbf{Pérez-Díez, V.}
\textit{``Full-Polarization Images from the VLBI Global Observing System (VGOS)''}.
Oral contribution, 52nd Young European Radio Astronomers Conference (YERAC),
Jodrell Bank Centre for Astrophysics, Manchester, United Kingdom,
12--15 September 2023.

\item \textbf{Pérez-Díez, V.}, Martí-Vidal, I., \& Albentosa-Ruiz, E.
\textit{``Towards an Astronomical Use of New Generation Geodetic Observations --
Imaging and Astronomical Results''}.
Contribution ID 562, XVI Scientific Meeting of the Spanish Astronomical Society,
Granada, Spain, 15--19 July 2024.

\item \textbf{Pérez-Díez, V.}
\textit{``Core-shift and Faraday rotation measurements from simultaneous
multi-frequency polarimetric VGOS observations: Evidence of a
non-equipartition energy regime in 1803+784''}.
Invited seminar, Scientific Seminar Astrogeodesy, GFZ Potsdam,
Potsdam, Germany, 22 July 2025.

\end{enumerate}

\subsubsection*{Co-Authored Contributions}

\begin{enumerate}

\item Albentosa-Ruiz, E., Martí-Vidal, I., \textbf{Pérez-Díez, V.}, et al.
\textit{``Towards an Astronomical Use of New Generation Geodetic Observations''}.
Poster contribution, XVI Scientific Meeting of the Spanish Astronomical Society,
Granada, Spain, 15--19 July 2024.

\item Jaron, F., Baldreich, L., Böhm, J., Charlot, P., Collioud, A.,
Gruber, J., Krásná, H., Martí-Vidal, I., Nothnagel, A., \&
\textbf{Pérez-Díez, V.}
\textit{``Mitigating Source Structure in Geodetic VLBI on the Visibility Level''}.
Poster contribution, 16th European VLBI Network Symposium,
Bonn, Germany, 2--6 September 2024.

\item Jaron, F., Martí-Vidal, I., Schartner, M., González-García, J.,
Albentosa-Ruiz, E., Bernhart, S., Böhm, J., Modiri, S.,
Nothnagel, A., \textbf{Pérez-Díez, V.}, Savolainen, T.,
Soja, B., Varenius, E., Xu, M. H., \& Cheung, C. C.
\textit{``Cross-polarization Gain Calibration of VGOS Antennas''}.
Poster contribution, 13th IVS Technical Operations Workshop,
MIT Haystack Observatory, United States, 4--8 May 2024. 

\item Jaron, F., Collioud, A., Nothnagel, A., Krásná, H.,
Martí-Vidal, I., Gruber, J., Böhm, J., Charlot, P., \&
\textbf{Pérez-Díez, V.}
\textit{``Removing source structure in DiFX visibility data''}.
Talk contribution, 10th International VLBI Technology Workshop,
Chalmers, Gothenburg, Sweden, 21--25 October 2025.
 
\item Martí-Vidal, I., \textbf{Pérez-Díez, V.}, \& Bachiller, R.
\textit{``Core-shifts and Faraday Rotation Measurements from VGOS Observations''}.
Poster contribution, IVS General Meeting 2026, Session 6: Crossroads VLBI,
Contribution ID 099, 12--17 Apr 2026.

\end{enumerate}

\subsection{Software}

\begin{enumerate}

\item Jaron, F. F. D., \& \textbf{Pérez-Díez, V.}
\textit{``VieSOFT -- Modeling and Removing Source Structure in VLBI Observations''}.
Software, Version v1, TU Wien, 2026.
doi: \href{https://doi.org/10.48436/mw2mg-9fy73}{10.48436/mw2mg-9fy73}.

\end{enumerate}

\subsection{Technical Reports}

\begin{enumerate}

\item González-García, J., García-Miró, C., \textbf{Pérez-Díez, V.},
Paredes, F., Martínez-Sánchez, E., Beltrán, F., de Vicente, P.,
\& Albo, C.
\textit{``Yebes Observatory Report''}.
In \textit{International VLBI Service for Geodesy and Astrometry
2021+2022 Biennial Report},
K. L. Armstrong, D. Behrend, \& K. D. Baver, eds.,
NASA/TP--20230014975, 2023, pp.~111--116.

\item \textbf{Pérez-Díez, V.}, Gómez-Garrido, M., González, J., de Vicente, P.,
Beltrán, F., Bolaño, R., \& García, S.
\textit{``Characterization of the Ny-Ålesund North (Nn) 13-m telescope''}.
Technical Report IT-CDT 2023-10, Centro de Desarrollos Tecnológicos,
Observatorio de Yebes, 2023.

\item Yebes VLBI Group.
\textit{``EVN Station Report -- 2021/2022 Yebes''}.
Station report, Yebes Observatory, 2022.
European VLBI Network, 2023, pp.~85--91.

\item Yebes VLBI Group.
\textit{``EVN Station Report -- 2023/2024 Yebes''}.
Station report, Yebes Observatory, 2024.
European VLBI Network.
\end{enumerate}

\subsection{Outreach Publications}

\begin{enumerate}

\item \textbf{Pérez-Díez, V.}
\textit{``Nuevo método para obtener imágenes ultraprecisas de galaxias activas''}.
Boletín de Actualidad IGN-CNIG, November 2022.

\item \textbf{Pérez-Díez, V.}
\textit{``La observación de cuásares para el estudio de los agujeros negros y de nuestro planeta''}.
Boletín de Actualidad IGN-CNIG, January 2025.

\item \textbf{Pérez-Díez, V.}
\textit{``Extragalactic research based on new-generation geodetic data''}.
EVN/JIVE Newsletter, Edition 70, January 2025.

\end{enumerate}

\pagebreak

\printbibliography[heading=bibintoc]

\end{document}